\documentclass[pdflatex,sn-mathphys-num]{sn-jnl}

\usepackage{graphicx}%
\usepackage{natbib}
\usepackage{multirow}%
\usepackage{amsmath,amssymb,amsfonts}%
\usepackage{amsthm}%
\usepackage{mathrsfs}%
\usepackage[title]{appendix}%
\usepackage{xcolor}%
\usepackage{textcomp}%
\usepackage{manyfoot}%
\usepackage{booktabs}%
\usepackage{algorithm}%
\usepackage{algorithmicx}%
\usepackage{algpseudocode}%
\usepackage{listings}%
\usepackage{slashed}

\theoremstyle{thmstyleone}%
\theoremstyle{thmstyletwo}%

\theoremstyle{thmstylethree}%
\newcommand{\nn}{\nonumber}
\newcommand{\vep}{\varepsilon}

\begin{document}

\title[Effective Field Theories for Neutron Stars Physics]{Effective Field Theories for Neutron Stars Physics}


\author*[1]{\fnm{J.~M.~Alarc\'on}}\email{jmanuel.alarcon@uah.es}

\author[2]{\fnm{E. Lope-Oter}}\email{mariaevl@ucm.es}
\equalcont{These authors contributed equally to this work.}

\author[1]{\fnm{Y.~Cano}}\email{ycano01@ucm.es}
\equalcont{All authors contributed equally to this work.}


\affil*[1]{Universidad de Alcal\'a, Grupo de F\'isica Nuclear,  Part\'iculas y Astrof\'isica, Departamento de F\'isica y
Matem\'aticas, 28805 Alcal\'a de Henares (Madrid), Spain}

\affil[2]{Departamento de F\'{\i}sica Te\'orica \& IPARCOS,\\ Universidad Complutense de Madrid, E-28040 Madrid, Spain}




\abstract{There is an increasing interest in the community for the Neutron Stars and what we can learn from them. 
In this review we show how chiral effective field theory, combined with many-body methods, can provide important results that connect Neutron Star properties at zero temperature to nuclear physics and allows to use these compact objects as laboratories of new physics.}




\maketitle
\newpage
\tableofcontents
\newpage
\section{Introduction} 
\label{Sec:Introduction}

Neutron stars are compact objects resulting from the gravitational collapse of stars with masses between 8-25 $M_\odot$ during a type Ib, Ic, or II supernova. This remnant is an object whose mass is greater than the Sun's and whose radius is of the order of ten kilometers, leading to densities about fourteen orders of magnitude higher than the solar density. These properties make neutron stars one of the densest objects in the universe, providing the opportunity to study matter under conditions that are difficult to reproduce in terrestrial laboratories. Additionally, the intense gravitational fields generated by the neutron star also allow us to test fundamental theories of gravitation.

In all these studies, the equation of state (EoS), understood as the value of the pressure as a function of the energy density, plays a fundamental role. On one hand, the stability of the neutron star depends on this relation, and only certain combinations of masses and radii are allowed for each equation of state. 
Additionally, the equation of state is necessary to understand the magnetic fields generated by neutron stars and also governs the dynamics of neutron stars mergers in binary systems or core-collapse supernovae. Since the equation of state is obtained from nuclear interaction in the density range of the star, this allows us to test our knowledge of fundamental interactions under extreme conditions. From the possible values of resulting masses and radii, known as mass-radius diagrams, we can verify the validity of the theory with which the equation of state has been calculated.

From a theoretical point of view, the equation of state poses a challenge due to the wide range of densities it explores. On the one hand, the fundamental theory of strong interactions can be solved analytically only when the interaction occurs at high energies, which happens at large densities. On the other hand, for the low density region, one can employ low energy approaches for nuclear physics. But the intermediate density region is more complicated, and normally requires some modeling.

Historically, a wide variety of methods have been used to calculate the equation of state of a neutron star. We can distinguish them into two main categories: phenomenological methods and first-principles methods. Phenomenological methods aim to use effective interactions to treat the many-body interaction that occurs in nuclear matter as an independent effective interaction through the energy density functional theory \cite{Duguet:2013dga}. A very popular type of approach to this problem is based on Skyrme-type effective interactions. There are also other phenomenological models such as the mean-field model in which the interaction between nucleons is modeled through meson exchange \cite{Serot:1992ti}. On the other hand, first-principles methods try to construct the interaction from general principles and solve it through different techniques. Examples of this are variational methods \cite{Akmal:1998cf},  effective chiral theories \cite{Holt:2013fwa}, Quantum Monte Carlo methods \cite{Gandolfi:2009fj}, or nuclear lattice effective theory calculations \cite{Ren:2023ued}. 
Both, ab-initio and phenomenological calculations are usually done at zero temperature, since at the typical densities of (not young) neutron stars, the degeneracy pressure dominates over the thermal pressure. 

In this review, we focus on the effective field theory calculations for the equation of state of neutron stars at zero temperature and its applications. 
Effective field theories allow to derive systematically the interactions between nucleons based on the relevant symmetries of quantum chromodynamics (QCD) in the energy scale of interest. 
The main advantage of this approach is the possibility of deriving the nuclear forces in a systematic way, with a theoretical uncertainty estimation provided by a power counting.
An effective field theory based on the chiral symmetry of QCD for nucleons was first derived in \cite{Gasser:1987rb} for the single nucleon sector, and was extended to include few-nucleon interactions in vacuum later \cite{Weinberg:1990rz}.
The standard calculations of the equation of state with effective field theories combine the vacuum few-nucleon formulation with many-body techniques to provide the energy per particle at nuclear densities around nuclear saturation ($n_s \approx 0.16$~fm$^{-3}$) or even $2n_s$. 
Although formulations of chiral effective field theory in the nuclear medium \cite{Oller:2001sn} were able to provide also the equation of state of symmetric nuclear matter and pure neutron matter for different densities \cite{Lacour:2009ej,Alarcon:2021kpx,Alarcon:2022vtn}.  
Any of these effective field theory approaches provide the possibility of calculating the equation of state of neutron stars within a controlled accuracy, so they can be applied to study neutron stars properties like mass-radius diagram, tidal deformability or gravitational wave emission due to binary mergers. 

This review is organized as follows: In section \ref{Sec:Effective_theories_for_nuclear_matter} we introduce fundamental ideas on the construction of effective field theories for studying strong interactions at low energies, mentioning some particular problems when considering baryons. Later, we comment on the different approaches used to extend chiral effective theories to the many-body sector in order to study nuclear matter. 
In section \ref{Sec:EoS} we analyze the structure of neutron star, mentioning the particularities of each part. 
Later, we introduce the different observables used to characterize the star, like its mass, radius and tidal deformability. 
Afterwards we review the different types of methods used to determine the EoS of a neutron star using effective field theories. 
We also study phase transitions in neutron stars and show how to characterize them.
Later, we analyze how EoS obtained in EFT can be matched to crustal equations of state. 
These results are new in the literature and have not been published elsewhere. 
Finally, we study how these results can be applied to study new physics with neutron stars analyzing how dark matter modifies its properties. 
These modifications allows to set constraints on the dark matter particle mass and the capture rate of the star.

\section{Effective Theories for Nuclear Matter}
\label{Sec:Effective_theories_for_nuclear_matter}

Low energy theorems for strong interactions were formulated even before quantum chromodynamics was established as the fundamental theory of the strong interactions \cite{Adler:1964um,Adler:1965ga,Weinberg:1966kf,Gell-Mann:1968hlm}. 
They are based on the relevant symmetries of the strong interactions at low energies and provide useful results related to interactions between hadrons. 
These results are exact at the specific kinematical configurations where they are derived, but they can be extended to a kinematical configuration of interest using perturbation theory. 
Chiral effective field theory is just the quantum field theory constructed to provide results for hadronic interactions rooted on the low energy theorems of QCD. 
In this perturbative approach to the physical point from the low energy theorem result, the power counting plays an essential role, as establish the hierarchy of the infinite possible corrections to the low energy theorem according to an expansion parameter. 
The Goldstone theorem allows to assign a scaling behavior to the momentum of the Goldstone bosons ($p$) in the interaction between hadrons and the masses acquired through the explicit breaking of chiral symmetry of QCD ($M_\pi$). 
Therefore, in chiral effective field theory normally the scaling parameters are $p$ and $M_\pi$. 
Since the perturbative (loop) expansion is suppressed by a scale $\Lambda_\chi \sim 1$~GeV, we expect convergence for $p \sim M_\pi < \Lambda_\chi$.

\begin{figure}[H]
 \begin{center}
 \includegraphics[width=.49\textwidth]{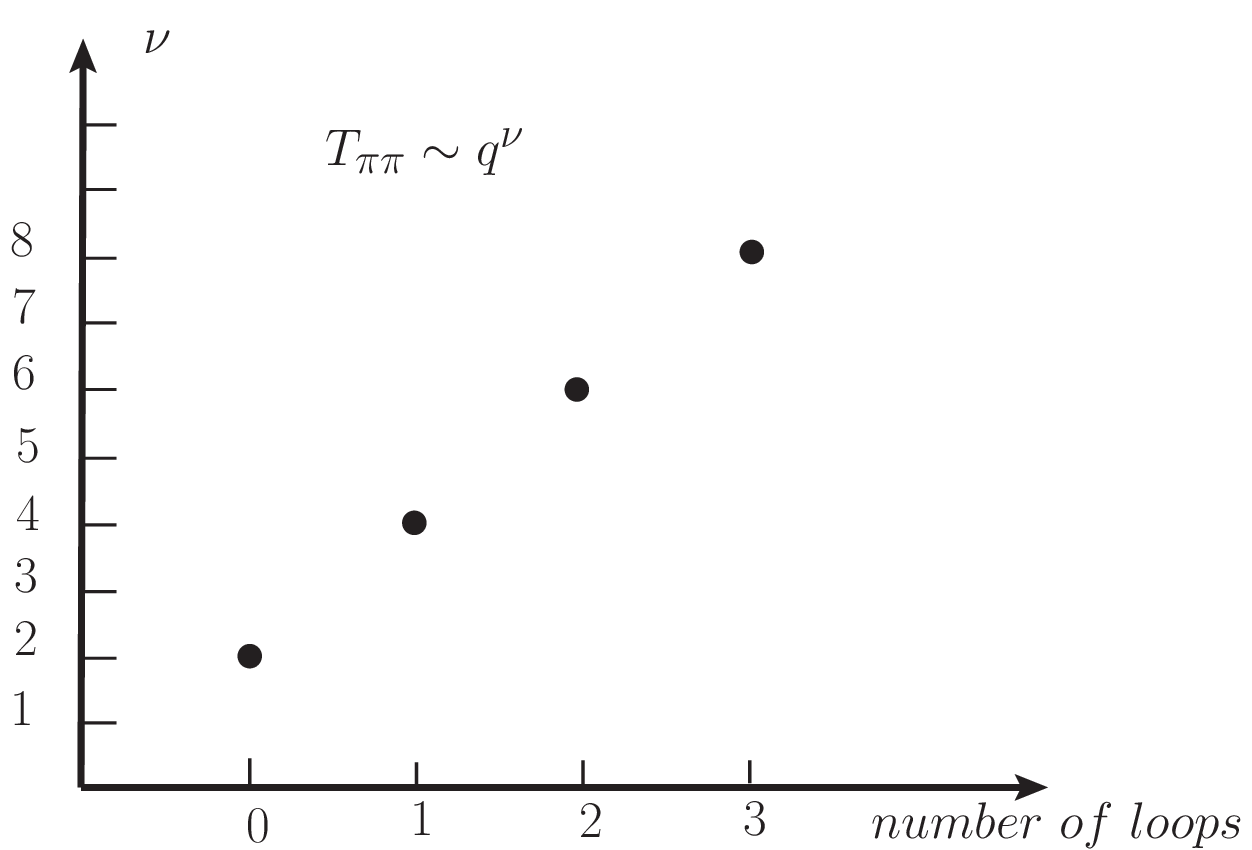}  \includegraphics[width=.5\textwidth]{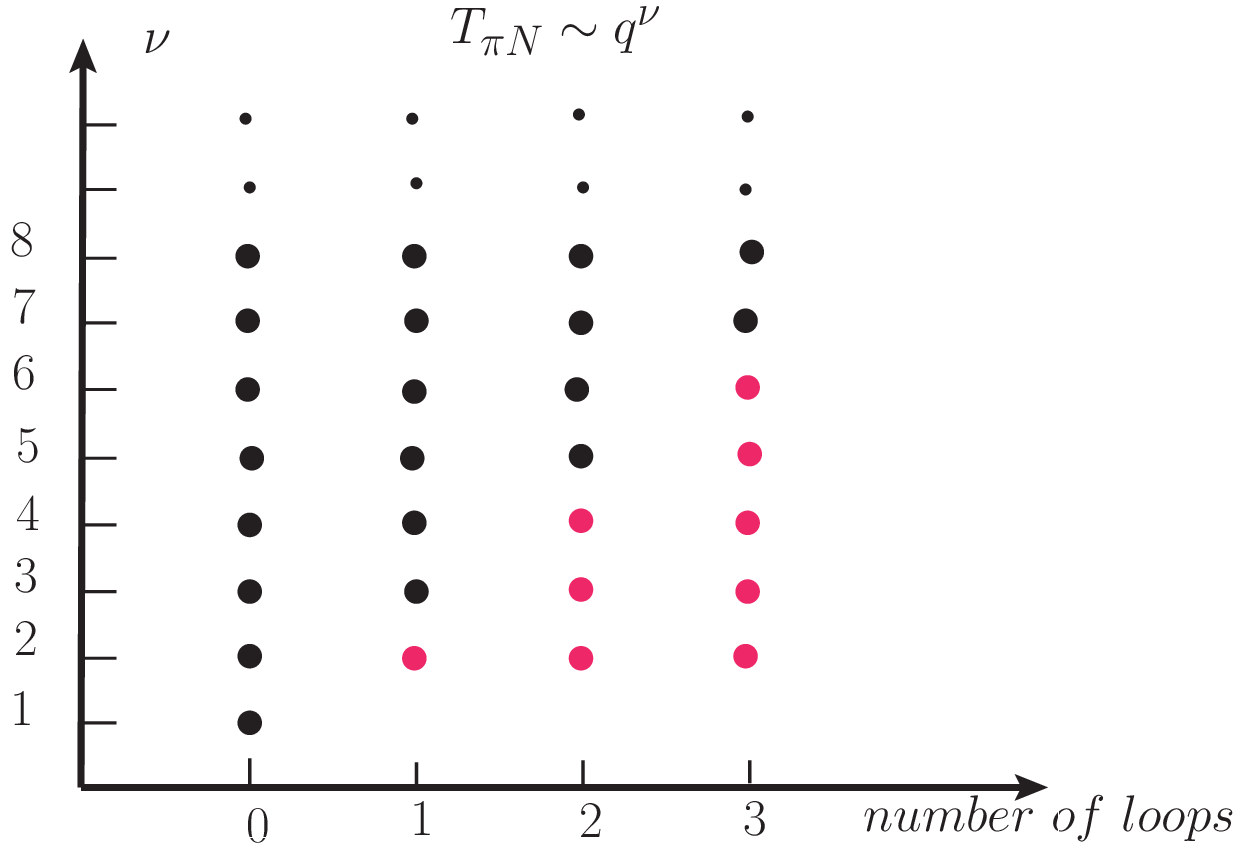} 
 \caption{Comparison of the naive scaling of the $\pi\pi$ scattering amplitude (left) with the $\pi N$ one (right). The red dots correspond to terms that break the power counting. One sees the contribution to all orders for any finite loop expansion.\label{Counting_nucleon_loops}}
 \end{center}
\end{figure}

When one introduces nucleons in the effective field theory, the counting becomes more involved, since the nucleon mass does not scale as $p$ or $M_\pi$ \cite{Gasser:1987rb}. 
A simple example that illustrates this issue is the leading one-loop correction to the nucleon mass that, according to the standard chiral counting, is of $\mathcal{O}(p^3)$. 
However, if one performs the calculation finds the following expression in dimensional regularization \cite{Alarcon:2012kn}

\begin{align}\label{Eq:nucleon_mass}
 m_N &= m - 4c_1 M_\pi^2 - \frac{3 g^2 m}{2 f^2}(2\bar{\lambda} (m^2+M_\pi^2)) + \frac{3 g^2 m M_\pi^2}{32 \pi^2 f^2} \nonumber\\
            &- \frac{3 g^2 M_\pi^3}{64 \pi^2 f^2} \left[ \frac{M_\pi}{m}�\log\left( \frac{M^2}{m^2} \right) - 4 \sqrt{1- \frac{M_\pi^2}{4m^2}} \arccos \left(\frac{M_\pi}{2m} \right) \right]
\end{align}

with $\bar{\lambda} = \frac{m^{d-4}}{16 \pi^2}\left\{ \frac{1}{d-4} - \frac{1}{2}[\log{4\pi}�+ \Gamma'(1) + 1] \right\}$, $M$ the pion mass, $d$ the space-time dimension, $f$, $m$ and $g$ the pion decay constant, the nucleon mass and the nucleon axial coupling in the chiral limit, respectively.
The result of Eq.~\ref{Eq:nucleon_mass} shows that the loop correction gives rise to terms of $\mathcal{O}(p^2)$ that break the power counting. 
In Fig.~\ref{Counting_nucleon_loops} it is shown what happens for an arbitrary number of loops.
A possible solution to this problem is to formulate the theory with baryons in a non-relativistic way so that the nucleon appears in the Lagrangian as a heavy static field. 
This is the so-called Heavy Baryon formulation \cite{Jenkins:1990jv}.
Another possibility is performing a renormalization of the Lagrangian that cancels both, the divergences and the power counting breaking terms, since the latter are always analytical in external momenta and quark masses.
This is the Extended-On-Mass-Shell renormalization scheme that preserves the covariance of the formulation of chiral effective field theory with nucleons \cite{Gegelia:1999gf,Fuchs:2003qc}.

When one considers more than one nucleon in the effective field theory, additional complications appear in the formulation due to the enhancement factors $2m_N/p$ that arise in the reducible two-nucleon diagrams of the $NN$ interaction \cite{Weinberg:1990rz} \footnote{In the vacuum $NN$ sector, the breakdown scale is estimated to be a bit lower, around $\Lambda_{NN} \approx 600$~MeV.}. 
A solution, proposed by Weinberg, consist in resuming this subclass of diagrams by solving a Lippmann-Schwinger (LS) equation, which is equivalent to resuming the right-hand cut of the $NN$ scattering matrix. Projected in partial waves, the LS equation for the $T$-matrix has the form

\begin{align}
T_{\ell \ell'} (p,p') = V_{\ell,\ell'} + \sum_{L} \int_{0}^{\infty} \frac{dk\, k^2}{2\pi^2} V_{\ell,L}(p,k)  \frac{1}{2\sqrt{p'^2+m_N^2}-2\sqrt{k^2+m_N^2} + i0^+}T_{L,\ell'}(k,p')
\end{align}

where $T_{\ell \ell'}$ is the $T$-matrix for the initial and final orbital angular momenta $\ell$ and $\ell'$ and $V_{\ell,\ell'}$ is the potential projected on these channels. 
This non-perturbative treatment requires to regularize the momentum integrals when using the $NN$ potentials derived from Chiral Effective Field Theory. 
In vacuum calculations is quite common to find combinations of local regulators and non-local regulators for different types of potentials derived from chiral symmetry \cite{Reinert:2017usi}. 
These regulators are introduced in the momentum space representation of the potentials in order to suppress the contribution of the high-energy modes in the momentum integrals. 
Historically, various types of regulator have been used for the same purpose \cite{Epelbaum:2004fk,Reinert:2017usi}.
The different outcome obtained in each analysis gives an idea of the systematics associated to the specific functional form chosen to regulate the divergent integrals. 
For a rigorous analysis of the artifacts introduced by the typical non-local regulators, see Ref.~\cite{Piarulli:2019cqu}.

Another issue of the vacuum nucleon two and few-nucleon forces is the inclusion of the $\Delta(1232)$ resonance as a dynamical degree of freedom. 
It is well known the importance of this resonance in EFT involving nucleons to describe and predict correctly the properties of the nucleon \cite{Alarcon:2011zs,Alarcon:2012kn,Alarcon:2012nr}.
From the EFT point of view, the inclusion of the $\Delta(1232)$ explicitly in the formulation allows to maintain the separation of scales required in effective theory and improves the convergence of the perturbative approach.
Unfortunately, it is quite standard to find EFT calculations without explicit $\Delta$ in the two and few-nucleon systems or nuclear matter calculations, although there are some exceptions \cite{Piarulli:2014bda,Piarulli:2016vel}. 

Chiral Effective Field theory allows to include interactions beyond the two-body forces, maintaining the hierarchy between the many-body forces. 
They are derived in the same way as the two-body forces, using the same counting but considering more nucleon bilinears than in the former case. 
These $n$-body forces are all interrelated, such that the contribution to one observable cannot be split into the different $n$-forces contributions, since they are representation-dependent and this separation is unphysical. 
A consistent treatment requires the inclusion of the few-body forces together with the two-body forces if they appear at a given order for a specific power counting in the EFT. 

So, in an effective field theory approach to nuclear forces one first constructs the most general Lagrangian satisfying the required symmetries up to the order consider in a particular power counting. 
Is quite common in the $NN$ sector to find results based on the Heavy Baryon formulation of chiral effective field theory, although there are recent publications using the covariant formulation as well \cite{Li:2016mln}.
With this Lagrangian one derives later the interaction potential that will be used to compute the $T$-matrix in a non-perturbative way. For example, if we are interested in a leading-order calculation of the $NN$ $T$-matrix, we consider fist the chiral Lagrangians that can be used to construct a $\mathcal{O}(p^0)$ (LO) diagram. These are the following $\pi N$ and $NN$ Lagrangians:

\begin{align}
&{\cal L}^{(1)}_{\pi N}  =  
 \bar{N} \left(i\gamma^\mu \partial_\mu - m_N
-\frac{1}{4f^2_\pi} \, \gamma^\mu
\mbox{\boldmath $\tau$} \cdot 
 ( \mbox{\boldmath $\pi$}
\times
 \partial_\mu \mbox{\boldmath $\pi$})
      - \frac{g_A}{2f_\pi} \gamma^\mu \gamma_5
        \mbox{\boldmath{$\tau$}} 
	\cdot \partial_\mu \mbox{\boldmath{$\pi$}}
  + \mathcal{O}(\mbox{\boldmath{$\pi$}}^3)
  \right) N  
\\
&{\cal L}^{(0)}_{NN} = -\frac{1}{2} C_S (\bar{N} N)(\bar{N} N) 
-\frac{1}{2} C_T (\bar{N} \overrightarrow{\sigma} N) \cdot (\bar{N} \overrightarrow{\sigma} N) \, ,
\end{align}

Where $\pi$ and $N$ represent the pion and nucleon fields, $\tau$ and $\sigma$ are the Pauli matrices in the isospin and spin space, respectively, $g_A$ the nucleon axial coupling and $f_\pi$ the pion weak decay constant. 
The ${\cal L}^{(1)}_{\pi N}$ Lagrangian contains nucleon bilinears with one and two pions.
At LO in $NN$ only the bilinear with one pion will contribute, while the one with two pions contributes to the NLO $NN$ potential. 
On the other hand, the ${\cal L}^{(0)}_{NN}$ Lagrangian contains only two terms with the product of two nucleon bilinears with no derivatives, which give rise to $\mathcal{O}(p^0)$ contact term contributions. 
These contact terms together with the diagram corresponding to the exchange of a pion between nucleons are all the $\mathcal{O}(p^0)$ contributions to the $NN$ interaction potential with chiral forces, that has the following simple form:

\begin{align}\label{Eq:VLO}
V_{LO} (q) = 
C_S + C_T \overrightarrow{\sigma}_1\cdot\overrightarrow{\sigma}_2 - 
\frac{g_A^2}{4f_\pi^2}
\mbox{\boldmath{$\tau$}}_1 \cdot \mbox{\boldmath{$\tau$}}_2 
\:
\frac{
\overrightarrow{\sigma_1} \cdot \overrightarrow{q} \,\, \overrightarrow{\sigma_2} \cdot \overrightarrow{q}}
{q^2 + M_\pi^2} 
\,.
\end{align}

\begin{figure}[H]
 \begin{center}
 \includegraphics[width=.99\textwidth]{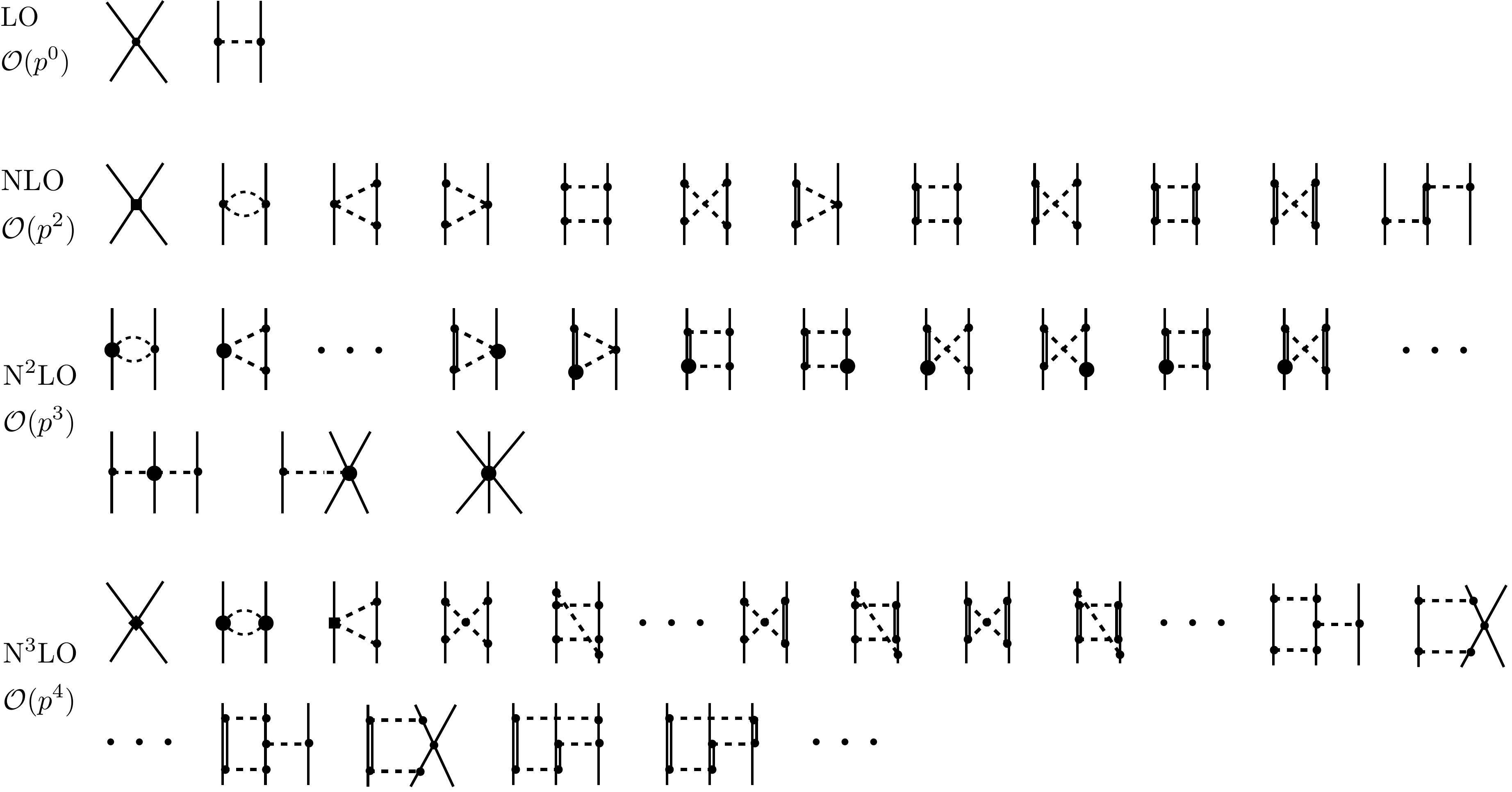} 
 \caption{Diagrammatic representation two- and three-nucleon forces obtained in chiral effective field theory according to the Weinberg power counting when the $\Delta$ is included. Solid, dashed and double lines correspond to nucleon, pion and $\Delta(1232)$, respectively. Figure adapted from Ref.~\cite{Piarulli:2019cqu}.  \label{NN_diagrams}}
 \end{center}
\end{figure}

In Fig.~\ref{NN_diagrams} is shown the diagrammatic representation of the different contributions to the $NN$ and $3N$ potentials derived from chiral forces up to $\mathcal{O}(p^4)$ considering also the $\Delta$ exchange.

There are many different approaches to treat the many-body problem using the nuclear forces derived from chiral effective field theory.
They are usually applied to calculate the energy per particle ($E/A$) in pure neutron matter (PNM) and symmetric nuclear matter (SNM), from where one can obtain the equation of state.
In SNM one considers an homogenous infinite system of nucleons with the same proton and neutron fractions. 
It is usually interesting to study the density at which the binding energy per nucleon is maximum, which is called the saturation density.
Note that the value of this binding energy per nucleon is different from the one obtained for the atomic nuclei, since in SNM there are no finite volume or surface corrections nor the electromagnetic interactions are usually considered.  
In PNM, on the other hand, one has an homogenous infinite system of nucleons with only neutrons.
This system is relevant to study neutron-rich matter and has a direct implication for neutron star physics, since one expects to have pure neutron matter in the outer core of neutron stars.
The difference between the energy per particle in PNM and SNM allows to compute the symmetry energy. 
This symmetry energy quantifies the energy needed to convert protons to neutrons, and it's dependence with the density is directly related to the neutron skin in heavy nuclei.  
In the following, we will summarize the different many-body methods used to study these systems with the chiral potentials derived form chiral effective field theory.

\begin{figure}[H]
 \begin{center}
 \includegraphics[width=.75\textwidth]{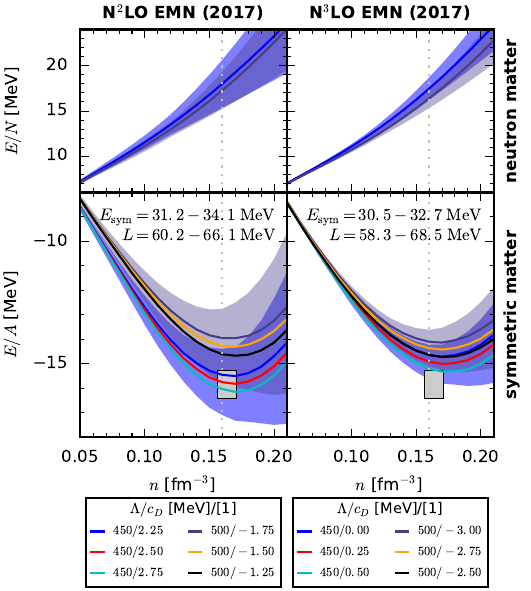} 
 \caption{Example of the results obtained in MBPT using chiral potentials. This plot corresponds to Ref.~\cite{Drischler:2017wtt}. Reprinted figure with permission from C.~Drischler, K.~Hebeler and A.~Schwenk,  Phys. Rev. Lett. \textbf{122}, no.4, 042501 (2019) Copyright 2019 by the American Physical Society \url{https://doi.org/10.1103/PhysRevLett.122.042501}. 
\label{MBPT_plot_EA}}
 \end{center}
\end{figure} 

\subsection{Many-Body Perturbation Theory}

One common approach is to use many-body perturbation theory (MBPT) \cite{Tichai:2020dna} with the $n$-body vacuum potentials to estimate the energy density of the many-nucleon system. 
The goal is to provide solutions to the time-independent Schr\"odinger equation for the many-body system given an Hamiltonian, which in this case is provided by Chiral EFT. 
The stating point is to write the Hamiltonian of the system ($H$) in terms of a one-body Hamiltonian $H_0 = T + U$ and a perturbation $H' = V - U$, where $T$ is the kinetic energy and $V$ the interaction potential taken from the EFT.
On the other hand, $U$ is an effective one-particle potential which is constructed commonly as

\begin{align}
U =  \int\frac{d^3 k'}{(2\pi)^3} \theta(k_F - k')\langle {\bf k} {\bf k'} | (1 - P_{12})V_{\text{2-body}} | {\bf k} {\bf k'} \rangle
\end{align}

where $\theta$ is the Fermi-Dirac distribution at zero temperature and $P_{12}$ is the Pauli exchange operator.
Given the non-perturbative nature of the nuclear many-body problem, it is necessary to choose an unperturbed state that captures the non-perturbative part of the problem at hand, since the perturbative approach won't reproduce it. 
In this framework it is possible to include the effects of the 3-body forces under some approximations that leads to a density-dependent modification of  $V_{\text{2-body}}$, although there are automated calculations that include 2-, 3- and 4-body forces without approximations for nuclear matter calculations in the literature \cite{Drischler:2017wtt}.
Many-Body Perturbation Theory allowed to compute the energy per particle in symmetric nuclear matter and pure neutron matter \cite{Hebeler:2010xb,Coraggio:2014nva,Drischler:2017wtt}. Some exapmles are shown in Fig.~\ref{MBPT_plot_EA}.

\subsection{Self Consisten Green Function Method}

\begin{figure}[H]
 \begin{center}
 \includegraphics[width=.45\textwidth]{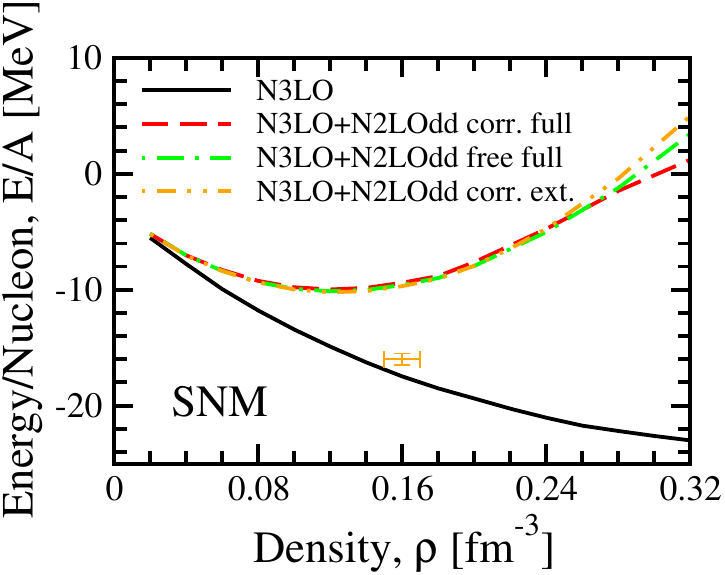} \includegraphics[width=.45\textwidth]{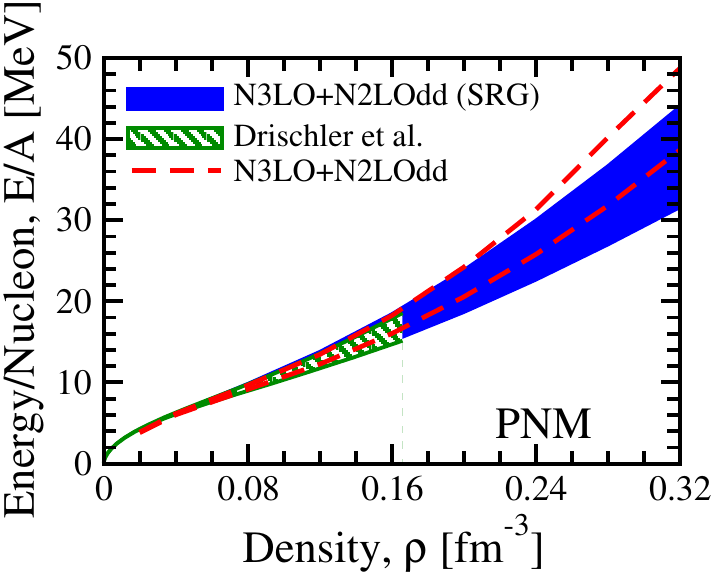} 
 \caption{Example of the results obtained in SCGF using chiral potentials. Left figure corresponds to SNM  and the right one to PNM. The plots are taken from Ref.~\cite{Carbone:2014mja}. Reprinted figures with permission from  A.~Carbone, A.~Rios and A.~Polls, Phys. Rev. C \textbf{90}, no.5, 054322 (2014). Copyright 2014 by the
American Physical Society.  \url{https://doi.org/10.1103/PhysRevC.90.054322}. \label{SCGF_plots}}
 \end{center}
\end{figure}

Another approach is the Self Consisten Green Function (SCGF) Method (see \cite{Rios:2020oad} for a recent review). 
As in MBPT, the stating point is to write the Hamiltonian as $H = H_0 + H'$, with the unperturbed Hamiltonian $H_0 = T + U$ and the perturbation $H' = V - U$, where $V$ contains the two- and few-body forces.  
With this decomposition, the one-body propagator has the following form

\begin{align}�\label{Eq:SCGF1}
 &G_{\alpha \beta}(t_{\alpha} - t_{\beta}) = \nonumber \\ 
 & - i \sum_{n = 0}^{\infty} \left( -i \right)^n�\frac{1}{n!} \int dt_1 \dots \int dt_n� \langle \Psi_0 |�T [H'(t_1) \dots H'(t_n)a_{\alpha}(t_{\alpha}) a^\dagger_{\beta}(t_\beta)]|\Psi_0\rangle_{\text{connected}}
\end{align}

where $T$ is the time order product operator and $\Psi_0$ is the $n$-body ground state wave function.
On the other hand $a$, $a^\dagger$ are one-body annihilation and creation operators and $H'$ is the perturbation, all of them in the interaction picture.
The different contributions in the perturbative expansion \eqref{Eq:SCGF1} can be reorganized to give rise to the Dyson's equation

\begin{align}�\label{Eq:SCGF3}
  G_{\alpha, \beta} (\omega)  =   G^{(0)}_{\alpha, \beta} (\omega) + \sum_{\alpha',\beta'}G^{(0)}_{\alpha,\alpha'} (\omega) \Sigma^*_{\alpha', \beta'}(\omega)G_{\beta',\beta}(\omega)
\end{align}

being $G^{(0)}_{\alpha, \beta} (\omega)$ the bare propagator and $\Sigma^*$ the irreducible self-energy.
The latter plays a central role in the SCGF method, since it encodes the information of the interaction with other particles. 
The calculation of $\Sigma^*$ requires the knowledge of the interaction potentials and it is where chiral forces are used.
The one-body Green function is computed iterating Eq.~\eqref{Eq:SCGF3} and is used to calculate the expectation value of any one-body operator $O^1 \equiv \sum_{\alpha, \beta} O^1_{\alpha,\beta} a_\alpha^\dagger a_\beta $ in the following way

\begin{align}
 \langle O^1 \rangle  = \sum_{\alpha ,\beta} \int_{C^+}\frac{d\omega}{2 \pi i}  O^1_{\alpha,\beta} G_{ \beta \alpha} (\omega).
\end{align}

where $C^+$ means that we close the contour integral from the upper part of the imaginary plane.
The one-body propagator also allows to compute the total energy of the system though the Galitski-Migdal-Koltun sum rule

\begin{align}
 E_0 = \sum_{\alpha ,\beta} \int_{C^+}\frac{d\omega}{2 \pi i}  [T_{\alpha \beta} + \omega \delta_{\alpha\beta}] G_{ \beta\alpha} (\omega).
\end{align}

being $T_{\alpha\beta}$ the matrix elements of the kinetic operator. 
This expression can be generalized to include the three-body forces \cite{Carbone:2013eqa}, and requires the calculation of the three-body propagator.
Results for SNM and PNM using this approach is show in Fig.~\ref{SCGF_plots}.

\subsection{Quantum Monte Carlo}

Quantum Monte Carlo (QMC) approaches are based on the path integral formalism, and there is a large variety of QMC methods that have been applied to study the many-body problem in nuclear physics. 
In this review, we will mention two of them, the Variational Monte Carlo (VMC) Method and the Green's Function Monte Carlo (GFMC) Method.

\subsubsection{Variational Monte Carlo}

In VMC one starts with a trial function $\psi$ that has a fixed functional form and some free parameters.
Since, for a given Hamiltonian $H$, the expected value 

\begin{align}
 E \equiv \frac{\langle  \psi | H | \psi \rangle }{\langle  \psi | \psi \rangle }
\end{align} 

is always equal or greater than the ground-state energy with the quantum numbers of $\psi$, one can find the ground state energy and wave function by minimizing $E$. 
This Hamiltonian contains the two and three-body potentials derived from Chiral EFT.
The Monte Carlo methods are applied to evaluate $E$ and minimize this function with respect to the free parameters of $\psi$.
The wave function is usually factorized into a long- and short-range parts

\begin{align}
 |\psi\rangle = O |\phi\rangle
\end{align} 

The operator $O$ is introduced to compute the short-range behavior of the wave function, and the state $\phi$ to capture the long-range one.
The short-range operator contains the interaction between two and three particles.
The long-range wave function $\phi$ can be chosen in different ways, and depends on the problem under consideration. 
Depending on the long-range correlation, one can use BCS form wave function or construct a wave function with cluster structure.
In practice, the calculations with VMC requiere a significant computational effort, since the computational cost growth exponentially with the number of particles.

\subsubsection{Green's Function Monte Carlo}

In GFMC one tries to reach the ground state $\psi_0$ by evolving a trial wave function $\phi$ according to

\begin{align}
 |\psi_0\rangle \propto \lim_{\tau�\to \infty} e^{-(H-\varepsilon_0)\tau} | \phi\rangle
\end{align}

where $\varepsilon_0$ is a parameter related to the normalization of the wave function. 
The trial wave function is commonly taken from the VMC results.
For nuclear physics applications, the evolution is computed in terms the product of short-time evolution operators inserting a complete set of states in the following way

\begin{align}
 |\psi_0 (r_N)\rangle =  \Pi_{1, \dots, N} \langle {\bf r_N}| e^{-(H-\varepsilon_0) \Delta \tau} |{\bf r_{N-1}}\rangle \dots  \langle {\bf r_1}| e^{-(H-\varepsilon_0) \Delta \tau} |{\bf r_{0}}\rangle |\phi (r_{0}) \rangle 
\end{align}

where $\Delta \tau$ is the shot-time interval. 
Monte Carlo methods are used to sample ${\bf r_i}$ in the propagation.
In the end, the energy of the system is calculated as a function of the imaginary time $\tau$ as

\begin{align}
\langle H(\tau)\rangle = \frac{\langle \psi(\tau) |H |\phi\rangle}{\langle \psi(\tau) |\phi\rangle}
\end{align}

For $\tau$ large enough the energy of the system fluctuates around a stable value which is identified with the ground state with the quantum numbers of the trial state. 
Matrix elements of other operators are also calculated through similar evaluation that require the knowledge of $|\psi (\tau)\rangle$ \cite{Pervin:2007sc}.

\begin{figure}[H]
 \begin{center}
 \includegraphics[width=.7\textwidth]{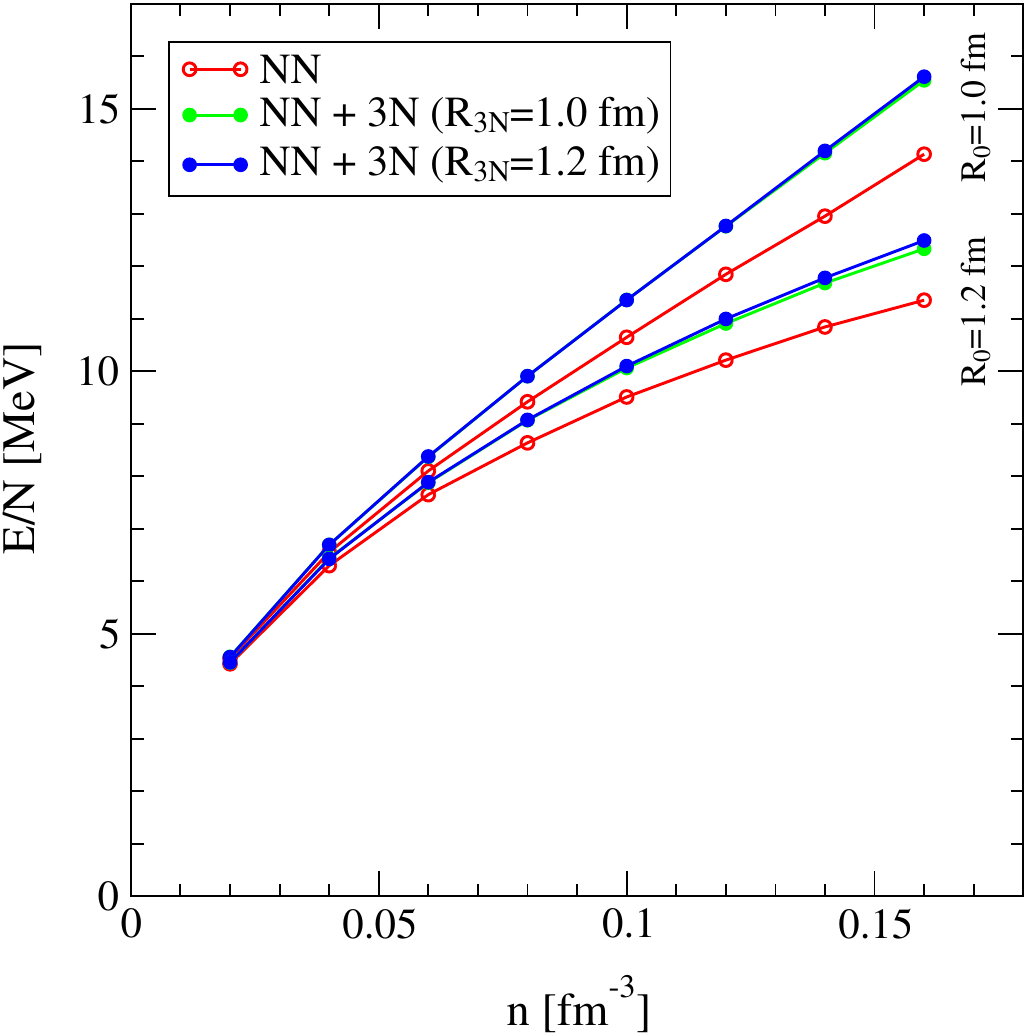} 
 \caption{Example of the calculation of the energy per particle in pure neutron matter with QMC calculations using chiral potentials. This plot corresponds to Ref.~\cite{Tews:2015ufa}. 
Reprinted figure with permission from I.~Tews, S.~Gandolfi, A.~Gezerlis and A.~Schwenk,
Phys. Rev. C \textbf{93}, no.2, 024305 (2016). Copyright 2016 by the
American Physical Society \url{https://doi.org/10.1103/PhysRevC.93.024305}. 
  \label{QMC_plots}}
 \end{center}
\end{figure} 

\subsection{In-medium Chiral Effective Field Theory}

Another possibility to study the many-body nuclear problem is to reformulate the EFT directly in the nuclear medium.
This has been done for the case of low density systems and zero temperature in \cite{Oller:2001sn}. 
By considering the transition amplitude for the finite density ground state $|\Omega\rangle$ at asymptotic times $t \to \pm \infty$ with external sources, one can derive a expression for the generating functional\footnote{This expression was derived for symmetric and spin-balanced systems. A generalization for asymmetric spin-polarized systems was also shown in \cite{Oller:2001sn}.} 

\begin{align}\label{Eq:generating_functional}
 &e^{i \mathcal{Z}} = \int [dU] (\text{det} D) \exp\left\{� i \int dx \mathcal{L}_{\pi \pi}  \right. \nonumber \\
 &�- i \int^{k_F}\!\! \frac{d{\bf p}}{(2\pi)^3 2E(p)} \int dx dy e^{ip(x-y)} \text{Tr} \left.(A[I_4 - D_0^{-1} A]^{-1}\right|_{(x,y)} (\slashed{p} + m_N) ) \nonumber \\
 &+ \frac{1}{2} \!\!\int^{k_F} \frac{d{\bf p}}{(2\pi)^3 2E(p)}\int^{k_F} \!\!\frac{d{\bf q}}{(2\pi)^3 2E(q)}  \int dx\, dx'\, dy\, dy'\, e^{ip(x-y)} e^{-iq(x'-y')}  \nonumber \\ 
 & \times \left. \text{Tr} \left.A[I_4 - D_0^{-1} A]^{-1}\right|_{(x,x')}  (\slashed{q} + m_N) \left.A[I_4 - D_0^{-1} A]^{-1})\right|_{(y,y')} (\slashed{p} + m_N) ) + \dots \right\}.
\end{align}

In this equation $I_4\equiv \delta(x_n-x_m)\delta_{n,m}$, with $n$ and $m$ labeling each of the incoming nucleons, $D_0$ is the Dirac operator $D_0 \equiv i \gamma^\mu \partial_\mu - m_N$, $A \equiv D - D_0$ with $D$ such that the bilinear terms in the nucleon fields of the chiral Lagrangian are expressed as $\mathcal{L}_{NN} = \bar{N}D N$, with $N$ the nucleon field. On the other hand, the symbol $[dU]$ under the integral represents the path integral over the pion fields fields and the trace is taken in the spin and isospin space.

\begin{figure}[H]
 \begin{center}
 \includegraphics[width=.9\textwidth]{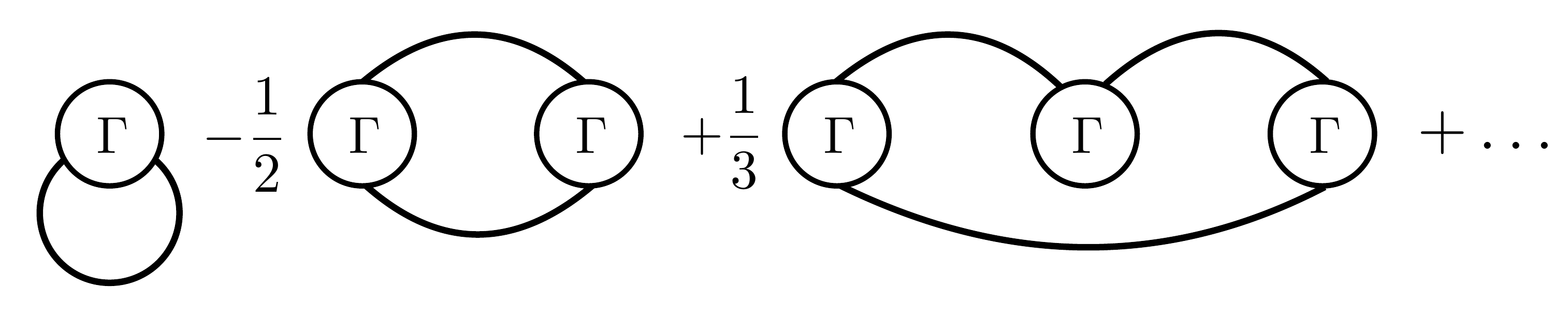} 
 \caption{Diagrammatic representation of the expansion in the number of Fermi seas (thick lines). \label{Fig:Fermi_sea_expansion}}
 \end{center}
\end{figure}

The Eq.~\eqref{Eq:generating_functional} involves two expansions.
One is the standard chiral expansion when considering a theory for nucleons in vacuum \cite{Gasser:1987rb}, which arise from the expansion of the operator $i \Gamma \equiv A[I_4 - D_0^{-1} A]^{-1}  = A + AD_0^{-1}A + AD_0^{-1}A D_0^{-1}A + \dots$ and $A = A^{(1)} +  A^{(2)} + \dots$, where the superscript in $A$ indicates the chiral order. 
The other expansion is related to the nuclear medium contribution, and is an expansion in the number of insertions of on-shell nucleons belonging to the Fermi sea. 
One can assign a counting $k_F \sim p$ to provide a unified counting for nuclear matter calculations at densities of the order of nuclear saturation or below \cite{Oller:2001sn}.
The expansion in the number of Fermi sea insertions allows the diagrammatic representation show in Fig.~\ref{Fig:Fermi_sea_expansion}.
So the EFT formulated in the nuclear medium permits the same kind of treatment as the vacuum counterpart.
There are, however, important differences with respect to the vacuum theory. 
One is the different counting of nucleon propagators in the nuclear medium. 
In vacuum they scale as $\mathcal{O}(p^{-1})$, but in the nuclear medium they may have a scaling of $\mathcal{O}(p^{-2})$ in certain configurations \cite{Meissner:2001gz}. 
This is not considered in many body approaches with vacuum $NN$ potentials and is important in establishing the hierarchy of the different contributions.

The non-perturbative nature of the nuclear interactions require non-perturbative techniques as in the vacuum case.
The counting provided by the in-medium formulation of the EFT allows to identify the diagrams that need to be resummed in this perturbative scheme. 
There are different methods available in the literature to perform this resummation, as we discuss in the following. 

\subsubsection{Unitary Chiral Perturbation Theory} 
One method is based on Unitary Chiral Perturbation Theory (UChPT), and allows to resum the unitary cut of each partial wave.
The starting point is that the inverse of a partial wave amplitude has to satisfy the following unitarity condition
\begin{align}
 \text{Im}T^{-1} = - \rho
\end{align}

where $\rho$ is the two-particle phase space.
Performing a once-subtracted dispersion relation of $T^{-1}$, allows to derive the main equation of UChPT, that can be written as

\begin{align}
  T = \left[I + N\cdot G \right]\cdot N
\end{align}
 
where $N$ is the interaction kernel, obtained from the perturbative scattering amplitude in EFT and $G$ is the unitarity loop function \cite{Oller:2000fj}.
This approach has been used to calculate the equation of state of pure neutron matter and symmetric nuclear matter in \cite{Lacour:2009ej} considering the in-medium contribution to the nucleon propagators.

\begin{figure}[H]
 \begin{center}
 \includegraphics[width=.7\textwidth]{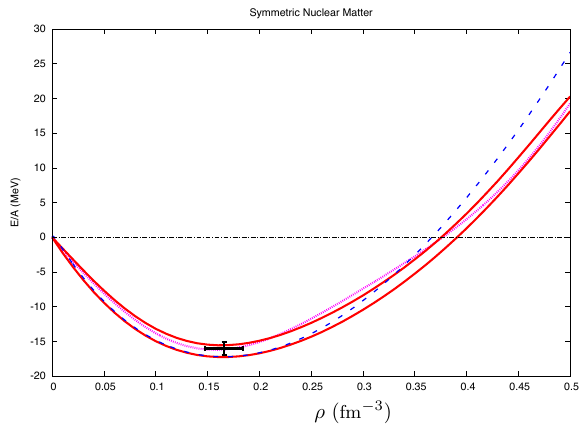} 
 \caption{Example of the results obtained for SNM in UChPT with in-medium chiral perturbation theory. This plot corresponds to Ref.~\cite{Lacour:2009ej}. Reprinted from Annals Phys. \textbf{326}, , A.~Lacour, J.~A.~Oller and U.~G.~Meissner,  "Non-perturbative methods for a chiral effective field theory of finite density nuclear systems'', 241-306 Copyright (2011), with permission from Elsevier�~\url{https://doi.org/10.1016/j.aop.2010.06.012}.
 \label{QMC_plots}}
 \end{center}
\end{figure}

\subsubsection{Ladder resummation} 
In Refs.~\cite{Lacour:2009ej,Kaiser:2012sr} Kaiser performed an algebraic resummation of the ladder diagrams with in-medium propagators for a pure contact S-wave interaction first, and S-wave effective range and P-wave scattering volume later. 
This work was extended to arbitrary partial wave content in a unified resumation scheme for any partial wave to any order in \cite{Alarcon:2021kpx}.
There, it was found that the contribution of the interaction to the energy density in the nuclear medium can be calculated, at low densities, from the in-medium $t$-matrix $t_m$. 
This matrix satisfies the following integral equation in partial waves

\begin{align}
 t_m(p',p) = v(p',p) + \frac{m}{2\pi^2}\int^\infty_0�dk \frac{k^2}{k^2 - p^2 - i \varepsilon} v(p',k)\cdot \mathcal{A}\cdot t_m(k,p)
\end{align}

where $v$ is the interaction potential in partial waves and $\mathcal{A}$ is a matrix constructed from the product of two spherical harmonics

\begin{align}
 \mathcal{A} \propto \int d{\bf�\hat{k}}\ Y_{\ell_2}^{m_3}({\bf \hat{k}})Y_{\ell_1}^{m_3}({\bf \hat{k}})\left[ 1 - 2\theta(k_F - |{\bf k} - a \hat{z} |)�\right]
\end{align}

\begin{figure}[H]
 \begin{center}
 \includegraphics[width=.48\textwidth]{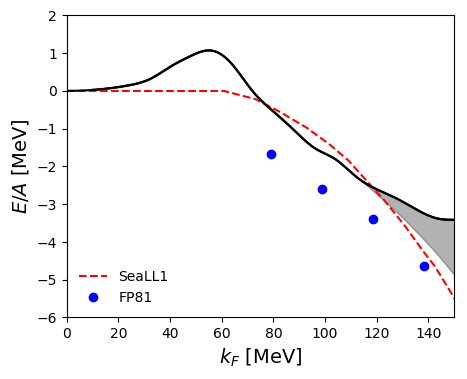} \includegraphics[width=.48\textwidth]{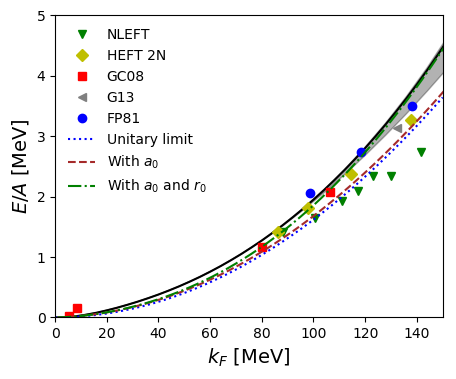} 
 \caption{Example of the results obtained with the ladder resummation method. Left plot shows the energy per particle in SNM and the right plot for PNM. Both figures are taken from Ref.~\cite{Alarcon:2022vtn}. 
Reprinted figures with permission from J.~M.~Alarc\'on and J.~A.~Oller,
Phys. Rev. C \textbf{107}, no.4, 044319 (2023) Copyright (2023) by the American Physical Society \url{https://doi.org/10.1103/PhysRevC.107.044319}.  \label{ladder_resummation_plots}}
 \end{center}
\end{figure}

Where ${\bf a} = ({\bf k_1} + {\bf k_2})/2$, being ${\bf k_1}$ and ${\bf k_2}$ the nucleons three-momenta, and $\ell_1$, $\ell_2$ the orbital angular momenta of the initial and final $NN$ state. 
If the interaction potential can be written as a polynomial in the off-shell momenta, the previous integral equation can be solved analytically. 
In \cite{Alarcon:2021kpx} it was shown that, for any partial wave, one can write the solution of this integral equation in terms of the effective range expansion plus an off-shell contribution that disappears when the cutoff is sent to infinity in a cutoff regularization scheme. 
This leaves the result in terms of the on-shell vacuum t-matrix and a two-point loop function with the insertion of one Fermi sea that can be calculated analytically. 
With this method, it has been calculated the Bertsch parameter in a unitary Fermi gas \cite{Alarcon:2021kpx}, studied the possibility of a liquid phase in a spin-balanced fermionic quantum system interacting in $P$-wave \cite{Alarcon:2021nwd} and the equation of state of symmetric nuclear matter and pure neutron matter \cite{Alarcon:2022vtn}.
The equation of state at low densities of symmetric nuclear matter and pure neutron matter were obtained in this approach directly from the $NN$ phase shifts and without the need of any regulator.
In SNM this approach is able to reproduce the so-called spinodal instability of homogeneous symmetric matter at low densities (see left plot in Fig.~\ref{ladder_resummation_plots}), which is not obtained by other chiral approaches.

\subsection{Nuclear Lattice Effective Field Theory}

Another possibility to face the many-body problem is using lattice techniques. 
In this approach, the vacuum $NN$ potentials obtained with a vacuum counting is put on a discrete time-space lattice. 
Normally, one first considers to fit the $NN$ observables with a lattice approach for a two-body problem to fix the $NN$ low-energy constants (LECs) \cite{Alarcon:2017zcv,Li:2018ymw}.
In this two-body calculation the calculation of the leading order is solved non-perturbatively and higher order corrections are calculated using perturbation theory.
Although the lattice regularize the divergences arising in a continuum calculation, the discretized potentials need to be modified with smearing functions and regulators \cite{Alarcon:2017zcv,Li:2018ymw}.
For example, if one considers the contact terms in the leading-order potential, the lattice implementation has the following form

\begin{align}
 V_{LO}^{\text{ct}} = \frac{C}{2}  :\sum_{\vec{n}} \rho(\vec{n}) \rho(\vec{n}):  + \frac{C_I}{2}  :\sum_{\vec{n}} \sum_I \rho_I^{}(\vec{n}) \rho_I^{}(\vec{n}):
\end{align}

where the symbol $:$ denotes normal ordering and the density operators $\rho(\vec{n})$ and $\rho_I(\vec{n})$ are defined in terms of the creation and annihilation operators $a^\dagger_{i,j}$ and $a_{i,j}$ as follows:

\begin{align}
&\rho(\vec{n}) \equiv \sum_{i,j} a_{i, j}^\dagger(\vec n) a_{i, j}^{}(\vec n), \\
&\rho_I^{}(\vec n) \equiv \sum_{i, j, j^\prime} a_{i, j}^\dag(\vec{n}) (\tau_I)_{j, j^\prime} a_{i, j^\prime}(\vec{n}), 
\end{align} 

The LO contact terms is smeared through the factor 

\begin{align}
f(\vec{q}\,) \equiv f_0^{-1} \exp\left(-b_s^{} \frac{\vec q\,^4}{4}\right),
\end{align}

where $b_s$ is a free parameter and $f_0$ is a normalization constant. 
All this is implemented in $V_{LO}^{\text{ct}}$ in the following way

\begin{align}
  V_{LO}^{\text{ct}} = \frac{C}{2L^3}  :\sum_{\vec{q}} f(\vec{q}\,) \rho(\vec{q}\,)\rho(-\vec{q}\,): +  \frac{C_I}{2L^3}  :\sum_{\vec{q}} f(\vec{q}\,)
\rho_I^{}(\vec{q}\,)\rho_I^{}(-\vec{q}\,):
\end{align}

The smearing of the LO contact terms potential is important to reproduce well the deuteron binding energy and the $NN$ phase shifts, and provides a reliable solution to apply perturbation theory later.

Once the LECs are fixed, the many-body problem is considered with the two- and three-body chiral potentials. 
In the case of nuclear matter, the observables are calculated using Monte Carlo techniques with auxiliary fields that transform the two-, three- and four-body forces into an interaction of a single nucleon with a fluctuating auxiliary field�\cite{Lu:2019nbg}. 
The first attempt to study nuclear matter on a lattice with effective theories was done in Ref.~\cite{Muller:1999cp}, and the method has been applied to study pure neutron matter�\cite{Epelbaum:2009rkz} and symmetric nuclear matter�\cite{Lu:2019nbg}.

\begin{figure}[H]
 \begin{center}
 \includegraphics[width=.8\textwidth]{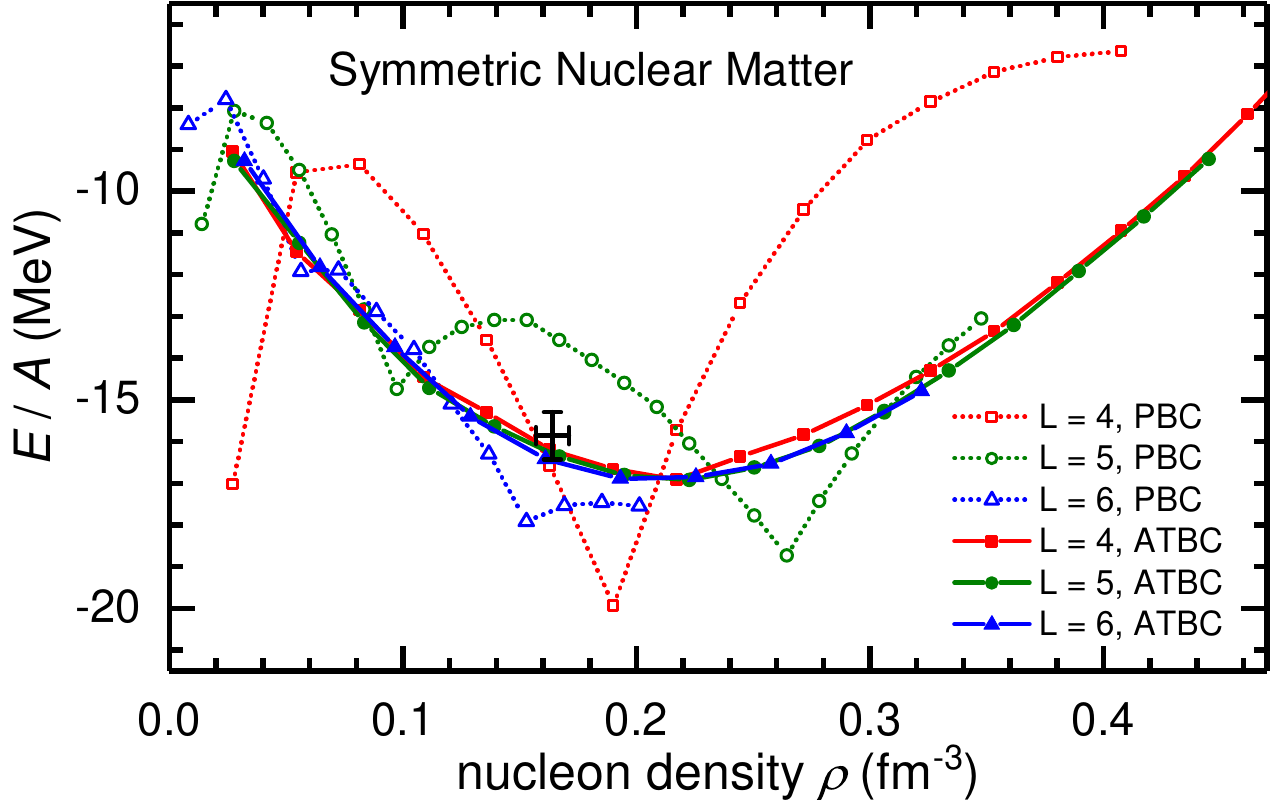} 
 \caption{Example of the results obtained with Nuclear Lattice Effective Field Theory. This plot corresponds to Ref.~\cite{Lu:2019nbg}.
 Reprinted figure with permission from B.~N.~Lu, N.~Li, S.~Elhatisari, D.~Lee, J.~E.~Drut, T.~A.~L\"ahde, E.~Epelbaum and U.~G.~Mei\ss{}ner,
Phys. Rev. Lett. \textbf{125}, no.19, 192502 (2020). Copyright (2020) by the
American Physical Society \url{https://doi.org/10.1103/PhysRevLett.125.192502}. \label{ladder_resummation_plots_2}}
 \end{center}
\end{figure}

\section{Equation of State of Neutron Stars}
\label{Sec:EoS}

Born in the aftermath of core collapse supernovae, neutron stars (NS) are supported against gravitational collapse by the neutron degeneracy pressure together with a repulsive strong force. Neutron stars are relatively cold systems for most of their lifetimes, with a temperature $T \leq 10^8$ K $\approx{10}$ keV \cite{Haensel:2007yy}, considered cold at the nuclear scale or generally from the perspective of nuclear and particle physics. So, an old NS can be treated as an object at zero temperature since its temperature is lower than the corresponding Fermi temperature of an immense gas of Fermions.

 The dynamical evolution of violent events such as core-collapse supernovae (CCS), leading to the formation of a NS or a black hole, is determined among others by the equation of state of matter. In addition, the EoS impacts the conditions for nucleosynthesis and the emerging neutrino spectra. Hence, the EoS is an essential ingredient in many astrophysical simulations and, therefore, the EoS becomes the main point of contact between macroscopic astrophysical studies and microscopic nuclear and particle physics works. Indeed, most of the static and dynamical properties of NSs can only be calculated once an equation of state describing the matter inside the stars is specified.
 In its most general form, the expression ��equation of state�� is used for any relationship between thermodynamic state variables and its construction assumes that the local system under consideration is in thermodynamic equilibrium. This implies that intensive thermodynamic variables such as temperature, pressure, or chemical potentials are well defined and that the conditions of thermal equilibrium (equivalent to a constant temperature $T$ throughout the chosen domain), mechanical equilibrium (constant pressure $P$) and chemical equilibrium (constant chemical potentials $\mu$) hold. 

 Assuming that full thermodynamic equilibrium is maintained throughout the different cooling stages \cite{HarrisonWheeler}, the interior of the resulting NS consists of "cold catalyzed matter", i.e., electrically neutral matter in its absolute ground state, with a density higher than that of the heaviest atomic nuclei. This scenario assumes that the reaction rates are much higher than the cooling rate. According to this cold catalyzed matter hypothesis, the interior of an old NS is in a state where thermal, nuclear and beta equilibrium prevail at a temperature $T$ low enough that thermal effects are negligible for the composition and pressure. The equilibrium conditions can reasonably be expected to be valid in any NS that is not accreting from a neighbour, but not for the crust of accreting NSs \cite{Pearson:2018tkr}. However, the core of a NS cools efficiently and, within minutes, thermal effects are no longer important in determining its structure and composition (with the exception of superfluidity).
 The composition is determined by the weak interaction equilibrium  ($\beta$-equilibrium), which means that neutron decay is balanced by the inverse process or electron capture dictating that the neutron, proton, and electron chemical potentials satisfy $\mu_n - \mu_p = \mu_e$. 
 
The microscopic property to describe the interior of a cold, mature NS is the one-parameter equation of state relating pressure $P$ to energy density $\varepsilon$. In contrast, other cases such as protoNS evolution and binary NS (BNS) mergers require an EoS of matter at non-zero temperature and out of (weak) $\beta$-equilibrium, i.e., the pressure becomes a function of three thermodynamic parameters (temperature, energy density and charge fraction). Furthermore,  the calculation of many of the properties of a NS (masses, radii, moments of inertia, binding energies, tidal deformabilities and oscillation frequencies) requires only the knowledge of the total energy density $\varepsilon$ and the pressure $P$ as a function of the baryon density, i.e., the one-parameter equation of state. Nuclear experiments provide important details of the equation of state up to $n_s$, and heavy-ion collisions add some constraints for densities up to about 5$n_s$, but only for hot, nearly symmetric nuclear matter (SNM). NS matter, by contrast, is cold and extremely neutron-rich with proton fractions of order a few percent near $n_s$, so a considerable extrapolation from hot, symmetric matter is required. 

NS mass and radius determinations have been made with increasing accuracy thanks to measurements of masses by Saphiro time delay \cite{Demorest:2010bx,Antoniadis:2013pzd,Fonseca:2021wxt}, NICER telescope of  masses and radii from X-ray data \cite{Riley:2019yda,Miller:2019cac,Riley:2021pdl,Miller:2021qha,Salmi:2022cgy}, and gravitational-wave detection from binary NS mergers by the LIGO and Virgo Collaborations \cite{LIGOScientific:2018hze,LIGOScientific:2018cki,LIGOScientific:2020aai}, which provide new constraints on the radii of typical (1.4$M_\odot$) stars via measurements of their tidal deformabilities.

\subsection{Neutron star structure}

A NS is typically divided into three main regions: the atmosphere, the crust (outer and inner), and the core (outer and inner). The central density of a NS is often expressed in terms of the baryon number density $n$ (nucleons per fm$^3$) related to the saturation density number $n_s$. The value of the saturation density is approximately $n_s=0.16$ fm$^{-3}$, which corresponds to approximately $2.7 \times 10^{14}$ g cm$^{-3}$. The central density of a NS can reach values that are 5 to 10 times higher than the saturation density.

 Moving towards the center of the NS, the density increases, and the nuclei in the crust progressively become more  neutron-rich. Eventually, at a certain point known as the neutron-drip point ($n\approx 2.6 \times 10^{-4}$ fm$^{-3}$), the density becomes high enough that neutrons start to leak out from the nuclei and the inner crust begins. Thus, in the inner region of the crust, neutron-rich nuclear clusters are supposed to be immersed in a neutron liquid and such matter cannot be reproduced in terrestrial laboratories and must be relied on theoretical models.
At approximately half of the saturation density ($n=n_s/2 \approx 0.08$ fm$^{-3}$),  the nuclei in the NS crust completely dissolve, and the core region begins. The core of a NS is much less understood than the crust and is also divided into outer and inner parts.
 
The outer core of a NS is believed to be composed of uniform matter at $\beta$-equilibrium, primarily neutrons with a small fraction of protons and electrons. The condition to fulfill beta-equilibrium (without neutrinos and without muons) is $\mu_n -\mu_p=\mu_e$. In cold catalyzed NS, neutrinos do not contribute to the chemical equilibrium. Muons can however appear with increasing densities when $\mu_\mu \geq m_\mu c^2$  and they contribute to $\beta$-equilibrium through the thermodynamic relation $\mu_e=\mu_\mu$. In addition, charge neutrality is also imposed: $n_e + n_\mu=n_p$. Thus, these three conditions complemented by the baryon conservation number $n = n_n+n_p$ lead to uniquely determine the composition of hadronic neutron-star matter called as $npe\mu$ matter \cite{Haensel:2007yy}.

Since the proton fraction in the uniform nucleonic matter is small, approximately $5\%$, neutron-star matter can be effectively described as pure neutron matter, whose properties are strongly related to the symmetry energy originating from the energy difference between pure neutron matter (PNM) and symmetric nuclear matter (SNM). Here, the degenerate neutron-rich gas, aided by repulsive interactions among the nucleons, provides pressure against the gravitational pull and keeps the star from collapsing to a black hole.






Low-density neutron matter (dilute neutron matter) is assumed be similar to a unitary Fermi gas, which is an idealized theoretical collection of fermions interacting only via pairwise s-wave interactions with an infinite scattering length and a vanishing effective range ($|a_0| \rightarrow \infty$ and $r_0=0$). 
In this limit, the
energy per particle $(E/A)_{\text{UG}}$ is proportional to the energy of a free Fermi gas $(E/A)_{\text{UG}}=\xi (E/A)_\text{F}$, $\xi$ being the Bertsch parameter, $\xi=0.370(5)(8)$ \cite{Zuern:2013}.
However, pure neutron matter at low densities has finite scattering length  and then both properties lead to larger energies than for a unitary gas. The Unitary Gas Conjecture (UGC) states that that $E/A\geq (E/A)_{\text{UG}}$ at all densities \cite{Tews:2016jhi,Lattimer:2023xjm} establishing a lower limit to the energy of pure neutron matter.

At higher densities, above approximately twice the saturation density ($n \approx 2n_s$), the composition of matter in the core becomes largely unknown.The inner core of NSs is the least understood part of the NS. Densities  can reach up to multiple times the saturation density, and it is unclear what degrees of freedom describe this region. There could be charged mesons, such as pions or kaons, that eventually form a Bose condensate, or other heavier baryons with strangeness, called hyperons. Finally, at higher densities, hadrons might not be relevant degrees of freedom and quark matter might appear \cite{Annala:2019puf}.
The outer regions of the NS, which are better understood, consist of a solid crust and a nuclear matter liquid just inside the crust, spanning baryon densities up to 1-2 $n_s$. The NS interior, for  baryon densities  $n \sim 2-10\, n_s$ is the most difficult to describe from first principles. Here, hyperons, quark matter in various phases (presumably unpaired at finite $T$, otherwise partly or totally condensed such as in the two-flavor color superconductivity (2SC) or Color-Flavor-Locked (CFL) phases respectively) may exist, but $\mu<\mu_{\text{pQCD}}$ (where pQCD refers to perturbative QCD). Only at much higher densities (n$_B \gtrsim  40n_s$, beyond those expected in neutron stars) can one apply perturbative QCD for dense matter directly. 

There are some approaches that propose to use the same nuclear model to describe the whole interior of NS, the core and the crust, which is known as unified EoS for NS.  The main challenge of this approach  is the difference between both regions: the crust has a crystalline structure difficult to model, while the core is homogeneous. These NS unified EoS are then model dependent, lead to different predictions depending on the nuclear interaction on which they are built, and do not provide an uncertainty band. 

\subsubsection{Outercrust}
 The region beneath the thin atmosphere (from milimetres to tens of centimetres) consists of a hundreds of meters of a thick solid layer what is called the crust. The crust is composed of inhomogeneous nucleonic matter and unlocalized $e^-$ in $\beta$-equilibrium. The crust can also be divided in outer crust, which is made of a body-centered cubic lattice of exotic nuclei in a charge neutralizing background of highly degenerate electrons, and an inner crust, consisting of neutron-proton clusters immersed in a neutron sea (possibly enriched with protons at sufficiently high densities). The internal constitution of the outermost layers of the crust is well known and the EoS of this region is determined by experimental atomic mass measurements. The pressure is given by that of a relativistic electron Fermi gas with corrections due to Coulomb interactions, exchange forces and charge screening  \cite{Blaschke:2018mqw}. 
Furthermore, modeling ions as finite-size particles instead of point-like has been shown via microscopic simulations to affect both the EoS and structural properties such as ion crystal formation. Both at low- and warm-temperature regions of the phase diagram, the binding energy of the ionic plasma is robustly reduced. Among other effects, 
this shifts the solid-liquid transition line to lower temperatures or higher densities, especially near or at inner crust densities. \cite{Barba-Gonzalez:2022pkn,Barba-Gonzalez:2023lln}.

Although the outer crust of a neutron star represents a small fraction of the stellar mass, it can be dynamically torn away by tidal forces and pressure during the collision of two neutron stars, or a neutron star and a black hole. There are two main approaches to describe the distribution of nucleons: either as a uniform interacting system or confined within discrete cells of specific shape and size. The common adopted method in the latter case is the Wigner-Seitz (WS) approximation, which models matter as being divided into charge-neutral cells. At lower densities, these cells are typically spherical and centered on a positively charged nucleus, immersed in a uniform background of electrons and, in some cases, free neutrons.
The crust is approached relying on phenomenological models. The classical way to determine the EoS is to use the so-called Baym-Pethick-Sutherland (BPS) model \cite{Baym:1971ax}, where the outer crust is supposed to be made of fully ionised atoms arranged in a body-centered cubic lattice at $T =$ 0  containing homogeneous crystalline structures made of one type of nuclides and coexisting with a degenerate electron gas. The EoS in each
layer of pressure $P$ is found by minimizing the Gibbs free energy per nucleon and the only microscopic
input is nuclear masses. Going deeper into the star, the inhomogeneous structures of matter are usually described using nuclear energy density functional \cite{Duguet:2013dga}. 

In recent works \cite{Chamel:2020hni}, the outer crust of a cold non accreting neutron star has been generally assumed to be stratified into different layers, each of which consists of a pure body-centered cubic ionic crystal in a charge
compensating background of highly degenerate electrons. Instead of performing the complete minimization of the Gibbs free energy per nucleon, they perform an  iterative minimization of the pressures between adjacent crustal layers, by using  very accurate analytical formulas for the pressure and baryon chemical potential at the interface between adjacent layers, with the  nuclear masses as the only microscopic inputs.
The accretion of matter onto a neutron star can radically change the internal constitution of the inner crust. The role of nuclear shell effects on the properties of accreted crusts has been studied in the framework of the nuclear energy density functional theory using the Extended Thomas Fermi Strutinsky Integral (EFTSI) method \cite{Blaschke:2018mqw}.

\subsubsection{Innercrust}
As density increases, nuclei become more neutron rich due to electron capture until neutrons start to drip out of nuclei at a threshold  density ($n_{\text{drip}}$), called neutron-drip transition, which marks the starting of the inner crust. The crust is supposed to be dissolved into a homogenous liquid at about half the saturation density $n_s$. This transition to the inner crust is found to occur at $n= 2.6 \times 10^{-4}$ fm$^{-3}$, although this density can be shifted due to the accretion of matter from a companion star. 
Such matter cannot be reproduced in experimental laboratories and requires theory. Some models rely on self-consistent mean-field (or Hartree-Fock) methods. Among those, we mention the Negele-Vautherin  \cite{Negele:1971vb} EoS, whose main drawback is that they are computationally very expensive. Thus, the pressure is generally not calculated directly, but it has to be evaluated numerically a posteriori by using a fit of the total energy per nucleon.  This fit may introduce systematic errors in the calculation of the global structure and in the dynamical evolution of NS. Moreover, this fit could induce loss of precision in the calculation of the adiabatic index \cite{Chamel:2015oqa}.

The crustal pressure is an essential ingredient entering the Tolman-Oppenheimer-Volkoff (TOV)
equations because of its implications for astrophysical phenomena such as pulsar glitches (sudden change in pulsar period)  \cite{Piekarewicz:2014lba}. The pressure in the inner crust is provided by the free gas of electrons and by the interacting dripped neutrons, aside from a correction from Coulomb exchange. 
 
A very popular model to describe the inner crust was first introduced by Baym, Bethe and Pethick
(BBP) \cite{Baym:1971ax}, based on the compressible liquid drop model (CLDM) to take into account the effect of the dripped neutrons. CLDM models parameterize the energy of the system in terms of global properties such as volume, asymmetry, surface, and Coulomb energy; their parameters are fitted phenomenologically. The main challenge here is that nucleons inside neutron-proton clusters and free neutrons outside cannot be treated separately, and be assumed to be uniformly distributed. Clusters have a sharp surface, and quantum-shell effects are neglected. 
There are currently two main groups to obtain the EoS of the inner crust: the first one uses phenomenological Skyrme interaction and the second one Chiral EFT Hamiltonians \cite{Grams:2022lci}. 

A partially phenomenological approach, based on the CLDM, was developed by \cite{Lattimer:1991ib}, who derived the EoS from a Skyrme nuclear effective force. Another EoS was developed by \cite{Shen:1998by} based on a nuclear Relativistic Mean Field (RMF) model. The crust was described in the Thomas-Fermi scheme using the variational method with trial profiles for the nucleon densities. Another common inner crust model in the CLDM approach was made by \cite{Douchin:2001sv}, formulating a unified EoS for NS on the basis of the SLy4 Skyrme nuclear effective force.  

Afterwards, the Brussels-Montreal group has derived unified EoS for NS \cite{Chamel:2008aa,Goriely:2009zzb,Goriely:2010bm,Goriely:2013xba,Goriely:2013nxa}, based on the BSk family of Skyrme nuclear effective forces. Each force
is fitted to the known masses of nuclei and adjusted among other constraints to reproduce a different
microscopic EoS of neutron matter with different stiffness at high density. The inner crust is treated
in the extended Thomas-Fermi approach with trial nucleon density profiles including perturbative shell corrections for protons via the Strutinsky integral method.
Skyrme forces are adjusted to the ground-state properties of finite nuclei, and therefore their predictions reproduce well SNM around $n_s$. However, their predictions in PNM differ largely and they represent the actual uncertainty for NM EoS based on phenomenological approaches \cite{Grams:2022lci}.

Another unified EoS from the outer crust to the core have been proposed by \cite{Sharma:2015bna} (BCPM), based on modern microscopic calculations using the Argonne V18 potential plus three body forces computed with the Urbana model. To deal with the inhomogeneous structures of matter in the NS crust, they used a nuclear energy density
functional that is directly based on the same microscopic calculations, and which is able to reproduce
the ground-state properties of nuclei along the periodic table. The EoS of the outer crust requires the
masses of neutron-rich nuclei, which are obtained through Hartree-Fock-Bogoliubov calculations with
the new functional when they are unknown experimentally. To compute the inner crust, Thomas-Fermi
calculations in Wigner-Seitz cells are performed with the same functional. We notice that the DH,
BSk and BCPM approaches are the only ones which construct a unified theory able to describe on a
microscopic level the complete structure of NS, from the outer crust to the inner core within the same
theoretical approach.
On the other hand, Chiral EFT Hamiltonians ($\chi$EFT) have been used for studying nuclear matter within various theoretical frameworks collected in the previous section, taking as experimental constraints the $NN$ scattering properties in vacuum complemented with the binding energy in the deuteron and $^3$He. In these approaches, $\chi$EFT band is much narrower than the dispersion among the Skyrme models \cite{Grams:2022lci}, because $\chi$EFT theory is well-suited to describe low-density NM, which is directly constrained by the nucleon-nucleon phase shifts and three-body forces.

At even higher densities, the extremely neutron-rich nuclei undergo a number of transitions, manifesting in
peculiar shapes and deformations in phases known as nuclear pasta \cite{Watanabe:2000rj}. In these phases, nuclei are no longer separated; instead protons and neutrons of neighboring nuclei organize themselves in different ��pasta�� geometries, such as spaghetti (rods) and lasagna (slabs). Eventually, the system becomes more and more neutron rich until the homogeneous matter phase, the outer core, is reached.

In summary, in the outer crust, the Skyrme models reproduce better the experimental nuclear binding in SNM and lead to tighter predictions than $\chi$EFT. However, in the inner crust, since a large amount of neutrons drip off clusters forming a neutron fluid, some properties such as the binding energy, the pressure, the speed of sound and the volume fraction are largely impacted by the properties of uniform neutron matter (NM) \cite{Grams:2022lci}. The experimental nuclear masses are, therefore, important constraints to accurately predict the EoS in NS outer crust, being essential to locate the neutron drip line.

\subsubsection{Core}
At about half the saturation density $n=n_s/2 = 0.08 $ fm$^3$, that is, $\rho \approx 1.8\times 10^{14}$ g/cm$^3$, $\varepsilon \approx 70$ MeV fm$^{-3}$ the nuclei completely melt and the NS core could begin. 
This region is much less understood than the crust and it is divided into outer and inner parts, as in the crust. 
The  outer core of NS is  supposedly composed of uniform matter at beta-equilibrium, whose properties are  strongly related to the symmetry energy originating from the energy difference between PNM and SNM.

At even higher densities $n\gtrsim 2n_s$, the composition of matter is basically unknown. In this inner core of the neutron star, many scenarios for the state of matter have been suggested, for example the formation of hyperons \cite{Lonardoni:2014bwa}, or quarks \cite{Alford:2007xm} or other more exotic condensates have been considered, such as kaons and their condensates, nuclear resonances, quarks, strange quark matter etc. \cite{Glendenning:1997wn}.
Although up to that point nuclear theory seems a reasonable guide, important features of the hadronic liquid interior remain unsolved. Different methods have been used to solve the EoS of the core:
 ab-initio many-body methods, based on first principles  \cite{Akmal:1998cf,Katayama:2015dga,Rijken:2016uon,Lonardoni:2014bwa} and phenomenological approaches, based on effective density-dependent interactions with parameters adjusted to reproduce nuclear observables and compact star properties \cite{Glendenning:1991es,Douchin:2001sv,Gaitanos:2003zg,Bonanno:2011ch}.

Ab-initio many-body methods are based on the underlying theory of strong interactions, Quantum Chromodynamics. However, in practice, calculations are only possible in limiting cases at low and high densities. At low densities ($n<2n_s$), the interactions can be described with effective theories of QCD (such as $\chi$EFT). Beyond these values of $n$ our knowledge come from astrophysical observations. However, at very high densities ($n\sim$ 40 $n_s$), QCD becomes asymptotically free and a phase transition to deconfined quark matter occurs. This is the perturbative pQCD regime, which can be used to impose robust constraints on the EoS of neutron stars. However, first-principle theoretical calculations at intermediate densities between $\chi$EFT and
pQCD limit are unavailable. Therefore, it is necessary to model the EoS in the density range between both regimes. A way to do that is the model-agnostic generation of EoSs, which are the EoS we will discuss in this review.



Phenomenological approaches commonly describe the structure of a NS by Unified EoS, which uses the same model to describe the crust and the core and, therefore, are model-dependent.
Other approaches (such as ab-initio models) need to use crustal EoS from phenomenological approaches to match to core EoS. However, using different EoS to connect crust and core could be not thermodynamically consistent, causing uncertainties in the calculation of some macrophysical parameters, such the radius, tidal deformability and moment of inertia. For instance, the resulting mass-radius diagrams, obtained  from the integration of the Tolman-Oppenheimer-Volkoff equations, depends on the procedure of matching the crust and core EoS segments. 
Ref.~\cite{Suleiman:2021hre} assesses the influence of the matching between an EoS for the core and another for the crust on the determination of radius, tidal deformability and moment of inertia.
This study concludes that a jump in the chemical potential at the core-crust interface should be as small as possible when employing non-unified EoS. This can be achieved by gluing core and crust at a density in the range of 0.08 to 0.1 fm$^{-3}$, due to the fact that nuclear models are adjusted to reproduce the results of laboratory experiments which constrain the property of matter up to roughly half the nuclear saturation density.
Matching the core and crust EoS at another density can create a relative difference with respect to the unified EoS as large as 5$\%$ for the radius, 20$\%$ for the deformability, and 10$\%$ for the moment of inertia.\\

\subsubsection{Static observables}

\textbf{Mass, Radius and TOV equations}
\\
\\
In order to examine the relation between the dense-matter EoS and the mass-radius relation of NS, we need the hydrostatic equilibrium relations in general relativity (GR). The basic equations for hydrostatic equilibrium were laid out in 1939 by Tolman, Oppenheimer and by Volkoff (TOV) as a generalization of the Newton hydrostatic equation from the point of view of the General Relativity (GR). This equation is a solution of Einstein'��s equations for a spherically symmetric, static body, with the approximation that an isolated star can be modeled as a perfect fluid. Thus,in equilibrium, the pressure needs to compensate for the weight of the upper layers and the increase in pressure is (using $c = 1$)
\begin{eqnarray}
\label{basicTOV}
   \frac{dP}{dr} &=& - \frac{G_N}{r^2}
\frac{(\varepsilon(r)+P(r))(M(r)+4\pi r^3P(r))}{1-\frac{2G_NM(r)}{r}}
    \\   
\color{black}\frac{dM}{dr} &=& 4\pi r^2 \color{black} \varepsilon(r) \ .
\end{eqnarray}
 The relativistic extension includes the Schwarzschild factor of the metric, weighs the energy density $\epsilon$, and includes the gravitational strength caused by the pressure $P(r)$.

The initial condition for these equations are $P(r = 0) = P_{central}$, $M(r = 0) = 0$. 
The only input required to solve TOV equation is the equation of state $P(\varepsilon)$. The solution of the TOV equation provides a sequence of neutron star masses $M$ and radii $R$ as a function of central pressure or density, which are plotted in the traditional M-R diagram. Each EoS generates a unique M-R diagram, characterized by a maximum mass ($M_{\text{TOV}}$). 

The stability condition for neutron  stars is determined by the sign of $dM/dn_c$, where $\varepsilon_c$ is the central energy density:
\begin{itemize}
    \item Stable branch: $dM/d\varepsilon_c > 0$
    \item Unstable branch: $dM/d\varepsilon_c < 0$ (leading to gravitational collapse into a black hole).
    \item $dM/d\varepsilon_c$ = 0, which marks the maximum stable mass $M_{\text{TOV}}$.    
\end{itemize}
\textbf{Tidal deformability}

From gravitational waves (GW) data can be used to extract information about neutron-star matter. The early inspiral phase of two coalescing neutron stars is affected by the
internal structure of NSs. To linear order, this impact can be characterized by a single
parameter �� the tidal deformability $\lambda$, which is defined as the ratio of the induced quadrupole moment of the star to the perturbing tidal field that causes the perturbation, i.e., how much the star deforms due to the external gravitational field of the companion.

A related quantity that is commonly also used is the dimensionless tidal deformability defined as 
\begin{equation}
\Lambda \equiv \frac{\lambda}{m^5} = \frac{2}{3} k_2\frac{R^5}{m^5} = \frac{2}{3} k_2 C^{-5}\label{Lambdadim},
\end{equation}
where $m$ is the mass of the star, $R$ being its radius, $C=m/R$ is its compactness. $k_2$ is the gravitational Love number with typical values around $0.2-0.3$ for different EoS. Thus, the information about neutron matter is then carried out by two dependences in compactness $C$ and in the second Love number $k_2$.
The second love number can be computed as \cite{Hinderer:2009ca}:
\begin{equation}
    \begin{split}
    k_2 &= \frac{8C^5}{5}(1-2C)^2[2+2C(y_R-1)-y_R]\times \\
    &[2C(6-3y_R+3C(5y_R-8))]+ \\ 
    & 4C^3(13-11y_R+C(3y_R-2)+2C^2(1+y_R))+\\
    & 3(1-2C)^2(2-y_R+2C(y_R-1))\log(1-2C)]^{-1}
    \end{split}
\end{equation}
\label{Sec:Neutron_Stars_for_Searches_of_New_Physics}
where $y_R=y(R) $ is solved together with the TOV from the following differential equation \cite{Zacchi:2020dxl}:
\begin{equation}
\label{eq:tidal}
    r\frac{dy(r)}{dr}+y^2(r)+y(r)F(r)+r^2Q(r)=0
\end{equation}
with
\begin{align}
    F(r) &= \frac{r-4\pi r^3(\varepsilon+P)}{r-2M}\\
    Q(r) &= \frac{4\pi r}{r-2M} \times\bigg[5\varepsilon+9P+\frac{\varepsilon+P}{c_{s}^2}  - \frac{6}{4\pi r^2} \bigg] -4\bigg[\frac{M+4\pi r^3P}{r^2-\big(1-\frac{2M}{r}\big)} \bigg]^2
\end{align}
where $c_s^2=dP/d\varepsilon$ is the squared speed of sound.
However, in an NS-NS merger, \emph{both} stars are distorted, so what the GW signal actually allows is the extraction of the binary tidal polarizability parameter $\widetilde{\Lambda}$, defined as a mass-weighted average of the individual $\Lambda_{1,2}$:
\begin{equation} \label{joint:tidal}
\widetilde{\Lambda} =\frac{16}{13}\left[\frac{(m_1+12m_2)m_1^{4}\Lambda_1}{(m_1+m_2)^{5}} + \frac{(m_2+12m_1)m_2^{4}\Lambda_2}{(m_1+m_2)^{5}}\right] \ .
\end{equation}

An upper bound on the dimensionless binary tidal deformability parameter $\bar{\Lambda}_{1.186}\leq$ 720 (low-spin priors) has been obtained from GW170817 \cite{LIGOScientific:2018hze}. Subsequently,  $1.4\,M_\odot$ NS tidal deformability was deduced in \cite{Abbott:2018exr} $\Lambda_{1.4}= 190^{+390}_{-120}$.

\subsection{EoS from chiral EFT}
From about one half to about twice nuclear saturation density, matter in the core of neutron can be modeled by considering nuclear matter, i.e., an infinite system of matter formed only by neutrons and protons. The properties of nuclear matter are very difficult to extract directly from the data because this system cannot be realized in terrestrial experiments, although experiments can provide constraints.

Thus, in general, at the lower densities  some crust models are used, typically from  \cite{Douchin:2001sv,Baym:1971pw,Negele:1971vb}, up to the crust-core transition density predicted by these models ($n\approx0.08$~fm$^{-3}=0.5n_s$. In the density range $0.5n_s\leq n \leq 2 n_s$, chiral effective field  theory ($\chi$EFT) with pion and nucleon degrees of freedom has become the dominant microscopic approach to describing nuclear interactions \cite{Epelbaum:2008ga,Machleidt:2011zz,Hammer:2019poc}. The application of $\chi$EFT to the EoS of infinite nuclear matter and neutron star structure \cite{Tews:2012fj,Hebeler:2013nza,Carbone:2013rca,Coraggio:2014nva,Holt:2016pjb,Drischler:2016djf,Drischler:2017wtt,Lonardoni:2019ypg,Piarulli:2019pfq} has also allowed significant progress.

At very high densities $n \gtrsim$  40 n$_s$, well above the densities reached in NSs, ab-initio calculations in QCD  offer significant and nontrivial information about the equation of state of matter in the cores of neutron stars \cite{Kurkela:2009gj,Gorda:2021kme}, due to the fact that pQCD constraints propagate from asymptotically high densities to lower densities, resulting in the softening of the EoS at densities higher than $\varepsilon \geq $750 MeV/fm$^3$. These features have been interpreted as the onset of a quark matter phase in \cite{Annala:2019puf}. However, this effect is not noticeable in those calculations that do not consider these pQCD constraints \cite{Gorda:2022jvk}.  

To summarize, the EoS of dense matter is known when the density is either very low or very high, but not at the energy density range that is expected to be reached inside the NS, between 2 and 10 times the saturation density. For such high densities, parametrization or interpolations methods are used.

\subsubsection{Interpolation Methods}

The complete information about the EoS is available through the thermodynamic gran canonical potential $\Omega(\mu)=-p(\mu)$ (at zero temperature, finite chemical potential and  in $\beta$- equilibrium).The basic conditions required from any EoS  are:
\begin{enumerate}
    \item The stability  condition ($\Omega(\mu)$ must be a concave function):
     \begin{equation}
        \frac{\partial^2 \Omega (\mu)}{\partial\mu^2} \leq 0 \rightarrow \frac{\partial n(\mu)}{\partial\mu} \geq 0
    \end{equation}
    \item The causality condition  between $\mu$ and $n$:
    \begin{equation}\label{slopen}
        c_s^{-2}=\frac{\mu}{n}\frac{\partial n}{\partial \mu} \geq 1  
    \end{equation}
    This imposes a minimal slope on the number density $\partial n/\partial \mu \geq n/ \mu$.
    \item The consistency condition: The EoS must simultaneous connect $n$,$\mu$ and $p$  of the two limits
        \begin{equation}\label{integralconstraint}
 \int_{\mu_1}^{\mu_2} n(\mu) d\mu= P_2-P_1= \Delta P   
\end{equation}
\end{enumerate}
and, therefore, with $\varepsilon$, related to these three parameters by Euler equation
\begin{equation}\label{eq:Euler}
    \mu= \frac{\varepsilon+ P}{n} \ .
\end{equation}

\vspace{0.5cm}
\begin{itemize}
    \item \textbf{pQCD constraints}
\end{itemize}

 At zero temperature but at asymptotically high density, the QCD coupling $\alpha_s$  is small  and a perturbative description of the bulk thermodynamics, i.e. perturbative QCD, becomes valid. This occurs when the baryon chemical potential  $\mu$
significantly exceeds the QCD energy scale, i.e., $\mu \gg \Lambda_{\text{QCD}}$.  The equation of state of dense deconfined quark matter (QM) can be  then determined in terms of a perturbative series in the strong coupling constant $\alpha_s$. 
The relevant results of pQCD calculations are constrained to zero temperature
matter composed of three flavors of massless quarks in beta-equilibrium. The details of the calculation of the QCD grand
canonical potential are shown in \cite{Kurkela:2009gj,Gorda:2021znl} for N$^2$LO and partial  N$^3$LO, respectively.

The resulting pQCD pressure depends on the chemical potential $\mu$ and the renormalization scale $\bar{\Lambda}$, which is related to a dimensionless parameter $X=3\bar{\Lambda}/2\mu$. The conventional approach for estimating theoretical uncertainties is $X \in$ [1/2,2]. A conventional choice for the lowest
chemical potential is set at $\mu \leq \mu_{\text{pQCD}}$ = 2.6 GeV.
To utilize the consistency requirement, it is necessary to determine the absolute maximum and minimum pressure differences between limits that an EoS can have when passing through a fixed point \cite{Komoltsev:2021jzg}:

\begin{equation}
\int_{\mu_{\rm L}}^{\mu_{\rm H}} n(\mu) d\mu =p_{\rm H}-p_{\rm L}\ .
\end{equation}

where the subscript $H$ denotes high-density limit entering pQCD and  $L$ denotes low-density limit at the end of chiral band. All possible interpolations of the full thermodynamic potential at zero temperature and in $\beta$-equilibrium, $\Omega (\mu) = -p(\mu)$ are then considered, between the low-density limit  of the chemical potential $\mu_L$ (corresponding to the end point of the chiral band, e.g., $n=1.3n_s$)  and the high-density limit  $\mu_H$ entering pQCD.

\begin{figure}
    \centering
    \includegraphics[width=0.39\linewidth]{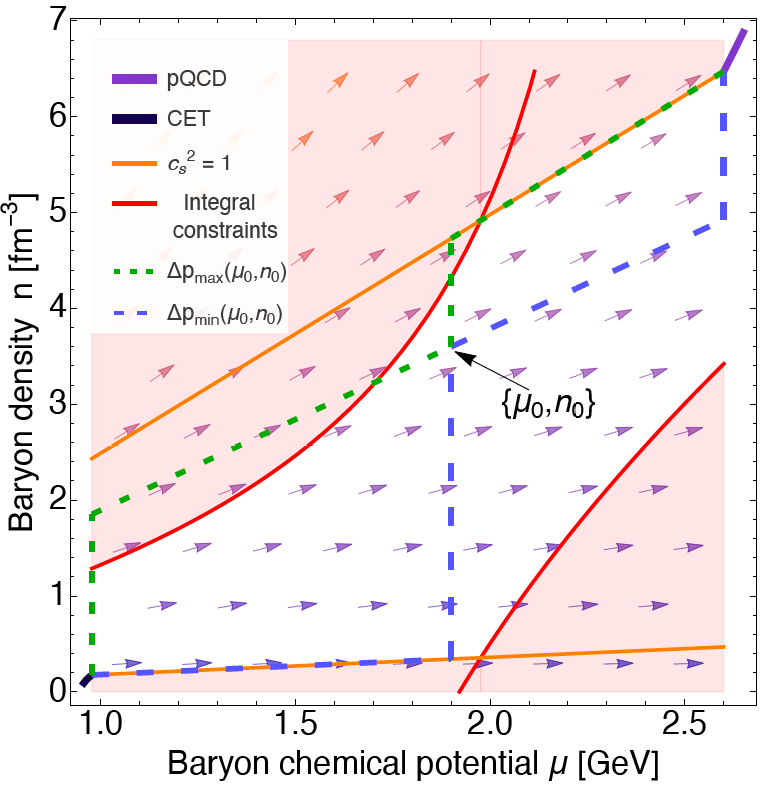}
        \includegraphics[width=0.59\linewidth]{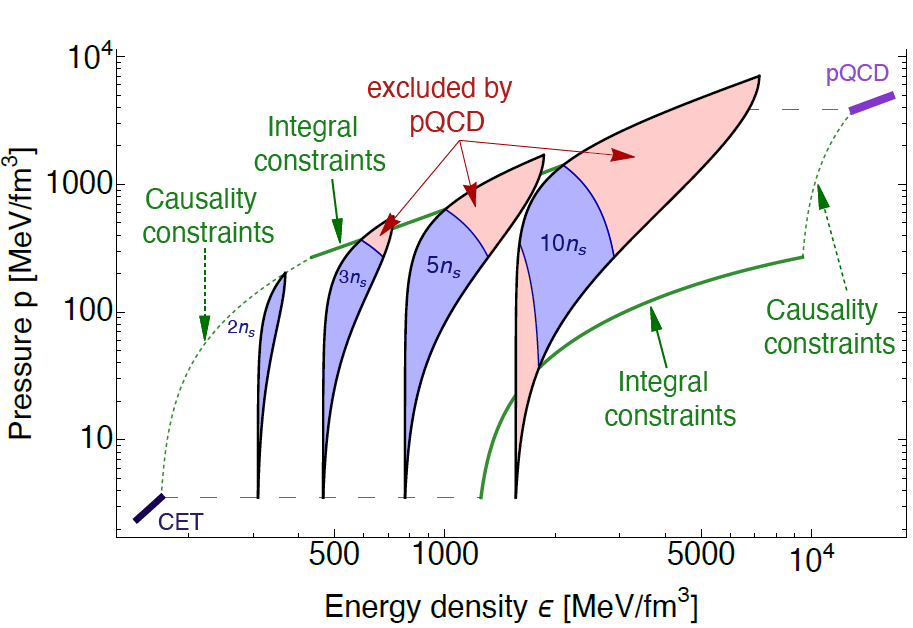}
 
    \caption{{\bf Left:} Baryon number density as a function of baryon chemical potential. Arrows indicate the minimal allowed slope dictated by causality, $\partial n (\mu)/\partial \mu \geq n/\mu$. The red
regions are excluded due to the simultaneous requirements of stability, causality, and
consistency. The constructions of $\Delta p_{min/max}$ for an arbitrary point $\mu_0$, n$_0$ are defined in \cite{Komoltsev:2021jzg}. {\bf Right:} The pQCD constraints in the $\varepsilon-P$ plane. The green envelope corresponds to the causality and integral constraints from the left figure. The black-outlined shapes represent the allowed $\varepsilon-P$ region when extrapolating a causal EoS from the low-density limit up to fixed densities n$_s=$ 2, 3, 5, and 10n$_s$. The blue regions correspond to the allowed areas where, in addition to stability and causality, consistency
with the high-density limit is imposed. The red regions are explicitly excluded by the pQCD limit. \emph{Reprinted  from \cite{Komoltsev:2021jzg} with permission of the American Physical Society http://dx.doi.org/10.1103/PhysRevLett.128.202701.} }
    \label{fig:IntegralconstraintsKomoltsev}
\end{figure}

Global constraints are derived taken into account these three requirements. Applying first stability and causality,the minimal slope $\partial n/\partial \mu \geq n/\mu$ can be visualized as a vector field on the $n �� \mu$ plane (by arrows at the left of Fig. \ref{fig:IntegralconstraintsKomoltsev}).
An important parameter is the so-called critical chemical potential, given by the intercept of the causal line and the integral constraint (orange line  an red line, respectively at the left of Fig. \ref{fig:IntegralconstraintsKomoltsev}):
\begin{equation}
\label{eq:muc}
\mu_c = \sqrt{\frac{\mu_L \mu_H ( \mu_H n_H-  \mu_L n_L-2 \Delta p )}{\mu_L n_H - \mu_H n_L} }.
\end{equation}
This parameter delimits the maximum and minimum values of $\varepsilon$ and $p$ at $c_s^2=$1 in the plane $\varepsilon- p$ shown at the right of Fig. \ref{fig:IntegralconstraintsKomoltsev}.

Therefore, if the maximum pressure difference $\Delta p_{max}$ for any EoS passing through a fixed point is smaller than $\Delta p = p_H �� p_L$ (fixed by low and high-density limits), such a point would be ruled out, as no causal and stable EoS can simultaneously connect $\mu$, $n$ and $p$. Similarly, if the minimum pressure difference $\Delta p_{min}$  is greater than $\Delta p$, the fixed point would be excluded. (For more details about all the expressions involved here see \cite{Komoltsev:2021jzg}).
This method propagate then the pQCD constraints to lower densities using only the
thermodynamic potential and the conditions of causality and mechanical stability. In practice, the method provides a necessary condition for an extrapolated EOS that
has to be fulfilled at all densities below the validity range of pQCD calculations.
Given all these limits, the next step is to construct the EoS from low to high densities. An ideal starting point for the EoS of neutron-star matter  at low densities is the EoS of pure neutron matter (PNM), which can be calculated using sophisticated many-body methods once a nuclear Hamiltonian is specified. 

\begin{itemize}
    \item \textbf{Meta model}
\end{itemize}

A common approach to model the homogeneous matter in the range $1n_s\leq n\leq 2n_s$ is the metamodel (MM) introduced in \cite{Margueron:2017lup,Margueron:2017eqc}. This approach is inspired from a Taylor expansion around the saturation density of symmetric nuclear matter, and proposes a parameterization in terms of the empirical parameters.  The MM is able to reproduce the EoSs predicted by a large number
of nucleonic models that exist in the literature \cite{Margueron:2017lup,Margueron:2017eqc}
including those from involved microscopic calculations
such as in the framework of $\chi$EFT \cite{Somasundaram:2020chb}. The MM is then a density  functional approach, similar to the Skyrme model \cite{Chabanat:1997un} that allows one to directly incorporate nuclear-physics knowledge encoded in terms of nuclear empirical parameters (NEPs) \cite{Koehn:2024set}. The model is based on that neutron-matter EoS up to a transition $n_{\text{break}}$ which vary in the range  $n_s$-$2n_s$.

In neutron matter, it is very common to use an asymmetry parameter, isospin asymmetry ($\alpha$), that quantifies the separation from SNM, as $\alpha\equiv \frac{n_n-n_p}{n_n+n_p}$,  in terms of the number densities for protons and neutrons. 
The two boundaries $\alpha=$ 0 and 1 correspond to symmetric nuclear matter (SNM) and pure neutron matter (PNM).  The saturation of symmetric nuclear matter is one of the hallmarks of the nuclear dynamics. 
One of the main feature of the EoS is the nuclear symmetry energy as a function of density. The nuclear symmetry energy characterizes the variation of the  binding energy when the ratio of neutrons to protons in a nuclear system is varied  \cite{Baldo:2016jhp} and is characterized in terms of $\alpha$ by
\begin{equation}\label{symmetryenergy0}
S (n=n_n+n_p) = \frac{1}{2} \left.\frac{\partial^2 E/N(n,\alpha)}{\partial \alpha^2}\right\vert_{\alpha=0}
\end{equation}
that can be experimentally determined from the binding energy per nucleon $E/N(n,\alpha)$.
However, there is another commonly used definition, which is  the  quadratic isospin-asymmetry expansion from symmetric nuclear matter (SNM, $\alpha = 0$) to pure neutron matter (PNM, $\alpha= 1$):  

\begin{equation}\label{symmetryenergy1}
    E/N(n,\alpha)= E/N(n,0) +S(n)\alpha^2 + \mathcal{O}(\alpha^4)
\end{equation}
The quadratic expansion is a reasonable approximation. Some studies, such as \cite{Wellenhofer:2016lnl}, indicate that higher-order terms in the expansion are relatively small, contributing with less than $1$-��$2$ MeV around $n_s$.
Therefore, for pure neutron matter:
\begin{equation}\label{symmetryenergy2}
 S (n=n_n+n_p)= E/N(n,\alpha=1) - E/N(n,\alpha=0)   
\end{equation}
Both expressions of the symmetry energies \eqref{symmetryenergy0} and \eqref{symmetryenergy2} are equal if the isospin dependence of the energy per particle is exactly quadratic.
Furthermore, the energy per nucleon of symmetric matter $E/N(n,0)$ and the symmetry energy $S(n)$ of \eqref{symmetryenergy2}, can be  expanded in the deviation  of the baryon density $n$ from the saturation density $n_s$, were the main coefficient of the Taylor expansion are known properties of the nuclear matter, which can be constrained by nuclear experiments.
These coefficients are defined as follow
\cite{Baldo:2016jhp}: 

\begin{align}
    &S_0=S (n=n_s) \\
    &L=3n_s\left(\frac{dS}{dn}\right)_{n_s}\\
    &K_{\rm sym}=9n_s^2\left(\frac{d^2S}{dn^2}\right)_{n_s}\\
    &Q_{\rm sym}=27n_s^3 \left(\frac{d^3S}{dn^3}\right)_{n_s}
\end{align}

where $S_0$ is the symmetry energy at saturation, $L$ the symmetry-energy slope, $K_{\rm sym}$ the isovector incompressibility, and $Q_{\rm sym}$   the skewness. If we define, in addition, $x\equiv\frac{n-n_s}{3n_s}$ \cite{Piekarewicz:2008nh}, then the  symmetry energy is given as

\begin{align}
S(n)= S_0+ L\,x + \frac{1}{2}K_{\rm sym}x^2+\frac{1}{6}Q_{\rm sym}x^3+ ...
\end{align}

If we focus on the first two terms of this expansion, the first term ($S_0$) represents the correction to the binding energy of symmetric nuclear matter, whereas the second term ($L$) states how rapidly the symmetry energy increases with density. Since symmetric nuclear matter saturates, that is, its pressure vanishes at saturation, the slope of the symmetry energy $L$ is closely related to the pressure of pure neutron matter (PNM) at saturation density. These nuclear parameters are then related to the neutron matter as follows:
\begin{equation}
    S_0 \equiv  S(n_s) =  E/N(n_s,1) - \epsilon_0
\end{equation}
\begin{equation}
    L = 3\frac{P_{\rm{NM}}}{n_s} 
   \end{equation}

In the same way, expanding  the energy per nucleon of symmetric nuclear matter $E/N(n,0)$ close to the saturation density $n_s$
\begin{equation}\label{expandingenergy} 
E/N(n,0)=\epsilon_0 +\frac{1}{2}K_{\rm sat} x^2+\frac{1}{6}Q_{\rm sat} x^3+ ..
\end{equation}
where  $\epsilon_0$ is the binding energy at saturation: $E_{\rm sat} \equiv \epsilon_0= -16.0 \pm 1.0$ MeV at $n_s = (0.16 \pm  0.01)$ fm$^{-3}$, from fits to nuclear binding energies. The $K$  coefficient is the incompressibility of symmetric nuclear matter, which controls the rate of change for deviations from the saturation density. From $^{208} {\rm Pb}$ data, this coefficient has been determined to be $K_{\rm sat}\simeq (240\pm 20)$ MeV.
The set ($S_0,L, K_{\rm sym},Q_{\rm sym}$) contains the parameters needed for pure neutron matter, with much larger uncertainties than those for symmetric matter in Eq. (\ref{expandingenergy}). They can be extracted from theoretical calculations of the binding energy per nucleon by taking appropriate derivatives.

Additionally, one can compute the energy density of nuclear matter $\varepsilon$
\begin{equation}\label{eq:densityenergy}
  \varepsilon= n\left(M_N+\frac{E}{N}\right)  
\end{equation}
 and the pressure $P$ at $T=0$
\begin{equation}\label{eq:firstlaw}
  P= n^2\,\frac{d(E/N) }{d n}  
\end{equation}

Furthermore, from the grand potential ensemble the baryon chemical potential $\mu$ can be calculated using the Euler equation \ref{eq:Euler}.
 These parameters of the nuclear EoS can be experimentally constrained  from different measurements, such as production in Heavy Ion Collisions (HICs) or measurements of the neutron skin thickness in heavy nuclei.
The slope $L$ can eventually be extracted from laboratory data of the the neutron-skin thickness \cite{Millerson:2019jkg}, so it is quite directly accessible to laboratory experiments such as the Lead Radius Experiments (PREX).


\begin{itemize}
    \item \textbf{Agnostic EoS models}
\end{itemize}

There are different ways to extend the new results of EFT to higher energies, the most popular being the so-called agnostic models. Agnostic EoS models can either be non-parametric, e.g., based on Gaussian processes, or parametric, e.g., based on a piece-wise polytrope or spectral representation,
or a parametrization of the speed of sound. All of them are characterized for constraining the EoS by imposing causality and mechanical stability. 
Some of these models take into account constraints on the EoS at NS densities arising from  calculations. Additionally, agnostic models constraint the EoSs taking into account astrophysical observations, and allow exploring the relationship between EoS properties and global NS properties, including  mass, radius, moment of inertia, tidal deformability, compactness, and Kepler frequency or maximum possible frequency of a rigidly rotating star.\\

\textit{Polytropic representation}\\

The neutron star model is performed  with the polytropic equation of state $P(\rho) = \kappa \rho^\Gamma$, with mass density $\rho = mn$,  where $\Gamma$ is the adiabatic index and $\kappa$ is known as a polytropic index. A first implementation of the polytropic representation was made by \cite{Hebeler:2013nza} and  also used in \cite{Kurkela:2014vha,Raithel:2016bux}. In this approach, to extend the EoS beyond  $\chi$EFT results at $n=1.1 n_s$, a general polytropic extension is used, usually with three segments, where the pressure of neutron star matter is piecewise given by $P(\rho) = \kappa \rho^\Gamma$, with mass density $\rho = mn$. The polytropic indices and the transition densities between the individual segments are
then varied to sample many possible EoS curves. A typical approach is to use use three polytropes with exponents $\Gamma_1$, $\Gamma_2$ and $\Gamma_3$, which make it possible to vary the softness or stiffness of the EoS in three regions of increasing density: a) $n_1\leq n \leq n_{12}$; b) $n_{12}\leq n \leq n_{23}$  and c) $n\geq n_{23}$. An example implementation is shown in Fig.~\ref{Polytropic hebeler} from \cite{Hebeler:2013nza}.
Thus, changing transition densities from $n_1=1.1 n_s$ the polytrope parameters over the ranges $1\leq \Gamma_1 \leq 4.5$,  $1.5 n_s\leq n \leq 8 n_s$, $0\leq \Gamma_2 \leq 8$, $n_{12}\leq n_{23}\leq 8 n_s$, and $0.5 \Gamma_3 \leq 8$. At densities  below $n \approx 0.5 n_s$, the $\chi$EFT band is connected to the Baym-Pethick-Sutherland (BPS) crust EoS \cite{Baym:1971pw}.
The main drawbacks of this approach are the introduction of discontinuities in the sound velocity and the lack of thermodynamic consistency, since it is only used the relation between pressure and energy density.

\vspace{0.5cm}
\textit{Speed-of-sound models}\\

One essential disadvantage of the previous models is the difficult to include phase transitions, as well as the discontinuities in the sound velocity introduced by polytropic representations.
To overcome these difficulties, another parametrization for the high-density EoS has been developed  in terms of the speed of sound $c_s$ (CS model) \cite{Greif:2018njt,Tews:2018kmu}. The speed of sound $c_s^2=\frac{\partial P}{\partial\varepsilon}$ can be obtained here in different ways. In this case, the functional form of such a parametrization is obtained taking into account constraints on the speed of sound. Causality imposes the constraint that $c_s$ cannot exceed the speed of light, mechanical stability imposes a non negative speed of sound, and at a sufficiently high density $c_s^2 \rightarrow 1/3$ since quarks at high momenta are asymptotically free. In the regime where the EoS is dominated by neutron matter $c_s^2$ grows with the density and, at higher densities, $c_s$
can have different possible behaviours, one exceeding the limit of 1/3 at intermediate densities and another one where this limit is valid, providing thus smoothness like the Gaussian parametrization. 

A second speed-of-sound model (CSM) to extend the EoS at higher densities  beyond MM was developed  by \cite{Tews:2018iwm}. The CSM constructs the EoS from a general parametrization of the speed of sound, 
with pressure $P$ and energy density $\varepsilon$, including phase transitions. Here they extend the parametrization to explore the full space for $c_s^2(n)$ by randomly sampling six reference points
 between $n_{\text{break}}$ and $12 n_s$  (or up to $25 n_s$ with 9 grid points in \cite{Koehn:2024set}) and connecting them by linear segments. 

\begin{figure} 
\centering
\includegraphics[width=5.127in]{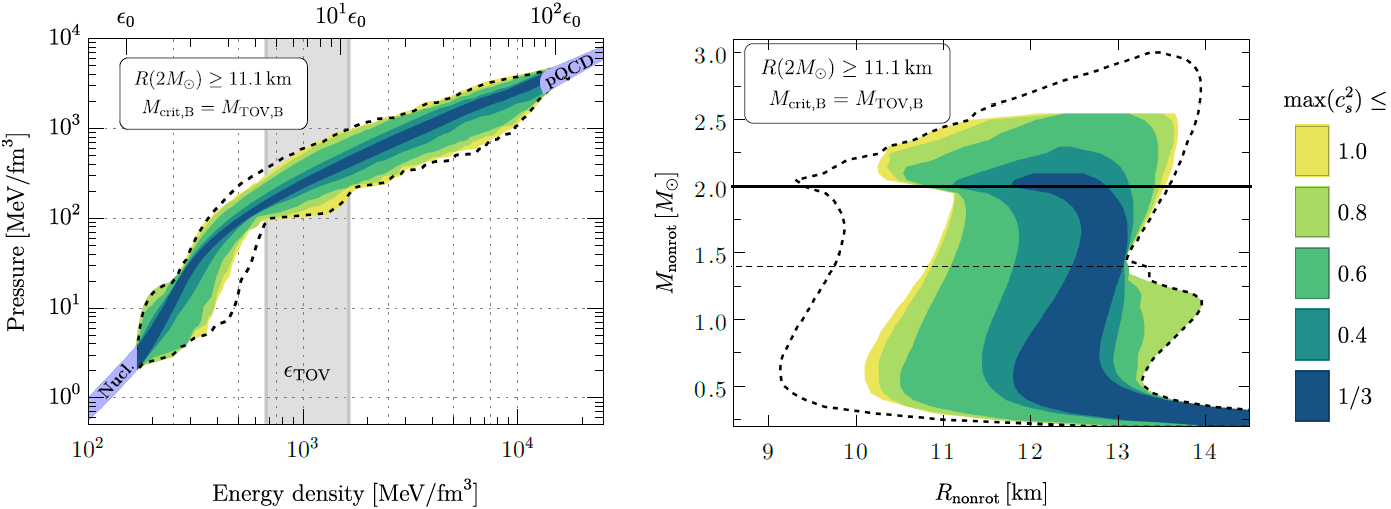}
 \caption{An example of the EoS and MR results from the analysis of \cite{Annala:2021gom}, where information from the NS radius constraints from the NICER collaboration and the
likely delayed gravitational collapse of the GW170817 merger remnant were taken
into account, assuming here the more conservative supramassive scenario. The
color coding refers to the maximal value the speed of sound squared attains at
any density, made feasible by the $c^2_s$ interpolation routine. \emph{Reprinted, with permission, from~\cite{Annala:2021gom} under Creative Commons License.}\vspace{0.5cm}\label{fig:counting} 
 \label{fig:Annalaband}
}
\end{figure}

A third speed-of-sound method  to overcome difficulties with politropic representations was provided by \cite{Annala:2019puf}. It uses a sound speed interpolation as function of the chemical potential $\mu $  typically for $N=5$ (seven segments), starting at the end of chiral regime: 

\begin{equation}
  c_s^2(\mu)=\frac{(\mu_{i+1}-\mu)c_{s,i}^2-(\mu-\mu_i)c_{s,i+1}^2}{\mu_{i+1}-\mu_{i-1}} 
\end{equation}
 and $0\leq c_s^2\leq 1$, with  $\mu_1 = \mu_{\chi EFT}$ (chemical potential at the end of chiral regime), $\mu_N =$ 2.6 GeV (chemical potential entering pQCD), and $\mu_{i-1} <  \mu_i < \mu_{i+1}$  for all other $i$. The following step is to impose the following two robust astrophysical constraints on the EoS: the requirement of supporting a 1.97M$_\odot$ NS \cite{Demorest:2010bx,Antoniadis:2013pzd}, and that the tidal deformability $\Lambda$ for a 1.4M$_\odot$ star obeys $70 < \Lambda(1.4M_\odot) <$ 580 \cite{TheLIGOScientific:2017qsa,Abbott:2018exr,LIGOScientific:2018hze}.
In this approach, the Baym-Pethick-Sutherland (BPS) crust model \cite{Baym:1971pw} is used and extend it up to $n=1.1n_s$ with random polytropes bounded by the soft and stiff EoSs of \cite{Hebeler:2013nza}. A similar model is used by \cite{Altiparmak:2022bke,Ecker:2022dlg}.
The results in \cite{Annala:2021gom} are shown in Fig. \ref{fig:Annalaband}, which represent the state-of-the-art of hard-cut type interpolation analyses. 

\vspace{0.5cm}
\textit{Grid interpolation}\\
\\
Another proposal has been reported in \cite{Oter:2019rqp} where a set of $\beta$-equilibrated NS matter EoS  are generated based on first principles, Chiral Perturbation Theory ($\chi$EFT) and pQCD, by interpolating in this intermediate area. The goal is to obtain an ensemble of EoS as model independent as possible and without considering astrophysical constraints in order to be used for testing GR an theories beyond GR such as modified gravity. 
The procedure is based on the chiral results, extending the EoS from the uncertainty region provided by $\chi$EFT up to the pQCD band. At lowest number densities ($n\leq 0.05n_s$)  nuclear data directly constrains the crustal EoS \cite{Baym:1971pw}. Low densities ($0.05n_s \leq n < 2 n_s$) are reasonably taken care of by $\chi$EFT predictions. The EoS are obtained from the chiral $E/A$ values  as a function of $n$ by using the Eq. \eqref{eq:densityenergy} and \eqref{eq:firstlaw}. Additionally, the chemical potential is obtained from the Euler Eq. \eqref{eq:Euler}.

Reciprocally, for each point of the EoS with $P(\varepsilon)$ assumed to be known, the corresponding values of $n$ and $E/A$ are obtained by using a discretized version of Eqs.~\eqref{eq:densityenergy} and \eqref{eq:firstlaw}, with the use of dense enough partitions (see \cite{Alarcon:2024hlj} for more details). On the other hand, at very high densities, the partial N$^3$LO results from \cite{Gorda:2021znl} are used, at chemical potential $\mu=$ 2.6 GeV, with their corresponding uncertainty for $X \in$ [1/2,2].

Starting from the obtained values for ($\varepsilon,P,E/A, n,\mu$) the EoS is interpolated to higher densities using the procedure developed in \cite{LopeOter:2019pcq}.
With this method, the first step is to establish the allowed region based on the conditions of causality and monotonic behavior, in terms of the sound velocity squared $c^2_s = dP/d\varepsilon \in$ [0,1].  Inside this area, a grid of $1000 \times 1000$ candidate points ($\varepsilon, P$) is constructed, representing potential EoS for the neutron matter. For each grid point $\varepsilon$, a value of $P$ is selected by applying different criteria to the slope, such as random variation, constant or continuous growth, phase transitions, etc., but always fulfilling stability and causality requirements.

Since for each point of the EoS the values ($\varepsilon,P,E/A, n,\mu$) are known, the causality and thermodynamic consistency conditions are ensured by imposing the causality condition in the $\mu-n$ plane by Eq. \eqref{slopen}.      
In addition, in each interpolation, the bands related to the integral constraints \cite{Komoltsev:2021jzg} are displayed, corresponding to the maximum and minimum $P(\vep)$ values of the pQCD regime. These bands are built according to the procedure of Ref.~\cite{Komoltsev:2021jzg}  (see this reference for the derivation and explicit expressions of the limiting curves). The most limiting band corresponds to the lowest pressure value of the pQCD band. However, given the uncertainty band of pQCD between the minimum and maximum $P$, the sum of the points of both bands are considered as potential pressure points, provided that the requirements of stability and causality in the $\mu-n$ plane are met.
In this way, the EoS is constructed up to pQCD, far away from the central densities expected in the NS, but convenient in order to be consistent with the pQCD band.\\
\begin{figure}
     \centering
     \includegraphics[width=0.43\textwidth]
     {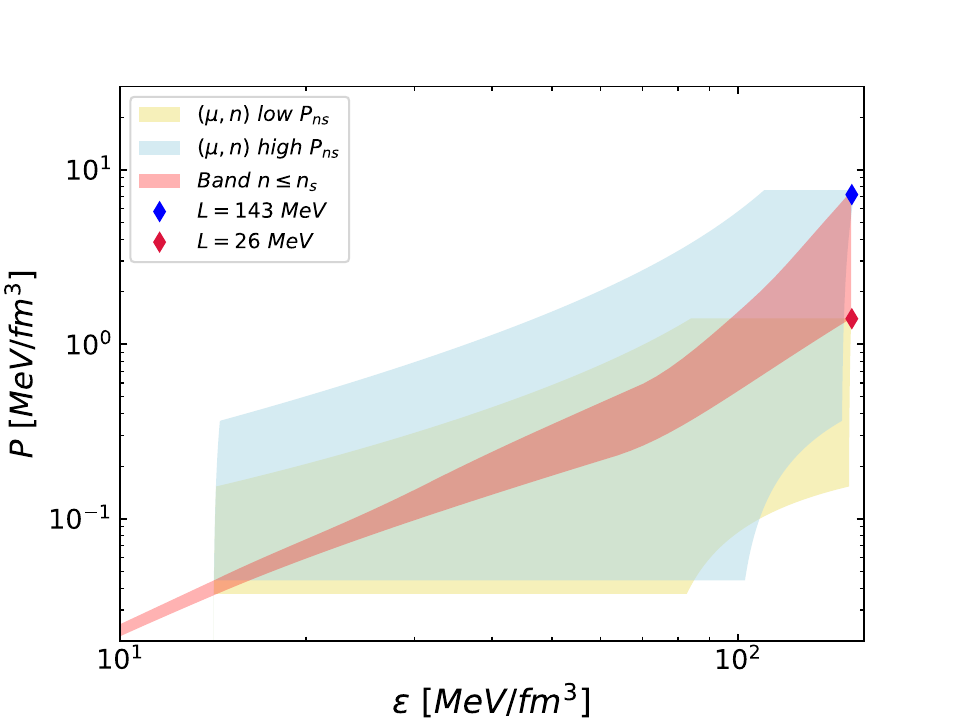}
     \includegraphics[width=0.48\textwidth]{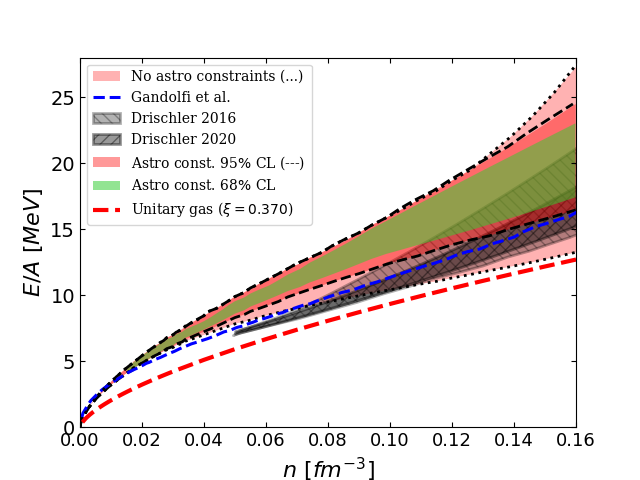}
 \caption{{\small {\bf Left:} Extrapolation from the upper limit of validity of the EoS \cite{Alarcon:2022vtn} up to $n_s$, $n_L<n<n_s$. Bounding areas of validity for PNM EoS by extrapolating the EoS of Ref.~\cite{Alarcon:2022vtn} (reddish band)  are shown from $n_L$ up to $n_s$. The blue and red diamonds correspond to the highest chemical potentials used in the construction of Ref.~\cite{Komoltsev:2021jzg} for $L=143~$MeV and 26~MeV (corresponding to the minimum EoS found compatible with the UGC, $L_{\text{UG}}\approx 25~$MeV ), respectively. They give rise to the palish blue and yellow areas, in this order. {\bf Right:}  Final band of  EoS's up to $n_s$ in three cases: The pink area corresponds to the direct extrapolation up to $n_s$, with its borders separated by dots. Once astrophysical constraints are taken into account the allowed area shrinks and it is depicted in green at 68\% confidence level (CL) and in orange at 95\% CL, with the borders for the latter signaled by dashed lines. We also compare with other results:  The blue dotted-dashed line is the Monte Carlo calculation of Ref.~\cite{Gandolfi:2011xu}. The many-body perturbation calculations from chiral potentials are given by the light  \cite{Drischler:2016djf} and darker \cite{Drischler:2020hwi}  gray areas. The dashed red one is the unitary limit for a Fermi gas.}
     \label{fig:LBandnolog} }
 \end{figure}

An example of this procedure can be seen in \cite{Alarcon:2024hlj}. In this work, interpolation from regulator-independent results have been developed, incorporating  an EoS for pure PNM at zero temperature and very low densities directly expressed in terms of the nucleon scattering data \cite{Alarcon:2022vtn}, i.e. phase shifts and mixing angles.
The results provided are renormalized and do not need to introduce any regulator, reducing the systematics.  
This EoS at low number density is interpolated up to known higher-density results from perturbative pQCD, following the grid methodology but applying here two steps.

i) A  first interpolation between the uncertainty band resulting for the aforementioned EoS \cite{Alarcon:2022vtn} for PNM  and  saturation density is performed, constrained  by nuclear experimental data regarding the values of the symmetry energy and its slope \cite{PREX:2021umo,CREX:2022kgg}. From PREX-II \cite{PREX:2021umo} Ref.~\cite{Reed:2021nqk} deduces the values $S_0= 38.1 \pm 4.7~{\text{MeV}}$, and $L = 106 \pm 37~{\text{MeV}}$. The maximum value for the symmetry slope is then $L=143$ MeV that fixes $P$ and $\mu$ at the upper band of this first interpolation. On the other hand, the minimum value is constraint by the Unitary Gas Conjecture, which states that $E/A \geq (E/A)_{\text{UG}}$  at all densities \cite{Tews:2016jhi,Lattimer:2023xjm} ($L_{\text{UG}}\approx 25~$MeV, $S_{\text{UG}}\approx 13~$MeV). The procedure is to generate first the EoS and filter them later, by ruling out all the EoS with a symmetry energy value  $S_0>42.8$ MeV (corresponding to the maximum value of $S_0$).
The band resulting from this first interpolation is shown on the left side of the Fig. \ref{fig:LBandnolog}.\\
ii) A second interpolation is used between this first band and the high-density pQCD regime \cite{Gorda:2021kme}. In this  second interpolation, phase transition are allowed only for $n\geq 2.5 n_s$. This constrain comes from the in-medium corrections of the quark condensate performed in \cite{Lacour:2010ci,Meissner:2001gz}. In addition, all the EoS must enter pQCD with $c_s^2=$ 1/3.

\begin{figure}
     \centering
     \includegraphics[width=0.45\textwidth]
     {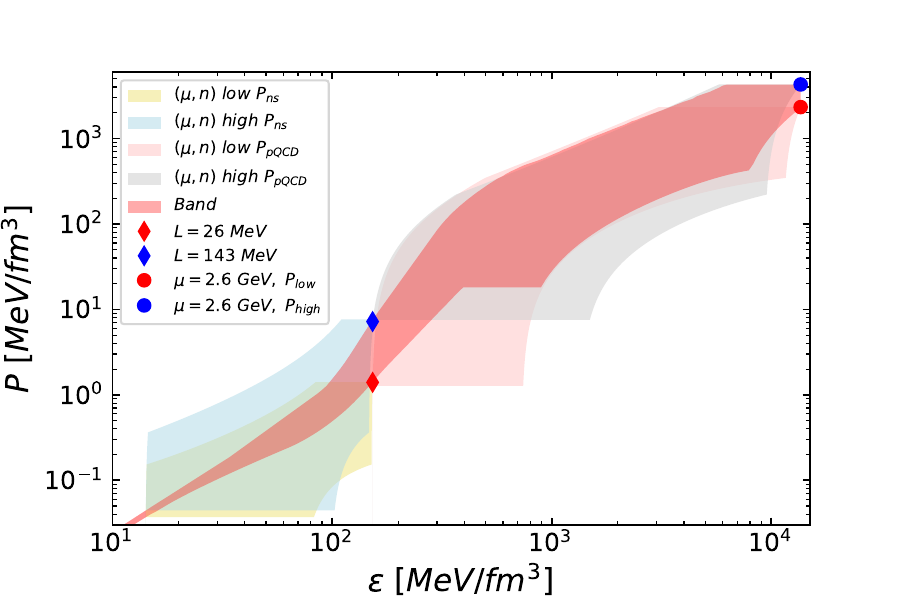}
   \includegraphics[width=0.45\textwidth]{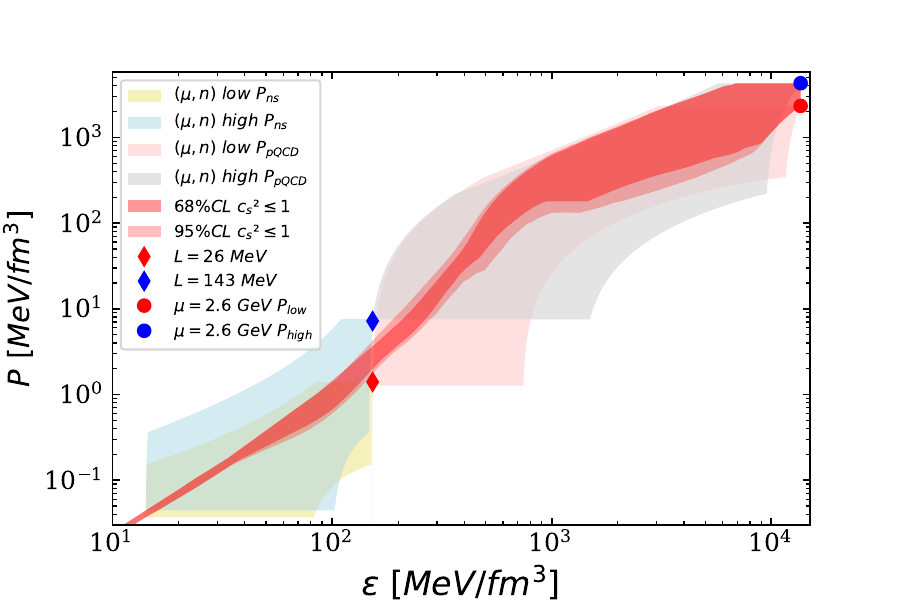}
     \includegraphics[width=0.482\textwidth]{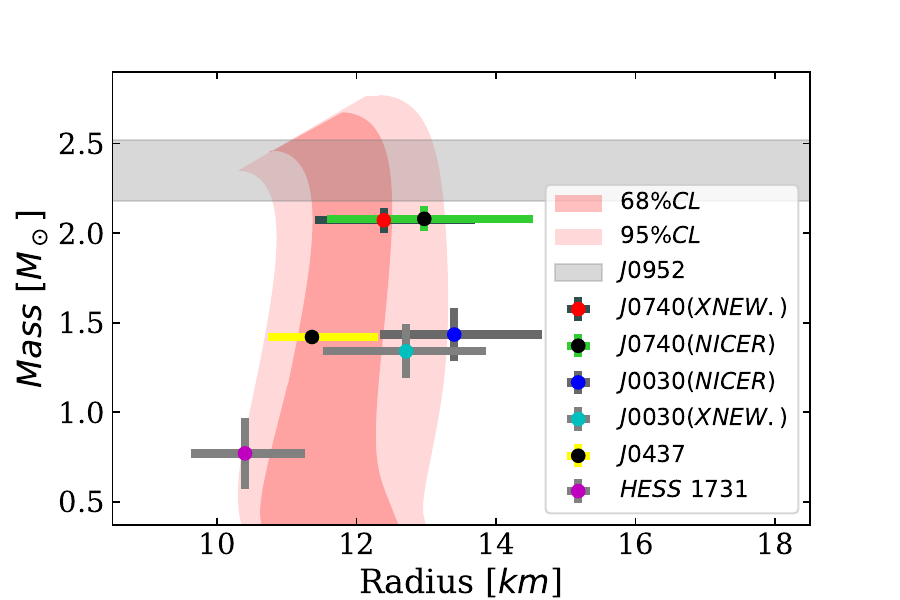}     
   \includegraphics[width=0.43\textwidth]{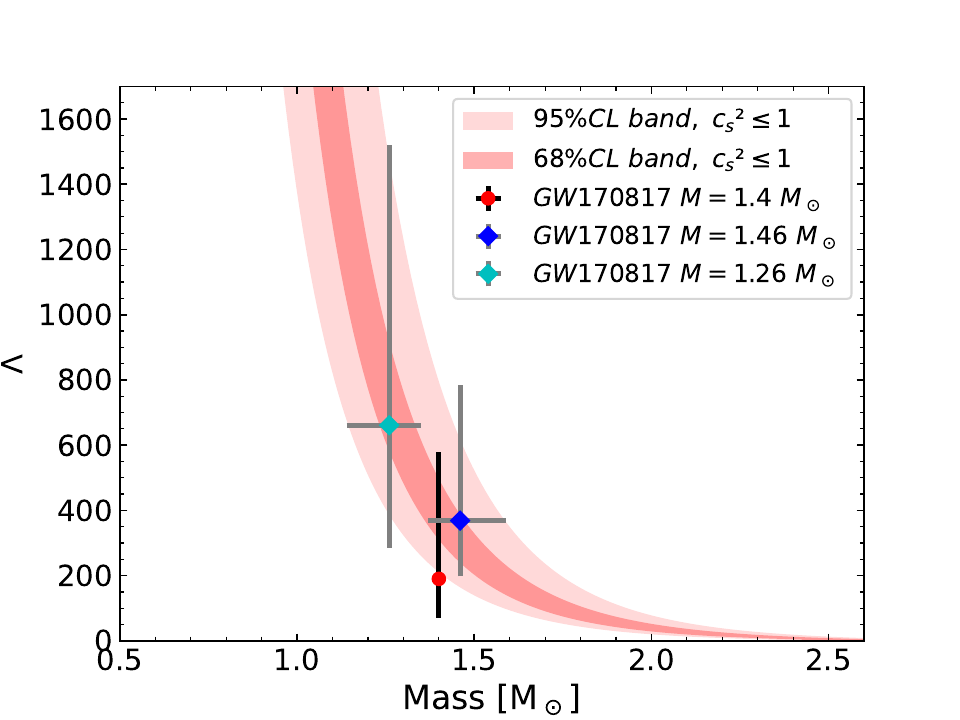}
\caption{{\small {\bf Top left:}   Band  obtained by interpolation from low to high densities, limited only by causality and thermodynamical consistency, pQCD and from measurements of nuclear parameters. {\bf Top right:} ($\varepsilon, p$) bands constrained  from astrophysical observables at 68$\%$ CL (red)  and 95$\%$ CL (pink) for $0\leq c_s^2 \leq 1$. {\bf Bottom left:}~ Mass-Radius diagrams for the EoS bands on the left. {\bf Bottom right:}~ Tidal deformability-Mass relation for the EoS bands on the left.}
\label{fig:totalband}}
 \end{figure}

Furthermore, in each interpolation, the bands related to the integral constraints \cite{Komoltsev:2021jzg} are displayed, corresponding to the maximum and minimum $P(\vep)$ values of the high matching band of interest, either at $n_s$ for the intermediate-density interpolation, or at the pQCD regime, for the higher-density part (see \cite{Alarcon:2022vtn} for more details).
In the second interpolation, the authors distinguish between two types of procedures, depending on whether we take into account astrophysical observables or not. In one case General Relativity (GR) is used to calculate mass and radii of NSs, while in the other one only experimental results from nuclear physics are accounted for. This is interesting  because one can then use  the resulting EoS in the latter case to test modified gravity theories \cite{LopeOter:2019pcq,Lope-Oter:2021vxl}. The total band resulting from the two interpolations without considering astrophysical constraints is shown at the top left of Fig. \ref{fig:totalband} (red area), taking into account possible phase transition (PT) for $n\geq 2.5 n_s$. The gray and pale pink areas display integral constraints from pQCD for the maximum pressure (blue circle) and the minimum pressure (red circle), respectively.

When taking into account astrophysical constraints, the method of least squares is used with six independent mass measurements $M_i$ at known radius plus the mass measurement of the most massive pulsar known. The astrophysical observables used are PSR J0952-0607 \cite{Romani:2022jhd} (M= 2.35 $\pm$ 0.17 M$_\odot$), HESS J1731-34 \cite{2022NatAs...6.1444D} [M=$0.77^{+0.20}_{-0.17}$ M$_\odot$, R=$10.4^{+0.86}_{-0.78}$ km], two independent radius measurements of  PSR J0030+045 \cite{Riley:2019yda} (M=($1.34^{+0.15}_{-0.16}$) M$_\odot$, R=$12.71^{+1.14}_{-1.19}$) km) and \cite{Miller:2019cac} [M=(1.4$\pm$ 0.05)M$_\odot$, R=$13.02^{+1.24}_{-1.06})$ km], two independent radius measurements of J0740+6620  \cite{Dittmann:2024mbo} [M=(2.08 $\pm$ 0.07) M$_\odot$, R=$12.92^{+2.09}_{-1.13}$ km] and \cite{Salmi:2024aum} [M=(2.073$\pm$ 0.069) M$_\odot$, R=$12.49^{+1.28}_{-0.88}$ km]
 and, finally, the recent PSR J0437-4715 \cite{Choudhury:2024xbk} [M=(1.418 $\pm$ 0.037) M$_\odot$, R=$11.36^{+0.95}_{-0.63}$ km].\\
 The results for the total band and M-R diagram and $\Lambda$-M diagram are shown at the top right, bottom left and bottom right, respectively.

All PTs that are found occur for $M\gtrsim 2.1\,M_\odot$ both at 68 and 95 $\%$ confidence level (CL), within the range of starting densities between $2.5 n_s \leq n \leq n_{c}$. Two kinds of PTs are found, one with short and the other with long phase-co-existence regions.  For $c_s^2 \leq 1$, short PTs appear for $2.5 n_s< n< 3.2n_s$ (68$\%$ CL) and they typically require afterwards a steep rise in $c_s^2$ to confront well with NS mass-radius data. Because of this they could also stand for a second PT that starts at densities less than or equal to the central density of the NS. The longest PTs found have a phase-coexistence region that extends over a range of densities $3.4 n_s\leq \Delta n \leq 19 n_s$ at 68$\%$ CL, such that  $1 \lesssim \Delta n/n\leq 3.8$ at 68$\%$ CL (for more details, see \cite{Alarcon:2024hlj}).
 The radius of the canonical 1.4 M$_\odot$ neutron star is found to be  $R_{1.4} \in$ [11.16,12.37] km ($68\%$ CL.) and $R_{1.4} \in$ [10.67,13.3] km ($95\%$ CL.). These values are slightly lower than those of other references, such as Refs.~\cite{Brandes:2023hma,Koehn:2024set}, due to the fact it is included the less massive known pulsar (HESS J1731-34) as well as the recently measured PSR J0437-4715 not included in those references. 
In addition, the values of the tidal deformability show very good agreement with the ones obtained from the GW170817 for three masses, 1.4 M$_\odot$ \cite{LIGOScientific:2018hze}, 1.26 and 1.36 M$_\odot$ \cite{Fasano:2019zwm}.\\
 The band of EoS's ($E/A$ versus $n$) is shown  at the right of Fig. \ref{fig:LBandnolog}, with the bands without considering astrophysical constraints (pink area), with astrophysical constraints at $68\%$ CL. (green area) and at $95\%$ CL. (red area) are compared with other calculations.
 Furthermore, astrophysical observables also constrain the nuclear observables, providing a more precise value for the energy symmetry and its slope,

\begin{align}
\label{240616.5}
&    32.9\leq S_0 \leq 39.5~\text{MeV}\,; ~  
    37.3 \leq L\leq 69.0~\text{MeV}~  (68\% \,\text{CL})~,\\
 &   32.1\leq S_0 \leq 40.6~\text{MeV}\,; ~34.6 \leq L\leq 80.0~\text{MeV}\  (95\% \ \text{CL})\,.\nn    
\end{align}

The interval for the obtained symmetry energy $S_0$ at 68~$\%$ remarkably lies inside 1$\sigma$ PREX-II value. The range  for $S_0$ at the 68\% CL is  compatible  with $30- 35~\text{MeV}$  \cite{Roca-Maza:2015eza}, $33^{+2.0}_{-1.8}$ \cite{Essick:2021kjb},   and $32\pm 1.7$~MeV  \cite{Lattimer:2023xjm}.

\vspace{0.5cm}
\textit{Bayesian analysis}\\

\begin{figure} 
\centering
\includegraphics[width=6.127in]{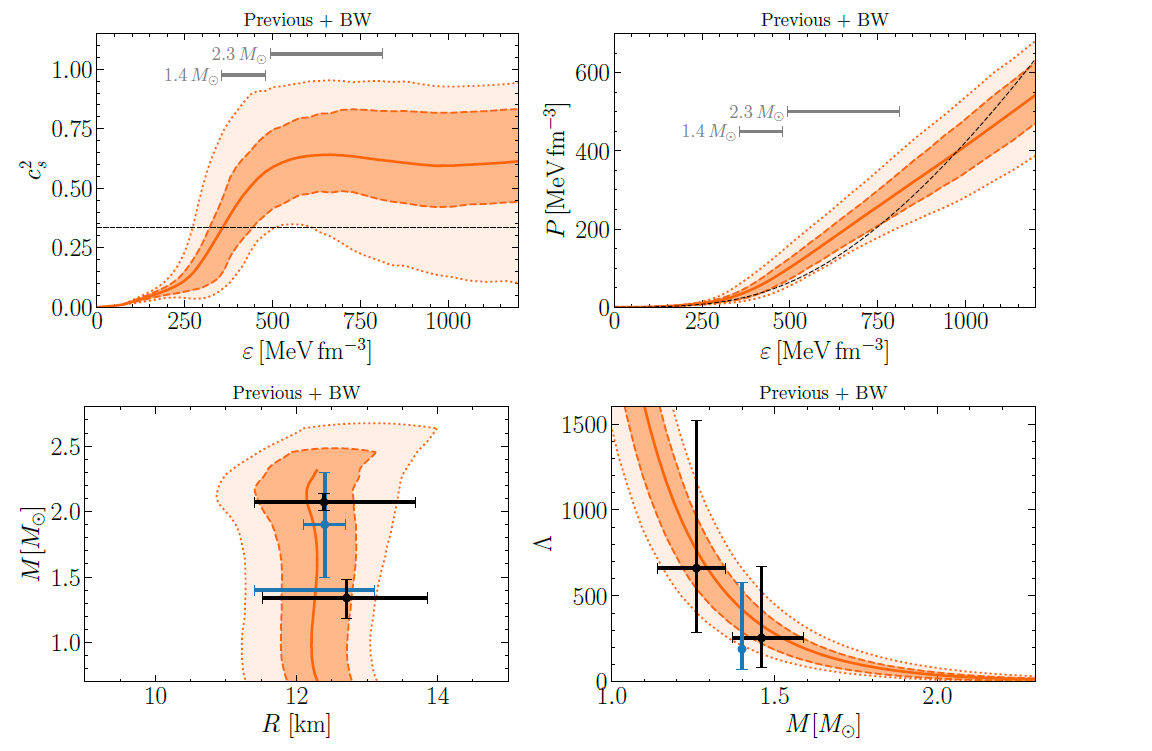}
\caption{{\bf Top:} Marginal posterior probability distributions at 95$\%$ and 68$\%$ level of $c_s^2$ and pressure $P$ as a function of energy density $\varepsilon$, inferred from the Table I of \cite{Brandes:2023bob}. The dashed black lines in the figure at right indicate the value of the conformal limit or represent the APR EoS \cite{Akmal:1998cf}. Solid lines show the medians of the posterior probability distributions at each $\varepsilon$ or M. Grey bars mark the 68$\%$ credible intervals of the central energy densities of neutron stars with masses M = 1.4M$_\odot$ and 2.3$_\odot$, respectively. {\bf Bottom:} Marginal posterior probability distributions for the mass-radius relation and the tidal deformability, $\Lambda$, as a function of neutron star mass M in units of the solar mass M$_\odot$.  The mass-radius relation is compared to the marginalized intervals at the 68$\%$ level from the NICER data analyses (black) \cite{Riley:2019yda,Riley:2021pdl} of PSR J0030+0451 and PSR J0740+6620. In addition the 68$\%$ mass-radius credible intervals of the thermonuclear burster 4U 1702-429 \cite{Al-Mamun:2020vzu} are displayed as well as the 68$\%$ credible interval of R(1.4M$_\odot$) extracted from quiescent low-mass x-ray binaries \cite{Nattila:2017wtj} (blue), both of which are not included in the Bayesian analysis. $\Lambda$(M) is compared to the masses and tidal deformabilities inferred in Ref. \cite{Fasano:2019zwm} for the two neutron stars in the merger event GW170817 at the 90$\%$level (black) as well as  $\Lambda$(1.4M$_\odot$) at the 90$\%$ level extracted from GW170817  \cite{LIGOScientific:2018cki} (blue). \emph{Reprinted from \cite{Brandes:2023hma,Brandes:2023bob} with permission from the American Physical Society http://dx.doi.org/10.1103/PhysRevD.108.094014.}  }
  \label{fig:BrandesBand}
\end{figure}

\begin{figure} 
\centering
\includegraphics[width=5.in]{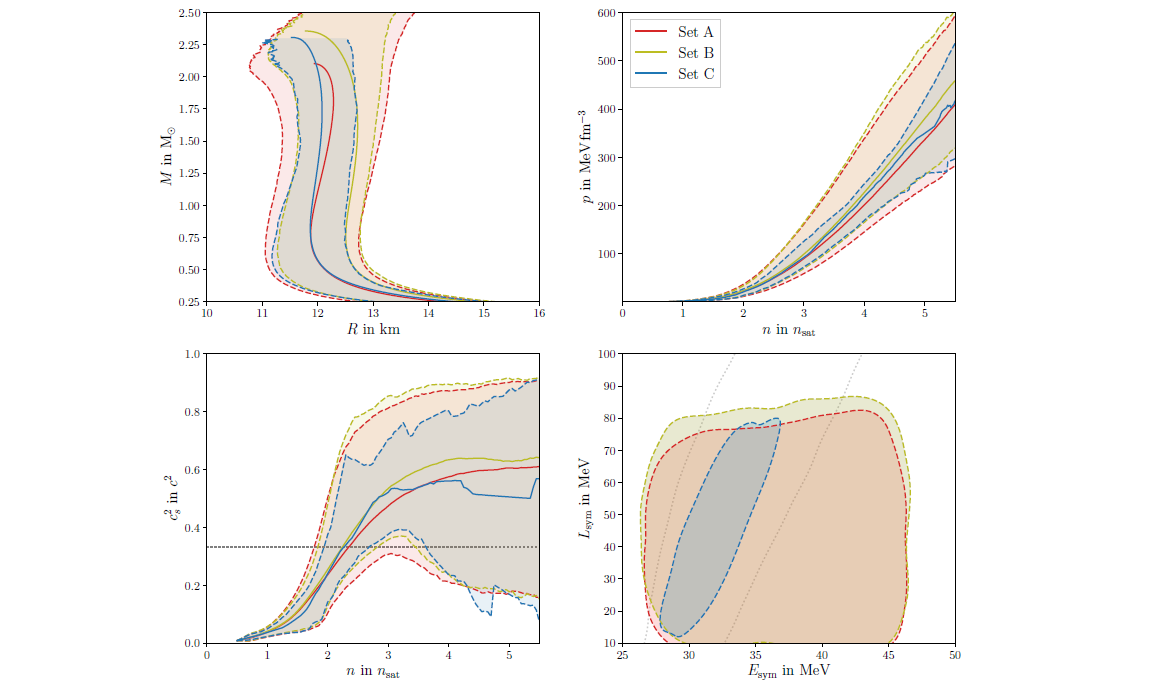}
 \caption{Final posterior estimate for the M-R and p-n relationship, speed of sound, and symmetry-energy parameters from the three different combinations of constraints A (red), B (yellow), C (blue). The solid lines in the top left-hand panel show the M-R curves with highest posterior likelihood, the dashed lines mark the 95$\%$ credibility intervals in radius at a given mass. Similarly, the dashed lines in the top right-hand panel indicate the 95$\%$ credible intervals for the pressure as a function of number density, the solid line there refers to the median. The bottom left-hand panel shows the same quantities for the speed of sound as a function of number density, with the medians drawn again as solid lines. For the calculation of these posterior
properties, we only include EoS samples below their TOV density, i.e., we consider $P(p|n, n < n_{\text{TOV}})$. The gray dotted line in the bottom left panel indicates the conformal limit $c^2_s$=1/3. The bottom right-hand panel shows the 95$\%$ credible regions for the nuclear symmetry parameters E$_{sym}$ and L$_{sym}$. The grey dotted line indicates the 95$\%$ credible region for these parameters from the combined results of PREX-II and CREX. \emph{Reprinted, with permission of ~\cite{Koehn:2024set} under Creative Commons License.} 
 \label{fig:Koehnband}
}
\end{figure}

In recent years, Bayesian analyses have multiplied considerably, with the objective of inferring a compatible range of EoS of neutron stars, with controlled uncertainties, by using information from astrophysical observations, nuclear experiments and nuclear theory \cite{Raaijmakers:2021uju,Huth:2021bsp,Gorda:2022jvk,Jiang:2022tps,Zhou:2023hzu,Brandes:2023hma,Prakash:2023afe,Annala:2023cwx,Essick:2023fso,Takatsy:2023xzf,Breschi:2024qlc,Koehn:2024set,Marquez:2024bzj,Rutherford:2024srk,Ecker:2024uqv,Beznogov:2024vcv,Somasundaram:2024ykk,Li:2025vhk,Magnall:2025zhm, Semposki:2025etb}. Bayes' theorem is stated mathematically as the following equation:

 \begin{equation}\label{eq:Bayes}
     P(EoS|data)= \frac{P(EoS)P(data|EoS)}{P(data)}
 \end{equation}
 where P(EoS) is the prior and $P(data | EoS)$ is the product of uncorrelated likelihoods of the data given an EoS.
Some of these results are shown in Figs. \ref{fig:BrandesBand} and \ref{fig:Koehnband}.\\ 

In  the results of Fig. \ref{fig:BrandesBand} from \cite{Brandes:2023bob,Brandes:2023hma}, the inference of the sound speed and equation-of-state for dense matter in neutron stars is
extended in view of recent new observational data. A second re-analysis is performed applying the perturbative QCD constraint at asymptotically high densities, in order to
clarify the influence of these constraints on the inference procedure.
The parameterisation here is represented by a set of N+1 points (N=5 and so seven segments) $\theta=(c_{s,i}^2,\varepsilon_i$), where $c_{s,i}^2$ is modeled as:

\begin{equation}  c_{s,i}^2(\varepsilon)=\frac{(\varepsilon_{i+1}-\varepsilon)c_{s,i}^2-(\varepsilon-\varepsilon_i)c_{s,i+1}^2}{\varepsilon_{i+1}-\varepsilon_{i-1}} 
\end{equation}
The next step is to compute likelihoods
for the different types of data considered. They add low-density constraint from $\chi$EFT as a likelihood instead
of a prior, permiting thus a balancing between the constraints from nuclear physics and astrophysics and allowing for a rigorous and statistically consistent Bayes factor analysis. Following the results of the analysis in \cite{Essick:2020flb}, they choose a conservative maximum density for the applicability of $\chi$EFT as $n = 1.3 n_s$. 

The astrophysical observables used are, for massive objects, the pulsars PSR J1614-��2230 \cite{Demorest:2010bx,Fonseca:2016tux}, PSR J0348+0432 \cite{Antoniadis:2013pzd},
and PSR J0740+6620 \cite{Romani:2022jhd}, PSR J0952-��0607; mass and radius data from PSR J0030+0451 \cite{Riley:2019yda} $M=(1.34^{+0.15}_{-0.16})\,M_\odot$, $R=12.71^{+1.14}_{-1.19}$ \text{km} and PSR J0740+6620 \cite{Riley:2021pdl} $M=(2.08 \pm 0.07)\,M_\odot$, $R=(12.92^{+2.09}_{-1.13})$\ \text{km}. For Binary Neutron Star Mergers (BNS), tidal deformability values from GW170817   \cite{Abbott:2018exr} ($\Lambda_{1.4}= 190^{+390}_{-120}$) and from GW190425 \cite{LIGOScientific:2020aai}  ($\widetilde{\Lambda}<$ 600).


The results of the inferred sound velocity are shown at the top left of Fig.\ref{fig:BrandesBand}. Within the $68\%$ highest posterior density credible bands, a rapid increase in $c_s^2$ is displayed beyond the conformal limit of $1/3$ at energy densities relevant for a wide range of neutron stars with masses $M \sim 1.4-��2.3\, M_\odot$. Quantification in terms of a corresponding Bayes factor demonstrates extreme evidence for $c_s^2$ exceeding the conformal bound within neutron stars.
At intermediate energy densities, $\varepsilon \sim 600$~MeV/fm$^{3}$, the speed of sound starts to form a plateau that extends up to higher energy densities. This behavior of the sound speed is reflected in the pressure. The plateau in $c_s^2$ corresponds to an approximately linear rise of the pressure with increasing energy density \cite{Brandes:2023bob}.
The results of including $\chi$EFT constraints show a moderate tension exists between $\chi$EFT  extrapolations of $c_s^2$ up to $n\simeq 2 n_s$ and the trend towards a stiffer EoS implied by the astrophysical data.

In this work the analysis of the pQCD constraints is also included as a likelihood.  The method developed by \cite{Komoltsev:2021jzg} requires that all the points of the EoS can be connected to the pQCD values by some causal and stable interpolation.  To implement these constraints, they take the logarithmic average over $X \in$ [1/2, 2] recommended by \cite{Gorda:2022jvk}. Since each EoS is constrained by neutron star data only up to the respective maximum central chemical potential $\mu_c$, density $n_c$ and pressure $P_c$ (by solving TOV equations), \cite{Brandes:2023bob} verifies at that point, ($\mu_{NS}$, $n_{NS}$, $P_{NS}$) = ($\mu_c$, $n_c$, $P_c$), whether a causal and thermodynamic interpolation to the asymptotic pQCD constraint exits. They find very little impact of the asymptotic pQCD and a strong evidence against minimum squared sound speeds smaller than $c_s^2 \leq$ 0.1, indicative of a possible first-order phase transition, in NS with masses up to $M \leq 2.1 M_\odot$.\\

Another example of a Bayesian analysis is shown in Fig. \ref{fig:Koehnband}  from Ref.~\cite{Koehn:2024set}. 
 They construct the EoS prior as follows. At the lowest densities $n\leq$ 0.076 fm$^{-��3}$, the authors use the crust EoS of \cite{Douchin:2001sv}  up to the crust-core transition density predicted by this model at  this density number. To combine the crust and core EoSs, they use a cubic spline in the speed of sound $c_s^2$ versus density plane between the crust-core transition density 0.076 fm$^{-��3}$ and the onset of the MM at 0.12 fm$^{-��3}$.
For the outer core, the authors assume that matter is composed of only nucleonic degrees of freedom in beta equilibrium up to a certain density $n_{\text{break}}$ that they randomly draw  from a uniform distribution in the range $1-��2 n_s$ to account for the possibility that non-nucleonic degrees of freedom appear at higher densities. To model the homogeneous matter  for 0.12 fm$^{-3}\leq n \leq n_{\text{break}}$, they employ the metamodel (MM) introduced in 
\cite{Margueron:2017lup,Margueron:2017eqc}.  

For $n>n_{\text{break}}$, they need to take into account the possibility that non-nucleonic degrees of freedom might appear \cite{Somasundaram:2022ztm}, while simultaneously allowing for an EoS prior that remains as conservative and broad as possible.
So, they employ a speed-of-sound approach which is a modified version of the scheme in \cite{Tews:2018iwm}. In this case,  they create a nonuniform grid in density between n$_{\text{break}}$ and $25 n_s$ with 9 grid points, where the ninth point is fixed at 25 $n_s$ and the other points are randomly distributed. Then, at each density grid point, the squared sound speed is varied uniformly between 0 and 1. Thus, Ref.~\cite{Koehn:2024set} creates a wide collection for different constraints on the EoS, where each collection utilizes a diverse set of data points, which they apply step by step to a wide, general prior for the dense matter EoS. Then, they determine the likelihood across all the EoS candidates for each given constraint, commenting on potential sources of systematic errors while applying a particular constraint, since different (statistical) interpretations of the data may affect the inferred results. For example, there is no specific a priori choice of how to implement theoretical constraints from $\chi$EFT or from pQCD in a Bayesian likelihood function \cite{Gorda:2022jvk,Brandes:2023bob,Drischler:2021kxf,Somasundaram:2022ztm}. Finally,  they group the constraints into different sets and study how the final estimates change, when they include tighter, but more model-dependent constraints.
They present three combinations of sets:
\begin{itemize}
    \item Set A: $\chi$EFT, pQCD, heavy pulsars with M$\sim$ 2M$_\odot$ (PSR J1614-2230, PSR J0348+0432, PSR J0740+6620), PSR 70030+451 (NICER), pSR J0740 (NICER), GW170817.
     \item Set B: HIC (heavy-ion collisions) , PSR J0952-0607, qLMXBs (quiescent low-mass X-ray binaries), GW170817+Kilonova+short gamma-ray burst (GRB170817A).
     \item Set C: PREXII, CREXII; $^{208}P_b$ dipole, thermonuclear accretion bursts in low-mass X-ray binaries (Burster 4U 1702-429, Burster J1808.8-3658), HESS J1731-347, GW190425,  GRB211211A, GW170817 postmerger.
\end{itemize}

To combine the individual constraints, they take the immediate dependencies into account by only multiplying independent likelihoods and omitting the dependent ones. When they combine two constraints that share a common prior, they reweigh the combination.
The authors find that $\chi$EFT softens the EoS and restricts the radii of a NS, while the pQCD constraint also softens the EoS but mostly impacts the maximum NS mass (TOV mass).
Regarding pQCD, the authors find that the constraining power of the QCD input is strongly dependent  on the termination density of the EoS. Thus, the more conservative choice $n_{\text{term}}=n_{\text{TOV}}$ offers little constraint, since only about 20$\%$ EoS are rejected (they cannot be connected to pQCD) \cite{Komoltsev:2023zor}. 
In the same reference the authors observe that the fact that this critical value of $n_{\text{term}}$ is so close to
n$_{\text{TOV}}$ is a coincidence, since if n$_{\text{pQCD}}$ were significantly lower, then it is possible that
this critical value of $n_{\text{term}}$ would also be lower.
On the other hand, the option $n_{\text{term}}=10 n_{s}$
shows stronger constraints from pQCD since now about 50-60$\%$ of the EoS cannot be connected to pQCD.

The results for the three sets are shown in Fig.~\ref{fig:Koehnband}, displaying the resulting constraints on the NS masses and radii, pressure and speed of sound, and the symmetry energy. The radii of NSs can be determined within an uncertainty of $\sim 0.5-��1.0$~km, depending on the number of constraints employed. 
With higher NS mass,
the uncertainty in the radius grows slightly. However, significant restrictions on the speed of sound and the empirical nuclear parameters are harder to obtain. Although the inclusion of $\chi$EFT allows them to restrict $L < 100$~MeV, narrower constraints can be achieved only when the results of PREX-II and CREX are directly included, especially on $S_0$. Regarding $c_s^2$, their inference implies that its value likely exceeds the conformal limit of 1/3 around 3 n$_s$, but for higher number densities the posterior distribution reaches a flat plateau and the upper limits on
$c_s^2$ remain weak and not particularly informative. 
Finally, including all the constraints mentioned, they get R$_{1.4} = 12.20^{+0.50}_{-��0.48}$~km and $M_{\text{TOV}} = 2.30^{+0.07}_{-��0.20}$ M$_\odot$.

All these analysis show that a key aspect to  be understood in NS core is the interplay between astrophysical observations, requiring a stiffening of the EoS, and pQCD calculations, requiring a softening of the EoS. This interplays yields a peak structure in the sound speed of NS
matter. This softening can be interpreted as a signature of a phase transition.

\vspace{0.5cm}
\textit{Non-parametric models}
\\



This non-parametric method for inferring the universal neutron star (NS) equation of state from GW and astrophysical observations was developed by \cite{Landry:2018prl,Landry:2020vaw,Miller:2019cac,Essick:2019ldf}. The prior consists of an ensemble of EoSs that extrapolate by using a Gauss proccess (GP) regression from the $\chi$EFT EoS to NS star densities $n \sim 5 - 10 n_s$. 
They generate GPs for an auxiliary variable

\begin{equation}\label{phiLE}    \phi(\varepsilon)=\log(c_s^2\frac{d\varepsilon}{dp}-1)
\end{equation}

conditioned on tabulated EoSs from the literature. 
In this way they construct generative models for synthetic EoSs and then compare them against the input data. The models are of different type, contain only hadronic matter, hadronic and hyperonic matter and hadronic and quark (and possibly hyperonic) matter. Finally, they condition the prior using NS observations.

In \cite{Essick:2019ldf},the work differs from  \cite{Landry:2018prl} in that they construct mixture models of GPs instead of relying on a single set of hyperparameters. That is,

\begin{equation}\label{phiEssick} 
\phi \sim \sum_i w_iP(\phi|\vec{\sigma_i})
\end{equation}
where $\vec{\sigma_i}$ and w$_i$ are the hyperparameters and weight associated with the ith element of the mixture model and $P(\phi|\vec{\sigma_i})$ is the normalized distribution.
 
\begin{equation}\label{P_Essick} 
P(\phi|\vec{\sigma_i}) = \mathcal{N}(\bar{\phi},\vec{\sigma_i})K(\varepsilon_i,\varepsilon_j)
\end{equation}
 and $K(\varepsilon_i,\varepsilon_j)=\sigma^2exp((\varepsilon_j-\varepsilon_i)/2l^2)$, in which the hyperparameters $\sigma$ and $l$ (correlation length) determine the behavior of the functions.\\
\\
\begin{figure}
    \centering
    \includegraphics[width=0.95\linewidth]{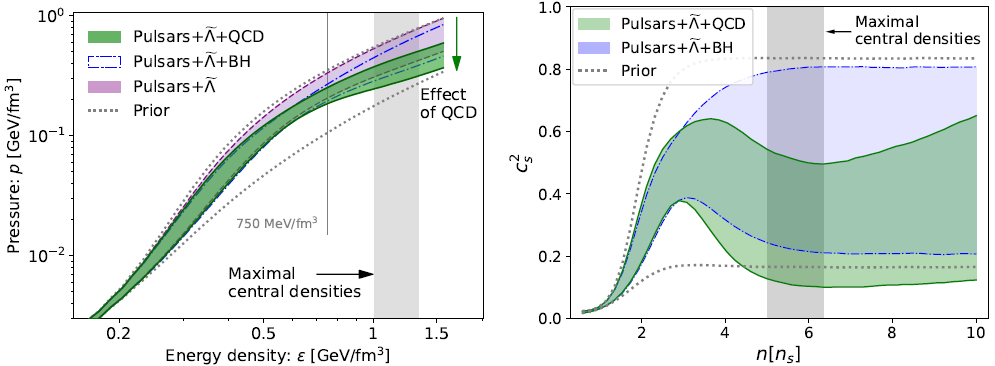}
    \caption{The impact of the QCD input on the EoS $P(\varepsilon)$ and $c_s^2(n)$. The bands represent 68$\%$ credible intervals conditioned on the different inputs. The label ��pulsars�� refers to the combined radio and X-ray posterior. The gray band indicates the 68$\%$ CL for the maximum densities reached in stable,
non-rotating NSs. \emph{Reprinted from ~\cite{Gorda:2022jvk} under Creative Commons License.} }
   \label{fig:abinitioGordaresults}
\end{figure}

 A different approach of  the non-parametric method is used in \cite{Gorda:2022jvk}. In this work, the authors construct a simple Bayesian-inference setup, which they use to study the interaction of the QCD input with the NS observations. They extrapolate the EoS from 
$\chi$EFT to $n = 10 n_s$ using a Gaussian-process regression and condition the prior using NS observations with or without QCD input. 
The question that arises is whether these features occurring at NS densities are genuine predictions of QCD or whether they arise as a consequence of the interpolation with a very restrictive interpolation function. To avoid the obstacles of the interpolation, \cite{Gorda:2022jvk} constructs a simple Bayesian-inference setup that they use to study the interaction of the QCD input with the NS observations.
 To do that, they construct a large ensemble of EoSs by using GP regression and incorporating it into a Bayesian inference of EoS, conditioning the prior thus obtained by using NS observations with or without QCD input.
Following Eq. \ref{eq:Bayes}, the 
 choice of prior $P(EoS|data)$ is based on the Gaussian processes  framework, $P(data|EoS)$ is the product of uncorrelated likelihoods of the data given an EoS, which can be expressed as $P(data|EoS)=P(QCD|EoS)P(Astro|EoS)$, when both observations and QCD input are used.
 
 To obtain the prior, they use the GP regression of \cite{Landry:2018prl,Landry:2020vaw} which is conditioned with an (beta-equilibrated) EoS computed in $\chi$EFT \cite{Hebeler:2013nza} up to $n = 1.1 n_s$. For densities below
$n = 0.57 n_s$ they merge the GP to the BPS crust EoS \cite{Baym:1971ax}.
 In contrast to the model~\cite{Landry:2018prl,Landry:2020vaw}, in this approach, they reconstruct the form of the EoS, pressure as a function of number density $P(n)$, instead of using only its reduced form as a function of energy density $P(\varepsilon)$, in order to use the ab-initio calculations in QCD. The auxiliary variable chosen to extrapolate related to the speed of sound is therefore
\begin{equation}\label{phiGorda}
    \phi (n)=-\log(1/c_s^2(n)-1)
\end{equation}

The choice of hyperparameters ($\bar{c_s^2}$, $l$, $\sigma$) determines the behavior of EoS in regions where no data are available. In that work, this choice is:
 $l\sim \mathcal{N}(1.0n_s,(0.25n_s)^2)$, $\sigma\sim \mathcal{N}(1.25,0.2^2$), $\bar{c_s^2}\sim \mathcal{N}(0,5,0.25^2$). The GP is then conditioned with the $\chi$EFT EoS. Solving then the TOV equations, the maximal masses supported by each EoS are found.
To obtain the astrophysical likelihoods, the authors consider the following astrophysical observations: radio measurements of PSR J0348+0432 with M =$2.01 \pm 0.04$ M$_\odot$ \cite{Antoniadis:2013pzd} and PSR
J1614-��2230 J1614-��2230 with M = $1.928 \pm 0.017$ M$_\odot$ \cite{Demorest:2010bx}; X-ray measurements of the mass and radius of PSR J0740+6620 \cite{Fonseca:2016tux,Miller:2021qha,Riley:2021pdl}; and multimessenger data from GW170817, including TD measurements \cite{TheLIGOScientific:2017qsa,LIGOScientific:2018cki,LIGOScientific:2018hze} and the Black Hole (BH) hypothesis (or maximum merger remnant mass to collapse to a BH) \cite{Margalit:2017dij,Rezzolla:2017aly,Ruiz:2017due,Shibata:2017xdx,Shibata:2019ctb}.

Regarding the construction of the QCD likelihood function, the essential issue is that the pQCD constraints are derived under the assumption of known values entering pQCD at a selected $\mu_{\text{pQCD}}$=2.6 GeV. In addition, the uncertainty band related to the normalization scale $X \in$ [0.5,2] should be taken into account. To do that, they marginalize over $X$ by sampling different values for $X$ from a log-uniform distribution between 1/2 and 2, evaluating the likelihood for each, and then taking the average.
Different results from analysis with and without the QCD input differ and are shown in Fig. \ref{fig:abinitioGordaresults}. The works with the QCD input report a strong softening of the EoS and lower speed of sounds at energy densities of the order $\varepsilon \sim$ 750 MeV/fm$^3$, interpreting them as the onset of a quark matter \cite{Annala:2019puf,Annala:2023cwx}. However, these features are not discernible in works without the QCD input nor present in nuclear models.

Regarding the constraining power of the QCD input on EoS
inference, there are conflicting conclusions between \cite{Gorda:2022jvk} and \cite{Somasundaram:2022ztm}. In \cite{Gorda:2022jvk}, the effect of QCD is significant for the termination density $n_{\text{term}}=10n_s$. In contrast, in \cite{Somasundaram:2022ztm}, with n$_{\text{term}}=n_{\text{TOV}}$, the impact appeared marginal.

\begin{figure}[H]
     \centering
      \includegraphics[width=0.49\textwidth]{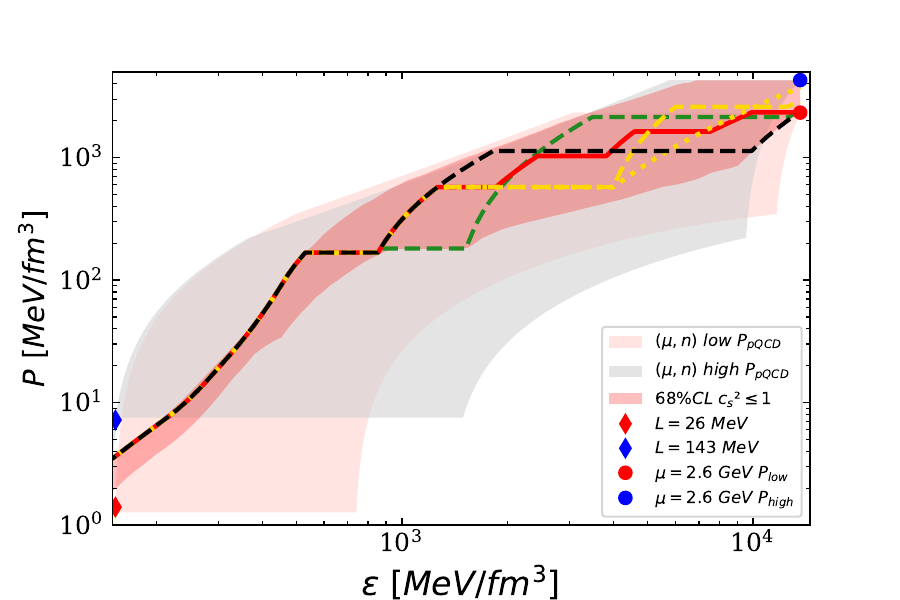}
     \includegraphics[width=0.49\textwidth]{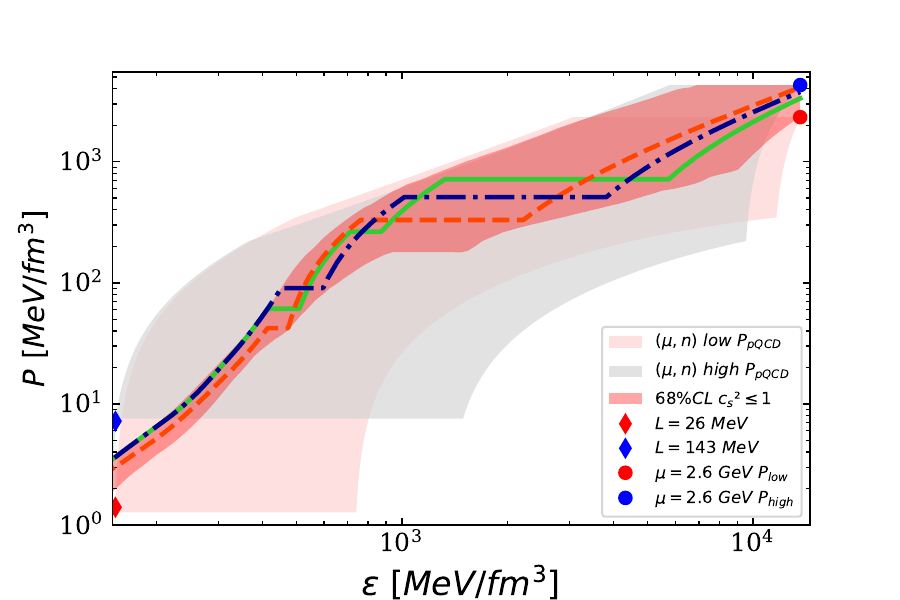}     

\caption{ {\small {\bf Left:} Four EoS's at 68 $\%$ CL of the band of Fig. \ref{fig:totalband} (top right), with the same first PT at $P=167~$MeV/fm$^3$ but reaching pQCD in distinct ways involving different number of PTs. 
 {\bf Right:}~  3 EoS's at 68 $\%$ CL of the band of Fig. \ref{fig:totalband}(top right) with several PTs, all of them starting at densities inside the NS.  For the two panels see the text for the meaning of the lines and more details.} }
     \label{fig:PTs}
 \end{figure}

\subsection{Phase transitions}

Different phases of matter are characterized by different exact symmetries. Between them, we can find in NS the following exact symmetries: quark phase invariance, leading to baryon number conservation (spontaneously broken in superfluids), spatial symmetry (whose breaking leads to inhomogeneous condensates and crystals), EM gauge invariance (spontaneously broken in semiconductors) or flavor/color internal quantum numbers. There are other approximate symmetries such as chiral symmetry, whose breaking is manifest in anomalously light pseudo-Goldstone bosons�~\cite{Alford:2019oge}.
Particularly, there is much variety of possible quark matter phases intermediate between hadrons and the asymptotic Color-Flavor Locked phase, due to the nine fermion species (u,d,s flavors in three colors) that can form a variety of Cooper pairs.

In the interior of a NS, at densities above $\approx$ 2$n_s$, the composition of matter is basically unknown.  Other particles than nucleons and electrons are expected to appear such as hyperon matter or quark matter. However, their presence softens the EoS, making it difficult to reach maximum masses higher than $2M_\odot$. This creates the so-called hyperon puzzle, which is the challenge of reconciling the measured masses of NSs with the presence of hyperons in their interior, and applies to any other degree of freedom in the NS core \cite{Chatterjee:2015pua,Blaschke:2018mqw,Baym:2017whm,Annala:2019puf,McLerran:2018hbz,Leonhardt:2019fua}. These challenges motivate the search for alternative schemes in the physics of compact objects, such as Extended Theories of Gravity (ETG) \cite{Joyce:2016vqv,Olmo:2019flu}.

In a recent paper \cite{Ferreira:2025dat}, the authors  analyse the slope of the NS mass radius curve $dM/dR$  from two large sets of relativistic mean field equations of state for nucleonic and hyperonic neutron star matter in order to identify hyperons in NS matter. They find that if the mass-radius curve always has a negative slope the probability that the star has
hyperons is very small. The appearance of other non-nucleon degrees of freedom, such as delta-baryons or kaon condensates, could have similar effects to those of the appearance of hyperons.\\

Phase transitions are divided into two main categories,
\begin{enumerate}
    \item First-order phase transitions (FOPT), which involve a latent heat. During such a transition, a system either absorbs or releases a fixed (and typically large) amount of energy per volume. In addition, the temperature (or the pressure) of the system will stays constant as heat is added: the system is in a "mixed-phase regime" in which some parts of the system have completed the transition and others have not. 
    \item Second-order phase transitions,  also called  continuous phase transitions or crossover.
   
\end{enumerate}

To describe the hadron-quark transition in a neutron star, the most commonly used construction is the Maxwell construction   describing the first-order transition while  the Gibbs construction is used to describe the crossover transitions.
In the Maxwell construction the pressure as well as the chemical potential are constant in the density interval of the mixed phase defined by the Gibbs thermodynamic-equilibrium condition:
 \begin{equation}\label{Gibbscondition}
      T_H = T_E = 0 \ , \ \ \ \ 
      \mu  = \mu_c \ , \ \ \ \ 
      P_H = P_E := P_c\ .
\end{equation}

Where $H$ and $E$ denotes hadronic and exotic phases, respectively.
What actually changes during the phase transition is the energy density, so they appear as a horizontal segment in the ($\varepsilon,P$) plane.
The Gibbs thermodynamic-equilibrium condition determines at what critical chemical potential $\mu_c$  will the pressure of the two phases  be equal.
In the Maxwell construction, the hadron-quark interface tension is regarded as being infinite.

The other  FOPT approach  used is the Gibbs construction (or rapid crossover), which corresponds to a vanishing interface tension, where the pressure in the mixed phase increases with the baryon density \cite{Glendenning:1992vb}, although $\partial P/\partial \varepsilon$ still changes discontinuously. The Gibbs construction also corresponds to global charge neutrality instead of local charge neutrality, appropriate for matter with two conserved charges (baryon number and electric charge) \cite{Glendenning:2001pe}.

Possible first-order phase transitions would leave distinct observable traces, such as a kink in the mass-radius diagram, or  possible shifts in the peak frequency of the post-merger GW signal when a quark-hadron phase transition is present  \cite{Bauswein:2018bma,Most:2018eaw}. 
Furthermore, for rotating neutron stars, the authors of \cite{Moreno:2023xez} extract three-dimensional sections of the ellipticity or dynamical angular momentum as a function of the star's mass and angular velocity.
A possible first-order phase transition in the equation of state leaves leaves a clear ridge (nonanalyticity) in these observables, similar to the sudden curvature in popular mass-radius diagrams for static stars.
So revealing the presence of a possible phase transition is
one of the goals of current and future detectors.  Although it is impossible to establish the existence of FOPT with current neutron-star observations, future observations using next-generation  gravitational-wave observatories, such as the Einstein telescope \cite{Abac:2025saz}, are expected to provide the necessary sensitivity and bandwidth to detect these signals.

First-order phase transitions (FoPT) to quark matter or other exotic forms of matter inside NS would be characterized by a jump in energy density $\Delta \varepsilon$ between hadronic and exotic phases. Furthermore, if the PT is sufficiently strong, meaning that the jump in energy density is significant, it could introduce a discontinuity in the mass-radius relation. After the hadronic branch, this discontinuity could reveal a new stable branch for exotic matter (a hybrid branch), which is called the "third family".  This second stable branch would invariably contain a hybrid star with the same mass but a smaller radius than at least one neutron star from the original branch. These two stars, commonly referred to as "twin stars" \cite{Glendenning:1998ag,Alford:2004pf,Benic:2014jia,Christian:2017jni,Christian:2021uhd}.\\


\subsubsection{Characterizing a first-order phase transition}

FOPT is characterized by a phase-coexistence region that extends over a range of densities $\Delta n$. The width of this region, $\Delta n$/$n$ ($n$ being the density at which the phase-coexistence region starts), is a measure of the strength of the phase transition \cite{Brandes:2023bob}. These authors refer to a "strong" first-order phase transition if $\Delta n$/$n > 1$ and a "weak" first-order phase transition if $\Delta n$/$n < 1$. Moreover, there are other criteria to characterize a FOPT, such as the Seidov criterium and the latent heat.


\vspace{0.5cm}
\textit{Seidov criterium}
\\
\\
When the EoS contains a phase transition with a discontinuity in the energy density $\Delta\varepsilon$,  the mass-radius sequence can become unstable close to the point of the transition. The Seidov-limit \cite{Seidov:1971sv} establishes a threshold value $\Delta \varepsilon_{crit}$ below which there is always a stable hybrid star branch connected to the neutron star branch:
\begin{equation}\label{Seidovsjump}
\Delta \varepsilon_{\text{crit}} := \varepsilon_E-\varepsilon_H = \varepsilon_H \left( \frac{1}{2} +\frac{3}{2}\frac{P_H}{\varepsilon_H} \right),
\end{equation}

which indicates the maximum "critical" discontinuity in the energy density (or jump). It was obtained by performing an expansion in powers of
the size of the core (small-core approximation).
When $\Delta \varepsilon < \Delta \varepsilon_{\text{crit}}$, a stable connected hybrid branch continues from the hadronic branch. However, if $\Delta \varepsilon > \Delta \varepsilon_{\text{crit}}$, there is no stable connected branch and the mass-radius sequence can become unstable if the transition from the hadronic to the exotic phase occurs faster than the oscillation period of the perturbation or the fundamental mode period that governs the star's stability \cite{Pereira:2017rmp}. However, stability can be regained for slower transitions \cite{Rau:2022ofy,Goncalves:2022ymr,Ranea-Sandoval:2022izm}. If the central pressure rises sharply enough with energy density following the phase transition, the mass-radius (M-R) diagram can recover stability.
The phase transition described by this EoS results in a new stable branch populated by hybrid stars, which may or may not be separated from the nucleonic branch by an instability region. 
The condition $\partial M/\partial n_c \geq 0$ is employed to evaluate dynamic stability, with the assumption that the conversion timescale at the phase transition is rapid.

\vspace{0.5cm}
\textit{Latent heat}
\\
\\
From the thermodynamic point of view, the way to characterize a first-order phase transition is by the specific latent heat of the transition.  It is defined as energy released or absorbed, by a body or a thermodynamic system, during a constant-temperature process or first-order phase transition.  A stronger transition will involve a larger amount of latent heat.

A FOPT is usually characterized by a jump in energy density between two phases, which is often referred to as latent heat. 
As an intensive property, the authors of \cite{Lope-Oter:2021mjp} use the specific latent heat $L$ normalized to the unit mass, which is a dimensionless quantity in natural units with $c=1$. To calculate $L$ in neutron star matter the starting point in the low-density regime is
\begin{equation}\label{def:Ln}
L|_n:=\frac{\Delta E}{NM_N}\ ,
\end{equation}
which is the latent heat per nucleon normalized to the vacuum neutron mass of $940$ MeV, computed from $\Delta E :=  E_E- E_H$, and to which they have added a subindex $n$ to distinguish it from $L|_\varepsilon$ defined shortly in Eq.~(\ref{LperA}).

To obtain $L|_n$ from the EoS they integrate  the first law of thermodynamics, Eq. \eqref{eq:firstlaw}, in terms of the pressure, the number density $n$ (baryon number is conserved) and the energy per nucleon ($E/N$), with limits extending from the transition point $n_{\rm tr}=n_{\rm H}$ (where the phase $H$ is pure) to that $n_{E}$ where the medium is completely in the presumed exotic phase.
Since the pressure $P$ is constant over the transition, $L|_n$  is obtained  directly by integrating from Eq.~(\ref{eq:firstlaw}). Then,
\begin{equation}\label{diff}
     L|_n=\frac{\Delta (E/N)}{M_n} =\frac{ P_H}{Mn} \frac{(n_E-n_H)}{n_E n_H}
\end{equation}
Some values of $L|_n$ are shown in  the center of Fig.
\ref{fig:LatentheatEoS}  for the EoS examples shown in the left. These EoSs were obtained by interpolating between $\chi$EFT EoS of \cite{Drischler:2020yad} at $n=1.3n_s$ and the pQCD bounds \cite{Kurkela:2009gj} at $\mu_{\text{pQCD}} \approx 2680$~MeV, displaying the largest
possible jump in energy density compatible with pQCD constraints, following the causality limit $c_s^2$ = 1 to reach the pQCD regime.  The largest jumps are found in the red area, which corresponds to the constraints from the lower pressure of the uncertainty band to enter pQCD regime.

Taking  into account the Eq.~\eqref{eq:densityenergy}  and Eq.~(\ref{diff}), an alternative  definition for the latent heat  directly evaluated from the Equation of State $P(\varepsilon)$ is \cite{Lope-Oter:2021mjp}:
\begin{equation}\label{LperA}
     L|_\varepsilon = P_H \frac{(\varepsilon_E-\varepsilon_H)}{\varepsilon_E  \varepsilon_H} \ .
\end{equation}
At low $\varepsilon$, Eq.~(\ref{LperA}) is equivalent to Eq.~(\ref{def:Ln}) except for the binding energy $0\simeq (E/A) \ll M_N$, introducing a few percent error quantified below. 
Near the density allowing maximum $L$, Eq.~(\ref{LperA}) is a very practical definition of $L$.
Nevertheless they will generally refer to  $L|_n$ from Eq.~(\ref{def:Ln}) with Eq.~(\ref{diff}) substituted therein.
The difference between computing $L|_n$ and  $L|_\varepsilon$ (shown at the right of  Fig.~\ref{fig:LatentheatEoS}) grows with energy density. Around $\varepsilon\simeq 2.6 \varepsilon_{\rm s}$ (with $\varepsilon_s = \varepsilon(n_s)$), the stiffest EoS leads to a 15\% change relative to  $L|_\varepsilon$; with the softest EoS at the bottom of the allowed band, the two heats agree within 15\% all the way to $\varepsilon\simeq 5.8 \varepsilon_{\rm s}=5.8 \times 153$~MeV/fm$^3$.
The corresponding binding energy per nucleon $E/A$ up to which the relative difference between $L|_n$ and  $L|_\varepsilon$  remains within 15\% is about $ 20\% M_N$ (with the precise figure depending on the EoS), still in a regime where the total energy is mass-dominated.
Beyond such energy densities it becomes really necessary to distinguish
$L|_\varepsilon$ from $L|_n$. 
To compute this second  one, it is necessary to use the relation between $n$ and $E$ in Eq.~(\ref{eq:densityenergy}), that requires the binding energy per nucleon at each point of the grid, $(E/A)_i$.

\subsubsection{Maximum Latent heat}

In Ref.~\cite{Lope-Oter:2021mjp} the authors aim to obtain the maximum latent heat in a FOPT, they generate specific EoS in order to look for high  $L|_n$ values. 
At lowest number densities $n\leq 0.05 n_{\rm s}$ nuclear data directly  constrains the crustal EoS ~\cite{Baym:1971pw,Negele:1971vb}. At low densities, $0.05 \leq n \leq 1.31 n_{\rm s}$, they use the low $\chi$EFT ($\Lambda=$500 MeV) from data of \cite{Drischler:2020yad} at  N$^3$LO. At $1.31\leq n \lesssim$ 40n$_s$, they extend the EoS by using grid interpolation and taking into account pQCD constraints \cite{Kurkela:2009gj}. 
Since the goal is to find the maximum latent heat, different EoS are constructed, including extreme EoS. Thus, any $P(\varepsilon)$ in this region is matched to $\chi$EFT at a number density $n_m$, with  slope $c^2_{s,m} \in (0,1)$ (given in the figure legends). In this case, they end the EoS at the matching pQCD points (blue and red points represent high and low pQCD limits, respectively). 
Checks have been conducted employing $n_m=2 n_s$ to explore the systematic, and almost no difference in $L|_n$ was found, although the energy density $\varepsilon$, at which the maximum transition occurs, shifts to higher values.

\begin{figure}[t]\vspace{-0.3cm}
\centering
\includegraphics[width=0.32\columnwidth]{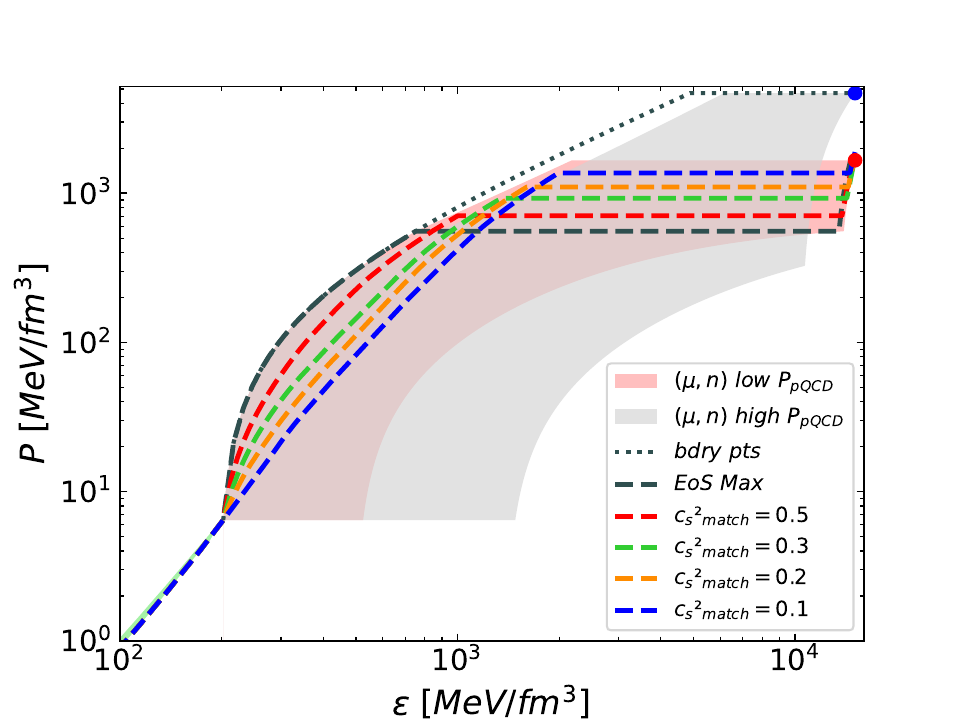}
\includegraphics[width=0.33\columnwidth]{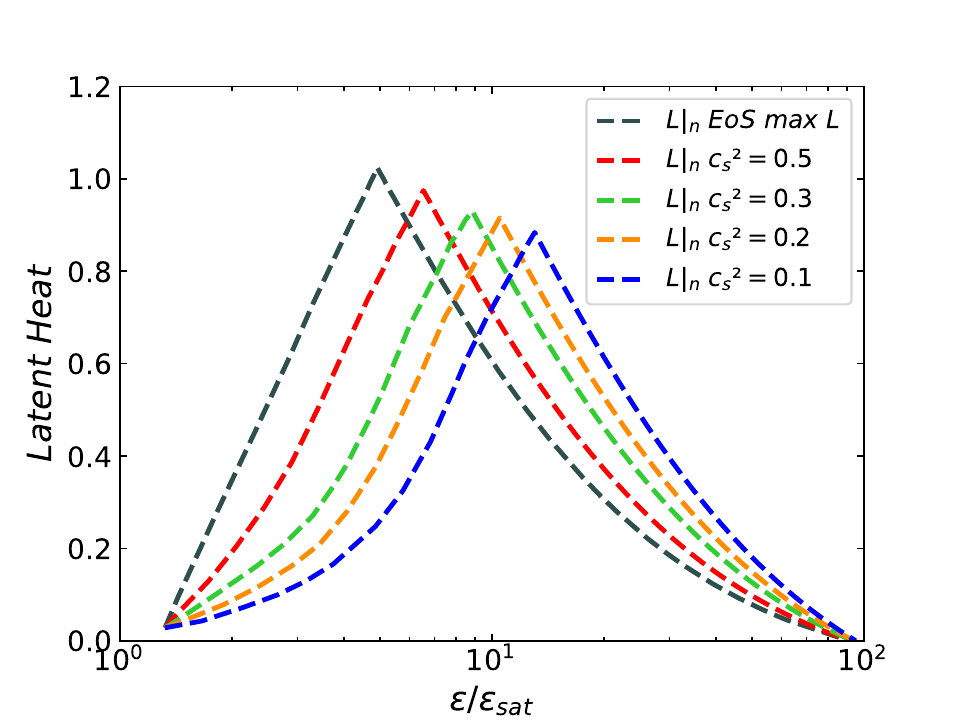}
\includegraphics[width=0.33\columnwidth]{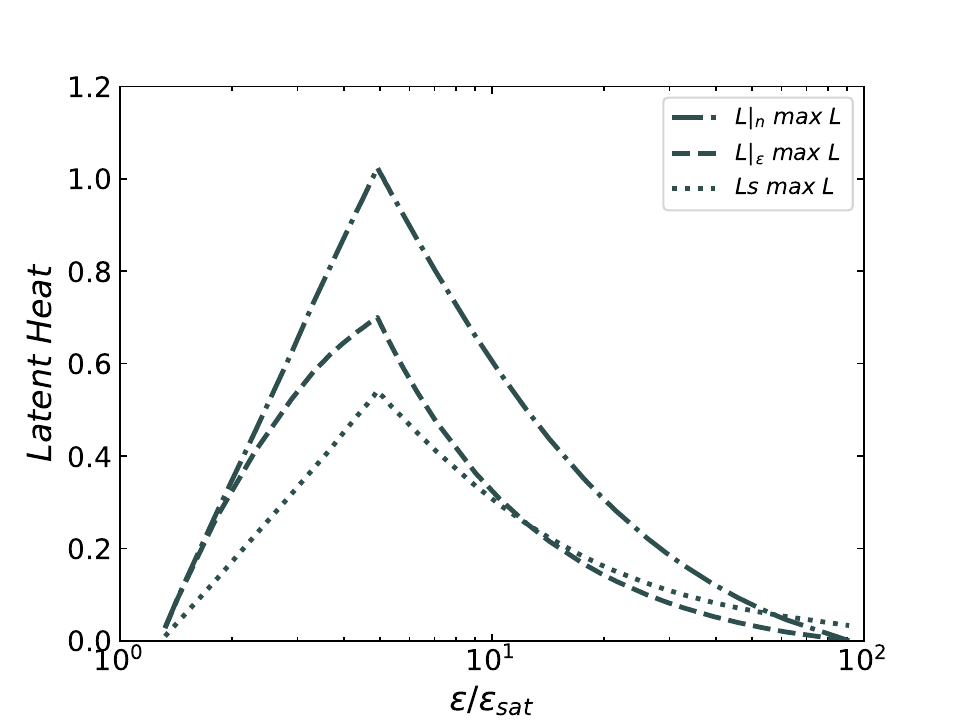}

\caption[Example EoS and maximum specific  $L|_\varepsilon$  of these EoS.]{\label{fig:band}  
{\bf Left}: Example EoS inside the band are ordered upwards by increasing $c_s^2$ at the $\chi$EFT matching point. The gray and red areas correspond to the constraints from imposing simultaneously the limit for  $n$ and $\mu$  at the high-$\varepsilon$ pQCD and the low-$\varepsilon$ pQCD  point, respectively.
 The low-density $\chi$EFT approximation is used up to $1.3n_s$, just above the nuclear saturation density.
{\bf Center}: Maximum specific  $L|_\nu$  with equation of state fitting within the grey nEoS band of the top of the figure. The $OX$ axis (energy density at which the transition triggers) extends to 100$n_{\rm s}$  (at chemical potential $\mu_B=2.6$ GeV) where  pQCD is matched. An absolute maximum $L|_\varepsilon \simeq 1$ is reached for
$\varepsilon \simeq 5 \varepsilon_{\rm s}$.
{\bf Right}: Comparing the bound on the latent heat $L|_n$ (upper line, dashed) from microscopic hadron physics alone  with the maximum latent heat related to the Seidov limit~\cite{Seidov:1971sv}.
\label{fig:LatentheatEoS}
}
\end{figure}

The $L|_n$ and $L|_\varepsilon$ are calculated from Eqs.  \eqref{diff}, \eqref{LperA}  for such phase transitions.  The results $L|_n$ are displayed in the center of Fig. \ref{fig:LatentheatEoS} for the EoS to the left of this figure. 
The largest maximum latent heat that they find compatible with the QCD-based results are $L|_n \simeq 1$ and $L\simeq 0.7$, both of which would be reached for $\varepsilon \simeq 5 \varepsilon_{\rm s}$, (for an EoS matched to nuclear matter with maximum slope $c^2_s\simeq 1$ at 1.3 $n_{\rm s}$).
It is important to note that the maximum latent heat is not reached for the maximum density jump at $\varepsilon=3.5 \varepsilon_s$ as it is generally assumed, but at a higher $\varepsilon$. However, solving the TOV equations, the maximum mass (M=3.6 M$_\odot$) for the corresponding EoS is reached at an intermediate $\varepsilon_c= 3.9\varepsilon_s$ between them,  with  $L|_n$=0.85 and $L|_\varepsilon$=0.62.

In \cite{Lope-Oter:2021mjp} the authors also compare this bound, obtained from hadron physics alone, with  the traditional Seidov bound, assuming the validity of GR and a perturbative assumption about a small size of the exotic core of a NS.  To do that, it is necessary to transform the Seidov limit for $\Delta \varepsilon$ in Eq.~(\ref{Seidovsjump}) to a limit on the latent heat $L|_n$ and therefore it is also necessary to express the number density $n = \varepsilon_{\rm max}^{\rm GR} /(M_N+E/A|_{\varepsilon_{\rm max}^{\rm GR}})$ in terms of the maximum energy density computed within General Relativity, an equation which is likewise iteratively stepped forward. $L_{\rm Seidov}$ together with   $L|_n$ lines are shown to the right of Fig. \ref{fig:LatentheatEoS}, where  $L|_n$, $L|_\varepsilon$ and $L_s$ are displayed for the stiffest EoS. 
It shows a big difference between $L|_n$, constrained by $\chi$EFT and pQCD bounds as well as by first principles, and $L_s$, constraint by GR. 

The maximum latent heat depends on each EoS. In the case of the band of Fig. \ref{fig:totalband} (top right) \cite{Alarcon:2024hlj}, for the stiffest EoS within 68$\%$ CL, $L|_n=$0.746, $L|_\varepsilon=$0.57, and $L_s$=0.44. The values were obtained considering the longest FOPT at the central density and $c_s^2=$ 1/3 from the end of FOPT up to the entry of pQCD \cite{Gorda:2021kme}.

\subsubsection{Identifying phase transition}

\begin{figure}
    \centering
    \includegraphics[width=0.75\linewidth]{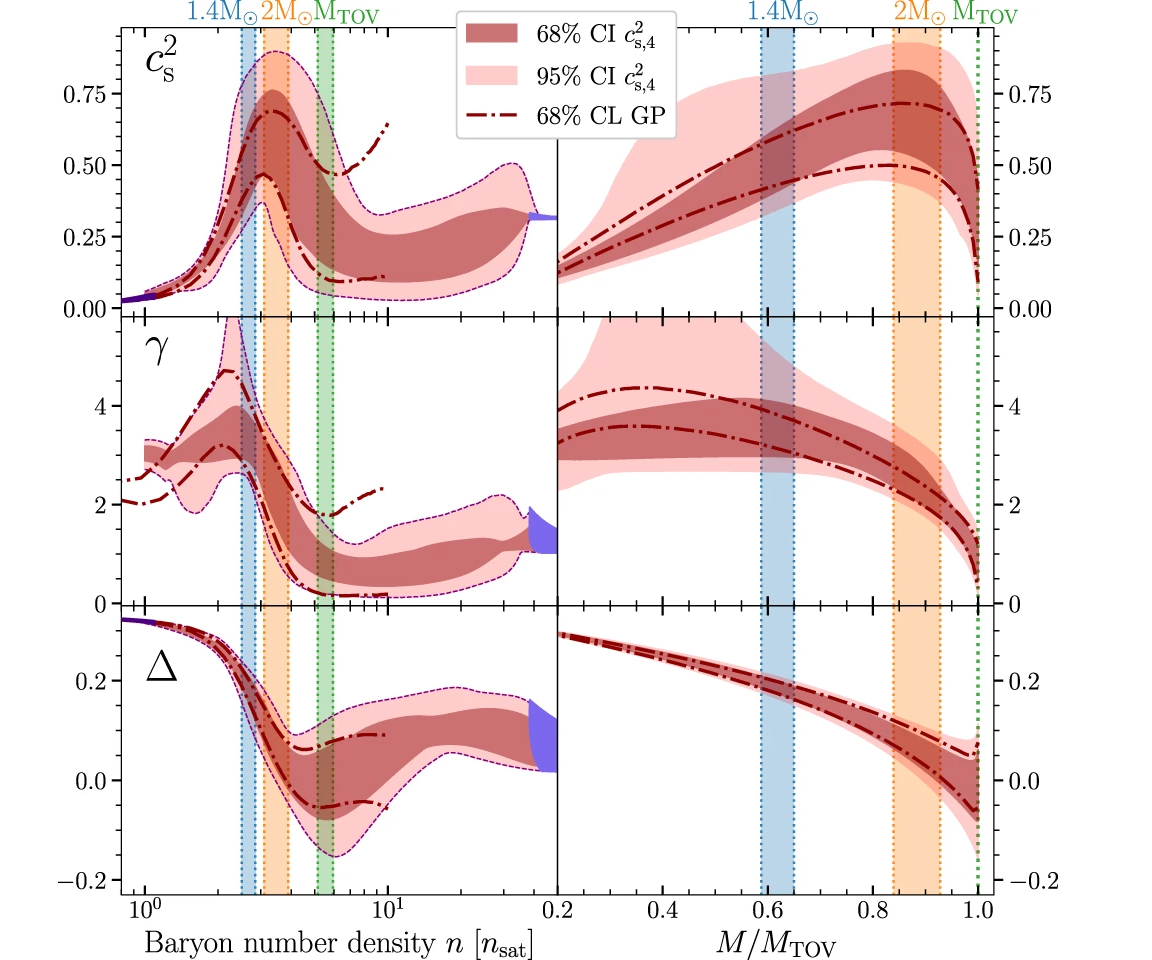}
    \caption{The normalized trace anomaly $\Delta$, polytropic index $\gamma$, and speed of sound squared $c_s^2$  as functions of (left) the baryon number density n and (right) M/M$_{MTOV}$. The dark and light bands represent the 68$\%$ and 95$\%$ CL obtained using $c_{s,4}^2$, while  the $68\%$ CL obtained from the GP ensemble from \cite{Gorda:2022jvk} is shown with a red dash-dotted line. \emph{Reprinted from \cite{Annala:2023cwx} under Creative Commons License.}}
    \label{fig:conformalparameters}
\end{figure}

It is known that QCD matter exhibit different properties from high density quark matter to nuclear matter densities. For quark matter the theory is approximately scale invariant and then there is conformal symmetry. However, for hadronic matter, the chiral symmetry is broken by a conformal anomaly introducing a scale that determines the size and mass of hadrons, except for the pseudoscalar mesons. Therefore, the EoSs of nuclear matter (NM) and QM are qualitatively distinct. This fundamental difference is reflected in some properties of the EoS, which are proposed as the signature of conformality in neutron stars. Examples of these  signatures are the normalized trace anomaly ($\Delta=(\varepsilon-3P)/3\varepsilon$) \cite{Fujimoto:2022ohj}, its logarithmic rate of change with respect to the energy density $\Delta'= d\Delta/dln\varepsilon$ \cite{Fujimoto:2022ohj}, $c_s^2$, the polytropic index $\gamma=dlnP/dln\varepsilon$ \cite{Annala:2019puf}, the change of the sign of the curvature of the  energy per particle $\beta$ \cite{Marczenko:2023txe} and the pressure normalized by that of a system of free quarks \cite{Annala:2023cwx}, $P/P_{free}$, among others. As the scale invariance becomes restored in QCD, $c_s^2 \rightarrow 1/3$, $\Delta \rightarrow 0$, $\Delta' \rightarrow 0$,$\gamma \rightarrow 1$ and $\beta$ changes the sign from positive to negative. Some of these properties are shown in Fig. \ref{fig:conformalparameters}.

Given all this variety of properties, in \cite{Annala:2023cwx} it is defined a new parameter called conformality ($d_c$) with the goal of tracking several of these properties at the same time while ensuring conformality at very high densities:
\begin{equation}\label{conformality}
    d_c= \sqrt{\Delta^2+ \Delta'^2}
\end{equation}
Since both $\Delta$ and $\Delta'$ are related to $c_s^2$ and $\gamma$,
\begin{equation}
    \Delta=\frac{1}{3}-\frac{c_s^2}{\gamma},\ \Delta'=c_s^2(\frac{1}{\gamma}-1)
\end{equation}
implying that when both $|\Delta|$ and $|\Delta'|$ are small, $\gamma$ and $c_s^2$ also approach their conformal limits 1 and 1/3, respectively. In
addition, a small value of $|\Delta'|$ naturally ensures that $\Delta$ will remain approximately constant at higher densities \cite{Annala:2023cwx}.
Therefore, a way to identify possible phase transition is by analyzing the shape of $c_s^2$,  $\Delta$ and $d_c$.

\vspace{0.5cm}
\textit{Speed-of-sound}
\\
\\
Inferred sound velocity results from different Bayesian analyses are displayed in Figs. ~\ref{fig:BrandesBand}, \ref{fig:Koehnband} and \ref{fig:abinitioGordaresults}, showing that astrophysical constraints require a stiffening of the EoS while
pQCD calculations require a softening of the EoS. This interplay between both constraints yields a peak structure in the sound speed of NS matter. In searching for this peak, a major challenge is how to extend the current EoS describing neutron stars (which terminates at the central density) to the pQCD regime, since the pQCD constraints require that all EoS must match these results. As the pQCD constraint becomes more restrictive at higher matching densities, two points have been proposed to extend the EoS to the pQCD regime: $n_{\text{term}} =n_{\text{TOV}}$ and $n_{\text{term}} = 10n_{s}$. The results obtained are different, clearly demonstrating how this termination density affects the constraining potential of the QCD input.

Thus, in \cite{Brandes:2023bob,Brandes:2023hma}, the authors use $n_{\text{term}} =n_{\text{TOV}}$ (considered as the most conservative choice) as the matching point and look for the possible appearance of a FOPT by using a very low squared sound velocity ($c_s^2 \leq$ 0.1), within the range of energy densities relevant for NS. These authors find that, at intermediate energy densities ($\varepsilon \sim 600$ MeV/fm$^{3}$), the speed of sound starts to form a plateau that extends up to higher energy densities. This behaviour of $c_s^2$ is reflected in the pressure, which  rises approximately linearly with increasing energy density. They quantify the evidence against FOPT by applying the Bayes factor obtaining strong evidence against it, not finding a peak in $c_s^2$ but the aforementioned plateau for $M\leq 2.1 M_\odot$ (taking into account the most massive pulsar, the black widow pulsar PSR J0952-0607). Moreover, the authors claim that small sound speeds $c_s^2 \leq$ 0.1 in the cores of NS with even higher masses, $M \gtrsim 2.1 M_\odot$, less constrained by the currently available astrophysical data, cannot be firmly excluded \cite{Brandes:2023bob}. These results agree with those from \cite{Alarcon:2024hlj}, where the EoS are first generated applying pQCD constraints and later constrained by astrophysical data, including explicitly FOPT and the pulsar PSR J0952-0607. Fig. \ref{fig:PTs} (left) shows four examples of EoS within $68\%$ CL \cite{Alarcon:2024hlj} with the same phase transition at $P= 167$ MeV/fm$^3$ and showing different ways to reach pQCD. The EoS represented by the yellow line gives rise to a twin-star configuration for a $M=2.17 M_\odot$. At the right plot of Fig. \ref{fig:PTs} three EoSs  (68$\%$ CL) are shown with several phase transitions that start at densities inside the NS. 

On the other hand, as a result of non-parametric interpolation, the authors of \cite{Gorda:2022jvk} find  that the effect of the QCD constraint becomes dominant at densities above $\varepsilon \sim$ 750 MeV/fm$^{3}$, including the black hole (BH) hypothesis (450 MeV/fm$^3$, if not including the BH hypothesis), indicating a softening of the EoS and finding a small reduction in the maximum TOV NS mass. These results have been obtained by extrapolating the $\chi$EFT EoS  to $n_{\text{term}} = 10n_s$ by a Gaussian process (GP) regression and conditioning the prior by NS observations and QCD input. The termination point $n_{\text{term}} = 10n_s$ is justified by \cite{Gorda:2022jvk,Komoltsev:2023zor} arguing that there are situations, such as binary neutron star mergers or differentially rotating neutron stars, where the maximum central density exceeds the TOV density. These results agree with \cite{Somasundaram:2021clp}, but both do not include FOPT. 

There is then a strong dependence of QCD constraining power on the choice of termination density. The reason for this strong dependence on the termination point is related to the way pQCD constraints are applied in Bayesian analyses that is as follows. 
If the condition $\Delta p_{min} \leq p_{High}- p_{\text{term}} \leq \Delta p_{max}$ is true, it is possible to connect this EoS with the pQCD prediction, and a likelihood of 1 is assigned to it. 
If it is not true, this EoS has a likelihood of zero. This likelihood gives equal weight to all possible ways of connecting the pre-existing EoS  with the pQCD prediction,  allowing for extreme EoS behaviors at $n_{\text{term}}=n_{\text{TOV}}$, such as long FOPT at the $n_{\text{term}}$ density.  This is considered the most conservative criterion for satisfying the pQCD constraints, since roughly 20$\%$  of the EoSs cannot be connected to pQCD for all values of $X$ \cite{Komoltsev:2023zor}. 
However, for $n_{\text{term}}=10n_s$, the pQCS constraints are more demanding (roughly 50  60$\%$ of EoS have a vanishing QCD likelihood) \cite{Komoltsev:2023zor}, and these extreme EoS behaviors are ruled out, due to smooth prior in Bayesian inferences. Furthermore, considering such a higher matching density beyond the range of control by data, the impact is expected to depend sensitively on choice of priors
in the unconstrained interpolation region \cite{Essick:2023fso} (depending in turn on the $\chi$EFT limits).
Another interesting point is that in the case of this strong FOPT at n$_{\text{TOV}}$, it does not destabilize the NS and then the location of the TOV density is not determined.
 
In summary, the pQCD likelihood can be integrated into different EoS inference models, but this approach requires an arbitrary selection of the density at which the constraints are applied. The EoS behavior is also treated differently on either side of the chosen density. Thus, the results strongly depend on the termination density of the EoS, with EoS requiring drastic softening beyond the termination density, followed by a high sound speed segment to reach the high density limit, being particularly sensitive. Methods that penalize such extreme behavior above the TOV density introduce additional model dependence.  
Therefore, observations of stars near their maximum mass are required to constrain this area effectively as well as from the post-merger GW signal.

\vspace{0.45cm}
\begin{figure}
    \centering
    \includegraphics[width=0.54\linewidth]{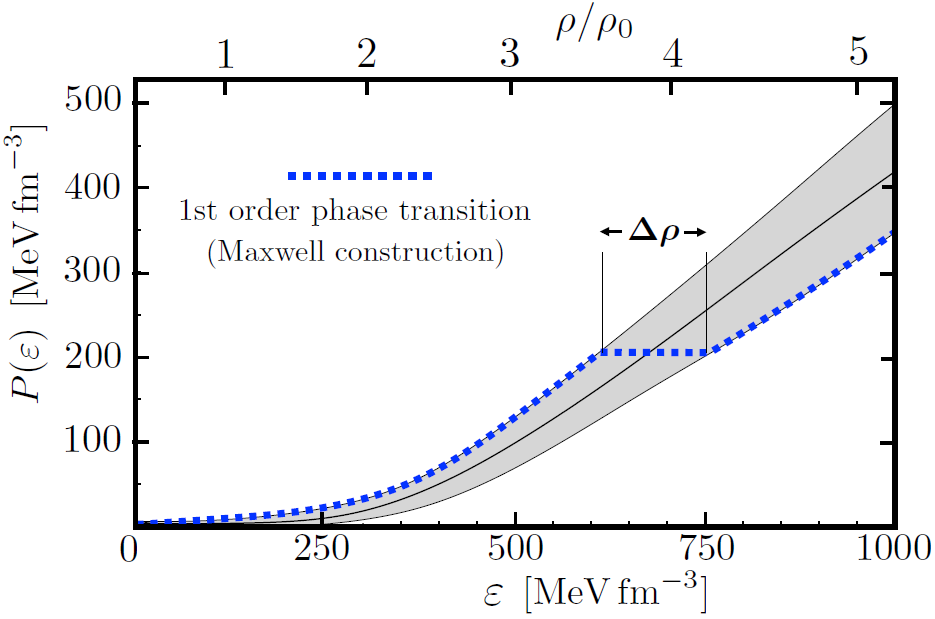}
    \includegraphics[width=0.45\linewidth]{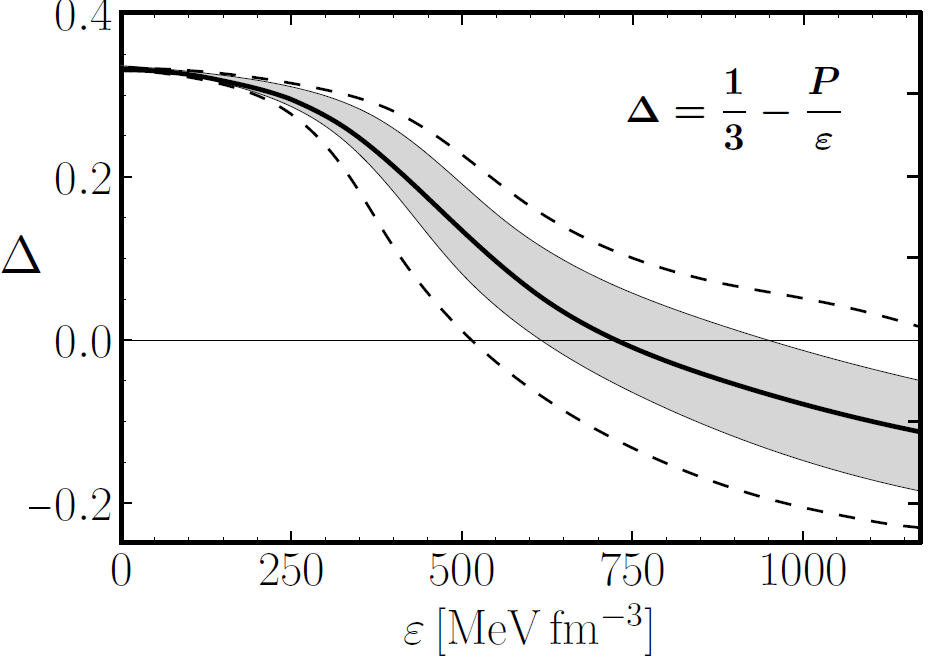}
    \caption{{\bf Left:} Illustration of the constraint on the maximum width of a Maxwell-constructed coexistence
region for a FOPT within the 68$\%$ CL of $P(\varepsilon)$ \cite{Brandes:2023bob}. The upper (baryon
density) scale refers to the median of the $P(\varepsilon)$ distribution. {\bf Right:} Median and confidence bands \cite{Brandes:2023bob} of the trace anomaly measure $\Delta$, at the 68$\%$ level
(gray band) and 95$\%$level (dashed lines)\emph{Reprinted from \cite{Brandes:2023bob} under Creative Commons License.}}. 
    \label{fig:FOPTBrandes}
\end{figure}

\textit{Trace anomaly}
\\
\\
The conformal symmetry also leaves a distinct signature on the trace anomaly $\Delta=\frac{1}{3} -\frac{P}{\varepsilon}$ of the EoS. Conformal matter is characterized by an EoS $P=\varepsilon/3$, so $\Delta \rightarrow 0$. 
In general, requiring the trace anomaly to satisfy causality ($P \leq \varepsilon$) and thermodynamic stability ($P > 0$) leads to the following allowed range for $\Delta$: \begin{equation} 
- \frac{2}{3} \leq \Delta \leq \frac{1}{3}. 
\end{equation}
The maximum value, $\Delta = 1/3$, is reached for $P = 0$ at the surface of the neutron star (NS), while the minimum value depends on the equation of state and on the central values ($\varepsilon_c, P_c$) for the maximum mass obtained by solving the Tolman-Oppenheimer-Volkoff (TOV) equations in General Relativity (GR).\\
The trace anomaly can become negative for very repulsive interactions with $c_s^2 > 1/3$, a consequence of the increased stiffness necessary to achieve massive NSs in GR. Recent statistical and Bayesian analyses \cite{Ecker:2022dlg,Marczenko:2022jhl,Annala:2023cwx,Brandes:2023hma,Takatsy:2023xzf,Jimenez:2024hib} have found evidence of a negative trace anomaly at high densities, challenging the conjecture $\Delta \geq 0$ within NSs \cite{Fujimoto:2022ohj} (see Ref.~\cite{Cai:2024oom} for another kind of approach). An illustrative example of these results is shown to the right of Fig. \ref{fig:FOPTBrandes}. Starting from $\Delta$ = 1/3 with zero energy density, the trace anomaly decreases with increasing energy density, its median becomes negative at $\varepsilon \sim$ 700 MeV/fm$^3$, entering a high-pressure domain with $P > \varepsilon/3$. This crossover occurs within the range of energy densities possibly reached in NS cores. Applying the Bayes factor, the authors of \cite{Brandes:2023bob} find moderate evidence that $\Delta$ becomes negative within neutron stars. In \cite{Brandes:2023hma}, an additional study is realized for first-order phase transitions with Maxwell construction. Such FOPT are characterized by a domain of phase coexistence that extends over a
certain range of baryon densities, $\Delta n$. The width of the domain of phase coexistence, $\Delta n/n$ (with $n$ the density at which the coexistence interval starts), is a measure of the strength of the phase transition, i.e., the magnitude of surface tension between the two coexisting phases. They refer to a
"strong" FOPT if $\Delta n/n >$ 1. In contrast, a "weak" FOPT has a value of $\Delta n/n < 1$. 
The posterior credible bands inferred from NS data, as displayed in the top right of Fig. \ref{fig:BrandesBand}. These results constrain the maximum possible widths of mixed-phase domains, $\Delta n/n$, showing an example in the left of Fig. \ref{fig:FOPTBrandes}.  They find that these maximum possible phase coexistence regions is narrow $(\Delta n/n)_{\text{max}} \approx$ 0.2 
 at the 68$\%$ level and  $(\Delta n/n)_{\text{max}} \lesssim$ 0.3 at the 95$\%$ level. In fact, they find that $(\Delta n/n)_{\text{max}}$ stays nearly constant as a function of $n$  (taken at the starting point of the mixed-phase interval) over the whole region of
densities $n \approx 2 - 5 n_s$ that are relevant for neutron stars. Therefore, a strong first-order phase transition is very unlikely to occur.
The discrepancies between \cite{Brandes:2023bob} and \cite{Gorda:2022jvk,Annala:2023cwx} again arise both from the way the pQCD constraints are applied and from the values of the maximum masses used.\\

\vspace{0.5cm}
\textit{Conformality and pressure normalised by free pressure}
\\
\\
\begin{figure}
    \centering
    \includegraphics[width=0.45\linewidth]{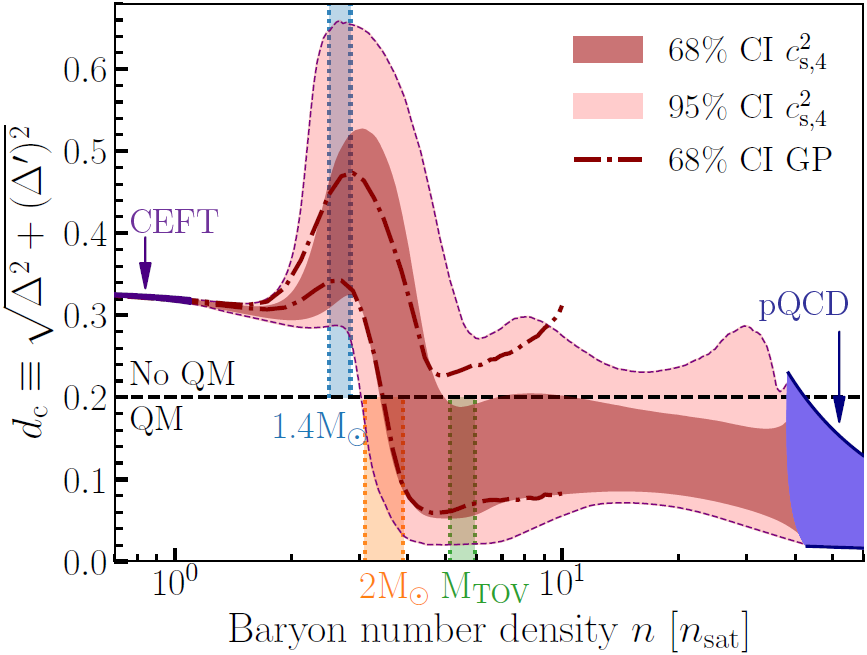}    \includegraphics[width=0.48\linewidth]{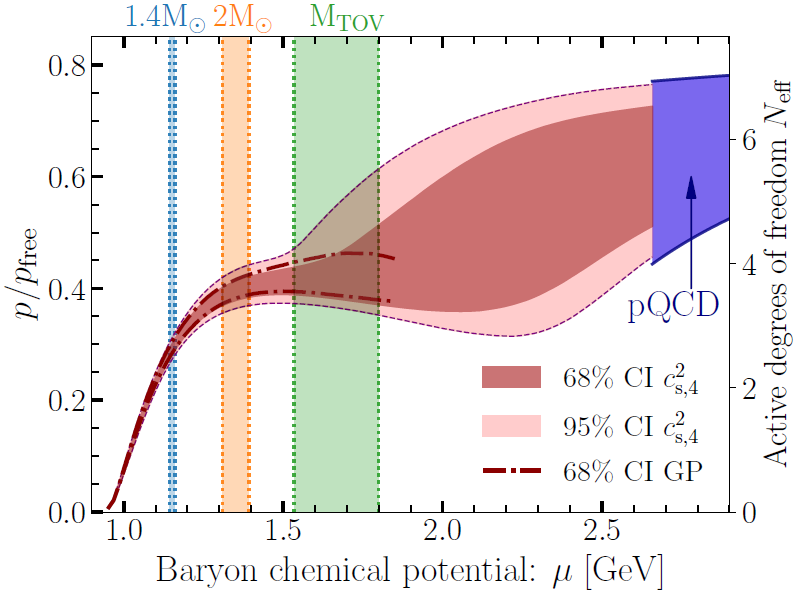}    
    \caption{{\bf Left:} The conformal parameter $d_c$, with a value of 0.2 shown as a black dashed line, plotted as a function of number density. The dark and light
bands represent the 68$\%$ and 95$\%$ confidence intervals (CIs) obtained using a four-segment sound speed interpolation ($c_{s,4}^2$). In addition, the 68$\%$ CI obtained from the GP ensemble is shown with a red dash-dotted line. The colored bands correspond to the 68$\%$ CI for the central densities of different masses. {\bf Right:} Pressure normalized to that of a free Fermi gas of quarks  as a
function of chemical potential. The dark and light bands represent the 68$\%$ and 95$\%$ CIs obtained using $c_{s,4}^2$. Additionally, the $68\%$ CI obtained from the GP ensemble is
shown with a red dash-dotted line. The colored bands correspond to the $68\%$ CI for the central densities of different masses. \emph{Reprinted from \cite{Annala:2023cwx} under Creative Commons License.}}
    \label{fig:conformality}
\end{figure}

Regarding the conformality parameter $d_c$ the authors of \cite{Annala:2023cwx} adopt the criteria $d_c <2$, justifying this value taking into account a FOPT ($c_s^2=0=\gamma$ and $\Delta'=1/3-\Delta$). The values for this parameter using a GP regression are shown in the left of Fig.~\ref{fig:conformality}, exhibiting a qualitative change in its behavior around $n\sim  2 - 3n_s$, indicating a softening of the EoSs.
However, not all matter that is near-conformal is QM and the conformalization does not ensure the presence of the deconfined phase. Thus, in \cite{Annala:2023cwx}, the authors also use the pressure normalized by the free Fermi gas, $P_{\text{free}}=(3/4\pi^2) (\mu_B/3)^4$, where $\mu_B$ is the barionic chemical potential), which is not fixed by conformal symmetry. This normalized pressure, which is proportional to the number of active degrees of freedom, $N_{\text{eff}}$, can be used to differentiate deconfined quark matter from other types of near-conformal behavior. Fig. \ref{fig:conformality} (right) shows the results for this for the normalized pressure, which is flattened at approximately $P/P_{\text{free}} =0.4\pm 0.03$ (about two-thirds of the pQCD value, or $N_{\text{eff}}=$1). This value is consistent with that of weakly interacting quark matter.

In summary, \cite{Annala:2023cwx,Komoltsev:2025vwn} finds that the matter in the cores of the most massive NSs exhibits near-conformal behavior as well as the effective number of active degrees of freedom flattens out around the TOV density, consistent with weakly coupled quark matter. These results are obtained by using speed-of-sound interpolation (4 segments) and  nonparametric GP regression between $n=1.2n_s$ ($\chi$EFT limit) and $10n_{\text{TOV}}$.

\subsection{Matching chiral equations of state with crustal equations of state.}

\begin{figure}
\centering 
\includegraphics[width=0.47\linewidth]{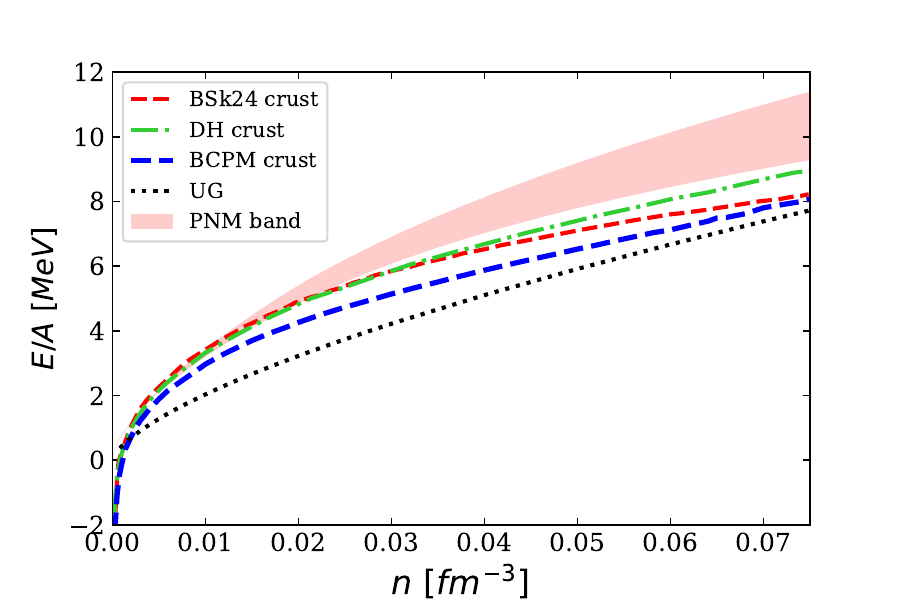} \includegraphics[width=0.47\linewidth]{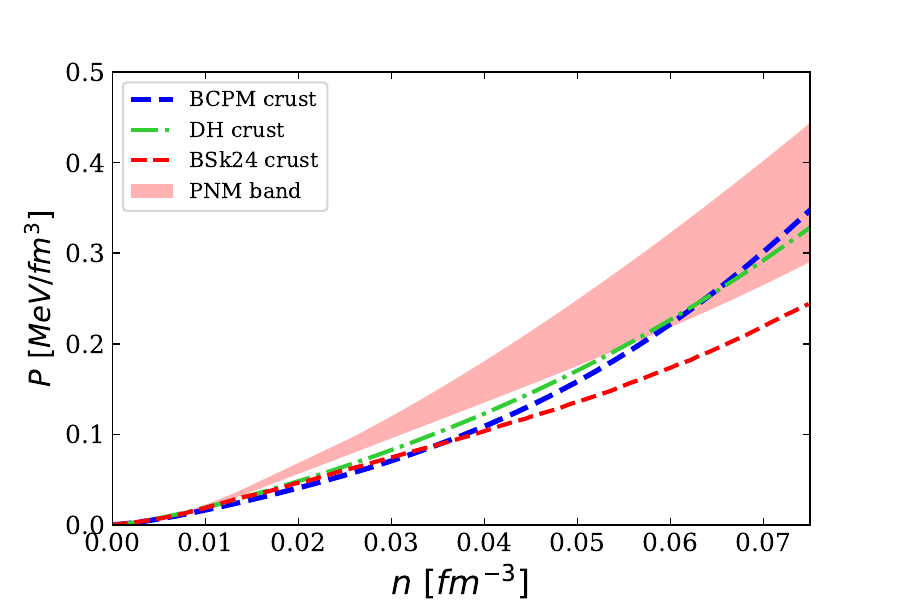}
    \caption{{\bf Left:} $E/A (n)$ EoS of the crust of DH, BSk24 and BPCM  EoS at densities up to half the saturation density ($n \leq 0.08$ fm$^{-3}$), as well as the PNM band at the same densities. {\bf Right:} $P(n)$ for the same EoS and PNM band.}
    \label{fig:CRusthalfdensity}
\end{figure}

\begin{figure}
\centering
\includegraphics[width=0.48\linewidth]{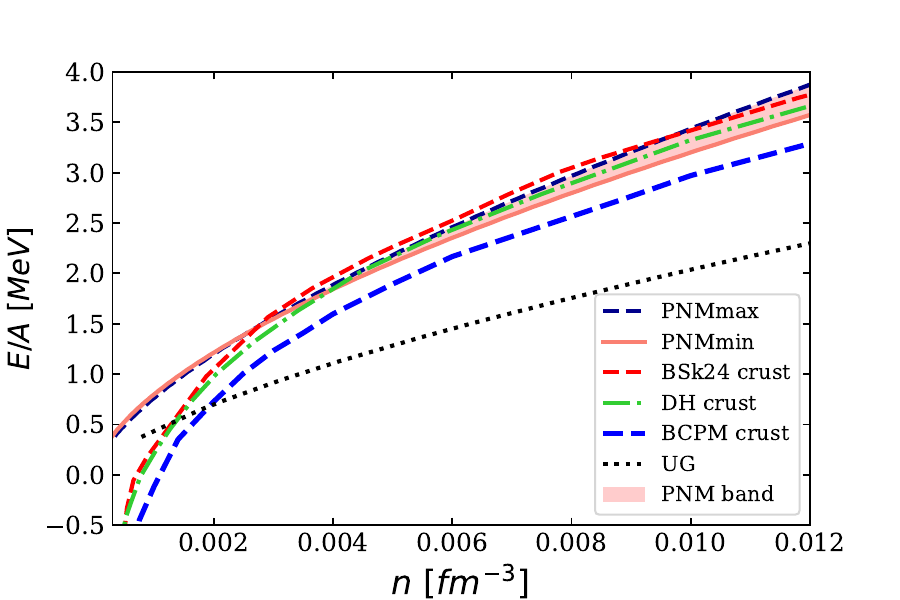}
\includegraphics[width=0.48\linewidth]{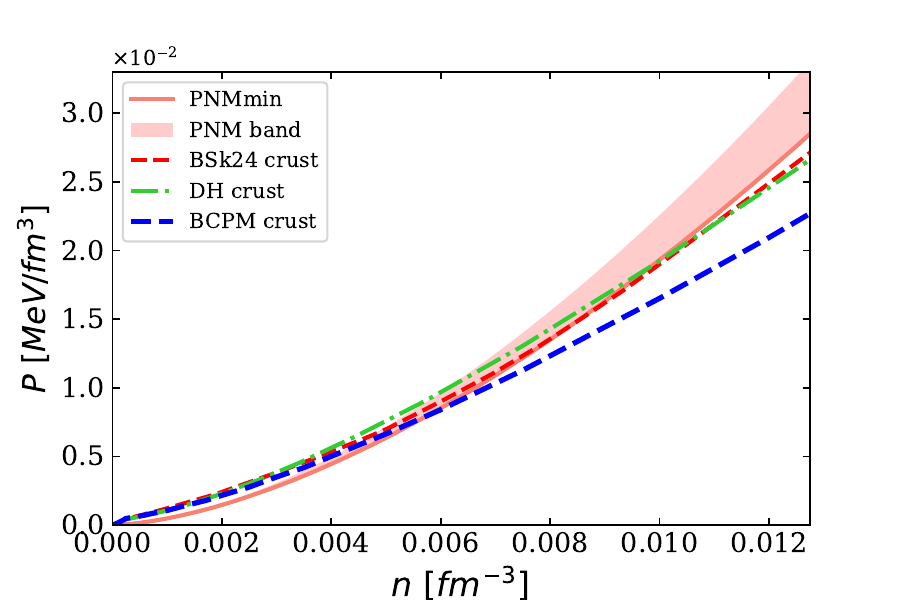}
    \caption{{\bf Left:} $E/A (n)$ of the crust of DH, BSk24 and BPCM EoS, as well as the PNM band at low densities. {\bf Right:} P(n) for the same EoS and PNM band.}
    \label{fig:comparacrust}
\end{figure}

EoS based on ab-initio many-body methods, such as $\chi$EFT EoS, describe pure neutron matter and so they are limited at low and moderate densities. To extend these EoS at higher densities different interpolation methods are used. However, the question arises about how to extend these EoS to densities where matter is not homogeneous. A common way to do that is by adding a crustal EoS.
To glue two EoS, all thermodynamic quantities should be matched: pressure $P$, energy density $\varepsilon$, and baryonic density $n$, so that the thermodynamic consistency is fulfilled \cite{Fortin:2016hny}.

The main obstacle one encounters when addressing the matching of a crustal EoS to a core EoS is that both EoS use different models. While the chiral EoSs use ab initio methos and have an uncertainty band, the crustal EoS use phenomenological models and provide no uncertainty band. Furthermore, the matching between the two EoS must be thermodynamically consistent, otherwise it would lead to artificial uncertainty in the radius, which can be as large as $4\%$, as the expected accuracy of current and future X-ray telescopes \cite{Suleiman:2021hre}.

Our proposal here is to add the crust to the EoS by taking the PNM result obtained at densities $n \leq n_s$ and that were constrained by nuclear experiments \cite{Alarcon:2024hlj}. For the crust, we consider the following EoSs:
\begin{itemize}
    \item The DH-SLy4 calculation  \cite{Douchin:2001sv}: It is constructed using the Skyrme effective nuclear forces SLy4 \cite{Chabanat:1997un}. This force was parametrized, among other constraints, to be consistent with the microscopic variational calculation of neutron matter of \cite{Wiringa:1988tp} above the nuclear saturation density. The inner crust is described using the Compressible Liquid Drop Model (CLDM)  \cite{Douchin:2001sv}. The parameters of the model are calculated using the SLy effective interaction. Their calculations have been limited to $\rho > 10^6$ g cm$^{-��3}$ and adjusted to experimental masses of neutron-rich nuclei.

\item BSkP24 (from Brussels-Montreal group) \cite{Pearson:2018tkr}: The matter is described using  the energy density functional family BsK. These functionals are based on generalized Skyrme-type forces and contact pairing forces. The parameters of these functionals were determined by fitting to known masses of nuclei, with the nuclear masses being calculated using the Hartree-��Fock-��Bogoliubov (HFB) method. To calculate the EoS of the inner crust they  use the Extended Thomas-Fermi approach  with the same functional as in the outer crust, including perturbatively shell corrections for protons via the Strutinsky integral method. The functional BSk24 was fitted to a symmetry energy coefficient S = 30 MeV.

\item 	BPCM [Barcelona-Paris-Catania-Madrid group] \cite{Sharma:2015bna}. The BCPM  nuclear energy density functional is used in the crust calculations. This functional has been obtained from the ab initio Brueckner-Hartree-Fock calculations (BHF) in nuclear matter within an approximation inspired by the Kohn-Sham formulation \cite{Kohn:1965zzb}.
The NS crust is modeled in the Wigner-Seitz (WS) approximation. To compute the outer crust, the masses of neutron-rich nuclei are taken from experiment (if they are measured), and deformed Hartree-Fock-Bogoliubov (HFB) calculations are performed with the BCPM energy density functional when the masses are unknown. To describe the inner crust, Thomas-Fermi calculations in Wigner-Seitz cells are performed with the same functional in different periodic configurations (spheres, cylinders, slabs, cylindrical holes, and spherical bubbles). 
\end{itemize}

\begin{figure}
    \centering
    \includegraphics[width=0.8\linewidth]{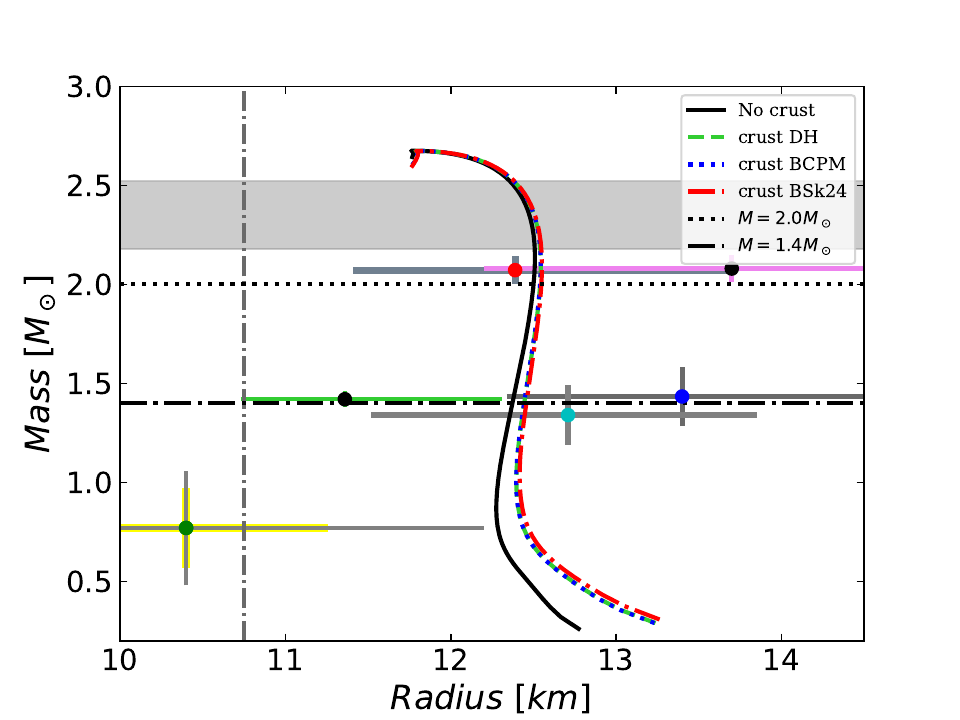}
    \caption{Mass-Radius diagram of the stiffest EoS (68$\%$ CL) without crustal EoS (black solid line) and matching with the three cases of crustal Eos: DH (green dashed line), BPCM (blue dotted line) and BsK24 (red dash-dotted line)}.  
    \label{fig:MvRthreecrust}
\end{figure}

Fig. \ref{fig:CRusthalfdensity} shows the EoS ($E/A(n)$ left plot, $P(n)$ right plot) of these three EoS as well as the PNM band from \cite{Alarcon:2024hlj}. Some authors \cite{Fortin:2016hny,Suleiman:2021hre} recommend matching the crustal and core EoS to half the saturation density ($n=0.08 ~$fm$^{-3}$), as the best region to make the different EoS thermodynamically consistent. However, Fig. \ref{fig:CRusthalfdensity}  shows the incompatibility of the three EoS with the PNM band, so it is not possible to glue the crustal EoS with the PNM band at the recommended density values.
The next step is to find a region of compatibility of the $n$, $E/A$ and $P$ values for both crustal and core EoS. This region is finally found in the innercrust, at lower densities $n<0.01$ fm$^{-3}= 6\times 10^{-2} n_s$, below the limit of the original PNM band of \cite{Alarcon:2022vtn}, as shown in Fig. \ref{fig:comparacrust}. The figure on the left shows that the $E/A$ values of the three crustal EoS are lower than those of the PNM band up to densities corresponding to the neutron drip line ($n\sim 3 \times 10^{-3}$ fm$^{-3}$), showing a $P(n)$ dependence stiffer than the corresponding PNM EoSs. Above this point there is a clear common region of compatility for the crustal EoS DH and BSk24 with the PNM band for $4\times 10^{-3}\leq n \leq 1.2 \times 10^{-2}$ fm$^{-3}$, while for the BPCM EoS, the $E/A$ values are lower than those of the PNM band over the whole range of densities.

To compare the three crust EoSs, a good common point to match these EoS to the minimum PNM EoS (PNM$_{Min}$) is $n \sim 5\times 10^{-3}$ fm$^{-3}$, although gluing with BPCM EoS requires a small increase in $E/A$. The results for the Mass-Radius diagrams corresponding to the three crustal EoS at this matching point are shown in Fig. \ref{fig:MvRthreecrust}. Since the crustal EoSs are stiffer than the PNM EoS up to the drip line point, the corresponding mass-radius diagrams of the resulting crust-PNM EoS are shifted to the right, showing slightly larger radii at low masses (for $M= M_\odot$, $\Delta R \leq  130$~m, for BSk24-PNM$_{Min}$ EoS and $111$~m for BPCM-PNM$_{Min}$ EoS and so, there are no significant differences between the three crust EoS.

\begin{figure}
    \centering
    \includegraphics[width=0.45\linewidth]{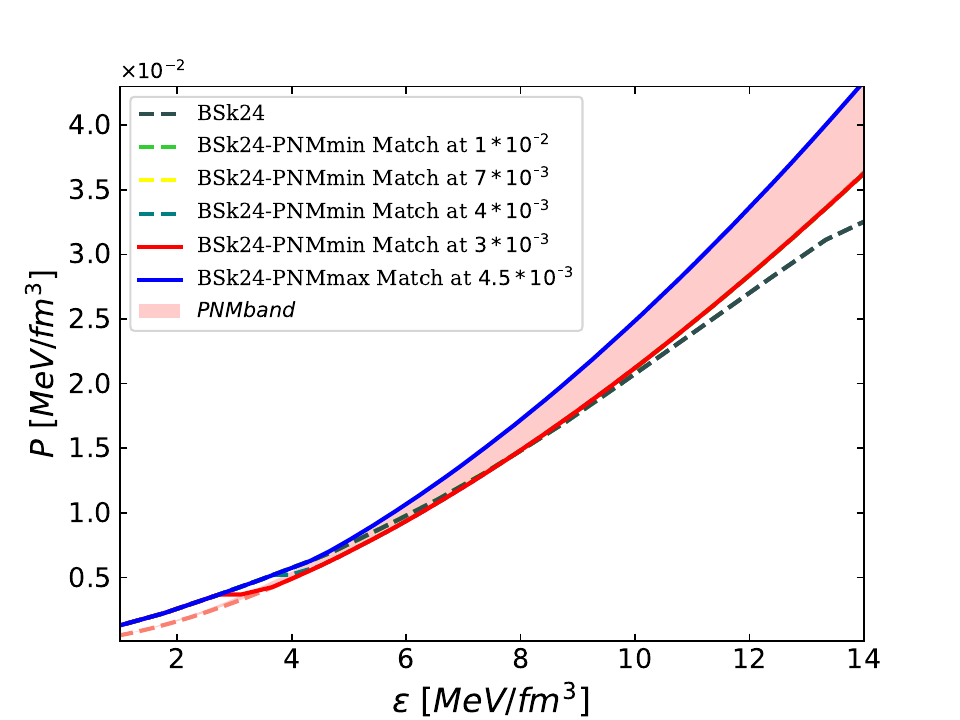}
    \includegraphics[width=0.45\linewidth]
{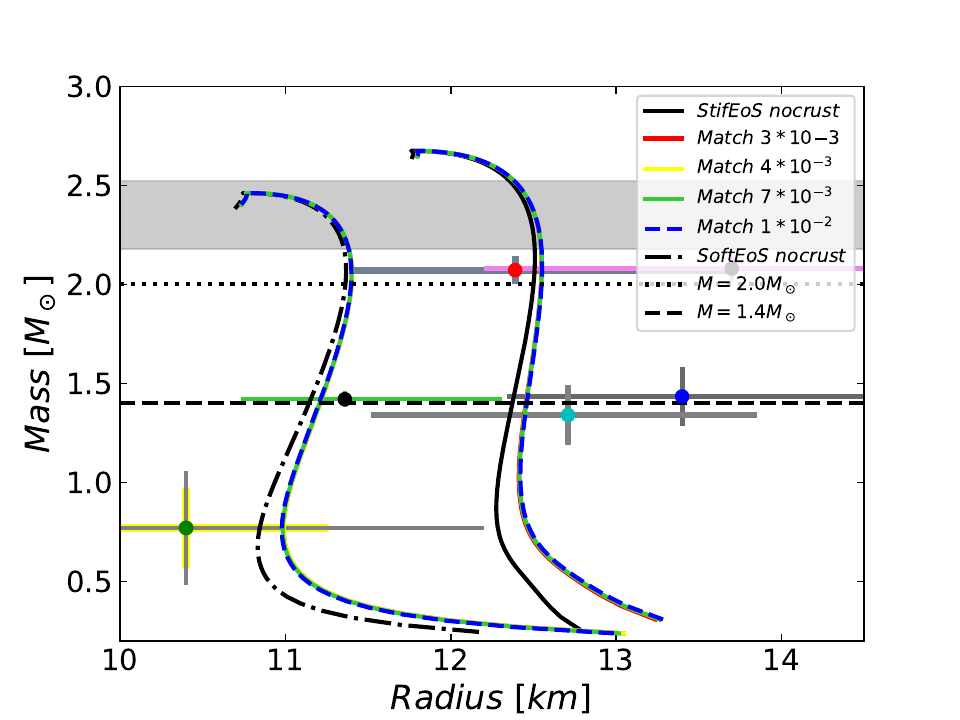}
    \caption{The resulting band (left), between the red and blue lines, and various M-R diagrams (right) when matching  the BSk24 crustal EoS, with the maximum EoS of the PNM band and with the minimum EoS of the PNM band for a density range $ 3 \times 10^{-3} \leq n\leq 10^{-2}$ fm$^{-3}$.}
    \label{fig:crustOllerband}
\end{figure}

Focusing on the most consistent EoS (DH and BSk24), we can see a common region of compatibility with the maximum PNM EoS (PNM$_{Max}$) at $4\times 10^{-3} \leq n \leq 6\times 10^{-3}$ fm$^{-3}$, and at $4\times 10^{-3} \leq n < 10^{-2}$ fm$^{-3}$  with PNM$_{Min}$.   
Fig. \ref{fig:crustOllerband} (left) shows the resulting crust band by matching BSk24 with the PNM band (left). The matching with the maximum EoS of the PNM band has been performed at $n=4.5\times 10^{-3}$ fm$^{-3}$, while the matching with the minimum EoS of PNM band has been performed at $3\times 10^{-3} \leq n\leq 10^{-2}$~fm$^{-3}$. Fig. \ref{fig:crustOllerband} (right) shows the Mass-Radius diagrams applying the crust BsK24 of various matching points to the stiffest EoS (solid black line) and the softest EoS (dashdotted black line) at 68$\%$ of the band of Fig. \ref{fig:totalband} (top right), with no significant differences between radius.
The results of matching DH EoS with PNM EoS are very similar to those obtained for the matching of BsK24-PNM EoS.

\section{Neutron Stars for Searches of New Physics}
\label{Sec:Neutron_Stars_for_Searches_of_New_Physics}
Dark matter is an hypothetical form of invisible matter that exerts gravitational effects on light and ordinary matter. 
Due to its weak interaction with other Standard-Model particles, experiments of both direct detection (attempting to register 
scattering of dark matter particles off target nuclei) and indirect detection (relying on the products of dark matter interactions) 
have so far been unsuccessful. Because of their extreme densities with conditions unattainable in Earth laboratories, neutron stars
have proved to be excellent laboratories for searches for new physics. These compact objects (as well as white dwarfs and black holes) 
are said to be able to accrete dark matter particles. The presence of dark matter inside neutron stars modifies the properties of a pure 
baryonic object, such as surface temperature (\cite{Kouvaris:2007ay,deLavallaz:2010wp}), changes in the rotational pattern (\cite{Kouvaris:2014rja}), 
X-ray pulse profiles (\cite{Miao:2022rqj,Yadav:2024xob}),  gravitational wave waveform (\cite{Kopp:2018jom,Giangrandi:2025rko}), and changes in the 
mass-radius relation and tidal deformabilities (\cite{Zhang:2020pfh,Ellis:2018bkr,Routaray:2022acz}), which gives us an opportunity to study different 
scenarios of dark matter, the two last observables being the most sought after effects by most authors. It is, however, important to note a few challenges that 
studying these two observables present. For changes in the mass-radius relation to be noticeable, the dark matter content bounded to the star would have to be at 
least of the order of $1\%$, which requires an anomalously high dark matter ambient density. The minimum
 amount of dark matter required to significantly modify the mass-radius diagram is dependent on the proposed dark matter model. Studying tidal 
deformabilities presents a similar challenge: heavy non-repulsive dark matter particles tend to form a dark core which  would decrease 
tidal deformability (with a high enough dark matter content). However, self-repulsive light particles can form an extended dark 
matter halo able to increase $\Lambda$ of the hybrid star even with a dark matter content low enough to be allowed by 
capture constraints\cite{Nelson:2018xtr}.\\

\subsection{Static observables: Two fluid generalization}

To study the properties of the resulting star formed by the admixture of dark and baryonic matter, the Tolmann-Oppenheimer-Volkoff 
in the two-fluid formalism was introduced by Sandin \& Ciarcelluti \cite{Sandin:2008db}. Due to the feeble interactions between 
dark and Standard-Model particles, it is possible to only consider the gravitational one, so that each fluid satisfies conservation 
of energy-momentum separately. At hydrostatic equilibrium the star has to satisfy the following conditions:
\begin{gather} \label{TOVpressure}
    \frac{dP_{BM} }{dr} = -(P_{BM}+\varepsilon_{BM})\frac{4\pi r^3(P_{BM} + P_{{DM}})+m(r)}{r(r-2m(r))} \\
    \frac{dP_{DM}}{dr} = -(P_{DM}+\varepsilon_{DM})\frac{4\pi r^3(P_{BM} + P_{{DM}})+m(r)}{r(r-2m(r))}
\end{gather}
where $P$ is the pressure, $\varepsilon$ the energy density, and the mass of each component and the total mass are given by
\begin{gather} 
    \frac{dm_i(r)}{dr} = 4\pi\varepsilon_i(r)r^2\\
    \label{TOVmass}
    m(r) = m_{BM}(r)+m_{DM}(r)
\end{gather}
where the subindex $i$ refers to one of the fluids. As well as in the one-fluid formalism, in order to solve this system of coupled differential equations (usually with the 4-th order Runge-Kutta algorithm) it is necessary 
to establish the initial conditions of the pressure and mass of each component at the center of the star, as well as to provide 
both equations of state.\\
\\
For a Dark Matter Admixed Neutron Star (DANS), the TOV equations do not yield a mass radius curve, but instead multiple star configurations (each for one combination of central pressures) for which one can plot different curves based on different criteria (the curve for a fixed ratio of central pressures would be different to the curve for a fixed fraction of dark matter) so one cannot naively use the maximum mass criteria for the onset of instability. For a two fluid star, this change from stable to unstable configurations happens when the number of particles for each component remains stationary under variations of both central energy densities $\varepsilon_i^c$:
\begin{equation}
\label{stability}
    \begin{pmatrix}
        \delta N_{BM}\\
        \delta N_{DM}
    \end{pmatrix}= 
    \begin{pmatrix}
        \partial N_{BM}/\partial\varepsilon_{BM}^c & \partial N_{BM}/\partial\varepsilon_{DM}^c \\
        \partial N_{DM}/\partial\varepsilon_{BM}^c & \partial N_{DM}/\partial\varepsilon_{DM}^c
    \end{pmatrix}
    \begin{pmatrix}
        \delta\varepsilon_{BM}^c \\
        \delta\varepsilon_{DM}^c
    \end{pmatrix}=0
\end{equation}
For this equation to have non-trivial solutions it is necessary that the matrix's determinant equals zero, so stable configurations are only allowed when both eigenvalues are positive. In the case of a single fluid, the determinant condition gives $\partial N/\partial \varepsilon^c = 0$, which, if the TOV equations are satisfied, is equivalent to the known $\partial M/\partial \varepsilon^c = 0$ criteria for the onset of instability.\\
\\
The expression of the tidal deformability \eqref{eq:tidal} can also be extended to include a second fluid with the below modifications \cite{Barbat:2024yvi}:
\begin{align}
    F(r) &= \frac{r-4\pi r^3[(\varepsilon_{BM}+\varepsilon_{DM})-(P_{BM}+P_{DM})]}{r-2(m_{BM}+m_{DM)}}\\
    \begin{split}
        Q(r) &= \frac{4\pi r}{r-2(m_{BM}+m_{DM})}\\
          &\times\bigg[5(\varepsilon_{BM}+\varepsilon_{DM})+9(P_{BM}+P_{DM})+\frac{\varepsilon_{BM}+P_{BM}}{c_{s,BM}^2} +\frac{\varepsilon_{DM}+P_{DM}}{c_{s,DM}^2} - \frac{6}{4\pi r^2} \bigg] \\
        &-4\bigg[\frac{(m_{BM}+m_{DM})+4\pi r^3(P_{BM}+P_{DM})}{r^2-\big(1-\frac{2(m_{BM}+m_{DM})}{r}\big)} \bigg]^2
    \end{split} 
\end{align}
where $c_{s,i}^2(r)=dP_i/d\varepsilon_i $ is the squared speed of sound for i=BM,DM.\\

\subsection{Dark matter effects on static observables}

\begin{figure}[h!]
    \centering
    \includegraphics[width=0.6\linewidth]{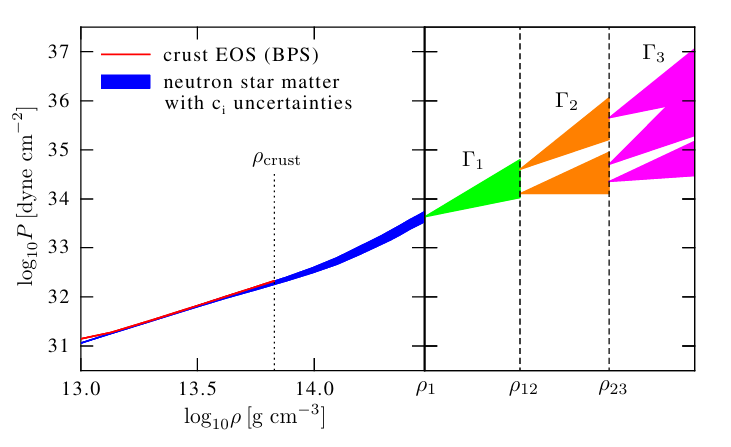}
    \caption{The left part compares the pressure band predicted by neutron matter results to the BPS EoS \cite{Baym:1971pw}. 
    The right part shows the piecewise polytropic extension to higher densities. This plot corresponds to \cite{Hebeler:2013nza}. Reproduced by permission of the AAS.}
    \label{Polytropic hebeler}
\end{figure}

\begin{figure}[h!]
    \centering
    \includegraphics[width=0.8\linewidth]{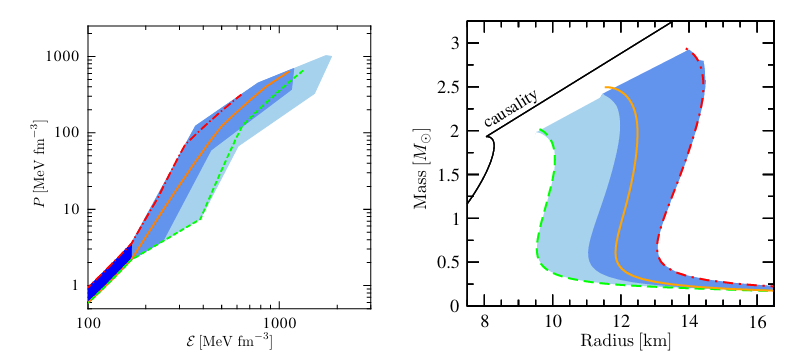}
    \caption{On the left panel the three representative EoS compared to the uncertainty band. 
    On the right panel the corresponding neutron star mass-radius results. Figures taken from \cite{Hebeler:2013nza}. Reproduced by permission of the AAS.}.
    \label{eos hebeler}
\end{figure}

In order to study new physics scenarios, it is necessary to employ Standard Model results with a controlled systematics. For that reason, in this review we focus on the EoS based on chiral effective field theory and pQCD.
One example is the three tabulated EoS (soft, intermediate and stiff) obtained in Ref.~\cite{Hebeler:2013nza}.
To construct these equation of state the authors perform microscopic matter calculations based on chiral 
$NN$ and 3$N$ interactions based heavily on their earlier work \cite{Hebeler:2009iv}. This gives the nuclear equation of 
state up to $n=1.1n_s$. They extend these results to higher energies with a piecewise polytropic extrapolation (previously explained in the \textit{Polytropic representation} of section 2) as shown in 
Fig.~\ref{Polytropic hebeler} which leads to a large number of EoS from which are selected those that remain causal and are able to support masses of $1.97M_\odot$ 
(the heaviest neutron star known at the time). The tabulated EoS are also consisted with astrophysical observation constraints. 
Figure \ref{eos hebeler} (left) shows the different EoS obtained in that work (soft, intermediate and stiff), while the right plot shows the mass-radius diagram obtained with these three types of equations.

\begin{figure}[h!]
    \centering
    \includegraphics[width=0.6\linewidth]{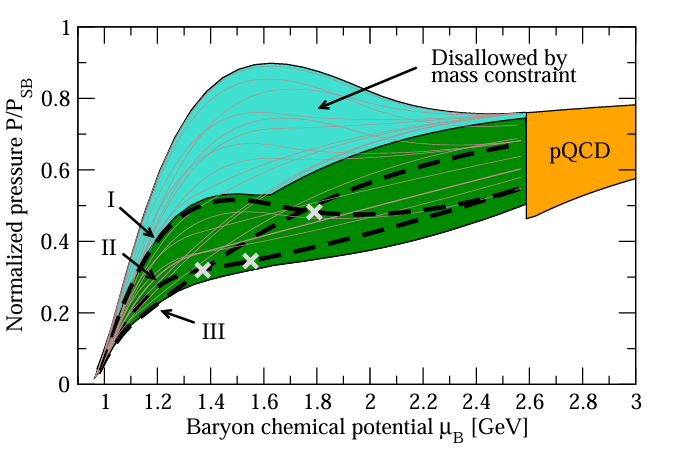}
    \caption{The green area supports a star of $2M_\odot$. The cross on each representative EoS denotes the maximal chemical 
    potential reached at the center of the star. Taken form \cite{Kurkela:2014vha}. Reproduced by permission of the AAS.}
    \label{fig:enter-label-2}
\end{figure}

In \cite{Deliyergiyev:2019vti} baryonic matter is represented by the EoSI obtained in \cite{Kurkela:2014vha} mapped with an inner crust 
\cite{Negele:1971vb} and outer crust \cite{Ruester:2005fm}. For even lower densities ($\rho<3.3\times10^3 $g/cm$^3$) the Harrison-Wheeler EoS \cite{HarrisonWheeler} is used. The EoS from \cite{Kurkela:2014vha} (EoSI, EoSII and EoSIII, see Fig.~\ref{fig:enter-label-2}) use the soft and stiff Chiral EFT
EoS for densities below $1.1n_s$ of \cite{Hebeler:2013nza} described above, and \cite{Fraga:2013qra} for the high energy regime.
\begin{figure}[h!]
    \centering
    \includegraphics[width=0.6\linewidth]{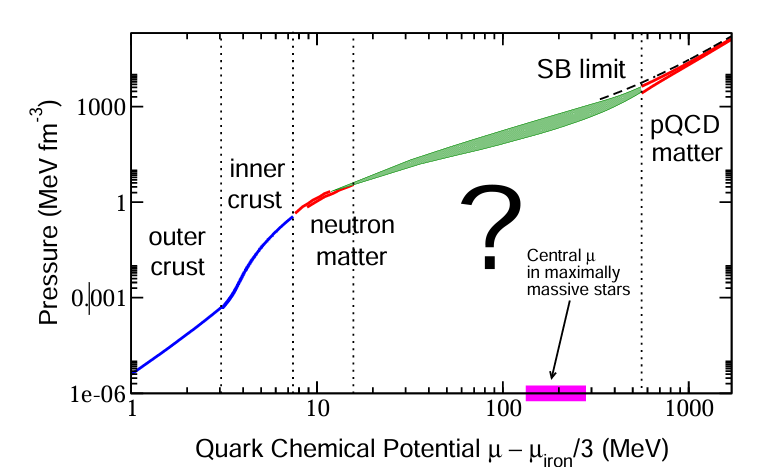}
    \caption{Known limits of the stellar EoS. The green band corresponds to the intepolating polytropic EoS. Results taken form \cite{Kurkela:2014vha}. Reproduced by permission of the AAS.}
    \label{fig:enter-label-1}
\end{figure}
To transit between these two extreme density regimes, they interpolate with two monotropes of the form $P(n)=\kappa n^\Gamma$ matched 
first in a smooth way but later with a first-order phase transition. In Fig.~\ref{fig:enter-label-1} is shown a schematic plot of the different density regimes.
They further constrain their EoS by requiring that they support two solar mass stars. Finally, the crust EoS from \cite{Baym:1971pw} is employed for densities below $0.6n_s$. 
EoSI gives the smallest radius, while EoSII and EoSIII give the maximal mass and radius, respectively.

For the dark matter component, in \cite{Deliyergiyev:2019vti} they model DM 
as a  non self annihilating, self-interacting Fermi gas with particle masses $m_f$ between 1 and 500 GeV with equation of state:
\begin{gather}
    p = \frac{m_f^4}{24\pi^2}[(2z^3-3z)(\sqrt{1+z^2})+3\text{sinh}^{-1}(z)] \\
    \epsilon = \frac{m_f^4}{24\pi^2}[(2z^3-3z)(\sqrt{1+z^2})-\text{sinh}^{-1}(z)]
\end{gather}
taken from \cite{Narain:2006kx} where $z=k_F/m_f$, accounting for weakly and strongly interacting dark matter. Using these combinations of EoS for the DANS, it is shown that strongly interacting particles allow for neutron stars heavier than $2M_\odot$. 

Later, in \cite{Dengler:2021qcq} The authors add the analysis of the tidal deformability and the dimensionless second 
Love number. They show that, with a high enough content of dark matter inside neutron stars, the second love number 
is significantly different compared to ordinary neutron stars. In agreement with \cite{Nelson:2018xtr} lower dark 
matter particles masses (MeV-GeV range) with strong interactions form a halo increasing the tidal deformability, 
while weakly interacting particles form a core that reduces the tidal deformability, consistent with the results of \cite{Ellis:2018bkr}.

This is studied furthermore in \cite{Barbat:2024yvi}, where the two limiting cases of EoSI and EoSII 
from \cite{Kurkela:2014vha} are taken. These two EoSs are representative of all the smooth EoS that fall between the them.
EoSI and EoSII are again mapped with \cite{Negele:1971vb}, \cite{Ruester:2005fm}, and \cite{HarrisonWheeler} for lower densities. 
They use the same model of non self-annihilating self-interacting fermion dark matter with masses $m_f\in(0.1,1)$GeV 
and multiple interaction strengths. Now it is also performed an analysis of the stable two-fluid configuration, following (\cite{Hippert:2022snq}) and
requiring that the number of particles for ordinary matter (OM) and DM is stationary under variations of each central density.
\begin{figure}[h!]
    \centering
    \includegraphics[width=0.7\linewidth]{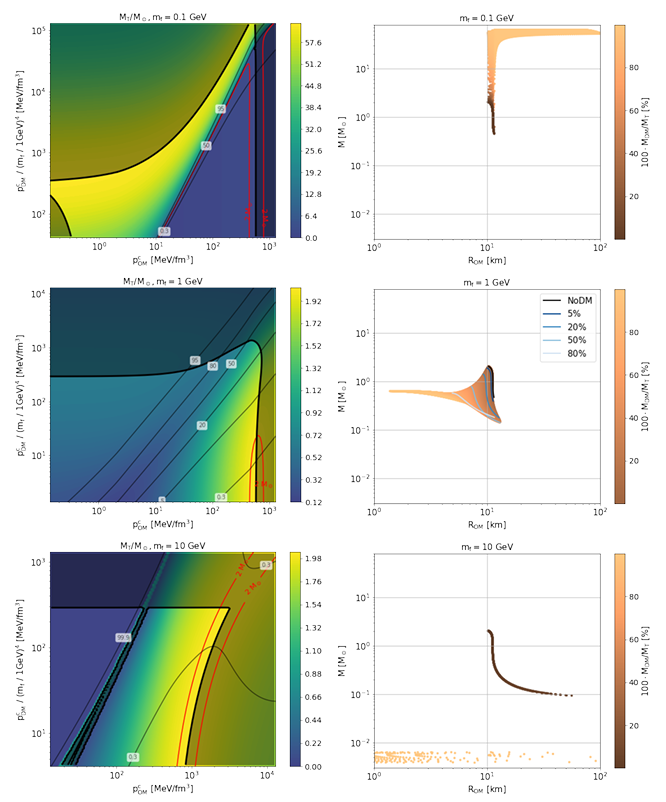}
    \caption{Configurations of DM admixed with OM. The black curves in the left panels represent the critical curves 
    with the unstable regions shaded. In the right panels are shown the mass-radius for stable configurations. This figure corresponds 
    to \cite{Barbat:2024yvi}. "Reprinted figure with permission from Mikel F. Barbat, Jurgen Schaffner-Bielich, and Laura Tolos. 
    Comprehensive study of compact stars with dark matter. Phys. Rev. D, 110, 2, 2024. Copyright (2024) by the American Physical Society.}
    \label{fig:enter-label-3}
\end{figure}
It is shown that for low masses ($m_f=0.1$ GeV) the DM forms a dark halo increasing the maximum mass which doesn't compromise 
the stability of the star. For intermediate masses ($m_f=1$~GeV) a stable DM core forms that reduces the maximum mass but for higher masses 
($m_f=10$~GeV) the dark core makes the configuration unstable. In Fig.~\ref{fig:enter-label-3} are shown the central pressure combinations that form stable stars, as well a their mass and composition. This results are similar for different OM EoSs and interaction strengths, only 
depending on the dark matter particle mass. They also study the change on tidal deformabilities, further suggesting that while for total 
masses around $1.4M_\odot$ the tidal deformabilty depends both and the DM particle mass and the ordinary matter EoS. Using the constrain of 
the tidal deformability from GW170817 they rule out configurations with a DM particle mass less than 0.1~GeV with more than $1\%$ of self-interacting DM.\\

In \cite{Dengler:2025ntz} the authors use the three EoSs of \cite{Kurkela:2014vha} characterized above. The dark matter component is modeled as three dark-quark state made of one flavor of fermions in the fundamental representation of $G_2$-QCD. $G_2$-QCD replaces the group SU(3) with $G_2$. 
The dark quarks 
equation of state is described in \cite{Wellegehausen:2013cya} as lattice data. In order to compute the TOV equations, they interpolate between 
points with piecewise polytropes ensuring thermodynamic consistency. The DM EoS is extrapolated to small densities by a free Fermi fluid and to 
high densities by a Fermi-Dirac distribution. They investigate dark matter candidates from 250~MeV to 4~GeV. Performing a similar stability analysis 
to the above described, they observe that the addition of dark matter allows to sustain larger central pressures for ordinary matter. It is also found 
that light dark matter has a stronger effect on the properties of the NS, making it possible to have stable NS of $8M_\odot$ (for high enough DM fractions).\\

The work of \cite{Hebeler:2013nza} has also been used by \cite{Rutherford:2022xeb} In this case, instead of using one of the three tabulated EoS 
of the original paper, the band is fitted to a single polytrope. This EoS is mapped to the BPS EoS \cite{Baym:1971pw} for the crust (below $0.5n_s$)  to a piecewise polytropic parameterization for high densities
high densities. In this paper the authors use the 
asymmetric dark matter model of \cite{Nelson:2018xtr} (self-repulsive bosonic particles) with the EoS:
\begin{equation}
    \epsilon_\chi(r) = m_\chi c^2n_\chi + \frac{g_\chi^2}{2m_\phi^2}\frac{\hbar^3}{c}n_\chi^2 \qquad p_\chi(r)=-\epsilon_\chi\tilde{\mu}_\chi=\frac{g_\chi^2}{2m_\phi^2}\frac{\hbar^3}{c}n_\chi^2 \qquad \tilde{\mu}_\chi=\frac{\partial\epsilon_\chi}{\partial n_\chi}
\end{equation}
where $m_\phi$ is the mass of the $\phi_\mu$ vector field $Z_{\mu\nu}=\nabla_\mu\phi_\nu-\nabla_\nu\phi_\mu $, $g_B$ the interaction strength, $n_\chi$ 
is the number density of asymmetric dark matter (ADM) and $\tilde{\mu}_\chi$ the ADM chemical potential for masses $m_\chi \in(10^{-2},10^{8}) $~MeV, self repulsion 
\begin{equation}
    10^{-2}\leq\frac{g_\chi}{m_\phi/\text{MeV}}\leq10^3
\end{equation}
and dark matter fraction $F_\chi\leq20\%$.
Performing the Bayesian analysis of \cite{Raaijmakers:2019qny}, it is suggested that if two 
neutron stars with the same mass but different radii were measured at the $2\%$ uncertainty level, it could be concluded that one neutron star has ADM 
and the other one doesn't. It is moreover shown that high masses ($m_\chi>10^6$~MeV) and low self-interaction strength   ($g_\chi/m_\phi<0.1$~MeV$^{-1} $) is 
disfavored by the neutron stars having to be more that $1M_\odot$. The authors also suggest that if the baryonic EoS uncertainties are tightened without 
using ADM, NICER and STROBE-X could constrain the ratio of $m_\chi$ and $g_\chi/m_\phi$ but no the individual 
quantities for bosonic ADM.

In \cite{Rutherford:2024bli} the previous work is extended using fermionic ADM cores, thus allowing the interaction strength to be zero. The fermionic dark 
matter EoS is then:
\begin{align}
    &\epsilon_\chi=\frac{c^5m_\chi^4}{8\pi^2\hbar^3}\left[\sqrt{1+z^2}(2z^3+z)-\log(z+\sqrt{1+z^2}) \right]+\frac{g_\chi^2}{2m_\phi^2}\frac{c^5(m_\chi z)^6}{\hbar^3(3\pi^2)^2} \\
    &p_\chi=\frac{c^5m_\chi^4}{8\pi^2\hbar^3}\left[\sqrt{1+z^2}\left(\frac{2}{3}z^3-z \right)+\log(z+\sqrt{1+z^2}) \right]+\frac{g_\chi^2}{2m_\phi^2}\frac{c^5(m_\chi z)^6}{\hbar^3(3\pi^2)^2}
\end{align}
where $z=\hbar k_\chi/m_\chi c$ is the relativity parameter in terms of the ADM Fermi momentum. They study fermionic dark matter with parameters 
$m_\chi=15$~GeV, $g_\chi/m_\phi=0.01$~MeV$^{-1} $ and fixing the fraction of dark matter $F_\chi=1.5\%$. Following the same Bayesian framework as before, this 
study shows that for the mass fraction considered, the change on the mass-radius diagram is so feeble that neither NICER nor STROBE-X will be able to differentiate 
between a ADM core with a stiffer baryonic EoS from a softer baryonic EoS without the core.\\

The use of Chiral EFT theories for the baryonic EoS of neutron star has confirmed the previously suggested results that heavier DM particles would form a DM core inside the neutron stars, while lighter DM forms a halo surrounding the star, which increases the tidal deformability. They've also shown that a small fraction of DM would yield a mass-radius curve almost undistinguishable from another similar curve given by a pure baryonic EoS. 
So to study the effects of DM on the mass-radius curve it would be necessary to have both a significant amount of DM and really tight constraints on the measurements of the NSs. Moreover, although most of the papers referenced here have included a stability analysis for the two fluid configuration of the DANS, these are still fairly uncommon (specially in older papers) and so it would be useful to look into it with further detail.   

Recently, it has been discovered a central compact object in the HESS J1731-347 supernova remnant. This compact object is supposed to be a strangely light 
neutron star ($M=0.77_{-0.17}^{+0.20}M_\odot $, $R=10.4_{-0.78}^{+0.86} $km), which has motivated several works studying the possibility of it being a DANS. In \cite{Lopes:2024ixl} 
its been shown (using the $\sigma\omega\rho\phi$ nuclear model with the $L1w^4$ parametrizations and the one fluid formalism of the TOV equations) that 
a DANS with DM particles of Lagrangian:
\begin{equation}
\label{ldm}
    \mathcal{L}_{DM}=\bar{\chi}(i\gamma^\mu\partial_\mu-(m_\chi-g_Hh))\chi+\frac{1}{2}(\partial^\mu h\partial_\mu h-m_H^2h^2)
\end{equation}
with Fermi momenta 0.03~GeV is compatible with the compact object measurements. In Eq.~\eqref{ldm} $\chi$ stands for the Dirac field that represents the 
dark fermion that self interacts through the exchange of the Higgs boson with mass $m_H=125$~GeV, and $g_H=0.1$ stands for the coupling constant. In \cite{Routaray:2023txs} 
they use the NITR-I EoS and also show that a dark matter particle of Fermi momenta $k=0.03$~GeV allows this NS to be a DANS. The employed DM model considers coupling between 
baryons and DM through Higgs exchange and the corresponding DM EoS is:

\begin{gather}
    \varepsilon_{DM}=\frac{2}{(2\pi)^3}\int_0^{k_f^{DM}}d^3k\sqrt{k^2+{M_{\chi}^\star}^2}+ \frac{1}{2}M_h^2h_0^2 \\
    P_{DM}=\frac{2}{3(2\pi)^3}\int_0^{k_f^{DM}}d^3k \frac{k^2}{\sqrt{k^2+{M_{\chi}^\star}^2}} - \frac{1}{2}M_h^2h_0^2
\end{gather}
Where $M_{\chi}^\star$ represents the effective mass of the DM, $k_f^{DM} $ its Fermi momentum, $M_h$ the Higgs mass , and $h$ the Higgs field. Again, using the single 
fluid approach the final EoS is considered to be the sum of both contributions. Also with the one-fluid formalism the work of \cite{Routaray:2022acz} (using for nuclear matter 
the RMF formalism with IOPB-I , G3 and QMC-RMF parameter set, where IOPB allows for larger radius and QMC-RMF for the larger maximum mass) shows that a DM particle mass of 24~GeV allows the resulting mass-radius curve to be consistent with the HESS measurements. 
In \cite{Sagun:2023rzp}, using both soft IST EoS and stiff BigApple EoS, the authors show that relativistic free fermi gas DM particles heavier than 1.15~GeV and a dark matter fraction 
above $4.2\%$ could explain the HESS measurement. 

Although all these works have shed light on the possible dark matter component in HESS J1731-347, the choice of a particular EoS could bias the possible new physics scenario. 
For that reason, it is interesting to sudy the possible presence of DM in this compact object using EoS based on first principles.
 The work of Ref.~\cite{Alarcon2025} uses some of the EoSs obtained in \cite{Alarcon:2024hlj}. In the latter reference, they use regulator independent pure neutron matter EoS for the low energy density regime  that is expressed in terms of data of nucleon-nucleon scattering. This result is matched to pQCD for the high-density regime. To interpolate between these two extreme regions, stability, thermodynamic consistency and causality are employed. The EoS are constrained by symmetry energy measurements of PREX-II and CREX, as well as with astrophysical observations of mass, radii and tidal deformabilities of known neutron stars.
Modelling dark matter as a free Fermi gas with masses between 1~MeV and 10~GeV the authors look for the combination of dark matter particle mass and dark matter fraction that will allow the 
HESS J1731-347 compact object to lay within the $1\sigma$ confidence level ($1\sigma$CL) band.

\begin{figure}[h!]

\centering
\includegraphics[width=.45\textwidth]{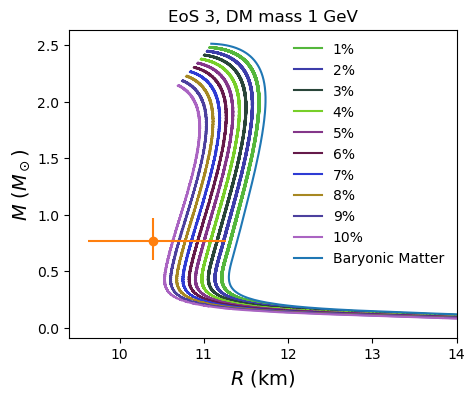}\hfill
\includegraphics[width=.45\textwidth]{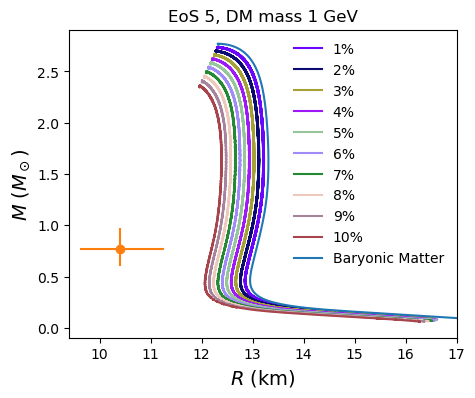}

\caption{Effects of 1~GeV dark matter particle on M-R curves for different EoS. Results taken from \cite{Alarcon2025}.}
\label{difeos}
\end{figure}

In Fig.~\ref{difeos} it is shown that, depending on the choice of baryonic EoS, it would be needed to have a DM fraction of as little as $2\%$ or way more than $10\%$. This results highlight 
the importance of narrowing the uncertainty of the HESS J1731-347 compact object as it will either highly constraint the allowed NS EoS or the properties of dark matter.

\section{Summary and Conclusions}

In this review we showed how chiral effective field theory can contribute to the physics of neutron stars. 
Rooted on the low-energy theorems of the strong interactions effective field theory allows to compute in a systematic manner the equation of state of nuclear matter at densities corresponding to the crust. 
This requires the use of the many-body techniques explained in Sec.~\ref{Sec:Effective_theories_for_nuclear_matter}. 
These methods normally introduce some additional systematic error with respect to the vacuum calculation, and requires additional considerations.
Nevertheless, EFT calculations with many-body methods are able to determine the EoS of the crust of a neutron star with a controlled accuracy, 
providing a sound starting point for the continuation to higher densities in the star. 
In Sec.~\ref{Sec:EoS} we explained how this can be done by imposing general constrains related to causality, mechanical stability and thermodynamic consistency. Also, matching the EoS to the pQCD results at very high densities provides strong constrains to the EoS that impacts on the softening of the EoS.  
With these considerations, some EoS provided by EFT where able to explain the different measurements of neutron star masses and radii, as well as the tidal deformabilities with a controlled systematics.  
The reliability in the determination of the systematics is one of the strongest points of EFT calculations for searches of new physics, as indicated in Sec.~\ref{Sec:Neutron_Stars_for_Searches_of_New_Physics}.
As shown there, EFT results allows to quantify the effects of possible dark matter content inside a supernova remnant with anomalous mass and radii, and put constrains on the dark matter particle and content in the compact object.
Appart from the modification of the mass-radius diagram, rigorous new physics searches in neutron stars involving changes of frecuencies in the X-ray emision or gravitaional wave signal are possible thanks to the controlled systematics.
This will open new possibilities for searches of new physics in compact objects in the multimessenger era.

\section{Acknowledgements}
We acknowledge partial financial support to the Grant PID2022-136510NB-C32 funded by
MCIN/AEI/10.13039/501100011033/ and FEDER, UE,
and to the EU Horizon 2020 research and innovation program, STRONG-2020 project, under grant agreement no. 824093. We also acknowledge financial support by the "L\'inea de Excelencia para el Profesorado Universitario de la UAH", project number EPU-INV-UAH/2023/003. 
The EoSs with crust of Sect. 3.4 are available from the corresponding author on reasonable request.



\bibliography{biblio}


\begin{thebibliography}{245}
\ifx \bisbn   \undefined \def \bisbn  #1{ISBN #1}\fi
\ifx \binits  \undefined \def \binits#1{#1}\fi
\ifx \bauthor  \undefined \def \bauthor#1{#1}\fi
\ifx \batitle  \undefined \def \batitle#1{#1}\fi
\ifx \bjtitle  \undefined \def \bjtitle#1{#1}\fi
\ifx \bvolume  \undefined \def \bvolume#1{\textbf{#1}}\fi
\ifx \byear  \undefined \def \byear#1{#1}\fi
\ifx \bissue  \undefined \def \bissue#1{#1}\fi
\ifx \bfpage  \undefined \def \bfpage#1{#1}\fi
\ifx \blpage  \undefined \def \blpage #1{#1}\fi
\ifx \burl  \undefined \def \burl#1{\textsf{#1}}\fi
\ifx \doiurl  \undefined \def \doiurl#1{\url{https://doi.org/#1}}\fi
\ifx \betal  \undefined \def \betal{\textit{et al.}}\fi
\ifx \binstitute  \undefined \def \binstitute#1{#1}\fi
\ifx \binstitutionaled  \undefined \def \binstitutionaled#1{#1}\fi
\ifx \bctitle  \undefined \def \bctitle#1{#1}\fi
\ifx \beditor  \undefined \def \beditor#1{#1}\fi
\ifx \bpublisher  \undefined \def \bpublisher#1{#1}\fi
\ifx \bbtitle  \undefined \def \bbtitle#1{#1}\fi
\ifx \bedition  \undefined \def \bedition#1{#1}\fi
\ifx \bseriesno  \undefined \def \bseriesno#1{#1}\fi
\ifx \blocation  \undefined \def \blocation#1{#1}\fi
\ifx \bsertitle  \undefined \def \bsertitle#1{#1}\fi
\ifx \bsnm \undefined \def \bsnm#1{#1}\fi
\ifx \bsuffix \undefined \def \bsuffix#1{#1}\fi
\ifx \bparticle \undefined \def \bparticle#1{#1}\fi
\ifx \barticle \undefined \def \barticle#1{#1}\fi
\bibcommenthead
\ifx \bconfdate \undefined \def \bconfdate #1{#1}\fi
\ifx \botherref \undefined \def \botherref #1{#1}\fi
\ifx \url \undefined \def \url#1{\textsf{#1}}\fi
\ifx \bchapter \undefined \def \bchapter#1{#1}\fi
\ifx \bbook \undefined \def \bbook#1{#1}\fi
\ifx \bcomment \undefined \def \bcomment#1{#1}\fi
\ifx \oauthor \undefined \def \oauthor#1{#1}\fi
\ifx \citeauthoryear \undefined \def \citeauthoryear#1{#1}\fi
\ifx \endbibitem  \undefined \def \endbibitem {}\fi
\ifx \bconflocation  \undefined \def \bconflocation#1{#1}\fi
\ifx \arxivurl  \undefined \def \arxivurl#1{\textsf{#1}}\fi
\csname PreBibitemsHook\endcsname

\bibitem[\protect\citeauthoryear{Duguet}{2014}]{Duguet:2013dga}
\begin{barticle}
\bauthor{\bsnm{Duguet}, \binits{T.}}:
\batitle{{The nuclear energy density functional formalism}}.
\bjtitle{Lect. Notes Phys.}
\bvolume{879},
\bfpage{293}--\blpage{350}
(\byear{2014})
\doiurl{10.1007/978-3-642-45141-6_7}
{\href{https://arxiv.org/abs/1309.0440}{{arXiv:1309.0440}}}
{[nucl-th]}
\end{barticle}
\endbibitem

\bibitem[\protect\citeauthoryear{Serot}{1992}]{Serot:1992ti}
\begin{barticle}
\bauthor{\bsnm{Serot}, \binits{B.D.}}:
\batitle{{Quantum hadrodynamics}}.
\bjtitle{Rept. Prog. Phys.}
\bvolume{55},
\bfpage{1855}--\blpage{1946}
(\byear{1992})
\doiurl{10.1088/0034-4885/55/11/001}
\end{barticle}
\endbibitem

\bibitem[\protect\citeauthoryear{Akmal et~al.}{1998}]{Akmal:1998cf}
\begin{barticle}
\bauthor{\bsnm{Akmal}, \binits{A.}},
\bauthor{\bsnm{Pandharipande}, \binits{V.R.}},
\bauthor{\bsnm{Ravenhall}, \binits{D.G.}}:
\batitle{{The Equation of state of nucleon matter and neutron star structure}}.
\bjtitle{Phys. Rev. C}
\bvolume{58},
\bfpage{1804}--\blpage{1828}
(\byear{1998})
\doiurl{10.1103/PhysRevC.58.1804}
{\href{https://arxiv.org/abs/nucl-th/9804027}{{arXiv:nucl-th/9804027}}}
\end{barticle}
\endbibitem

\bibitem[\protect\citeauthoryear{Holt et~al.}{2013}]{Holt:2013fwa}
\begin{barticle}
\bauthor{\bsnm{Holt}, \binits{J.W.}},
\bauthor{\bsnm{Kaiser}, \binits{N.}},
\bauthor{\bsnm{Weise}, \binits{W.}}:
\batitle{{Nuclear chiral dynamics and thermodynamics}}.
\bjtitle{Prog. Part. Nucl. Phys.}
\bvolume{73},
\bfpage{35}--\blpage{83}
(\byear{2013})
\doiurl{10.1016/j.ppnp.2013.08.001}
{\href{https://arxiv.org/abs/1304.6350}{{arXiv:1304.6350}}}
{[nucl-th]}
\end{barticle}
\endbibitem

\bibitem[\protect\citeauthoryear{Gandolfi et~al.}{2009}]{Gandolfi:2009fj}
\begin{barticle}
\bauthor{\bsnm{Gandolfi}, \binits{S.}},
\bauthor{\bsnm{Illarionov}, \binits{A.Y.}},
\bauthor{\bsnm{Schmidt}, \binits{K.E.}},
\bauthor{\bsnm{Pederiva}, \binits{F.}},
\bauthor{\bsnm{Fantoni}, \binits{S.}}:
\batitle{{Quantum Monte Carlo calculation of the equation of state of neutron
  matter}}.
\bjtitle{Phys. Rev. C}
\bvolume{79},
\bfpage{054005}
(\byear{2009})
\doiurl{10.1103/PhysRevC.79.054005}
{\href{https://arxiv.org/abs/0903.2610}{{arXiv:0903.2610}}}
{[nucl-th]}
\end{barticle}
\endbibitem

\bibitem[\protect\citeauthoryear{Ren et~al.}{2024}]{Ren:2023ued}
\begin{barticle}
\bauthor{\bsnm{Ren}, \binits{Z.}},
\bauthor{\bsnm{Elhatisari}, \binits{S.}},
\bauthor{\bsnm{L\"ahde}, \binits{T.A.}},
\bauthor{\bsnm{Lee}, \binits{D.}},
\bauthor{\bsnm{Mei\ss{}ner}, \binits{U.-G.}}:
\batitle{{Ab initio study of nuclear clustering in hot dilute nuclear matter}}.
\bjtitle{Phys. Lett. B}
\bvolume{850},
\bfpage{138463}
(\byear{2024})
\doiurl{10.1016/j.physletb.2024.138463}
{\href{https://arxiv.org/abs/2305.15037}{{arXiv:2305.15037}}}
{[nucl-th]}
\end{barticle}
\endbibitem

\bibitem[\protect\citeauthoryear{Gasser et~al.}{1988}]{Gasser:1987rb}
\begin{barticle}
\bauthor{\bsnm{Gasser}, \binits{J.}},
\bauthor{\bsnm{Sainio}, \binits{M.E.}},
\bauthor{\bsnm{Svarc}, \binits{A.}}:
\batitle{{Nucleons with chiral loops}}.
\bjtitle{Nucl. Phys. B}
\bvolume{307},
\bfpage{779}--\blpage{853}
(\byear{1988})
\doiurl{10.1016/0550-3213(88)90108-3}
\end{barticle}
\endbibitem

\bibitem[\protect\citeauthoryear{Weinberg}{1990}]{Weinberg:1990rz}
\begin{barticle}
\bauthor{\bsnm{Weinberg}, \binits{S.}}:
\batitle{{Nuclear forces from chiral Lagrangians}}.
\bjtitle{Phys. Lett. B}
\bvolume{251},
\bfpage{288}--\blpage{292}
(\byear{1990})
\doiurl{10.1016/0370-2693(90)90938-3}
\end{barticle}
\endbibitem

\bibitem[\protect\citeauthoryear{Oller}{2002}]{Oller:2001sn}
\begin{barticle}
\bauthor{\bsnm{Oller}, \binits{J.A.}}:
\batitle{{Chiral Lagrangians at finite density}}.
\bjtitle{Phys. Rev. C}
\bvolume{65},
\bfpage{025204}
(\byear{2002})
\doiurl{10.1103/PhysRevC.65.025204}
{\href{https://arxiv.org/abs/hep-ph/0101204}{{arXiv:hep-ph/0101204}}}
\end{barticle}
\endbibitem

\bibitem[\protect\citeauthoryear{Lacour et~al.}{2011}]{Lacour:2009ej}
\begin{barticle}
\bauthor{\bsnm{Lacour}, \binits{A.}},
\bauthor{\bsnm{Oller}, \binits{J.A.}},
\bauthor{\bsnm{Mei{\ss}ner}, \binits{U.-G.}}:
\batitle{{Non-perturbative methods for a chiral effective field theory of
  finite density nuclear systems}}.
\bjtitle{Annals Phys.}
\bvolume{326},
\bfpage{241}--\blpage{306}
(\byear{2011})
\doiurl{10.1016/j.aop.2010.06.012}
{\href{https://arxiv.org/abs/0906.2349}{{arXiv:0906.2349}}}
{[nucl-th]}
\end{barticle}
\endbibitem

\bibitem[\protect\citeauthoryear{Alarcon and Oller}{2022}]{Alarcon:2021kpx}
\begin{barticle}
\bauthor{\bsnm{Alarcon}, \binits{J.M.}},
\bauthor{\bsnm{Oller}, \binits{J.A.}}:
\batitle{{Ladder resummation of spin 1/2 fermion many-body systems with
  arbitrary partial-wave content}}.
\bjtitle{Annals Phys.}
\bvolume{437},
\bfpage{168741}
(\byear{2022})
\doiurl{10.1016/j.aop.2021.168741}
{\href{https://arxiv.org/abs/2106.02652}{{arXiv:2106.02652}}}
{[nucl-th]}
\end{barticle}
\endbibitem

\bibitem[\protect\citeauthoryear{Alarc\'on and Oller}{2023}]{Alarcon:2022vtn}
\begin{barticle}
\bauthor{\bsnm{Alarc\'on}, \binits{J.M.}},
\bauthor{\bsnm{Oller}, \binits{J.A.}}:
\batitle{{Nuclear matter from the ladder resummation in terms of the
  experimental nucleon-nucleon scattering amplitudes}}.
\bjtitle{Phys. Rev. C}
\bvolume{107}(\bissue{4}),
\bfpage{044319}
(\byear{2023})
\doiurl{10.1103/PhysRevC.107.044319}
{\href{https://arxiv.org/abs/2212.05092}{{arXiv:2212.05092}}}
{[nucl-th]}
\end{barticle}
\endbibitem

\bibitem[\protect\citeauthoryear{Adler}{1965a}]{Adler:1964um}
\begin{barticle}
\bauthor{\bsnm{Adler}, \binits{S.L.}}:
\batitle{{Consistency conditions on the strong interactions implied by a
  partially conserved axial vector current}}.
\bjtitle{Phys. Rev.}
\bvolume{137},
\bfpage{1022}--\blpage{1033}
(\byear{1965})
\doiurl{10.1103/PhysRev.137.B1022}
\end{barticle}
\endbibitem

\bibitem[\protect\citeauthoryear{Adler}{1965b}]{Adler:1965ga}
\begin{barticle}
\bauthor{\bsnm{Adler}, \binits{S.L.}}:
\batitle{{Consistency conditions on the strong interactions implied by a
  partially conserved axial-vector current. II}}.
\bjtitle{Phys. Rev.}
\bvolume{139},
\bfpage{1638}--\blpage{1643}
(\byear{1965})
\doiurl{10.1103/PhysRev.139.B1638}
\end{barticle}
\endbibitem

\bibitem[\protect\citeauthoryear{Weinberg}{1966}]{Weinberg:1966kf}
\begin{barticle}
\bauthor{\bsnm{Weinberg}, \binits{S.}}:
\batitle{{Pion scattering lengths}}.
\bjtitle{Phys. Rev. Lett.}
\bvolume{17},
\bfpage{616}--\blpage{621}
(\byear{1966})
\doiurl{10.1103/PhysRevLett.17.616}
\end{barticle}
\endbibitem

\bibitem[\protect\citeauthoryear{Alarcon et~al.}{2013}]{Alarcon:2012kn}
\begin{barticle}
\bauthor{\bsnm{Alarcon}, \binits{J.M.}},
\bauthor{\bsnm{Martin~Camalich}, \binits{J.}},
\bauthor{\bsnm{Oller}, \binits{J.A.}}:
\batitle{{Improved description of the $\pi N$-scattering phenomenology in
  covariant baryon chiral perturbation theory}}.
\bjtitle{Annals Phys.}
\bvolume{336},
\bfpage{413}--\blpage{461}
(\byear{2013})
\doiurl{10.1016/j.aop.2013.06.001}
{\href{https://arxiv.org/abs/1210.4450}{{arXiv:1210.4450}}}
{[hep-ph]}
\end{barticle}
\endbibitem

\bibitem[\protect\citeauthoryear{Jenkins and Manohar}{1991}]{Jenkins:1990jv}
\begin{barticle}
\bauthor{\bsnm{Jenkins}, \binits{E.E.}},
\bauthor{\bsnm{Manohar}, \binits{A.V.}}:
\batitle{{Baryon chiral perturbation theory using a heavy fermion Lagrangian}}.
\bjtitle{Phys. Lett. B}
\bvolume{255},
\bfpage{558}--\blpage{562}
(\byear{1991})
\doiurl{10.1016/0370-2693(91)90266-S}
\end{barticle}
\endbibitem

\bibitem[\protect\citeauthoryear{Gegelia and Japaridze}{1999}]{Gegelia:1999gf}
\begin{barticle}
\bauthor{\bsnm{Gegelia}, \binits{J.}},
\bauthor{\bsnm{Japaridze}, \binits{G.}}:
\batitle{{Matching heavy particle approach to relativistic theory}}.
\bjtitle{Phys. Rev. D}
\bvolume{60},
\bfpage{114038}
(\byear{1999})
\doiurl{10.1103/PhysRevD.60.114038}
{\href{https://arxiv.org/abs/hep-ph/9908377}{{arXiv:hep-ph/9908377}}}
\end{barticle}
\endbibitem

\bibitem[\protect\citeauthoryear{Fuchs et~al.}{2003}]{Fuchs:2003qc}
\begin{barticle}
\bauthor{\bsnm{Fuchs}, \binits{T.}},
\bauthor{\bsnm{Gegelia}, \binits{J.}},
\bauthor{\bsnm{Japaridze}, \binits{G.}},
\bauthor{\bsnm{Scherer}, \binits{S.}}:
\batitle{{Renormalization of relativistic baryon chiral perturbation theory and
  power counting}}.
\bjtitle{Phys. Rev. D}
\bvolume{68},
\bfpage{056005}
(\byear{2003})
\doiurl{10.1103/PhysRevD.68.056005}
{\href{https://arxiv.org/abs/hep-ph/0302117}{{arXiv:hep-ph/0302117}}}
\end{barticle}
\endbibitem

\bibitem[\protect\citeauthoryear{Reinert et~al.}{2018}]{Reinert:2017usi}
\begin{barticle}
\bauthor{\bsnm{Reinert}, \binits{P.}},
\bauthor{\bsnm{Krebs}, \binits{H.}},
\bauthor{\bsnm{Epelbaum}, \binits{E.}}:
\batitle{{Semilocal momentum-space regularized chiral two-nucleon potentials up
  to fifth order}}.
\bjtitle{Eur. Phys. J. A}
\bvolume{54}(\bissue{5}),
\bfpage{86}
(\byear{2018})
\doiurl{10.1140/epja/i2018-12516-4}
{\href{https://arxiv.org/abs/1711.08821}{{arXiv:1711.08821}}}
{[nucl-th]}
\end{barticle}
\endbibitem

\bibitem[\protect\citeauthoryear{Epelbaum et~al.}{2005}]{Epelbaum:2004fk}
\begin{barticle}
\bauthor{\bsnm{Epelbaum}, \binits{E.}},
\bauthor{\bsnm{Glockle}, \binits{W.}},
\bauthor{\bsnm{Meissner}, \binits{U.-G.}}:
\batitle{{The Two-nucleon system at next-to-next-to-next-to-leading order}}.
\bjtitle{Nucl. Phys. A}
\bvolume{747},
\bfpage{362}--\blpage{424}
(\byear{2005})
\doiurl{10.1016/j.nuclphysa.2004.09.107}
{\href{https://arxiv.org/abs/nucl-th/0405048}{{arXiv:nucl-th/0405048}}}
\end{barticle}
\endbibitem

\bibitem[\protect\citeauthoryear{Piarulli and Tews}{2020}]{Piarulli:2019cqu}
\begin{barticle}
\bauthor{\bsnm{Piarulli}, \binits{M.}},
\bauthor{\bsnm{Tews}, \binits{I.}}:
\batitle{{Local Nucleon-Nucleon and Three-Nucleon Interactions Within Chiral
  Effective Field Theory}}.
\bjtitle{Front. in Phys.}
\bvolume{7},
\bfpage{245}
(\byear{2020})
\doiurl{10.3389/fphy.2019.00245}
{\href{https://arxiv.org/abs/2002.00032}{{arXiv:2002.00032}}}
{[nucl-th]}
\end{barticle}
\endbibitem

\bibitem[\protect\citeauthoryear{Alarcon et~al.}{2012}]{Alarcon:2011zs}
\begin{barticle}
\bauthor{\bsnm{Alarcon}, \binits{J.M.}},
\bauthor{\bsnm{Martin~Camalich}, \binits{J.}},
\bauthor{\bsnm{Oller}, \binits{J.A.}}:
\batitle{{The chiral representation of the $\pi N$ scattering amplitude and the
  pion-nucleon sigma term}}.
\bjtitle{Phys. Rev. D}
\bvolume{85},
\bfpage{051503}
(\byear{2012})
\doiurl{10.1103/PhysRevD.85.051503}
{\href{https://arxiv.org/abs/1110.3797}{{arXiv:1110.3797}}}
{[hep-ph]}
\end{barticle}
\endbibitem

\bibitem[\protect\citeauthoryear{Alarcon et~al.}{2014}]{Alarcon:2012nr}
\begin{barticle}
\bauthor{\bsnm{Alarcon}, \binits{J.M.}},
\bauthor{\bsnm{Geng}, \binits{L.S.}},
\bauthor{\bsnm{Martin~Camalich}, \binits{J.}},
\bauthor{\bsnm{Oller}, \binits{J.A.}}:
\batitle{{The strangeness content of the nucleon from effective field theory
  and phenomenology}}.
\bjtitle{Phys. Lett. B}
\bvolume{730},
\bfpage{342}--\blpage{346}
(\byear{2014})
\doiurl{10.1016/j.physletb.2014.01.065}
{\href{https://arxiv.org/abs/1209.2870}{{arXiv:1209.2870}}}
{[hep-ph]}
\end{barticle}
\endbibitem

\bibitem[\protect\citeauthoryear{Piarulli et~al.}{2015}]{Piarulli:2014bda}
\begin{barticle}
\bauthor{\bsnm{Piarulli}, \binits{M.}},
\bauthor{\bsnm{Girlanda}, \binits{L.}},
\bauthor{\bsnm{Schiavilla}, \binits{R.}},
\bauthor{\bsnm{Navarro~P\'erez}, \binits{R.}},
\bauthor{\bsnm{Amaro}, \binits{J.E.}},
\bauthor{\bsnm{Ruiz~Arriola}, \binits{E.}}:
\batitle{{Minimally nonlocal nucleon-nucleon potentials with chiral two-pion
  exchange including $\Delta$ resonances}}.
\bjtitle{Phys. Rev. C}
\bvolume{91}(\bissue{2}),
\bfpage{024003}
(\byear{2015})
\doiurl{10.1103/PhysRevC.91.024003}
{\href{https://arxiv.org/abs/1412.6446}{{arXiv:1412.6446}}}
{[nucl-th]}
\end{barticle}
\endbibitem

\bibitem[\protect\citeauthoryear{Piarulli et~al.}{2016}]{Piarulli:2016vel}
\begin{barticle}
\bauthor{\bsnm{Piarulli}, \binits{M.}},
\bauthor{\bsnm{Girlanda}, \binits{L.}},
\bauthor{\bsnm{Schiavilla}, \binits{R.}},
\bauthor{\bsnm{Kievsky}, \binits{A.}},
\bauthor{\bsnm{Lovato}, \binits{A.}},
\bauthor{\bsnm{Marcucci}, \binits{L.E.}},
\bauthor{\bsnm{Pieper}, \binits{S.C.}},
\bauthor{\bsnm{Viviani}, \binits{M.}},
\bauthor{\bsnm{Wiringa}, \binits{R.B.}}:
\batitle{{Local chiral potentials with $\Delta$-intermediate states and the
  structure of light nuclei}}.
\bjtitle{Phys. Rev. C}
\bvolume{94}(\bissue{5}),
\bfpage{054007}
(\byear{2016})
\doiurl{10.1103/PhysRevC.94.054007}
{\href{https://arxiv.org/abs/1606.06335}{{arXiv:1606.06335}}}
{[nucl-th]}
\end{barticle}
\endbibitem

\bibitem[\protect\citeauthoryear{Li et~al.}{2018}]{Li:2016mln}
\begin{barticle}
\bauthor{\bsnm{Li}, \binits{K.-W.}},
\bauthor{\bsnm{Ren}, \binits{X.-L.}},
\bauthor{\bsnm{Geng}, \binits{L.-S.}},
\bauthor{\bsnm{Long}, \binits{B.-W.}}:
\batitle{{Leading order relativistic hyperon-nucleon interactions in chiral
  effective field theory}}.
\bjtitle{Chin. Phys. C}
\bvolume{42}(\bissue{1}),
\bfpage{014105}
(\byear{2018})
\doiurl{10.1088/1674-1137/42/1/014105}
{\href{https://arxiv.org/abs/1612.08482}{{arXiv:1612.08482}}}
{[nucl-th]}
\end{barticle}
\endbibitem

\bibitem[\protect\citeauthoryear{Drischler et~al.}{2019}]{Drischler:2017wtt}
\begin{barticle}
\bauthor{\bsnm{Drischler}, \binits{C.}},
\bauthor{\bsnm{Hebeler}, \binits{K.}},
\bauthor{\bsnm{Schwenk}, \binits{A.}}:
\batitle{{Chiral interactions up to next-to-next-to-next-to-leading order and
  nuclear saturation}}.
\bjtitle{Phys. Rev. Lett.}
\bvolume{122}(\bissue{4}),
\bfpage{042501}
(\byear{2019})
\doiurl{10.1103/PhysRevLett.122.042501}
{\href{https://arxiv.org/abs/1710.08220}{{arXiv:1710.08220}}}
{[nucl-th]}
\end{barticle}
\endbibitem

\bibitem[\protect\citeauthoryear{Tichai et~al.}{2020}]{Tichai:2020dna}
\begin{barticle}
\bauthor{\bsnm{Tichai}, \binits{A.}},
\bauthor{\bsnm{Roth}, \binits{R.}},
\bauthor{\bsnm{Duguet}, \binits{T.}}:
\batitle{{Many-body perturbation theories for finite nuclei}}.
\bjtitle{Front. in Phys.}
\bvolume{8},
\bfpage{164}
(\byear{2020})
\doiurl{10.3389/fphy.2020.00164}
{\href{https://arxiv.org/abs/2001.10433}{{arXiv:2001.10433}}}
{[nucl-th]}
\end{barticle}
\endbibitem

\bibitem[\protect\citeauthoryear{Hebeler et~al.}{2011}]{Hebeler:2010xb}
\begin{barticle}
\bauthor{\bsnm{Hebeler}, \binits{K.}},
\bauthor{\bsnm{Bogner}, \binits{S.K.}},
\bauthor{\bsnm{Furnstahl}, \binits{R.J.}},
\bauthor{\bsnm{Nogga}, \binits{A.}},
\bauthor{\bsnm{Schwenk}, \binits{A.}}:
\batitle{{Improved nuclear matter calculations from chiral low-momentum
  interactions}}.
\bjtitle{Phys. Rev. C}
\bvolume{83},
\bfpage{031301}
(\byear{2011})
\doiurl{10.1103/PhysRevC.83.031301}
{\href{https://arxiv.org/abs/1012.3381}{{arXiv:1012.3381}}}
{[nucl-th]}
\end{barticle}
\endbibitem

\bibitem[\protect\citeauthoryear{Coraggio et~al.}{2014}]{Coraggio:2014nva}
\begin{barticle}
\bauthor{\bsnm{Coraggio}, \binits{L.}},
\bauthor{\bsnm{Holt}, \binits{J.W.}},
\bauthor{\bsnm{Itaco}, \binits{N.}},
\bauthor{\bsnm{Machleidt}, \binits{R.}},
\bauthor{\bsnm{Marcucci}, \binits{L.E.}},
\bauthor{\bsnm{Sammarruca}, \binits{F.}}:
\batitle{{Nuclear-matter equation of state with consistent two- and three-body
  perturbative chiral interactions}}.
\bjtitle{Phys. Rev. C}
\bvolume{89}(\bissue{4}),
\bfpage{044321}
(\byear{2014})
\doiurl{10.1103/PhysRevC.89.044321}
{\href{https://arxiv.org/abs/1402.0965}{{arXiv:1402.0965}}}
{[nucl-th]}
\end{barticle}
\endbibitem

\bibitem[\protect\citeauthoryear{Rios}{2020}]{Rios:2020oad}
\begin{barticle}
\bauthor{\bsnm{Rios}, \binits{A.}}:
\batitle{{Green's Function Techniques for Infinite Nuclear Systems}}.
\bjtitle{Front. in Phys.}
\bvolume{8},
\bfpage{387}
(\byear{2020})
\doiurl{10.3389/fphy.2020.00387}
{\href{https://arxiv.org/abs/2006.10610}{{arXiv:2006.10610}}}
{[nucl-th]}
\end{barticle}
\endbibitem

\bibitem[\protect\citeauthoryear{Carbone et~al.}{2013}]{Carbone:2013eqa}
\begin{barticle}
\bauthor{\bsnm{Carbone}, \binits{A.}},
\bauthor{\bsnm{Cipollone}, \binits{A.}},
\bauthor{\bsnm{Barbieri}, \binits{C.}},
\bauthor{\bsnm{Rios}, \binits{A.}},
\bauthor{\bsnm{Polls}, \binits{A.}}:
\batitle{{Self-consistent Green's functions formalism with three-body
  interactions}}.
\bjtitle{Phys. Rev. C}
\bvolume{88}(\bissue{5}),
\bfpage{054326}
(\byear{2013})
\doiurl{10.1103/PhysRevC.88.054326}
{\href{https://arxiv.org/abs/1310.3688}{{arXiv:1310.3688}}}
{[nucl-th]}
\end{barticle}
\endbibitem

\bibitem[\protect\citeauthoryear{Carbone et~al.}{2014}]{Carbone:2014mja}
\begin{barticle}
\bauthor{\bsnm{Carbone}, \binits{A.}},
\bauthor{\bsnm{Rios}, \binits{A.}},
\bauthor{\bsnm{Polls}, \binits{A.}}:
\batitle{{Correlated density-dependent chiral forces for infinite matter
  calculations within the Green's function approach}}.
\bjtitle{Phys. Rev. C}
\bvolume{90}(\bissue{5}),
\bfpage{054322}
(\byear{2014})
\doiurl{10.1103/PhysRevC.90.054322}
{\href{https://arxiv.org/abs/1408.0717}{{arXiv:1408.0717}}}
{[nucl-th]}
\end{barticle}
\endbibitem

\bibitem[\protect\citeauthoryear{Pervin et~al.}{2007}]{Pervin:2007sc}
\begin{barticle}
\bauthor{\bsnm{Pervin}, \binits{M.}},
\bauthor{\bsnm{Pieper}, \binits{S.C.}},
\bauthor{\bsnm{Wiringa}, \binits{R.B.}}:
\batitle{{Quantum Monte Carlo calculations of electroweak transition matrix
  elements in A = 6,7 nuclei}}.
\bjtitle{Phys. Rev. C}
\bvolume{76},
\bfpage{064319}
(\byear{2007})
\doiurl{10.1103/PhysRevC.76.064319}
{\href{https://arxiv.org/abs/0710.1265}{{arXiv:0710.1265}}}
{[nucl-th]}
\end{barticle}
\endbibitem

\bibitem[\protect\citeauthoryear{Tews et~al.}{2016}]{Tews:2015ufa}
\begin{barticle}
\bauthor{\bsnm{Tews}, \binits{I.}},
\bauthor{\bsnm{Gandolfi}, \binits{S.}},
\bauthor{\bsnm{Gezerlis}, \binits{A.}},
\bauthor{\bsnm{Schwenk}, \binits{A.}}:
\batitle{{Quantum Monte Carlo calculations of neutron matter with chiral
  three-body forces}}.
\bjtitle{Phys. Rev. C}
\bvolume{93}(\bissue{2}),
\bfpage{024305}
(\byear{2016})
\doiurl{10.1103/PhysRevC.93.024305}
{\href{https://arxiv.org/abs/1507.05561}{{arXiv:1507.05561}}}
{[nucl-th]}
\end{barticle}
\endbibitem

\bibitem[\protect\citeauthoryear{Mei{\ss}ner et~al.}{2002}]{Meissner:2001gz}
\begin{barticle}
\bauthor{\bsnm{Mei{\ss}ner}, \binits{U.G.}},
\bauthor{\bsnm{Oller}, \binits{J.A.}},
\bauthor{\bsnm{Wirzba}, \binits{A.}}:
\batitle{{In-medium chiral perturbation theory beyond the mean field
  approximation}}.
\bjtitle{Annals Phys.}
\bvolume{297},
\bfpage{27}--\blpage{66}
(\byear{2002})
\doiurl{10.1006/aphy.2002.6244}
{\href{https://arxiv.org/abs/nucl-th/0109026}{{arXiv:nucl-th/0109026}}}
\end{barticle}
\endbibitem

\bibitem[\protect\citeauthoryear{Oller and Meissner}{2001}]{Oller:2000fj}
\begin{barticle}
\bauthor{\bsnm{Oller}, \binits{J.A.}},
\bauthor{\bsnm{Meissner}, \binits{U.G.}}:
\batitle{{Chiral dynamics in the presence of bound states: Kaon nucleon
  interactions revisited}}.
\bjtitle{Phys. Lett. B}
\bvolume{500},
\bfpage{263}--\blpage{272}
(\byear{2001})
\doiurl{10.1016/S0370-2693(01)00078-8}
{\href{https://arxiv.org/abs/hep-ph/0011146}{{arXiv:hep-ph/0011146}}}
\end{barticle}
\endbibitem

\bibitem[\protect\citeauthoryear{Kaiser}{2012}]{Kaiser:2012sr}
\begin{barticle}
\bauthor{\bsnm{Kaiser}, \binits{N.}}:
\batitle{{Resummation of in-medium ladder diagrams: s-wave effective range and
  p-wave interaction}}.
\bjtitle{Eur. Phys. J. A}
\bvolume{48},
\bfpage{148}
(\byear{2012})
\doiurl{10.1140/epja/i2012-12148-8}
{\href{https://arxiv.org/abs/1210.0783}{{arXiv:1210.0783}}}
{[nucl-th]}
\end{barticle}
\endbibitem

\bibitem[\protect\citeauthoryear{Alarc\'on and Oller}{2022}]{Alarcon:2021nwd}
\begin{barticle}
\bauthor{\bsnm{Alarc\'on}, \binits{J.M.}},
\bauthor{\bsnm{Oller}, \binits{J.A.}}:
\batitle{{Ultracold spin-balanced fermionic quantum liquids with renormalized
  P-wave interactions}}.
\bjtitle{Phys. Rev. C}
\bvolume{106}(\bissue{5}),
\bfpage{054003}
(\byear{2022})
\doiurl{10.1103/PhysRevC.106.054003}
{\href{https://arxiv.org/abs/2107.08051}{{arXiv:2107.08051}}}
{[cond-mat.quant-gas]}
\end{barticle}
\endbibitem

\bibitem[\protect\citeauthoryear{Alarc\'on et~al.}{2017}]{Alarcon:2017zcv}
\begin{barticle}
\bauthor{\bsnm{Alarc\'on}, \binits{J.M.}},
\bauthor{\bsnm{Du}, \binits{D.}},
\bauthor{\bsnm{Klein}, \binits{N.}},
\bauthor{\bsnm{L\"ahde}, \binits{T.A.}},
\bauthor{\bsnm{Lee}, \binits{D.}},
\bauthor{\bsnm{Li}, \binits{N.}},
\bauthor{\bsnm{Lu}, \binits{B.-N.}},
\bauthor{\bsnm{Luu}, \binits{T.}},
\bauthor{\bsnm{Mei\ss{}ner}, \binits{U.-G.}}:
\batitle{{Neutron-proton scattering at next-to-next-to-leading order in Nuclear
  Lattice Effective Field Theory}}.
\bjtitle{Eur. Phys. J. A}
\bvolume{53}(\bissue{5}),
\bfpage{83}
(\byear{2017})
\doiurl{10.1140/epja/i2017-12273-x}
{\href{https://arxiv.org/abs/1702.05319}{{arXiv:1702.05319}}}
{[nucl-th]}
\end{barticle}
\endbibitem

\bibitem[\protect\citeauthoryear{Li et~al.}{2018}]{Li:2018ymw}
\begin{barticle}
\bauthor{\bsnm{Li}, \binits{N.}},
\bauthor{\bsnm{Elhatisari}, \binits{S.}},
\bauthor{\bsnm{Epelbaum}, \binits{E.}},
\bauthor{\bsnm{Lee}, \binits{D.}},
\bauthor{\bsnm{Lu}, \binits{B.-N.}},
\bauthor{\bsnm{Mei\ss{}ner}, \binits{U.-G.}}:
\batitle{{Neutron-proton scattering with lattice chiral effective field theory
  at next-to-next-to-next-to-leading order}}.
\bjtitle{Phys. Rev. C}
\bvolume{98}(\bissue{4}),
\bfpage{044002}
(\byear{2018})
\doiurl{10.1103/PhysRevC.98.044002}
{\href{https://arxiv.org/abs/1806.07994}{{arXiv:1806.07994}}}
{[nucl-th]}
\end{barticle}
\endbibitem

\bibitem[\protect\citeauthoryear{Lu et~al.}{2020}]{Lu:2019nbg}
\begin{barticle}
\bauthor{\bsnm{Lu}, \binits{B.-N.}},
\bauthor{\bsnm{Li}, \binits{N.}},
\bauthor{\bsnm{Elhatisari}, \binits{S.}},
\bauthor{\bsnm{Lee}, \binits{D.}},
\bauthor{\bsnm{Drut}, \binits{J.E.}},
\bauthor{\bsnm{L\"ahde}, \binits{T.A.}},
\bauthor{\bsnm{Epelbaum}, \binits{E.}},
\bauthor{\bsnm{Mei\ss{}ner}, \binits{U.-G.}}:
\batitle{{$Ab Initio$ Nuclear Thermodynamics}}.
\bjtitle{Phys. Rev. Lett.}
\bvolume{125}(\bissue{19}),
\bfpage{192502}
(\byear{2020})
\doiurl{10.1103/PhysRevLett.125.192502}
{\href{https://arxiv.org/abs/1912.05105}{{arXiv:1912.05105}}}
{[nucl-th]}
\end{barticle}
\endbibitem

\bibitem[\protect\citeauthoryear{Muller et~al.}{2000}]{Muller:1999cp}
\begin{barticle}
\bauthor{\bsnm{Muller}, \binits{H.M.}},
\bauthor{\bsnm{Koonin}, \binits{S.E.}},
\bauthor{\bsnm{Seki}, \binits{R.}},
\bauthor{\bsnm{Kolck}, \binits{U.}}:
\batitle{{Nuclear matter on a lattice}}.
\bjtitle{Phys. Rev. C}
\bvolume{61},
\bfpage{044320}
(\byear{2000})
\doiurl{10.1103/PhysRevC.61.044320}
{\href{https://arxiv.org/abs/nucl-th/9910038}{{arXiv:nucl-th/9910038}}}
\end{barticle}
\endbibitem

\bibitem[\protect\citeauthoryear{Epelbaum et~al.}{2009}]{Epelbaum:2009rkz}
\begin{barticle}
\bauthor{\bsnm{Epelbaum}, \binits{E.}},
\bauthor{\bsnm{Krebs}, \binits{H.}},
\bauthor{\bsnm{Lee}, \binits{D.}},
\bauthor{\bsnm{Mei{\ss}ner}, \binits{U.-G.}}:
\batitle{{Ground state energy of dilute neutron matter at next-to-leading order
  in lattice chiral effective field theory}}.
\bjtitle{Eur. Phys. J. A}
\bvolume{40},
\bfpage{199}--\blpage{213}
(\byear{2009})
\doiurl{10.1140/epja/i2009-10755-0}
{\href{https://arxiv.org/abs/0812.3653}{{arXiv:0812.3653}}}
{[nucl-th]}
\end{barticle}
\endbibitem

\bibitem[\protect\citeauthoryear{Haensel et~al.}{2007}]{Haensel:2007yy}
\begin{bbook}
\bauthor{\bsnm{Haensel}, \binits{P.}},
\bauthor{\bsnm{Potekhin}, \binits{A.Y.}},
\bauthor{\bsnm{Yakovlev}, \binits{D.G.}}:
\bbtitle{{Neutron Stars 1: Equations of State and Structure}}
vol. \bseriesno{326}.
\bpublisher{Springer},
\blocation{New York, USA}
(\byear{2007}).
\doiurl{10.1007/978-0-387-47301-7}
\end{bbook}
\endbibitem

\bibitem[\protect\citeauthoryear{Harrison et~al.}{1965}]{HarrisonWheeler}
\begin{bbook}
\bauthor{\bsnm{Harrison}, \binits{B.K.}},
\bauthor{\bsnm{Thorne}, \binits{K.S.}},
\bauthor{\bsnm{Wakano}, \binits{M.}},
\bauthor{\bsnm{Wheeler}, \binits{J.A.}}:
\bbtitle{Gravitation Theory and Gravitational Collapse}.
\bpublisher{The University of Chicago Press},
\blocation{Chicago}
(\byear{1965})
\end{bbook}
\endbibitem

\bibitem[\protect\citeauthoryear{Pearson et~al.}{2018}]{Pearson:2018tkr}
\begin{barticle}
\bauthor{\bsnm{Pearson}, \binits{J.M.}},
\bauthor{\bsnm{Chamel}, \binits{N.}},
\bauthor{\bsnm{Potekhin}, \binits{A.Y.}},
\bauthor{\bsnm{Fantina}, \binits{A.F.}},
\bauthor{\bsnm{Ducoin}, \binits{C.}},
\bauthor{\bsnm{Dutta}, \binits{A.K.}},
\bauthor{\bsnm{Goriely}, \binits{S.}}:
\batitle{{Unified equations of state for cold non-accreting neutron stars with
  Brussels-Montreal functionals - I. Role of symmetry energy}}.
\bjtitle{Mon. Not. Roy. Astron. Soc.}
\bvolume{481}(\bissue{3}),
\bfpage{2994}--\blpage{3026}
(\byear{2018})
\doiurl{10.1093/mnras/sty2413}
{\href{https://arxiv.org/abs/1903.04981}{{arXiv:1903.04981}}}
{[astro-ph.HE]}.
\bcomment{[Erratum: Mon.Not.Roy.Astron.Soc. 486, 768 (2019)]}
\end{barticle}
\endbibitem

\bibitem[\protect\citeauthoryear{Demorest et~al.}{2010}]{Demorest:2010bx}
\begin{barticle}
\bauthor{\bsnm{Demorest}, \binits{P.B.}},
\bauthor{\bsnm{Pennucci}, \binits{T.}},
\bauthor{\bsnm{Ransom}, \binits{S.M.}},
\bauthor{\bsnm{Roberts}, \binits{M.S.E.}},
\bauthor{\bsnm{Hessels}, \binits{J.H.T.}}:
\batitle{{A two-solar-mass neutron star measured using Shapiro delay}}.
\bjtitle{Nature}
\bvolume{467},
\bfpage{1081}--\blpage{1083}
(\byear{2010})
\doiurl{10.1038/nature09466}
{\href{https://arxiv.org/abs/1010.5788}{{arXiv:1010.5788}}}
{[astro-ph.HE]}
\end{barticle}
\endbibitem

\bibitem[\protect\citeauthoryear{Antoniadis et~al.}{2013}]{Antoniadis:2013pzd}
\begin{barticle}
\bauthor{\bsnm{Antoniadis}, \binits{J.}}, \betal:
\batitle{{A massive pulsar in a compact relativistic binary}}.
\bjtitle{Science}
\bvolume{340}(\bissue{6131}),
\bfpage{1233232}
(\byear{2013})
\doiurl{10.1126/science.1233232}
{\href{https://arxiv.org/abs/1304.6875}{{arXiv:1304.6875}}}
{[astro-ph.HE]}
\end{barticle}
\endbibitem

\bibitem[\protect\citeauthoryear{Fonseca et~al.}{2021}]{Fonseca:2021wxt}
\begin{barticle}
\bauthor{\bsnm{Fonseca}, \binits{E.}}, \betal:
\batitle{{Refined Mass and Geometric Measurements of the High-mass PSR
  J0740+6620}}.
\bjtitle{Astrophys. J. Lett.}
\bvolume{915}(\bissue{1}),
\bfpage{12}
(\byear{2021})
\doiurl{10.3847/2041-8213/ac03b8}
{\href{https://arxiv.org/abs/2104.00880}{{arXiv:2104.00880}}}
{[astro-ph.HE]}
\end{barticle}
\endbibitem

\bibitem[\protect\citeauthoryear{Riley et~al.}{2019}]{Riley:2019yda}
\begin{barticle}
\bauthor{\bsnm{Riley}, \binits{T.E.}}, \betal:
\batitle{{A $NICER$ view of PSR J0030+0451: Millisecond pulsar parameter
  estimation}}.
\bjtitle{Astrophys. J. Lett.}
\bvolume{887}(\bissue{1}),
\bfpage{21}
(\byear{2019})
\doiurl{10.3847/2041-8213/ab481c}
{\href{https://arxiv.org/abs/1912.05702}{{arXiv:1912.05702}}}
{[astro-ph.HE]}
\end{barticle}
\endbibitem

\bibitem[\protect\citeauthoryear{Miller et~al.}{2019}]{Miller:2019cac}
\begin{barticle}
\bauthor{\bsnm{Miller}, \binits{M.C.}}, \betal:
\batitle{{PSR J0030+0451 mass and radius from $NICER$ data and implications for
  the properties of neutron star matter}}.
\bjtitle{Astrophys. J. Lett.}
\bvolume{887}(\bissue{1}),
\bfpage{24}
(\byear{2019})
\doiurl{10.3847/2041-8213/ab50c5}
{\href{https://arxiv.org/abs/1912.05705}{{arXiv:1912.05705}}}
{[astro-ph.HE]}
\end{barticle}
\endbibitem

\bibitem[\protect\citeauthoryear{Riley et~al.}{2021}]{Riley:2021pdl}
\begin{barticle}
\bauthor{\bsnm{Riley}, \binits{T.E.}}, \betal:
\batitle{{A NICER View of the Massive Pulsar PSR J0740+6620 Informed by Radio
  Timing and XMM-Newton Spectroscopy}}.
\bjtitle{Astrophys. J. Lett.}
\bvolume{918}(\bissue{2}),
\bfpage{27}
(\byear{2021})
\doiurl{10.3847/2041-8213/ac0a81}
{\href{https://arxiv.org/abs/2105.06980}{{arXiv:2105.06980}}}
{[astro-ph.HE]}
\end{barticle}
\endbibitem

\bibitem[\protect\citeauthoryear{Miller et~al.}{2021}]{Miller:2021qha}
\begin{barticle}
\bauthor{\bsnm{Miller}, \binits{M.C.}}, \betal:
\batitle{{The Radius of PSR J0740+6620 from NICER and XMM-Newton Data}}.
\bjtitle{Astrophys. J. Lett.}
\bvolume{918}(\bissue{2}),
\bfpage{28}
(\byear{2021})
\doiurl{10.3847/2041-8213/ac089b}
{\href{https://arxiv.org/abs/2105.06979}{{arXiv:2105.06979}}}
{[astro-ph.HE]}
\end{barticle}
\endbibitem

\bibitem[\protect\citeauthoryear{Salmi et~al.}{2022}]{Salmi:2022cgy}
\begin{barticle}
\bauthor{\bsnm{Salmi}, \binits{T.}}, \betal:
\batitle{{The Radius of PSR J0740+6620 from NICER with NICER Background
  Estimates}}.
\bjtitle{Astrophys. J.}
\bvolume{941}(\bissue{2}),
\bfpage{150}
(\byear{2022})
\doiurl{10.3847/1538-4357/ac983d}
{\href{https://arxiv.org/abs/2209.12840}{{arXiv:2209.12840}}}
{[astro-ph.HE]}
\end{barticle}
\endbibitem

\bibitem[\protect\citeauthoryear{Abbott et~al.}{2019}]{LIGOScientific:2018hze}
\begin{barticle}
\bauthor{\bsnm{Abbott}, \binits{B.P.}}, \betal:
\batitle{{Properties of the binary neutron star merger GW170817}}.
\bjtitle{Phys. Rev. X}
\bvolume{9}(\bissue{1}),
\bfpage{011001}
(\byear{2019})
\doiurl{10.1103/PhysRevX.9.011001}
{\href{https://arxiv.org/abs/1805.11579}{{arXiv:1805.11579}}}
{[gr-qc]}
\end{barticle}
\endbibitem

\bibitem[\protect\citeauthoryear{Abbott et~al.}{2018}]{LIGOScientific:2018cki}
\begin{barticle}
\bauthor{\bsnm{Abbott}, \binits{B.P.}}, \betal:
\batitle{{GW170817: Measurements of neutron star radii and equation of state}}.
\bjtitle{Phys. Rev. Lett.}
\bvolume{121}(\bissue{16}),
\bfpage{161101}
(\byear{2018})
\doiurl{10.1103/PhysRevLett.121.161101}
{\href{https://arxiv.org/abs/1805.11581}{{arXiv:1805.11581}}}
{[gr-qc]}
\end{barticle}
\endbibitem

\bibitem[\protect\citeauthoryear{Abbott et~al.}{2020}]{LIGOScientific:2020aai}
\begin{barticle}
\bauthor{\bsnm{Abbott}, \binits{B.P.}}, \betal:
\batitle{{GW190425: Observation of a Compact Binary Coalescence with Total Mass
  $\sim 3.4 M_{\odot}$}}.
\bjtitle{Astrophys. J. Lett.}
\bvolume{892}(\bissue{1}),
\bfpage{3}
(\byear{2020})
\doiurl{10.3847/2041-8213/ab75f5}
{\href{https://arxiv.org/abs/2001.01761}{{arXiv:2001.01761}}}
{[astro-ph.HE]}
\end{barticle}
\endbibitem

\bibitem[\protect\citeauthoryear{Zuern et~al.}{2013}]{Zuern:2013}
\begin{barticle}
\bauthor{\bsnm{Zuern}, \binits{G.}},
\bauthor{\bsnm{Lompe}, \binits{T.}},
\bauthor{\bsnm{Wenz}, \binits{A.}},
\bauthor{\bsnm{Jochim}, \binits{S.}},
\bauthor{\bsnm{Julienne}, \binits{P.}},
\bauthor{\bsnm{Hutson}, \binits{J.}}:
\batitle{{Precise Characterization of $^6\text{Li}$ Feshbach Resonances Using
  Trap-Sideband-Resolved RF Spectroscopy of Weakly Bound Molecules}}.
\bjtitle{Phys. Rev. Lett.}
\bvolume{110},
\bfpage{135301}
(\byear{2013})
\doiurl{10.1103/PhysRevLett.110.135301}
\end{barticle}
\endbibitem

\bibitem[\protect\citeauthoryear{Tews et~al.}{2017}]{Tews:2016jhi}
\begin{barticle}
\bauthor{\bsnm{Tews}, \binits{I.}},
\bauthor{\bsnm{Lattimer}, \binits{J.M.}},
\bauthor{\bsnm{Ohnishi}, \binits{A.}},
\bauthor{\bsnm{Kolomeitsev}, \binits{E.E.}}:
\batitle{{Symmetry Parameter Constraints from a Lower Bound on Neutron-matter
  Energy}}.
\bjtitle{Astrophys. J.}
\bvolume{848}(\bissue{2}),
\bfpage{105}
(\byear{2017})
\doiurl{10.3847/1538-4357/aa8db9}
{\href{https://arxiv.org/abs/1611.07133}{{arXiv:1611.07133}}}
{[nucl-th]}
\end{barticle}
\endbibitem

\bibitem[\protect\citeauthoryear{Lattimer}{2023}]{Lattimer:2023xjm}
\begin{barticle}
\bauthor{\bsnm{Lattimer}, \binits{J.M.}}:
\batitle{{Constraints on the Nuclear Symmetry Energy from Experiments, Theory
  and Observations}}.
\bjtitle{J. Phys. Conf. Ser.}
\bvolume{2536}(\bissue{1}),
\bfpage{012009}
(\byear{2023})
\doiurl{10.1088/1742-6596/2536/1/012009}
{\href{https://arxiv.org/abs/2308.08001}{{arXiv:2308.08001}}}
{[nucl-th]}
\end{barticle}
\endbibitem

\bibitem[\protect\citeauthoryear{Annala et~al.}{2020}]{Annala:2019puf}
\begin{barticle}
\bauthor{\bsnm{Annala}, \binits{E.}},
\bauthor{\bsnm{Gorda}, \binits{T.}},
\bauthor{\bsnm{Kurkela}, \binits{A.}},
\bauthor{\bsnm{N\"attil\"a}, \binits{J.}},
\bauthor{\bsnm{Vuorinen}, \binits{A.}}:
\batitle{{Evidence for quark-matter cores in massive neutron stars}}.
\bjtitle{Nat. Phys.}
\bvolume{16}(\bissue{9}),
\bfpage{907}--\blpage{910}
(\byear{2020})
\doiurl{10.1038/s41567-020-0914-9}
{\href{https://arxiv.org/abs/1903.09121}{{arXiv:1903.09121}}}
{[astro-ph.HE]}
\end{barticle}
\endbibitem

\bibitem[\protect\citeauthoryear{Blaschke and Chamel}{2018}]{Blaschke:2018mqw}
\begin{barticle}
\bauthor{\bsnm{Blaschke}, \binits{D.}},
\bauthor{\bsnm{Chamel}, \binits{N.}}:
\batitle{{Phases of dense matter in compact stars}}.
\bjtitle{Astrophys. Space Sci. Libr.}
\bvolume{457},
\bfpage{337}--\blpage{400}
(\byear{2018})
\doiurl{10.1007/978-3-319-97616-7_7}
{\href{https://arxiv.org/abs/1803.01836}{{arXiv:1803.01836}}}
{[nucl-th]}
\end{barticle}
\endbibitem

\bibitem[\protect\citeauthoryear{Barba-Gonz{\'a}lez
  et~al.}{2022}]{Barba-Gonzalez:2022pkn}
\begin{barticle}
\bauthor{\bsnm{Barba-Gonz{\'a}lez}, \binits{D.}},
\bauthor{\bsnm{Albertus}, \binits{C.}},
\bauthor{\bsnm{P{\'e}rez-Garc{\'\i}a}, \binits{M.{\'A}.}}:
\batitle{{Crystallization in single- and multicomponent neutron star crusts}}.
\bjtitle{Phys. Rev. C}
\bvolume{106}(\bissue{6}),
\bfpage{065806}
(\byear{2022})
\doiurl{10.1103/PhysRevC.106.065806}
{\href{https://arxiv.org/abs/2207.14323}{{arXiv:2207.14323}}}
{[astro-ph.HE]}
\end{barticle}
\endbibitem

\bibitem[\protect\citeauthoryear{Barba-Gonz{\'a}lez
  et~al.}{2024}]{Barba-Gonzalez:2023lln}
\begin{barticle}
\bauthor{\bsnm{Barba-Gonz{\'a}lez}, \binits{D.}},
\bauthor{\bsnm{Albertus}, \binits{C.}},
\bauthor{\bsnm{P{\'e}rez-Garc{\'\i}a}, \binits{M.{\'A}.}}:
\batitle{{Virialized equation of state for warm and dense stellar plasmas in
  proto-neutron stars and Supernova matter}}.
\bjtitle{Mon. Not. Roy. Astron. Soc.}
\bvolume{528},
\bfpage{3498}--\blpage{3508}
(\byear{2024})
\doiurl{10.1093/mnras/stae235}
{\href{https://arxiv.org/abs/2312.16252}{{arXiv:2312.16252}}}
{[astro-ph.HE]}
\end{barticle}
\endbibitem

\bibitem[\protect\citeauthoryear{Baym et~al.}{1971}]{Baym:1971ax}
\begin{barticle}
\bauthor{\bsnm{Baym}, \binits{G.}},
\bauthor{\bsnm{Bethe}, \binits{H.A.}},
\bauthor{\bsnm{Pethick}, \binits{C.}}:
\batitle{{Neutron star matter}}.
\bjtitle{Nucl. Phys. A}
\bvolume{175},
\bfpage{225}--\blpage{271}
(\byear{1971})
\doiurl{10.1016/0375-9474(71)90281-8}
\end{barticle}
\endbibitem

\bibitem[\protect\citeauthoryear{Chamel}{2020}]{Chamel:2020hni}
\begin{barticle}
\bauthor{\bsnm{Chamel}, \binits{N.}}:
\batitle{{Analytical determination of the structure of the outer crust of a
  cold nonaccreted neutron star}}.
\bjtitle{Phys. Rev. C}
\bvolume{101}(\bissue{3}),
\bfpage{032801}
(\byear{2020})
\doiurl{10.1103/PhysRevC.101.032801}
{\href{https://arxiv.org/abs/2003.00983}{{arXiv:2003.00983}}}
{[astro-ph.HE]}
\end{barticle}
\endbibitem

\bibitem[\protect\citeauthoryear{Negele and Vautherin}{1973}]{Negele:1971vb}
\begin{barticle}
\bauthor{\bsnm{Negele}, \binits{J.W.}},
\bauthor{\bsnm{Vautherin}, \binits{D.}}:
\batitle{{Neutron star matter at subnuclear densities}}.
\bjtitle{Nucl. Phys. A}
\bvolume{207},
\bfpage{298}--\blpage{320}
(\byear{1973})
\doiurl{10.1016/0375-9474(73)90349-7}
\end{barticle}
\endbibitem

\bibitem[\protect\citeauthoryear{Chamel et~al.}{2015}]{Chamel:2015oqa}
\begin{barticle}
\bauthor{\bsnm{Chamel}, \binits{N.}},
\bauthor{\bsnm{Fantina}, \binits{A.F.}},
\bauthor{\bsnm{Zdunik}, \binits{J.L.}},
\bauthor{\bsnm{Haensel}, \binits{P.}}:
\batitle{{Neutron drip transition in accreting and nonaccreting neutron star
  crusts}}.
\bjtitle{Phys. Rev. C}
\bvolume{91}(\bissue{5}),
\bfpage{055803}
(\byear{2015})
\doiurl{10.1103/PhysRevC.91.055803}
{\href{https://arxiv.org/abs/1504.04537}{{arXiv:1504.04537}}}
{[astro-ph.HE]}
\end{barticle}
\endbibitem

\bibitem[\protect\citeauthoryear{Piekarewicz
  et~al.}{2014}]{Piekarewicz:2014lba}
\begin{barticle}
\bauthor{\bsnm{Piekarewicz}, \binits{J.}},
\bauthor{\bsnm{Fattoyev}, \binits{F.J.}},
\bauthor{\bsnm{Horowitz}, \binits{C.J.}}:
\batitle{{Pulsar Glitches: The Crust may be Enough}}.
\bjtitle{Phys. Rev. C}
\bvolume{90}(\bissue{1}),
\bfpage{015803}
(\byear{2014})
\doiurl{10.1103/PhysRevC.90.015803}
{\href{https://arxiv.org/abs/1404.2660}{{arXiv:1404.2660}}}
{[nucl-th]}
\end{barticle}
\endbibitem

\bibitem[\protect\citeauthoryear{Grams et~al.}{2022}]{Grams:2022lci}
\begin{barticle}
\bauthor{\bsnm{Grams}, \binits{G.}},
\bauthor{\bsnm{Margueron}, \binits{J.}},
\bauthor{\bsnm{Somasundaram}, \binits{R.}},
\bauthor{\bsnm{Reddy}, \binits{S.}}:
\batitle{{Confronting a set of Skyrme and $\chi _{EFT}$ predictions for the
  crust of neutron stars: On the origin of uncertainties in model
  predictions}}.
\bjtitle{Eur. Phys. J. A}
\bvolume{58}(\bissue{3}),
\bfpage{56}
(\byear{2022})
\doiurl{10.1140/epja/s10050-022-00706-w}
{\href{https://arxiv.org/abs/2203.11645}{{arXiv:2203.11645}}}
{[nucl-th]}
\end{barticle}
\endbibitem

\bibitem[\protect\citeauthoryear{Lattimer et~al.}{1991}]{Lattimer:1991ib}
\begin{barticle}
\bauthor{\bsnm{Lattimer}, \binits{J.M.}},
\bauthor{\bsnm{Prakash}, \binits{M.}},
\bauthor{\bsnm{Pethick}, \binits{C.J.}},
\bauthor{\bsnm{Haensel}, \binits{P.}}:
\batitle{{Direct URCA process in neutron stars}}.
\bjtitle{Phys. Rev. Lett.}
\bvolume{66},
\bfpage{2701}--\blpage{2704}
(\byear{1991})
\doiurl{10.1103/PhysRevLett.66.2701}
\end{barticle}
\endbibitem

\bibitem[\protect\citeauthoryear{Shen et~al.}{1998}]{Shen:1998by}
\begin{barticle}
\bauthor{\bsnm{Shen}, \binits{H.}},
\bauthor{\bsnm{Toki}, \binits{H.}},
\bauthor{\bsnm{Oyamatsu}, \binits{K.}},
\bauthor{\bsnm{Sumiyoshi}, \binits{K.}}:
\batitle{{Relativistic equation of state of nuclear matter for supernova
  explosion}}.
\bjtitle{Prog. Theor. Phys.}
\bvolume{100},
\bfpage{1013}
(\byear{1998})
\doiurl{10.1143/PTP.100.1013}
{\href{https://arxiv.org/abs/nucl-th/9806095}{{arXiv:nucl-th/9806095}}}
\end{barticle}
\endbibitem

\bibitem[\protect\citeauthoryear{Douchin and Haensel}{2001}]{Douchin:2001sv}
\begin{barticle}
\bauthor{\bsnm{Douchin}, \binits{F.}},
\bauthor{\bsnm{Haensel}, \binits{P.}}:
\batitle{{A unified equation of state of dense matter and neutron star
  structure}}.
\bjtitle{Astron. Astrophys.}
\bvolume{380},
\bfpage{151}
(\byear{2001})
\doiurl{10.1051/0004-6361:20011402}
{\href{https://arxiv.org/abs/astro-ph/0111092}{{arXiv:astro-ph/0111092}}}
\end{barticle}
\endbibitem

\bibitem[\protect\citeauthoryear{Chamel et~al.}{2008}]{Chamel:2008aa}
\begin{barticle}
\bauthor{\bsnm{Chamel}, \binits{N.}},
\bauthor{\bsnm{Goriely}, \binits{S.}},
\bauthor{\bsnm{Pearson}, \binits{J.M.}}:
\batitle{{Further explorations of Skyrme-Hartree-Fock-Bogoliubov mass formulas.
  IX. Constraint of pairing force to S(0)-1 neutron-matter gap}}.
\bjtitle{Nucl. Phys. A}
\bvolume{812},
\bfpage{72}--\blpage{98}
(\byear{2008})
\doiurl{10.1016/j.nuclphysa.2008.08.015}
{\href{https://arxiv.org/abs/0809.0447}{{arXiv:0809.0447}}}
{[nucl-th]}
\end{barticle}
\endbibitem

\bibitem[\protect\citeauthoryear{Goriely et~al.}{2009}]{Goriely:2009zzb}
\begin{barticle}
\bauthor{\bsnm{Goriely}, \binits{S.}},
\bauthor{\bsnm{Chamel}, \binits{N.}},
\bauthor{\bsnm{Pearson}, \binits{J.M.}}:
\batitle{{Skyrme-Hartree-Fock-Bogoliubov nuclear mass formulas: Crossing the
  0.6 MeV threshold with microscopically deduced pairing}}.
\bjtitle{Phys. Rev. Lett.}
\bvolume{102},
\bfpage{152503}
(\byear{2009})
\doiurl{10.1103/PhysRevLett.102.152503}
{\href{https://arxiv.org/abs/0906.2607}{{arXiv:0906.2607}}}
{[nucl-th]}
\end{barticle}
\endbibitem

\bibitem[\protect\citeauthoryear{Goriely et~al.}{2010}]{Goriely:2010bm}
\begin{barticle}
\bauthor{\bsnm{Goriely}, \binits{S.}},
\bauthor{\bsnm{Chamel}, \binits{N.}},
\bauthor{\bsnm{Pearson}, \binits{J.M.}}:
\batitle{{Further explorations of Skyrme-Hartree-Fock-Bogoliubov mass formulas.
  XII: Stiffness and stability of neutron-star matter}}.
\bjtitle{Phys. Rev. C}
\bvolume{82},
\bfpage{035804}
(\byear{2010})
\doiurl{10.1103/PhysRevC.82.035804}
{\href{https://arxiv.org/abs/1009.3840}{{arXiv:1009.3840}}}
{[nucl-th]}
\end{barticle}
\endbibitem

\bibitem[\protect\citeauthoryear{Goriely et~al.}{2013a}]{Goriely:2013xba}
\begin{barticle}
\bauthor{\bsnm{Goriely}, \binits{S.}},
\bauthor{\bsnm{Chamel}, \binits{N.}},
\bauthor{\bsnm{Pearson}, \binits{J.M.}}:
\batitle{{Further explorations of Skyrme-Hartree-Fock-Bogoliubov mass formulas.
  13. The 2012 atomic mass evaluation and the symmetry coefficient}}.
\bjtitle{Phys. Rev. C}
\bvolume{88}(\bissue{2}),
\bfpage{024308}
(\byear{2013})
\doiurl{10.1103/PhysRevC.88.024308}
\end{barticle}
\endbibitem

\bibitem[\protect\citeauthoryear{Goriely et~al.}{2013b}]{Goriely:2013nxa}
\begin{barticle}
\bauthor{\bsnm{Goriely}, \binits{S.}},
\bauthor{\bsnm{Chamel}, \binits{N.}},
\bauthor{\bsnm{Pearson}, \binits{J.M.}}:
\batitle{{Hartree-Fock-Bogoliubov nuclear mass model with 0.50 MeV accuracy
  based on standard forms of Skyrme and pairing functionals}}.
\bjtitle{Phys. Rev. C}
\bvolume{88}(\bissue{6}),
\bfpage{061302}
(\byear{2013})
\doiurl{10.1103/PhysRevC.88.061302}
\end{barticle}
\endbibitem

\bibitem[\protect\citeauthoryear{Sharma et~al.}{2015}]{Sharma:2015bna}
\begin{barticle}
\bauthor{\bsnm{Sharma}, \binits{B.K.}},
\bauthor{\bsnm{Centelles}, \binits{M.}},
\bauthor{\bsnm{Vi\~nas}, \binits{X.}},
\bauthor{\bsnm{Baldo}, \binits{M.}},
\bauthor{\bsnm{Burgio}, \binits{G.F.}}:
\batitle{{Unified equation of state for neutron stars on a microscopic basis}}.
\bjtitle{Astron. Astrophys.}
\bvolume{584},
\bfpage{103}
(\byear{2015})
\doiurl{10.1051/0004-6361/201526642}
{\href{https://arxiv.org/abs/1506.00375}{{arXiv:1506.00375}}}
{[nucl-th]}
\end{barticle}
\endbibitem

\bibitem[\protect\citeauthoryear{Watanabe et~al.}{2000}]{Watanabe:2000rj}
\begin{barticle}
\bauthor{\bsnm{Watanabe}, \binits{G.}},
\bauthor{\bsnm{Iida}, \binits{K.}},
\bauthor{\bsnm{Sato}, \binits{K.}}:
\batitle{{Thermodynamic properties of nuclear 'pasta' in neutron star crusts}}.
\bjtitle{Nucl. Phys. A}
\bvolume{676},
\bfpage{455}--\blpage{473}
(\byear{2000})
\doiurl{10.1016/S0375-9474(00)00197-4}
{\href{https://arxiv.org/abs/astro-ph/0001273}{{arXiv:astro-ph/0001273}}}.
\bcomment{[Erratum: Nucl.Phys.A 726, 357--365 (2003)]}
\end{barticle}
\endbibitem

\bibitem[\protect\citeauthoryear{Lonardoni et~al.}{2015}]{Lonardoni:2014bwa}
\begin{barticle}
\bauthor{\bsnm{Lonardoni}, \binits{D.}},
\bauthor{\bsnm{Lovato}, \binits{A.}},
\bauthor{\bsnm{Gandolfi}, \binits{S.}},
\bauthor{\bsnm{Pederiva}, \binits{F.}}:
\batitle{{Hyperon Puzzle: Hints from Quantum Monte Carlo Calculations}}.
\bjtitle{Phys. Rev. Lett.}
\bvolume{114}(\bissue{9}),
\bfpage{092301}
(\byear{2015})
\doiurl{10.1103/PhysRevLett.114.092301}
{\href{https://arxiv.org/abs/1407.4448}{{arXiv:1407.4448}}}
{[nucl-th]}
\end{barticle}
\endbibitem

\bibitem[\protect\citeauthoryear{Alford et~al.}{2008}]{Alford:2007xm}
\begin{barticle}
\bauthor{\bsnm{Alford}, \binits{M.G.}},
\bauthor{\bsnm{Schmitt}, \binits{A.}},
\bauthor{\bsnm{Rajagopal}, \binits{K.}},
\bauthor{\bsnm{Sch\"afer}, \binits{T.}}:
\batitle{{Color superconductivity in dense quark matter}}.
\bjtitle{Rev. Mod. Phys.}
\bvolume{80},
\bfpage{1455}--\blpage{1515}
(\byear{2008})
\doiurl{10.1103/RevModPhys.80.1455}
{\href{https://arxiv.org/abs/0709.4635}{{arXiv:0709.4635}}}
{[hep-ph]}
\end{barticle}
\endbibitem

\bibitem[\protect\citeauthoryear{Glendenning}{1997}]{Glendenning:1997wn}
\begin{bbook}
\bauthor{\bsnm{Glendenning}, \binits{N.K.}}:
\bbtitle{{Compact Stars: Nuclear Physics, Particle Physics, and General
  Relativity}}.
\bpublisher{Springer},
\blocation{Germany}
(\byear{1997})
\end{bbook}
\endbibitem

\bibitem[\protect\citeauthoryear{Katayama and Saito}{2015}]{Katayama:2015dga}
\begin{barticle}
\bauthor{\bsnm{Katayama}, \binits{T.}},
\bauthor{\bsnm{Saito}, \binits{K.}}:
\batitle{{Hyperons in neutron stars}}.
\bjtitle{Phys. Lett. B}
\bvolume{747},
\bfpage{43}--\blpage{47}
(\byear{2015})
\doiurl{10.1016/j.physletb.2015.03.039}
{\href{https://arxiv.org/abs/1501.05419}{{arXiv:1501.05419}}}
{[nucl-th]}
\end{barticle}
\endbibitem

\bibitem[\protect\citeauthoryear{Rijken and Schulze}{2016}]{Rijken:2016uon}
\begin{barticle}
\bauthor{\bsnm{Rijken}, \binits{T.A.}},
\bauthor{\bsnm{Schulze}, \binits{H.J.}}:
\batitle{{Hyperon-hyperon interactions with the Nijmegen ESC08 model}}.
\bjtitle{Eur. Phys. J. A}
\bvolume{52}(\bissue{2}),
\bfpage{21}
(\byear{2016})
\doiurl{10.1140/epja/i2016-16021-6}
\end{barticle}
\endbibitem

\bibitem[\protect\citeauthoryear{Glendenning and
  Moszkowski}{1991}]{Glendenning:1991es}
\begin{barticle}
\bauthor{\bsnm{Glendenning}, \binits{N.K.}},
\bauthor{\bsnm{Moszkowski}, \binits{S.A.}}:
\batitle{{Reconciliation of neutron star masses and binding of the lambda in
  hypernuclei}}.
\bjtitle{Phys. Rev. Lett.}
\bvolume{67},
\bfpage{2414}--\blpage{2417}
(\byear{1991})
\doiurl{10.1103/PhysRevLett.67.2414}
\end{barticle}
\endbibitem

\bibitem[\protect\citeauthoryear{Gaitanos et~al.}{2004}]{Gaitanos:2003zg}
\begin{barticle}
\bauthor{\bsnm{Gaitanos}, \binits{T.}},
\bauthor{\bsnm{Di~Toro}, \binits{M.}},
\bauthor{\bsnm{Typel}, \binits{S.}},
\bauthor{\bsnm{Baran}, \binits{V.}},
\bauthor{\bsnm{Fuchs}, \binits{C.}},
\bauthor{\bsnm{Greco}, \binits{V.}},
\bauthor{\bsnm{Wolter}, \binits{H.H.}}:
\batitle{{On the Lorentz structure of the symmetry energy}}.
\bjtitle{Nucl. Phys. A}
\bvolume{732},
\bfpage{24}--\blpage{48}
(\byear{2004})
\doiurl{10.1016/j.nuclphysa.2003.12.001}
{\href{https://arxiv.org/abs/nucl-th/0309021}{{arXiv:nucl-th/0309021}}}
\end{barticle}
\endbibitem

\bibitem[\protect\citeauthoryear{Bonanno and Sedrakian}{2012}]{Bonanno:2011ch}
\begin{barticle}
\bauthor{\bsnm{Bonanno}, \binits{L.}},
\bauthor{\bsnm{Sedrakian}, \binits{A.}}:
\batitle{{Composition and stability of hybrid stars with hyperons and quark
  color-superconductivity}}.
\bjtitle{Astron. Astrophys.}
\bvolume{539},
\bfpage{16}
(\byear{2012})
\doiurl{10.1051/0004-6361/201117832}
{\href{https://arxiv.org/abs/1108.0559}{{arXiv:1108.0559}}}
{[astro-ph.SR]}
\end{barticle}
\endbibitem

\bibitem[\protect\citeauthoryear{Suleiman et~al.}{2021}]{Suleiman:2021hre}
\begin{barticle}
\bauthor{\bsnm{Suleiman}, \binits{L.}},
\bauthor{\bsnm{Fortin}, \binits{M.}},
\bauthor{\bsnm{Zdunik}, \binits{J.L.}},
\bauthor{\bsnm{Haensel}, \binits{P.}}:
\batitle{{Influence of the crust on the neutron star macrophysical quantities
  and universal relations}}.
\bjtitle{Phys. Rev. C}
\bvolume{104}(\bissue{1}),
\bfpage{015801}
(\byear{2021})
\doiurl{10.1103/PhysRevC.104.015801}
{\href{https://arxiv.org/abs/2106.12845}{{arXiv:2106.12845}}}
{[astro-ph.HE]}
\end{barticle}
\endbibitem

\bibitem[\protect\citeauthoryear{Hinderer et~al.}{2010}]{Hinderer:2009ca}
\begin{barticle}
\bauthor{\bsnm{Hinderer}, \binits{T.}},
\bauthor{\bsnm{Lackey}, \binits{B.D.}},
\bauthor{\bsnm{Lang}, \binits{R.N.}},
\bauthor{\bsnm{Read}, \binits{J.S.}}:
\batitle{{Tidal deformability of neutron stars with realistic equations of
  state and their gravitational wave signatures in binary inspiral}}.
\bjtitle{Phys. Rev. D}
\bvolume{81},
\bfpage{123016}
(\byear{2010})
\doiurl{10.1103/PhysRevD.81.123016}
{\href{https://arxiv.org/abs/0911.3535}{{arXiv:0911.3535}}}
{[astro-ph.HE]}
\end{barticle}
\endbibitem

\bibitem[\protect\citeauthoryear{Zacchi}{2020}]{Zacchi:2020dxl}
\begin{botherref}
\oauthor{\bsnm{Zacchi}, \binits{A.}}:
{Gravitational quadrupole deformation and the tidal deformability for stellar
  systems: (The number of) Love for undergraduates}
(2020)
{\href{https://arxiv.org/abs/2007.00423}{{arXiv:2007.00423}}}
{[astro-ph.HE]}
\end{botherref}
\endbibitem

\bibitem[\protect\citeauthoryear{Abbott et~al.}{2018}]{Abbott:2018exr}
\begin{barticle}
\bauthor{\bsnm{Abbott}, \binits{B.P.}}, \betal:
\batitle{{GW170817: Measurements of neutron star radii and equation of state}}.
\bjtitle{Phys. Rev. Lett.}
\bvolume{121}(\bissue{16}),
\bfpage{161101}
(\byear{2018})
\doiurl{10.1103/PhysRevLett.121.161101}
{\href{https://arxiv.org/abs/1805.11581}{{arXiv:1805.11581}}}
{[gr-qc]}
\end{barticle}
\endbibitem

\bibitem[\protect\citeauthoryear{Baym et~al.}{1971}]{Baym:1971pw}
\begin{barticle}
\bauthor{\bsnm{Baym}, \binits{G.}},
\bauthor{\bsnm{Pethick}, \binits{C.}},
\bauthor{\bsnm{Sutherland}, \binits{P.}}:
\batitle{{The Ground state of matter at high densities: Equation of state and
  stellar models}}.
\bjtitle{Astrophys. J.}
\bvolume{170},
\bfpage{299}--\blpage{317}
(\byear{1971})
\doiurl{10.1086/151216}
\end{barticle}
\endbibitem

\bibitem[\protect\citeauthoryear{Epelbaum et~al.}{2009}]{Epelbaum:2008ga}
\begin{barticle}
\bauthor{\bsnm{Epelbaum}, \binits{E.}},
\bauthor{\bsnm{Hammer}, \binits{H.-W.}},
\bauthor{\bsnm{Meissner}, \binits{U.-G.}}:
\batitle{{Modern Theory of Nuclear Forces}}.
\bjtitle{Rev. Mod. Phys.}
\bvolume{81},
\bfpage{1773}--\blpage{1825}
(\byear{2009})
\doiurl{10.1103/RevModPhys.81.1773}
{\href{https://arxiv.org/abs/0811.1338}{{arXiv:0811.1338}}}
{[nucl-th]}
\end{barticle}
\endbibitem

\bibitem[\protect\citeauthoryear{Machleidt and Entem}{2011}]{Machleidt:2011zz}
\begin{barticle}
\bauthor{\bsnm{Machleidt}, \binits{R.}},
\bauthor{\bsnm{Entem}, \binits{D.R.}}:
\batitle{{Chiral effective field theory and nuclear forces}}.
\bjtitle{Phys. Rept.}
\bvolume{503},
\bfpage{1}--\blpage{75}
(\byear{2011})
\doiurl{10.1016/j.physrep.2011.02.001}
{\href{https://arxiv.org/abs/1105.2919}{{arXiv:1105.2919}}}
{[nucl-th]}
\end{barticle}
\endbibitem

\bibitem[\protect\citeauthoryear{Hammer et~al.}{2020}]{Hammer:2019poc}
\begin{barticle}
\bauthor{\bsnm{Hammer}, \binits{H.-W.}},
\bauthor{\bsnm{K\"onig}, \binits{S.}},
\bauthor{\bsnm{Kolck}, \binits{U.}}:
\batitle{{Nuclear effective field theory: status and perspectives}}.
\bjtitle{Rev. Mod. Phys.}
\bvolume{92}(\bissue{2}),
\bfpage{025004}
(\byear{2020})
\doiurl{10.1103/RevModPhys.92.025004}
{\href{https://arxiv.org/abs/1906.12122}{{arXiv:1906.12122}}}
{[nucl-th]}
\end{barticle}
\endbibitem

\bibitem[\protect\citeauthoryear{Tews et~al.}{2013}]{Tews:2012fj}
\begin{barticle}
\bauthor{\bsnm{Tews}, \binits{I.}},
\bauthor{\bsnm{Kr\"uger}, \binits{T.}},
\bauthor{\bsnm{Hebeler}, \binits{K.}},
\bauthor{\bsnm{Schwenk}, \binits{A.}}:
\batitle{{Neutron matter at next-to-next-to-next-to-leading order in chiral
  effective field theory}}.
\bjtitle{Phys. Rev. Lett.}
\bvolume{110}(\bissue{3}),
\bfpage{032504}
(\byear{2013})
\doiurl{10.1103/PhysRevLett.110.032504}
{\href{https://arxiv.org/abs/1206.0025}{{arXiv:1206.0025}}}
{[nucl-th]}
\end{barticle}
\endbibitem

\bibitem[\protect\citeauthoryear{Hebeler et~al.}{2013}]{Hebeler:2013nza}
\begin{barticle}
\bauthor{\bsnm{Hebeler}, \binits{K.}},
\bauthor{\bsnm{Lattimer}, \binits{J.M.}},
\bauthor{\bsnm{Pethick}, \binits{C.J.}},
\bauthor{\bsnm{Schwenk}, \binits{A.}}:
\batitle{{Equation of state and neutron star properties constrained by nuclear
  physics and observation}}.
\bjtitle{Astrophys. J.}
\bvolume{773}(\bissue{1}),
\bfpage{11}
(\byear{2013})
\doiurl{10.1088/0004-637X/773/1/11}
{\href{https://arxiv.org/abs/1303.4662}{{arXiv:1303.4662}}}
{[astro-ph.SR]}
\end{barticle}
\endbibitem

\bibitem[\protect\citeauthoryear{Carbone et~al.}{2013}]{Carbone:2013rca}
\begin{barticle}
\bauthor{\bsnm{Carbone}, \binits{A.}},
\bauthor{\bsnm{Rios}, \binits{A.}},
\bauthor{\bsnm{Polls}, \binits{A.}}:
\batitle{{Symmetric nuclear matter with chiral three-nucleon forces in the
  self-consistent Green's functions approach}}.
\bjtitle{Phys. Rev. C}
\bvolume{88},
\bfpage{044302}
(\byear{2013})
\doiurl{10.1103/PhysRevC.88.044302}
{\href{https://arxiv.org/abs/1307.1889}{{arXiv:1307.1889}}}
{[nucl-th]}
\end{barticle}
\endbibitem

\bibitem[\protect\citeauthoryear{Holt and Kaiser}{2017}]{Holt:2016pjb}
\begin{barticle}
\bauthor{\bsnm{Holt}, \binits{J.W.}},
\bauthor{\bsnm{Kaiser}, \binits{N.}}:
\batitle{{Equation of state of nuclear and neutron matter at third-order in
  perturbation theory from chiral effective field theory}}.
\bjtitle{Phys. Rev. C}
\bvolume{95}(\bissue{3}),
\bfpage{034326}
(\byear{2017})
\doiurl{10.1103/PhysRevC.95.034326}
{\href{https://arxiv.org/abs/1612.04309}{{arXiv:1612.04309}}}
{[nucl-th]}
\end{barticle}
\endbibitem

\bibitem[\protect\citeauthoryear{Drischler et~al.}{2016}]{Drischler:2016djf}
\begin{barticle}
\bauthor{\bsnm{Drischler}, \binits{C.}},
\bauthor{\bsnm{Carbone}, \binits{A.}},
\bauthor{\bsnm{Hebeler}, \binits{K.}},
\bauthor{\bsnm{Schwenk}, \binits{A.}}:
\batitle{{Neutron matter from chiral two- and three-nucleon calculations up to
  N$^3$LO}}.
\bjtitle{Phys. Rev. C}
\bvolume{94}(\bissue{5}),
\bfpage{054307}
(\byear{2016})
\doiurl{10.1103/PhysRevC.94.054307}
{\href{https://arxiv.org/abs/1608.05615}{{arXiv:1608.05615}}}
{[nucl-th]}
\end{barticle}
\endbibitem

\bibitem[\protect\citeauthoryear{Lonardoni et~al.}{2020}]{Lonardoni:2019ypg}
\begin{barticle}
\bauthor{\bsnm{Lonardoni}, \binits{D.}},
\bauthor{\bsnm{Tews}, \binits{I.}},
\bauthor{\bsnm{Gandolfi}, \binits{S.}},
\bauthor{\bsnm{Carlson}, \binits{J.}}:
\batitle{{Nuclear and neutron-star matter from local chiral interactions}}.
\bjtitle{Phys. Rev. Res.}
\bvolume{2}(\bissue{2}),
\bfpage{022033}
(\byear{2020})
\doiurl{10.1103/PhysRevResearch.2.022033}
{\href{https://arxiv.org/abs/1912.09411}{{arXiv:1912.09411}}}
{[nucl-th]}
\end{barticle}
\endbibitem

\bibitem[\protect\citeauthoryear{Piarulli et~al.}{2020}]{Piarulli:2019pfq}
\begin{barticle}
\bauthor{\bsnm{Piarulli}, \binits{M.}},
\bauthor{\bsnm{Bombaci}, \binits{I.}},
\bauthor{\bsnm{Logoteta}, \binits{D.}},
\bauthor{\bsnm{Lovato}, \binits{A.}},
\bauthor{\bsnm{Wiringa}, \binits{R.B.}}:
\batitle{{Benchmark calculations of pure neutron matter with realistic
  nucleon-nucleon interactions}}.
\bjtitle{Phys. Rev. C}
\bvolume{101}(\bissue{4}),
\bfpage{045801}
(\byear{2020})
\doiurl{10.1103/PhysRevC.101.045801}
{\href{https://arxiv.org/abs/1908.04426}{{arXiv:1908.04426}}}
{[nucl-th]}
\end{barticle}
\endbibitem

\bibitem[\protect\citeauthoryear{Kurkela et~al.}{2010}]{Kurkela:2009gj}
\begin{barticle}
\bauthor{\bsnm{Kurkela}, \binits{A.}},
\bauthor{\bsnm{Romatschke}, \binits{P.}},
\bauthor{\bsnm{Vuorinen}, \binits{A.}}:
\batitle{{Cold quark matter}}.
\bjtitle{Phys. Rev. D}
\bvolume{81}(\bissue{10}),
\bfpage{105021}
(\byear{2010})
\doiurl{10.1103/PhysRevD.81.105021}
{\href{https://arxiv.org/abs/0912.1856}{{arXiv:0912.1856}}}
{[hep-ph]}
\end{barticle}
\endbibitem

\bibitem[\protect\citeauthoryear{Gorda et~al.}{2021}]{Gorda:2021kme}
\begin{barticle}
\bauthor{\bsnm{Gorda}, \binits{T.}},
\bauthor{\bsnm{Kurkela}, \binits{A.}},
\bauthor{\bsnm{Paatelainen}, \binits{R.}},
\bauthor{\bsnm{S\"appi}, \binits{S.}},
\bauthor{\bsnm{Vuorinen}, \binits{A.}}:
\batitle{{Cold quark matter at N3LO: Soft contributions}}.
\bjtitle{Phys. Rev. D}
\bvolume{104}(\bissue{7}),
\bfpage{074015}
(\byear{2021})
\doiurl{10.1103/PhysRevD.104.074015}
{\href{https://arxiv.org/abs/2103.07427}{{arXiv:2103.07427}}}
{[hep-ph]}
\end{barticle}
\endbibitem

\bibitem[\protect\citeauthoryear{Gorda et~al.}{2023}]{Gorda:2022jvk}
\begin{barticle}
\bauthor{\bsnm{Gorda}, \binits{T.}},
\bauthor{\bsnm{Komoltsev}, \binits{O.}},
\bauthor{\bsnm{Kurkela}, \binits{A.}}:
\batitle{{Ab-initio QCD Calculations Impact the Inference of the
  Neutron-star-matter Equation of State}}.
\bjtitle{Astrophys. J.}
\bvolume{950}(\bissue{2}),
\bfpage{107}
(\byear{2023})
\doiurl{10.3847/1538-4357/acce3a}
{\href{https://arxiv.org/abs/2204.11877}{{arXiv:2204.11877}}}
{[nucl-th]}
\end{barticle}
\endbibitem

\bibitem[\protect\citeauthoryear{Gorda et~al.}{2021}]{Gorda:2021znl}
\begin{barticle}
\bauthor{\bsnm{Gorda}, \binits{T.}},
\bauthor{\bsnm{Kurkela}, \binits{A.}},
\bauthor{\bsnm{Paatelainen}, \binits{R.}},
\bauthor{\bsnm{S\"appi}, \binits{S.}},
\bauthor{\bsnm{Vuorinen}, \binits{A.}}:
\batitle{{Soft Interactions in Cold Quark Matter}}.
\bjtitle{Phys. Rev. Lett.}
\bvolume{127}(\bissue{16}),
\bfpage{162003}
(\byear{2021})
\doiurl{10.1103/PhysRevLett.127.162003}
{\href{https://arxiv.org/abs/2103.05658}{{arXiv:2103.05658}}}
{[hep-ph]}
\end{barticle}
\endbibitem

\bibitem[\protect\citeauthoryear{Komoltsev and
  Kurkela}{2022}]{Komoltsev:2021jzg}
\begin{barticle}
\bauthor{\bsnm{Komoltsev}, \binits{O.}},
\bauthor{\bsnm{Kurkela}, \binits{A.}}:
\batitle{{How Perturbative QCD Constrains the Equation of State at Neutron-Star
  Densities}}.
\bjtitle{Phys. Rev. Lett.}
\bvolume{128}(\bissue{20}),
\bfpage{202701}
(\byear{2022})
\doiurl{10.1103/PhysRevLett.128.202701}
{\href{https://arxiv.org/abs/2111.05350}{{arXiv:2111.05350}}}
{[nucl-th]}
\end{barticle}
\endbibitem

\bibitem[\protect\citeauthoryear{Margueron et~al.}{2018a}]{Margueron:2017lup}
\begin{barticle}
\bauthor{\bsnm{Margueron}, \binits{J.}},
\bauthor{\bsnm{Hoffmann~Casali}, \binits{R.}},
\bauthor{\bsnm{Gulminelli}, \binits{F.}}:
\batitle{{Equation of state for dense nucleonic matter from metamodeling. II.
  Predictions for neutron star properties}}.
\bjtitle{Phys. Rev. C}
\bvolume{97}(\bissue{2}),
\bfpage{025806}
(\byear{2018})
\doiurl{10.1103/PhysRevC.97.025806}
{\href{https://arxiv.org/abs/1708.06895}{{arXiv:1708.06895}}}
{[nucl-th]}
\end{barticle}
\endbibitem

\bibitem[\protect\citeauthoryear{Margueron et~al.}{2018b}]{Margueron:2017eqc}
\begin{barticle}
\bauthor{\bsnm{Margueron}, \binits{J.}},
\bauthor{\bsnm{Hoffmann~Casali}, \binits{R.}},
\bauthor{\bsnm{Gulminelli}, \binits{F.}}:
\batitle{{Equation of state for dense nucleonic matter from metamodeling. I.
  Foundational aspects}}.
\bjtitle{Phys. Rev. C}
\bvolume{97}(\bissue{2}),
\bfpage{025805}
(\byear{2018})
\doiurl{10.1103/PhysRevC.97.025805}
{\href{https://arxiv.org/abs/1708.06894}{{arXiv:1708.06894}}}
{[nucl-th]}
\end{barticle}
\endbibitem

\bibitem[\protect\citeauthoryear{Somasundaram
  et~al.}{2021}]{Somasundaram:2020chb}
\begin{barticle}
\bauthor{\bsnm{Somasundaram}, \binits{R.}},
\bauthor{\bsnm{Drischler}, \binits{C.}},
\bauthor{\bsnm{Tews}, \binits{I.}},
\bauthor{\bsnm{Margueron}, \binits{J.}}:
\batitle{{Constraints on the nuclear symmetry energy from asymmetric-matter
  calculations with chiral $NN$ and $3N$ interactions}}.
\bjtitle{Phys. Rev. C}
\bvolume{103}(\bissue{4}),
\bfpage{045803}
(\byear{2021})
\doiurl{10.1103/PhysRevC.103.045803}
{\href{https://arxiv.org/abs/2009.04737}{{arXiv:2009.04737}}}
{[nucl-th]}
\end{barticle}
\endbibitem

\bibitem[\protect\citeauthoryear{Chabanat et~al.}{1998}]{Chabanat:1997un}
\begin{barticle}
\bauthor{\bsnm{Chabanat}, \binits{E.}},
\bauthor{\bsnm{Bonche}, \binits{P.}},
\bauthor{\bsnm{Haensel}, \binits{P.}},
\bauthor{\bsnm{Meyer}, \binits{J.}},
\bauthor{\bsnm{Schaeffer}, \binits{R.}}:
\batitle{{A Skyrme parametrization from subnuclear to neutron star densities.
  2. Nuclei far from stablities}}.
\bjtitle{Nucl. Phys. A}
\bvolume{635},
\bfpage{231}--\blpage{256}
(\byear{1998})
\doiurl{10.1016/S0375-9474(98)00180-8} .
\bcomment{[Erratum: Nucl.Phys.A 643, 441--441 (1998)]}
\end{barticle}
\endbibitem

\bibitem[\protect\citeauthoryear{Koehn et~al.}{2024}]{Koehn:2024set}
\begin{botherref}
\oauthor{\bsnm{Koehn}, \binits{H.}}, et al.:
{An overview of existing and new nuclear and astrophysical constraints on the
  equation of state of neutron-rich dense matter}
(2024)
{\href{https://arxiv.org/abs/2402.04172}{{arXiv:2402.04172}}}
{[astro-ph.HE]}
\end{botherref}
\endbibitem

\bibitem[\protect\citeauthoryear{Baldo and Burgio}{2016}]{Baldo:2016jhp}
\begin{barticle}
\bauthor{\bsnm{Baldo}, \binits{M.}},
\bauthor{\bsnm{Burgio}, \binits{G.F.}}:
\batitle{{The nuclear symmetry energy}}.
\bjtitle{Prog. Part. Nucl. Phys.}
\bvolume{91},
\bfpage{203}--\blpage{258}
(\byear{2016})
\doiurl{10.1016/j.ppnp.2016.06.006}
{\href{https://arxiv.org/abs/1606.08838}{{arXiv:1606.08838}}}
{[nucl-th]}
\end{barticle}
\endbibitem

\bibitem[\protect\citeauthoryear{Wellenhofer
  et~al.}{2016}]{Wellenhofer:2016lnl}
\begin{barticle}
\bauthor{\bsnm{Wellenhofer}, \binits{C.}},
\bauthor{\bsnm{Holt}, \binits{J.W.}},
\bauthor{\bsnm{Kaiser}, \binits{N.}}:
\batitle{{Divergence of the isospin-asymmetry expansion of the nuclear equation
  of state in many-body perturbation theory}}.
\bjtitle{Phys. Rev. C}
\bvolume{93}(\bissue{5}),
\bfpage{055802}
(\byear{2016})
\doiurl{10.1103/PhysRevC.93.055802}
{\href{https://arxiv.org/abs/1603.02935}{{arXiv:1603.02935}}}
{[nucl-th]}
\end{barticle}
\endbibitem

\bibitem[\protect\citeauthoryear{Piekarewicz and
  Centelles}{2009}]{Piekarewicz:2008nh}
\begin{barticle}
\bauthor{\bsnm{Piekarewicz}, \binits{J.}},
\bauthor{\bsnm{Centelles}, \binits{M.}}:
\batitle{{Incompressibility of neutron-rich matter}}.
\bjtitle{Phys. Rev. C}
\bvolume{79},
\bfpage{054311}
(\byear{2009})
\doiurl{10.1103/PhysRevC.79.054311}
{\href{https://arxiv.org/abs/0812.4499}{{arXiv:0812.4499}}}
{[nucl-th]}
\end{barticle}
\endbibitem

\bibitem[\protect\citeauthoryear{Millerson and
  Sammarruca}{2019}]{Millerson:2019jkg}
\begin{botherref}
\oauthor{\bsnm{Millerson}, \binits{R.}},
\oauthor{\bsnm{Sammarruca}, \binits{F.}}:
{Properties of isospin asymmetric matter derived from chiral effective field
  theory}
(2019)
{\href{https://arxiv.org/abs/1906.02905}{{arXiv:1906.02905}}}
{[nucl-th]}
\end{botherref}
\endbibitem

\bibitem[\protect\citeauthoryear{Kurkela et~al.}{2014}]{Kurkela:2014vha}
\begin{barticle}
\bauthor{\bsnm{Kurkela}, \binits{A.}},
\bauthor{\bsnm{Fraga}, \binits{E.S.}},
\bauthor{\bsnm{Schaffner-Bielich}, \binits{J.}},
\bauthor{\bsnm{Vuorinen}, \binits{A.}}:
\batitle{{Constraining neutron star matter with Quantum Chromodynamics}}.
\bjtitle{Astrophys. J.}
\bvolume{789}(\bissue{2}),
\bfpage{127}
(\byear{2014})
\doiurl{10.1088/0004-637X/789/2/127}
{\href{https://arxiv.org/abs/1402.6618}{{arXiv:1402.6618}}}
{[astro-ph.HE]}
\end{barticle}
\endbibitem

\bibitem[\protect\citeauthoryear{Raithel et~al.}{2016}]{Raithel:2016bux}
\begin{barticle}
\bauthor{\bsnm{Raithel}, \binits{C.A.}},
\bauthor{\bsnm{Ozel}, \binits{F.}},
\bauthor{\bsnm{Psaltis}, \binits{D.}}:
\batitle{{From Neutron Star Observables to the Equation of State: An Optimal
  Parametrization}}.
\bjtitle{Astrophys. J.}
\bvolume{831}(\bissue{1}),
\bfpage{44}
(\byear{2016})
\doiurl{10.3847/0004-637X/831/1/44}
{\href{https://arxiv.org/abs/1605.03591}{{arXiv:1605.03591}}}
{[astro-ph.HE]}
\end{barticle}
\endbibitem

\bibitem[\protect\citeauthoryear{Greif et~al.}{2019}]{Greif:2018njt}
\begin{barticle}
\bauthor{\bsnm{Greif}, \binits{S.K.}},
\bauthor{\bsnm{Raaijmakers}, \binits{G.}},
\bauthor{\bsnm{Hebeler}, \binits{K.}},
\bauthor{\bsnm{Schwenk}, \binits{A.}},
\bauthor{\bsnm{Watts}, \binits{A.L.}}:
\batitle{{Equation of state sensitivities when inferring neutron star and dense
  matter properties}}.
\bjtitle{Mon. Not. Roy. Astron. Soc.}
\bvolume{485}(\bissue{4}),
\bfpage{5363}--\blpage{5376}
(\byear{2019})
\doiurl{10.1093/mnras/stz654}
{\href{https://arxiv.org/abs/1812.08188}{{arXiv:1812.08188}}}
{[astro-ph.HE]}
\end{barticle}
\endbibitem

\bibitem[\protect\citeauthoryear{Tews et~al.}{2018a}]{Tews:2018kmu}
\begin{barticle}
\bauthor{\bsnm{Tews}, \binits{I.}},
\bauthor{\bsnm{Carlson}, \binits{J.}},
\bauthor{\bsnm{Gandolfi}, \binits{S.}},
\bauthor{\bsnm{Reddy}, \binits{S.}}:
\batitle{{Constraining the speed of sound inside neutron stars with chiral
  effective field theory interactions and observations}}.
\bjtitle{Astrophys. J.}
\bvolume{860}(\bissue{2}),
\bfpage{149}
(\byear{2018})
\doiurl{10.3847/1538-4357/aac267}
{\href{https://arxiv.org/abs/1801.01923}{{arXiv:1801.01923}}}
{[nucl-th]}
\end{barticle}
\endbibitem

\bibitem[\protect\citeauthoryear{Tews et~al.}{2018b}]{Tews:2018iwm}
\begin{barticle}
\bauthor{\bsnm{Tews}, \binits{I.}},
\bauthor{\bsnm{Margueron}, \binits{J.}},
\bauthor{\bsnm{Reddy}, \binits{S.}}:
\batitle{{Critical examination of constraints on the equation of state of dense
  matter obtained from GW170817}}.
\bjtitle{Phys. Rev. C}
\bvolume{98}(\bissue{4}),
\bfpage{045804}
(\byear{2018})
\doiurl{10.1103/PhysRevC.98.045804}
{\href{https://arxiv.org/abs/1804.02783}{{arXiv:1804.02783}}}
{[nucl-th]}
\end{barticle}
\endbibitem

\bibitem[\protect\citeauthoryear{Abbott
  et~al.}{2017}]{TheLIGOScientific:2017qsa}
\begin{barticle}
\bauthor{\bsnm{Abbott}, \binits{B.P.}}, \betal:
\batitle{{GW170817: Observation of gravitational waves from a binary neutron
  star inspiral}}.
\bjtitle{Phys. Rev. Lett.}
\bvolume{119}(\bissue{16}),
\bfpage{161101}
(\byear{2017})
\doiurl{10.1103/PhysRevLett.119.161101}
{\href{https://arxiv.org/abs/1710.05832}{{arXiv:1710.05832}}}
{[gr-qc]}
\end{barticle}
\endbibitem

\bibitem[\protect\citeauthoryear{Altiparmak et~al.}{2022}]{Altiparmak:2022bke}
\begin{barticle}
\bauthor{\bsnm{Altiparmak}, \binits{S.}},
\bauthor{\bsnm{Ecker}, \binits{C.}},
\bauthor{\bsnm{Rezzolla}, \binits{L.}}:
\batitle{{On the Sound Speed in Neutron Stars}}.
\bjtitle{Astrophys. J. Lett.}
\bvolume{939}(\bissue{2}),
\bfpage{34}
(\byear{2022})
\doiurl{10.3847/2041-8213/ac9b2a}
{\href{https://arxiv.org/abs/2203.14974}{{arXiv:2203.14974}}}
{[astro-ph.HE]}
\end{barticle}
\endbibitem

\bibitem[\protect\citeauthoryear{Ecker and Rezzolla}{2022}]{Ecker:2022dlg}
\begin{barticle}
\bauthor{\bsnm{Ecker}, \binits{C.}},
\bauthor{\bsnm{Rezzolla}, \binits{L.}}:
\batitle{{Impact of large-mass constraints on the properties of neutron
  stars}}.
\bjtitle{Mon. Not. Roy. Astron. Soc.}
\bvolume{519}(\bissue{2}),
\bfpage{2615}--\blpage{2622}
(\byear{2022})
\doiurl{10.1093/mnras/stac3755}
{\href{https://arxiv.org/abs/2209.08101}{{arXiv:2209.08101}}}
{[astro-ph.HE]}
\end{barticle}
\endbibitem

\bibitem[\protect\citeauthoryear{Annala et~al.}{2022}]{Annala:2021gom}
\begin{barticle}
\bauthor{\bsnm{Annala}, \binits{E.}},
\bauthor{\bsnm{Gorda}, \binits{T.}},
\bauthor{\bsnm{Katerini}, \binits{E.}},
\bauthor{\bsnm{Kurkela}, \binits{A.}},
\bauthor{\bsnm{N\"attil\"a}, \binits{J.}},
\bauthor{\bsnm{Paschalidis}, \binits{V.}},
\bauthor{\bsnm{Vuorinen}, \binits{A.}}:
\batitle{{Multimessenger Constraints for Ultradense Matter}}.
\bjtitle{Phys. Rev. X}
\bvolume{12}(\bissue{1}),
\bfpage{011058}
(\byear{2022})
\doiurl{10.1103/PhysRevX.12.011058}
{\href{https://arxiv.org/abs/2105.05132}{{arXiv:2105.05132}}}
{[astro-ph.HE]}
\end{barticle}
\endbibitem

\bibitem[\protect\citeauthoryear{Oter et~al.}{2019}]{Oter:2019rqp}
\begin{barticle}
\bauthor{\bsnm{Oter}, \binits{E.L.}},
\bauthor{\bsnm{Windisch}, \binits{A.}},
\bauthor{\bsnm{Llanes-Estrada}, \binits{F.J.}},
\bauthor{\bsnm{Alford}, \binits{M.}}:
\batitle{{nEoS: Neutron Star Equation of State from hadron physics alone}}.
\bjtitle{J. Phys. G}
\bvolume{46}(\bissue{8}),
\bfpage{084001}
(\byear{2019})
\doiurl{10.1088/1361-6471/ab2567}
{\href{https://arxiv.org/abs/1901.05271}{{arXiv:1901.05271}}}
{[gr-qc]}
\end{barticle}
\endbibitem

\bibitem[\protect\citeauthoryear{Alarc{\'o}n et~al.}{2025}]{Alarcon:2024hlj}
\begin{barticle}
\bauthor{\bsnm{Alarc{\'o}n}, \binits{J.M.}},
\bauthor{\bsnm{Lope-Oter}, \binits{E.}},
\bauthor{\bsnm{Oller}, \binits{J.A.}}:
\batitle{{Regulator-independent equations of state for neutron stars generated
  from first principles}}.
\bjtitle{Phys. Rev. D}
\bvolume{112}(\bissue{2}),
\bfpage{023034}
(\byear{2025})
\doiurl{10.1103/tzzc-1tq6}
{\href{https://arxiv.org/abs/2410.14776}{{arXiv:2410.14776}}}
{[nucl-th]}
\end{barticle}
\endbibitem

\bibitem[\protect\citeauthoryear{Lope~Oter et~al.}{2019}]{LopeOter:2019pcq}
\begin{barticle}
\bauthor{\bsnm{Lope~Oter}, \binits{E.}},
\bauthor{\bsnm{Windisch}, \binits{A.}},
\bauthor{\bsnm{Llanes-Estrada}, \binits{F.J.}},
\bauthor{\bsnm{Alford}, \binits{M.}}:
\batitle{{nEoS: Neutron Star Equation of State from hadron physics alone}}.
\bjtitle{J. Phys. G}
\bvolume{46}(\bissue{8}),
\bfpage{084001}
(\byear{2019})
\doiurl{10.1088/1361-6471/ab2567}
{\href{https://arxiv.org/abs/1901.05271}{{arXiv:1901.05271}}}
{[gr-qc]}
\end{barticle}
\endbibitem

\bibitem[\protect\citeauthoryear{Gandolfi et~al.}{2012}]{Gandolfi:2011xu}
\begin{barticle}
\bauthor{\bsnm{Gandolfi}, \binits{S.}},
\bauthor{\bsnm{Carlson}, \binits{J.}},
\bauthor{\bsnm{Reddy}, \binits{S.}}:
\batitle{{The maximum mass and radius of neutron stars and the nuclear symmetry
  energy}}.
\bjtitle{Phys. Rev. C}
\bvolume{85},
\bfpage{032801}
(\byear{2012})
\doiurl{10.1103/PhysRevC.85.032801}
{\href{https://arxiv.org/abs/1101.1921}{{arXiv:1101.1921}}}
{[nucl-th]}
\end{barticle}
\endbibitem

\bibitem[\protect\citeauthoryear{Drischler et~al.}{2020}]{Drischler:2020hwi}
\begin{barticle}
\bauthor{\bsnm{Drischler}, \binits{C.}},
\bauthor{\bsnm{Furnstahl}, \binits{R.J.}},
\bauthor{\bsnm{Melendez}, \binits{J.A.}},
\bauthor{\bsnm{Phillips}, \binits{D.R.}}:
\batitle{{How Well Do We Know the Neutron-Matter Equation of State at the
  Densities Inside Neutron Stars? A Bayesian Approach with Correlated
  Uncertainties}}.
\bjtitle{Phys. Rev. Lett.}
\bvolume{125}(\bissue{20}),
\bfpage{202702}
(\byear{2020})
\doiurl{10.1103/PhysRevLett.125.202702}
{\href{https://arxiv.org/abs/2004.07232}{{arXiv:2004.07232}}}
{[nucl-th]}
\end{barticle}
\endbibitem

\bibitem[\protect\citeauthoryear{Adhikari et~al.}{2021}]{PREX:2021umo}
\begin{barticle}
\bauthor{\bsnm{Adhikari}, \binits{D.}}, \betal:
\batitle{{Accurate Determination of the Neutron Skin Thickness of $^{208}$Pb
  through Parity-Violation in Electron Scattering}}.
\bjtitle{Phys. Rev. Lett.}
\bvolume{126}(\bissue{17}),
\bfpage{172502}
(\byear{2021})
\doiurl{10.1103/PhysRevLett.126.172502}
{\href{https://arxiv.org/abs/2102.10767}{{arXiv:2102.10767}}}
{[nucl-ex]}
\end{barticle}
\endbibitem

\bibitem[\protect\citeauthoryear{Adhikari et~al.}{2022}]{CREX:2022kgg}
\begin{barticle}
\bauthor{\bsnm{Adhikari}, \binits{D.}}, \betal:
\batitle{{Precision Determination of the Neutral Weak Form Factor of Ca48}}.
\bjtitle{Phys. Rev. Lett.}
\bvolume{129}(\bissue{4}),
\bfpage{042501}
(\byear{2022})
\doiurl{10.1103/PhysRevLett.129.042501}
{\href{https://arxiv.org/abs/2205.11593}{{arXiv:2205.11593}}}
{[nucl-ex]}
\end{barticle}
\endbibitem

\bibitem[\protect\citeauthoryear{Reed et~al.}{2021}]{Reed:2021nqk}
\begin{barticle}
\bauthor{\bsnm{Reed}, \binits{B.T.}},
\bauthor{\bsnm{Fattoyev}, \binits{F.J.}},
\bauthor{\bsnm{Horowitz}, \binits{C.J.}},
\bauthor{\bsnm{Piekarewicz}, \binits{J.}}:
\batitle{{Implications of PREX-2 on the Equation of State of Neutron-Rich
  Matter}}.
\bjtitle{Phys. Rev. Lett.}
\bvolume{126}(\bissue{17}),
\bfpage{172503}
(\byear{2021})
\doiurl{10.1103/PhysRevLett.126.172503}
{\href{https://arxiv.org/abs/2101.03193}{{arXiv:2101.03193}}}
{[nucl-th]}
\end{barticle}
\endbibitem

\bibitem[\protect\citeauthoryear{Lacour et~al.}{2010}]{Lacour:2010ci}
\begin{barticle}
\bauthor{\bsnm{Lacour}, \binits{A.}},
\bauthor{\bsnm{Oller}, \binits{J.A.}},
\bauthor{\bsnm{Mei{\ss}ner}, \binits{U.-G.}}:
\batitle{{The Chiral quark condensate and pion decay constant in nuclear matter
  at next-to-leading order}}.
\bjtitle{J. Phys. G}
\bvolume{37},
\bfpage{125002}
(\byear{2010})
\doiurl{10.1088/0954-3899/37/12/125002}
{\href{https://arxiv.org/abs/1007.2574}{{arXiv:1007.2574}}}
{[nucl-th]}
\end{barticle}
\endbibitem

\bibitem[\protect\citeauthoryear{Lope-Oter and
  Llanes-Estrada}{2022}]{Lope-Oter:2021vxl}
\begin{barticle}
\bauthor{\bsnm{Lope-Oter}, \binits{E.}},
\bauthor{\bsnm{Llanes-Estrada}, \binits{F.J.}}:
\batitle{{Unbiased interpolated neutron-star EoS at finite T for modified
  gravity studies}}.
\bjtitle{Eur. Phys. J. A}
\bvolume{58}(\bissue{1}),
\bfpage{9}
(\byear{2022})
\doiurl{10.1140/epja/s10050-021-00656-9}
{\href{https://arxiv.org/abs/2108.04027}{{arXiv:2108.04027}}}
{[hep-ph]}
\end{barticle}
\endbibitem

\bibitem[\protect\citeauthoryear{Romani et~al.}{2022}]{Romani:2022jhd}
\begin{barticle}
\bauthor{\bsnm{Romani}, \binits{R.W.}},
\bauthor{\bsnm{Kandel}, \binits{D.}},
\bauthor{\bsnm{Filippenko}, \binits{A.V.}},
\bauthor{\bsnm{Brink}, \binits{T.G.}},
\bauthor{\bsnm{Zheng}, \binits{W.}}:
\batitle{{PSR J0952\ensuremath{-}0607: The Fastest and Heaviest Known Galactic
  Neutron Star}}.
\bjtitle{Astrophys. J. Lett.}
\bvolume{934}(\bissue{2}),
\bfpage{18}
(\byear{2022})
\doiurl{10.3847/2041-8213/ac8007}
{\href{https://arxiv.org/abs/2207.05124}{{arXiv:2207.05124}}}
{[astro-ph.HE]}
\end{barticle}
\endbibitem

\bibitem[\protect\citeauthoryear{{Doroshenko}
  et~al.}{2022}]{2022NatAs...6.1444D}
\begin{barticle}
\bauthor{\bsnm{{Doroshenko}}, \binits{V.}},
\bauthor{\bsnm{{Suleimanov}}, \binits{V.}},
\bauthor{\bsnm{{P{\"u}hlhofer}}, \binits{G.}},
\bauthor{\bsnm{{Santangelo}}, \binits{A.}}:
\batitle{{A strangely light neutron star within a supernova remnant}}.
\bjtitle{Nature Astronomy}
\bvolume{6},
\bfpage{1444}--\blpage{1451}
(\byear{2022})
\doiurl{10.1038/s41550-022-01800-1}
\end{barticle}
\endbibitem

\bibitem[\protect\citeauthoryear{Dittmann et~al.}{2024}]{Dittmann:2024mbo}
\begin{botherref}
\oauthor{\bsnm{Dittmann}, \binits{A.J.}}, et al.:
{A More Precise Measurement of the Radius of PSR J0740+6620 Using Updated NICER
  Data}
(2024)
{\href{https://arxiv.org/abs/2406.14467}{{arXiv:2406.14467}}}
{[astro-ph.HE]}
\end{botherref}
\endbibitem

\bibitem[\protect\citeauthoryear{Salmi et~al.}{2024}]{Salmi:2024aum}
\begin{botherref}
\oauthor{\bsnm{Salmi}, \binits{T.}}, et al.:
{The Radius of the High Mass Pulsar PSR J0740+6620 With 3.6 Years of NICER
  Data}
(2024)
{\href{https://arxiv.org/abs/2406.14466}{{arXiv:2406.14466}}}
{[astro-ph.HE]}
\end{botherref}
\endbibitem

\bibitem[\protect\citeauthoryear{Choudhury et~al.}{2024}]{Choudhury:2024xbk}
\begin{barticle}
\bauthor{\bsnm{Choudhury}, \binits{D.}}, \betal:
\batitle{{A NICER View of the Nearest and Brightest Millisecond Pulsar: PSR
  J0437\textendash{}4715}}.
\bjtitle{Astrophys. J. Lett.}
\bvolume{971}(\bissue{1}),
\bfpage{20}
(\byear{2024})
\doiurl{10.3847/2041-8213/ad5a6f}
{\href{https://arxiv.org/abs/2407.06789}{{arXiv:2407.06789}}}
{[astro-ph.HE]}
\end{barticle}
\endbibitem

\bibitem[\protect\citeauthoryear{Brandes et~al.}{2023}]{Brandes:2023hma}
\begin{barticle}
\bauthor{\bsnm{Brandes}, \binits{L.}},
\bauthor{\bsnm{Weise}, \binits{W.}},
\bauthor{\bsnm{Kaiser}, \binits{N.}}:
\batitle{{Evidence against a strong first-order phase transition in neutron
  star cores: Impact of new data}}.
\bjtitle{Phys. Rev. D}
\bvolume{108}(\bissue{9}),
\bfpage{094014}
(\byear{2023})
\doiurl{10.1103/PhysRevD.108.094014}
{\href{https://arxiv.org/abs/2306.06218}{{arXiv:2306.06218}}}
{[nucl-th]}
\end{barticle}
\endbibitem

\bibitem[\protect\citeauthoryear{Fasano et~al.}{2019}]{Fasano:2019zwm}
\begin{barticle}
\bauthor{\bsnm{Fasano}, \binits{M.}},
\bauthor{\bsnm{Abdelsalhin}, \binits{T.}},
\bauthor{\bsnm{Maselli}, \binits{A.}},
\bauthor{\bsnm{Ferrari}, \binits{V.}}:
\batitle{{Constraining the Neutron Star Equation of State Using Multiband
  Independent Measurements of Radii and Tidal Deformabilities}}.
\bjtitle{Phys. Rev. Lett.}
\bvolume{123}(\bissue{14}),
\bfpage{141101}
(\byear{2019})
\doiurl{10.1103/PhysRevLett.123.141101}
{\href{https://arxiv.org/abs/1902.05078}{{arXiv:1902.05078}}}
{[astro-ph.HE]}
\end{barticle}
\endbibitem

\bibitem[\protect\citeauthoryear{Roca-Maza et~al.}{2015}]{Roca-Maza:2015eza}
\begin{barticle}
\bauthor{\bsnm{Roca-Maza}, \binits{X.}},
\bauthor{\bsnm{Vi\~nas}, \binits{X.}},
\bauthor{\bsnm{Centelles}, \binits{M.}},
\bauthor{\bsnm{Agrawal}, \binits{B.K.}},
\bauthor{\bsnm{Colo'}, \binits{G.}},
\bauthor{\bsnm{Paar}, \binits{N.}},
\bauthor{\bsnm{Piekarewicz}, \binits{J.}},
\bauthor{\bsnm{Vretenar}, \binits{D.}}:
\batitle{{The neutron skin thickness from the measured electric dipole
  polarizability in $^{68}$Ni, $^{120}$Sn, and $^{208}$Pb}}.
\bjtitle{Phys. Rev. C}
\bvolume{92},
\bfpage{064304}
(\byear{2015})
\doiurl{10.1103/PhysRevC.92.064304}
{\href{https://arxiv.org/abs/1510.01874}{{arXiv:1510.01874}}}
{[nucl-th]}
\end{barticle}
\endbibitem

\bibitem[\protect\citeauthoryear{Essick et~al.}{2021}]{Essick:2021kjb}
\begin{botherref}
\oauthor{\bsnm{Essick}, \binits{R.}},
\oauthor{\bsnm{Tews}, \binits{I.}},
\oauthor{\bsnm{Landry}, \binits{P.}},
\oauthor{\bsnm{Schwenk}, \binits{A.}}:
{Astrophysical constraints on the symmetry energy and the neutron skin of
  $^{208}$Pb with minimal modeling assumptions}
(2021)
{\href{https://arxiv.org/abs/2102.10074}{{arXiv:2102.10074}}}
{[nucl-th]}
\end{botherref}
\endbibitem

\bibitem[\protect\citeauthoryear{Raaijmakers
  et~al.}{2021}]{Raaijmakers:2021uju}
\begin{barticle}
\bauthor{\bsnm{Raaijmakers}, \binits{G.}},
\bauthor{\bsnm{Greif}, \binits{S.K.}},
\bauthor{\bsnm{Hebeler}, \binits{K.}},
\bauthor{\bsnm{Hinderer}, \binits{T.}},
\bauthor{\bsnm{Nissanke}, \binits{S.}},
\bauthor{\bsnm{Schwenk}, \binits{A.}},
\bauthor{\bsnm{Riley}, \binits{T.E.}},
\bauthor{\bsnm{Watts}, \binits{A.L.}},
\bauthor{\bsnm{Lattimer}, \binits{J.M.}},
\bauthor{\bsnm{Ho}, \binits{W.C.G.}}:
\batitle{{Constraints on the Dense Matter Equation of State and Neutron Star
  Properties from NICER\textquoteright{}s Mass\textendash{}Radius Estimate of
  PSR J0740+6620 and Multimessenger Observations}}.
\bjtitle{Astrophys. J. Lett.}
\bvolume{918}(\bissue{2}),
\bfpage{29}
(\byear{2021})
\doiurl{10.3847/2041-8213/ac089a}
{\href{https://arxiv.org/abs/2105.06981}{{arXiv:2105.06981}}}
{[astro-ph.HE]}
\end{barticle}
\endbibitem

\bibitem[\protect\citeauthoryear{Huth et~al.}{2022}]{Huth:2021bsp}
\begin{barticle}
\bauthor{\bsnm{Huth}, \binits{S.}}, \betal:
\batitle{{Constraining Neutron-Star Matter with Microscopic and Macroscopic
  Collisions}}.
\bjtitle{Nature}
\bvolume{606},
\bfpage{276}--\blpage{280}
(\byear{2022})
\doiurl{10.1038/s41586-022-04750-w}
{\href{https://arxiv.org/abs/2107.06229}{{arXiv:2107.06229}}}
{[nucl-th]}
\end{barticle}
\endbibitem

\bibitem[\protect\citeauthoryear{Jiang et~al.}{2023}]{Jiang:2022tps}
\begin{barticle}
\bauthor{\bsnm{Jiang}, \binits{J.-L.}},
\bauthor{\bsnm{Ecker}, \binits{C.}},
\bauthor{\bsnm{Rezzolla}, \binits{L.}}:
\batitle{{Bayesian Analysis of Neutron-star Properties with Parameterized
  Equations of State: The Role of the Likelihood Functions}}.
\bjtitle{Astrophys. J.}
\bvolume{949}(\bissue{1}),
\bfpage{11}
(\byear{2023})
\doiurl{10.3847/1538-4357/acc4be}
{\href{https://arxiv.org/abs/2211.00018}{{arXiv:2211.00018}}}
{[gr-qc]}
\end{barticle}
\endbibitem

\bibitem[\protect\citeauthoryear{Zhou et~al.}{2023}]{Zhou:2023hzu}
\begin{barticle}
\bauthor{\bsnm{Zhou}, \binits{J.}},
\bauthor{\bsnm{Xu}, \binits{J.}},
\bauthor{\bsnm{Papakonstantinou}, \binits{P.}}:
\batitle{{Bayesian inference of neutron-star observables based on effective
  nuclear interactions}}.
\bjtitle{Phys. Rev. C}
\bvolume{107}(\bissue{5}),
\bfpage{055803}
(\byear{2023})
\doiurl{10.1103/PhysRevC.107.055803}
{\href{https://arxiv.org/abs/2301.07904}{{arXiv:2301.07904}}}
{[nucl-th]}
\end{barticle}
\endbibitem

\bibitem[\protect\citeauthoryear{Prakash et~al.}{2024}]{Prakash:2023afe}
\begin{barticle}
\bauthor{\bsnm{Prakash}, \binits{A.}},
\bauthor{\bsnm{Gupta}, \binits{I.}},
\bauthor{\bsnm{Breschi}, \binits{M.}},
\bauthor{\bsnm{Kashyap}, \binits{R.}},
\bauthor{\bsnm{Radice}, \binits{D.}},
\bauthor{\bsnm{Bernuzzi}, \binits{S.}},
\bauthor{\bsnm{Logoteta}, \binits{D.}},
\bauthor{\bsnm{Sathyaprakash}, \binits{B.S.}}:
\batitle{{Detectability of QCD phase transitions in binary neutron star
  mergers: Bayesian inference with the next generation gravitational wave
  detectors}}.
\bjtitle{Phys. Rev. D}
\bvolume{109}(\bissue{10}),
\bfpage{103008}
(\byear{2024})
\doiurl{10.1103/PhysRevD.109.103008}
{\href{https://arxiv.org/abs/2310.06025}{{arXiv:2310.06025}}}
{[gr-qc]}
\end{barticle}
\endbibitem

\bibitem[\protect\citeauthoryear{Annala et~al.}{2023}]{Annala:2023cwx}
\begin{barticle}
\bauthor{\bsnm{Annala}, \binits{E.}},
\bauthor{\bsnm{Gorda}, \binits{T.}},
\bauthor{\bsnm{Hirvonen}, \binits{J.}},
\bauthor{\bsnm{Komoltsev}, \binits{O.}},
\bauthor{\bsnm{Kurkela}, \binits{A.}},
\bauthor{\bsnm{N\"attil\"a}, \binits{J.}},
\bauthor{\bsnm{Vuorinen}, \binits{A.}}:
\batitle{{Strongly interacting matter exhibits deconfined behavior in massive
  neutron stars}}.
\bjtitle{Nature Commun.}
\bvolume{14}(\bissue{1}),
\bfpage{8451}
(\byear{2023})
\doiurl{10.1038/s41467-023-44051-y}
{\href{https://arxiv.org/abs/2303.11356}{{arXiv:2303.11356}}}
{[astro-ph.HE]}
\end{barticle}
\endbibitem

\bibitem[\protect\citeauthoryear{Essick et~al.}{2023}]{Essick:2023fso}
\begin{barticle}
\bauthor{\bsnm{Essick}, \binits{R.}},
\bauthor{\bsnm{Legred}, \binits{I.}},
\bauthor{\bsnm{Chatziioannou}, \binits{K.}},
\bauthor{\bsnm{Han}, \binits{S.}},
\bauthor{\bsnm{Landry}, \binits{P.}}:
\batitle{{Phase transition phenomenology with nonparametric representations of
  the neutron star equation of state}}.
\bjtitle{Phys. Rev. D}
\bvolume{108}(\bissue{4}),
\bfpage{043013}
(\byear{2023})
\doiurl{10.1103/PhysRevD.108.043013}
{\href{https://arxiv.org/abs/2305.07411}{{arXiv:2305.07411}}}
{[astro-ph.HE]}
\end{barticle}
\endbibitem

\bibitem[\protect\citeauthoryear{Takatsy et~al.}{2023}]{Takatsy:2023xzf}
\begin{barticle}
\bauthor{\bsnm{Takatsy}, \binits{J.}},
\bauthor{\bsnm{Kovacs}, \binits{P.}},
\bauthor{\bsnm{Wolf}, \binits{G.}},
\bauthor{\bsnm{Schaffner-Bielich}, \binits{J.}}:
\batitle{{What neutron stars tell about the hadron-quark phase transition: A
  Bayesian study}}.
\bjtitle{Phys. Rev. D}
\bvolume{108}(\bissue{4}),
\bfpage{043002}
(\byear{2023})
\doiurl{10.1103/PhysRevD.108.043002}
{\href{https://arxiv.org/abs/2303.00013}{{arXiv:2303.00013}}}
{[astro-ph.HE]}
\end{barticle}
\endbibitem

\bibitem[\protect\citeauthoryear{Breschi et~al.}{2024}]{Breschi:2024qlc}
\begin{barticle}
\bauthor{\bsnm{Breschi}, \binits{M.}},
\bauthor{\bsnm{Gamba}, \binits{R.}},
\bauthor{\bsnm{Carullo}, \binits{G.}},
\bauthor{\bsnm{Godzieba}, \binits{D.}},
\bauthor{\bsnm{Bernuzzi}, \binits{S.}},
\bauthor{\bsnm{Perego}, \binits{A.}},
\bauthor{\bsnm{Radice}, \binits{D.}}:
\batitle{{Bayesian inference of multimessenger astrophysical data: Joint and
  coherent inference of gravitational waves and kilonovae}}.
\bjtitle{Astron. Astrophys.}
\bvolume{689},
\bfpage{51}
(\byear{2024})
\doiurl{10.1051/0004-6361/202449173}
{\href{https://arxiv.org/abs/2401.03750}{{arXiv:2401.03750}}}
{[astro-ph.HE]}
\end{barticle}
\endbibitem

\bibitem[\protect\citeauthoryear{Marquez et~al.}{2024}]{Marquez:2024bzj}
\begin{barticle}
\bauthor{\bsnm{Marquez}, \binits{K.D.}},
\bauthor{\bsnm{Malik}, \binits{T.}},
\bauthor{\bsnm{Pais}, \binits{H.}},
\bauthor{\bsnm{Menezes}, \binits{D.P.}},
\bauthor{\bsnm{Provid{\^e}ncia}, \binits{C.}}:
\batitle{{Nambu{\textendash}Jona-Lasinio description of hadronic matter from a
  Bayesian approach}}.
\bjtitle{Phys. Rev. D}
\bvolume{110}(\bissue{6}),
\bfpage{063040}
(\byear{2024})
\doiurl{10.1103/PhysRevD.110.063040}
{\href{https://arxiv.org/abs/2407.18452}{{arXiv:2407.18452}}}
{[nucl-th]}
\end{barticle}
\endbibitem

\bibitem[\protect\citeauthoryear{Rutherford et~al.}{2024}]{Rutherford:2024srk}
\begin{barticle}
\bauthor{\bsnm{Rutherford}, \binits{N.}}, \betal:
\batitle{{Constraining the Dense Matter Equation of State with New NICER
  Mass{\textendash}Radius Measurements and New Chiral Effective Field Theory
  Inputs}}.
\bjtitle{Astrophys. J. Lett.}
\bvolume{971}(\bissue{1}),
\bfpage{19}
(\byear{2024})
\doiurl{10.3847/2041-8213/ad5f02}
{\href{https://arxiv.org/abs/2407.06790}{{arXiv:2407.06790}}}
{[astro-ph.HE]}
\end{barticle}
\endbibitem

\bibitem[\protect\citeauthoryear{Ecker et~al.}{2025}]{Ecker:2024uqv}
\begin{barticle}
\bauthor{\bsnm{Ecker}, \binits{C.}},
\bauthor{\bsnm{Gorda}, \binits{T.}},
\bauthor{\bsnm{Kurkela}, \binits{A.}},
\bauthor{\bsnm{Rezzolla}, \binits{L.}}:
\batitle{{Constraining the equation of state in neutron-star cores via the
  long-ringdown signal}}.
\bjtitle{Nature Commun.}
\bvolume{16}(\bissue{1}),
\bfpage{1320}
(\byear{2025})
\doiurl{10.1038/s41467-025-56500-x}
{\href{https://arxiv.org/abs/2403.03246}{{arXiv:2403.03246}}}
{[astro-ph.HE]}
\end{barticle}
\endbibitem

\bibitem[\protect\citeauthoryear{Beznogov and Raduta}{2024}]{Beznogov:2024vcv}
\begin{barticle}
\bauthor{\bsnm{Beznogov}, \binits{M.V.}},
\bauthor{\bsnm{Raduta}, \binits{A.R.}}:
\batitle{{Bayesian inference of the dense matter equation~of state built upon
  extended Skyrme interactions}}.
\bjtitle{Phys. Rev. C}
\bvolume{110}(\bissue{3}),
\bfpage{035805}
(\byear{2024})
\doiurl{10.1103/PhysRevC.110.035805}
{\href{https://arxiv.org/abs/2403.19325}{{arXiv:2403.19325}}}
{[nucl-th]}
\end{barticle}
\endbibitem

\bibitem[\protect\citeauthoryear{Somasundaram
  et~al.}{2024}]{Somasundaram:2024ykk}
\begin{botherref}
\oauthor{\bsnm{Somasundaram}, \binits{R.}},
\oauthor{\bsnm{Svensson}, \binits{I.}},
\oauthor{\bsnm{De}, \binits{S.}},
\oauthor{\bsnm{Deneris}, \binits{A.E.}},
\oauthor{\bsnm{Dietz}, \binits{Y.}},
\oauthor{\bsnm{Landry}, \binits{P.}},
\oauthor{\bsnm{Schwenk}, \binits{A.}},
\oauthor{\bsnm{Tews}, \binits{I.}}:
{Inferring three-nucleon couplings from multi-messenger neutron-star
  observations}
(2024)
{\href{https://arxiv.org/abs/2410.00247}{{arXiv:2410.00247}}}
{[nucl-th]}
\end{botherref}
\endbibitem

\bibitem[\protect\citeauthoryear{Li and Sedrakian}{2025}]{Li:2025vhk}
\begin{barticle}
\bauthor{\bsnm{Li}, \binits{J.-J.}},
\bauthor{\bsnm{Sedrakian}, \binits{A.}}:
\batitle{{Bayesian inferences on covariant density functionals from
  multimessenger astrophysical data: The impacts of likelihood functions of low
  density matter constraints}}.
\bjtitle{Phys. Rev. C}
\bvolume{112}(\bissue{1}),
\bfpage{015802}
(\byear{2025})
\doiurl{10.1103/c1k3-k4l5}
{\href{https://arxiv.org/abs/2505.00911}{{arXiv:2505.00911}}}
{[nucl-th]}
\end{barticle}
\endbibitem

\bibitem[\protect\citeauthoryear{Magnall et~al.}{2025}]{Magnall:2025zhm}
\begin{botherref}
\oauthor{\bsnm{Magnall}, \binits{S.J.}},
\oauthor{\bsnm{Ecker}, \binits{C.}},
\oauthor{\bsnm{Rezzolla}, \binits{L.}},
\oauthor{\bsnm{Lasky}, \binits{P.D.}},
\oauthor{\bsnm{Goode}, \binits{S.R.}}:
{Physics-Informed Priors Improve Gravitational-Wave Constraints on Neutron-Star
  Matter}
(2025)
{\href{https://arxiv.org/abs/2504.21526}{{arXiv:2504.21526}}}
{[astro-ph.HE]}
\end{botherref}
\endbibitem

\bibitem[\protect\citeauthoryear{Semposki et~al.}{2025}]{Semposki:2025etb}
\begin{botherref}
\oauthor{\bsnm{Semposki}, \binits{A.C.}},
\oauthor{\bsnm{Drischler}, \binits{C.}},
\oauthor{\bsnm{Furnstahl}, \binits{R.J.}},
\oauthor{\bsnm{Phillips}, \binits{D.R.}}:
{Microscopic constraints for the equation of state and structure of neutron
  stars: a Bayesian model mixing framework}
(2025)
{\href{https://arxiv.org/abs/2505.18921}{{arXiv:2505.18921}}}
{[nucl-th]}
\end{botherref}
\endbibitem

\bibitem[\protect\citeauthoryear{Brandes and Weise}{2024}]{Brandes:2023bob}
\begin{barticle}
\bauthor{\bsnm{Brandes}, \binits{L.}},
\bauthor{\bsnm{Weise}, \binits{W.}}:
\batitle{{Constraints on Phase Transitions in Neutron Star Matter}}.
\bjtitle{Symmetry}
\bvolume{16}(\bissue{1}),
\bfpage{111}
(\byear{2024})
\doiurl{10.3390/sym16010111}
{\href{https://arxiv.org/abs/2312.11937}{{arXiv:2312.11937}}}
{[nucl-th]}
\end{barticle}
\endbibitem

\bibitem[\protect\citeauthoryear{Al-Mamun et~al.}{2021}]{Al-Mamun:2020vzu}
\begin{barticle}
\bauthor{\bsnm{Al-Mamun}, \binits{M.}},
\bauthor{\bsnm{Steiner}, \binits{A.W.}},
\bauthor{\bsnm{N\"attil\"a}, \binits{J.}},
\bauthor{\bsnm{Lange}, \binits{J.}},
\bauthor{\bsnm{O'Shaughnessy}, \binits{R.}},
\bauthor{\bsnm{Tews}, \binits{I.}},
\bauthor{\bsnm{Gandolfi}, \binits{S.}},
\bauthor{\bsnm{Heinke}, \binits{C.}},
\bauthor{\bsnm{Han}, \binits{S.}}:
\batitle{{Combining electromagnetic and gravitational-wave constraints on
  neutron-star masses and radii}}.
\bjtitle{Phys. Rev. Lett.}
\bvolume{126}(\bissue{6}),
\bfpage{061101}
(\byear{2021})
\doiurl{10.1103/PhysRevLett.126.061101}
{\href{https://arxiv.org/abs/2008.12817}{{arXiv:2008.12817}}}
{[astro-ph.HE]}
\end{barticle}
\endbibitem

\bibitem[\protect\citeauthoryear{N\"attil\"a et~al.}{2017}]{Nattila:2017wtj}
\begin{barticle}
\bauthor{\bsnm{N\"attil\"a}, \binits{J.}},
\bauthor{\bsnm{Miller}, \binits{M.C.}},
\bauthor{\bsnm{Steiner}, \binits{A.W.}},
\bauthor{\bsnm{Kajava}, \binits{J.J.E.}},
\bauthor{\bsnm{Suleimanov}, \binits{V.F.}},
\bauthor{\bsnm{Poutanen}, \binits{J.}}:
\batitle{{Neutron star mass and radius measurements from atmospheric model fits
  to X-ray burst cooling tail spectra}}.
\bjtitle{Astron. Astrophys.}
\bvolume{608},
\bfpage{31}
(\byear{2017})
\doiurl{10.1051/0004-6361/201731082}
{\href{https://arxiv.org/abs/1709.09120}{{arXiv:1709.09120}}}
{[astro-ph.HE]}
\end{barticle}
\endbibitem

\bibitem[\protect\citeauthoryear{Essick et~al.}{2020}]{Essick:2020flb}
\begin{barticle}
\bauthor{\bsnm{Essick}, \binits{R.}},
\bauthor{\bsnm{Tews}, \binits{I.}},
\bauthor{\bsnm{Landry}, \binits{P.}},
\bauthor{\bsnm{Reddy}, \binits{S.}},
\bauthor{\bsnm{Holz}, \binits{D.E.}}:
\batitle{{Direct astrophysical tests of chiral effective field theory at
  supranuclear densities}}.
\bjtitle{Phys. Rev. C}
\bvolume{102}(\bissue{5}),
\bfpage{055803}
(\byear{2020})
\doiurl{10.1103/PhysRevC.102.055803}
{\href{https://arxiv.org/abs/2004.07744}{{arXiv:2004.07744}}}
{[astro-ph.HE]}
\end{barticle}
\endbibitem

\bibitem[\protect\citeauthoryear{Fonseca et~al.}{2016}]{Fonseca:2016tux}
\begin{barticle}
\bauthor{\bsnm{Fonseca}, \binits{E.}}, \betal:
\batitle{{The NANOGrav nine-year data set: Mass and geometric measurements of
  binary millisecond pulsars}}.
\bjtitle{Astrophys. J.}
\bvolume{832}(\bissue{2}),
\bfpage{167}
(\byear{2016})
\doiurl{10.3847/0004-637X/832/2/167}
{\href{https://arxiv.org/abs/1603.00545}{{arXiv:1603.00545}}}
{[astro-ph.HE]}
\end{barticle}
\endbibitem

\bibitem[\protect\citeauthoryear{Somasundaram
  et~al.}{2023}]{Somasundaram:2022ztm}
\begin{barticle}
\bauthor{\bsnm{Somasundaram}, \binits{R.}},
\bauthor{\bsnm{Tews}, \binits{I.}},
\bauthor{\bsnm{Margueron}, \binits{J.}}:
\batitle{{Perturbative QCD and the neutron star equation~of state}}.
\bjtitle{Phys. Rev. C}
\bvolume{107}(\bissue{5}),
\bfpage{052801}
(\byear{2023})
\doiurl{10.1103/PhysRevC.107.L052801}
{\href{https://arxiv.org/abs/2204.14039}{{arXiv:2204.14039}}}
{[nucl-th]}
\end{barticle}
\endbibitem

\bibitem[\protect\citeauthoryear{Drischler et~al.}{2021}]{Drischler:2021kxf}
\begin{barticle}
\bauthor{\bsnm{Drischler}, \binits{C.}},
\bauthor{\bsnm{Holt}, \binits{J.W.}},
\bauthor{\bsnm{Wellenhofer}, \binits{C.}}:
\batitle{{Chiral Effective Field Theory and the High-Density Nuclear Equation
  of State}}.
\bjtitle{Ann. Rev. Nucl. Part. Sci.}
\bvolume{71},
\bfpage{403}--\blpage{432}
(\byear{2021})
\doiurl{10.1146/annurev-nucl-102419-041903}
{\href{https://arxiv.org/abs/2101.01709}{{arXiv:2101.01709}}}
{[nucl-th]}
\end{barticle}
\endbibitem

\bibitem[\protect\citeauthoryear{Komoltsev et~al.}{2023}]{Komoltsev:2023zor}
\begin{botherref}
\oauthor{\bsnm{Komoltsev}, \binits{O.}},
\oauthor{\bsnm{Somasundaram}, \binits{R.}},
\oauthor{\bsnm{Gorda}, \binits{T.}},
\oauthor{\bsnm{Kurkela}, \binits{A.}},
\oauthor{\bsnm{Margueron}, \binits{J.}},
\oauthor{\bsnm{Tews}, \binits{I.}}:
{Equation of state at neutron-star densities and beyond from perturbative QCD}
(2023)
{\href{https://arxiv.org/abs/2312.14127}{{arXiv:2312.14127}}}
{[nucl-th]}
\end{botherref}
\endbibitem

\bibitem[\protect\citeauthoryear{Landry and Essick}{2019}]{Landry:2018prl}
\begin{barticle}
\bauthor{\bsnm{Landry}, \binits{P.}},
\bauthor{\bsnm{Essick}, \binits{R.}}:
\batitle{{Nonparametric inference of the neutron star equation of state from
  gravitational wave observations}}.
\bjtitle{Phys. Rev. D}
\bvolume{99}(\bissue{8}),
\bfpage{084049}
(\byear{2019})
\doiurl{10.1103/PhysRevD.99.084049}
{\href{https://arxiv.org/abs/1811.12529}{{arXiv:1811.12529}}}
{[gr-qc]}
\end{barticle}
\endbibitem

\bibitem[\protect\citeauthoryear{Landry et~al.}{2020}]{Landry:2020vaw}
\begin{barticle}
\bauthor{\bsnm{Landry}, \binits{P.}},
\bauthor{\bsnm{Essick}, \binits{R.}},
\bauthor{\bsnm{Chatziioannou}, \binits{K.}}:
\batitle{{Nonparametric constraints on neutron star matter with existing and
  upcoming gravitational wave and pulsar observations}}.
\bjtitle{Phys. Rev. D}
\bvolume{101}(\bissue{12}),
\bfpage{123007}
(\byear{2020})
\doiurl{10.1103/PhysRevD.101.123007}
{\href{https://arxiv.org/abs/2003.04880}{{arXiv:2003.04880}}}
{[astro-ph.HE]}
\end{barticle}
\endbibitem

\bibitem[\protect\citeauthoryear{Essick et~al.}{2020}]{Essick:2019ldf}
\begin{barticle}
\bauthor{\bsnm{Essick}, \binits{R.}},
\bauthor{\bsnm{Landry}, \binits{P.}},
\bauthor{\bsnm{Holz}, \binits{D.E.}}:
\batitle{{Nonparametric inference of neutron star composition, equation of
  state, and maximum mass with GW170817}}.
\bjtitle{Phys. Rev. D}
\bvolume{101}(\bissue{6}),
\bfpage{063007}
(\byear{2020})
\doiurl{10.1103/PhysRevD.101.063007}
{\href{https://arxiv.org/abs/1910.09740}{{arXiv:1910.09740}}}
{[astro-ph.HE]}
\end{barticle}
\endbibitem

\bibitem[\protect\citeauthoryear{Margalit and Metzger}{2017}]{Margalit:2017dij}
\begin{barticle}
\bauthor{\bsnm{Margalit}, \binits{B.}},
\bauthor{\bsnm{Metzger}, \binits{B.D.}}:
\batitle{{Constraining the maximum mass of neutron stars from multi-messenger
  observations of GW170817}}.
\bjtitle{Astrophys. J. Lett.}
\bvolume{850}(\bissue{2}),
\bfpage{19}
(\byear{2017})
\doiurl{10.3847/2041-8213/aa991c}
{\href{https://arxiv.org/abs/1710.05938}{{arXiv:1710.05938}}}
{[astro-ph.HE]}
\end{barticle}
\endbibitem

\bibitem[\protect\citeauthoryear{Rezzolla et~al.}{2018}]{Rezzolla:2017aly}
\begin{barticle}
\bauthor{\bsnm{Rezzolla}, \binits{L.}},
\bauthor{\bsnm{Most}, \binits{E.R.}},
\bauthor{\bsnm{Weih}, \binits{L.R.}}:
\batitle{{Using gravitational-wave observations and quasi-universal relations
  to constrain the maximum mass of neutron stars}}.
\bjtitle{Astrophys. J. Lett.}
\bvolume{852}(\bissue{2}),
\bfpage{25}
(\byear{2018})
\doiurl{10.3847/2041-8213/aaa401}
{\href{https://arxiv.org/abs/1711.00314}{{arXiv:1711.00314}}}
{[astro-ph.HE]}
\end{barticle}
\endbibitem

\bibitem[\protect\citeauthoryear{Ruiz et~al.}{2018}]{Ruiz:2017due}
\begin{barticle}
\bauthor{\bsnm{Ruiz}, \binits{M.}},
\bauthor{\bsnm{Shapiro}, \binits{S.L.}},
\bauthor{\bsnm{Tsokaros}, \binits{A.}}:
\batitle{{GW170817, general relativistic magnetohydrodynamic simulations, and
  the neutron star maximum mass}}.
\bjtitle{Phys. Rev. D}
\bvolume{97}(\bissue{2}),
\bfpage{021501}
(\byear{2018})
\doiurl{10.1103/PhysRevD.97.021501}
{\href{https://arxiv.org/abs/1711.00473}{{arXiv:1711.00473}}}
{[astro-ph.HE]}
\end{barticle}
\endbibitem

\bibitem[\protect\citeauthoryear{Shibata et~al.}{2017}]{Shibata:2017xdx}
\begin{barticle}
\bauthor{\bsnm{Shibata}, \binits{M.}},
\bauthor{\bsnm{Fujibayashi}, \binits{S.}},
\bauthor{\bsnm{Hotokezaka}, \binits{K.}},
\bauthor{\bsnm{Kiuchi}, \binits{K.}},
\bauthor{\bsnm{Kyutoku}, \binits{K.}},
\bauthor{\bsnm{Sekiguchi}, \binits{Y.}},
\bauthor{\bsnm{Tanaka}, \binits{M.}}:
\batitle{{Modeling GW170817 based on numerical relativity and its
  implications}}.
\bjtitle{Phys. Rev. D}
\bvolume{96}(\bissue{12}),
\bfpage{123012}
(\byear{2017})
\doiurl{10.1103/PhysRevD.96.123012}
{\href{https://arxiv.org/abs/1710.07579}{{arXiv:1710.07579}}}
{[astro-ph.HE]}
\end{barticle}
\endbibitem

\bibitem[\protect\citeauthoryear{Shibata et~al.}{2019}]{Shibata:2019ctb}
\begin{barticle}
\bauthor{\bsnm{Shibata}, \binits{M.}},
\bauthor{\bsnm{Zhou}, \binits{E.}},
\bauthor{\bsnm{Kiuchi}, \binits{K.}},
\bauthor{\bsnm{Fujibayashi}, \binits{S.}}:
\batitle{{Constraint on the maximum mass of neutron stars using GW170817
  event}}.
\bjtitle{Phys. Rev. D}
\bvolume{100}(\bissue{2}),
\bfpage{023015}
(\byear{2019})
\doiurl{10.1103/PhysRevD.100.023015}
{\href{https://arxiv.org/abs/1905.03656}{{arXiv:1905.03656}}}
{[astro-ph.HE]}
\end{barticle}
\endbibitem

\bibitem[\protect\citeauthoryear{Alford et~al.}{2019}]{Alford:2019oge}
\begin{barticle}
\bauthor{\bsnm{Alford}, \binits{M.G.}},
\bauthor{\bsnm{Han}, \binits{S.}},
\bauthor{\bsnm{Schwenzer}, \binits{K.}}:
\batitle{{Signatures for quark matter from multi-messenger observations}}.
\bjtitle{J. Phys. G}
\bvolume{46}(\bissue{11}),
\bfpage{114001}
(\byear{2019})
\doiurl{10.1088/1361-6471/ab337a}
{\href{https://arxiv.org/abs/1904.05471}{{arXiv:1904.05471}}}
{[nucl-th]}
\end{barticle}
\endbibitem

\bibitem[\protect\citeauthoryear{Chatterjee and
  Vida\~na}{2016}]{Chatterjee:2015pua}
\begin{barticle}
\bauthor{\bsnm{Chatterjee}, \binits{D.}},
\bauthor{\bsnm{Vida\~na}, \binits{I.}}:
\batitle{{Do hyperons exist in the interior of neutron stars?}}
\bjtitle{Eur. Phys. J. A}
\bvolume{52}(\bissue{2}),
\bfpage{29}
(\byear{2016})
\doiurl{10.1140/epja/i2016-16029-x}
{\href{https://arxiv.org/abs/1510.06306}{{arXiv:1510.06306}}}
{[nucl-th]}
\end{barticle}
\endbibitem

\bibitem[\protect\citeauthoryear{Baym et~al.}{2018}]{Baym:2017whm}
\begin{barticle}
\bauthor{\bsnm{Baym}, \binits{G.}},
\bauthor{\bsnm{Hatsuda}, \binits{T.}},
\bauthor{\bsnm{Kojo}, \binits{T.}},
\bauthor{\bsnm{Powell}, \binits{P.D.}},
\bauthor{\bsnm{Song}, \binits{Y.}},
\bauthor{\bsnm{Takatsuka}, \binits{T.}}:
\batitle{{From hadrons to quarks in neutron stars: a review}}.
\bjtitle{Rept. Prog. Phys.}
\bvolume{81}(\bissue{5}),
\bfpage{056902}
(\byear{2018})
\doiurl{10.1088/1361-6633/aaae14}
{\href{https://arxiv.org/abs/1707.04966}{{arXiv:1707.04966}}}
{[astro-ph.HE]}
\end{barticle}
\endbibitem

\bibitem[\protect\citeauthoryear{McLerran and Reddy}{2019}]{McLerran:2018hbz}
\begin{barticle}
\bauthor{\bsnm{McLerran}, \binits{L.}},
\bauthor{\bsnm{Reddy}, \binits{S.}}:
\batitle{{Quarkyonic Matter and Neutron Stars}}.
\bjtitle{Phys. Rev. Lett.}
\bvolume{122}(\bissue{12}),
\bfpage{122701}
(\byear{2019})
\doiurl{10.1103/PhysRevLett.122.122701}
{\href{https://arxiv.org/abs/1811.12503}{{arXiv:1811.12503}}}
{[nucl-th]}
\end{barticle}
\endbibitem

\bibitem[\protect\citeauthoryear{Leonhardt et~al.}{2020}]{Leonhardt:2019fua}
\begin{barticle}
\bauthor{\bsnm{Leonhardt}, \binits{M.}},
\bauthor{\bsnm{Pospiech}, \binits{M.}},
\bauthor{\bsnm{Schallmo}, \binits{B.}},
\bauthor{\bsnm{Braun}, \binits{J.}},
\bauthor{\bsnm{Drischler}, \binits{C.}},
\bauthor{\bsnm{Hebeler}, \binits{K.}},
\bauthor{\bsnm{Schwenk}, \binits{A.}}:
\batitle{{Symmetric nuclear matter from the strong interaction}}.
\bjtitle{Phys. Rev. Lett.}
\bvolume{125}(\bissue{14}),
\bfpage{142502}
(\byear{2020})
\doiurl{10.1103/PhysRevLett.125.142502}
{\href{https://arxiv.org/abs/1907.05814}{{arXiv:1907.05814}}}
{[nucl-th]}
\end{barticle}
\endbibitem

\bibitem[\protect\citeauthoryear{Joyce et~al.}{2016}]{Joyce:2016vqv}
\begin{barticle}
\bauthor{\bsnm{Joyce}, \binits{A.}},
\bauthor{\bsnm{Lombriser}, \binits{L.}},
\bauthor{\bsnm{Schmidt}, \binits{F.}}:
\batitle{{Dark Energy Versus Modified Gravity}}.
\bjtitle{Ann. Rev. Nucl. Part. Sci.}
\bvolume{66},
\bfpage{95}--\blpage{122}
(\byear{2016})
\doiurl{10.1146/annurev-nucl-102115-044553}
{\href{https://arxiv.org/abs/1601.06133}{{arXiv:1601.06133}}}
{[astro-ph.CO]}
\end{barticle}
\endbibitem

\bibitem[\protect\citeauthoryear{Olmo et~al.}{2020}]{Olmo:2019flu}
\begin{barticle}
\bauthor{\bsnm{Olmo}, \binits{G.J.}},
\bauthor{\bsnm{Rubiera-Garcia}, \binits{D.}},
\bauthor{\bsnm{Wojnar}, \binits{A.}}:
\batitle{{Stellar structure models in modified theories of gravity: Lessons and
  challenges}}.
\bjtitle{Phys. Rept.}
\bvolume{876},
\bfpage{1}--\blpage{75}
(\byear{2020})
\doiurl{10.1016/j.physrep.2020.07.001}
{\href{https://arxiv.org/abs/1912.05202}{{arXiv:1912.05202}}}
{[gr-qc]}
\end{barticle}
\endbibitem

\bibitem[\protect\citeauthoryear{Ferreira and
  Provid{\^e}ncia}{2025}]{Ferreira:2025dat}
\begin{botherref}
\oauthor{\bsnm{Ferreira}, \binits{M.}},
\oauthor{\bsnm{Provid{\^e}ncia}, \binits{C.}}:
{Identifying hyperons in neutron star matter from the slope of the mass-radius
  diagram}
(2025)
{\href{https://arxiv.org/abs/2506.00550}{{arXiv:2506.00550}}}
{[nucl-th]}
\end{botherref}
\endbibitem

\bibitem[\protect\citeauthoryear{Glendenning}{1992}]{Glendenning:1992vb}
\begin{barticle}
\bauthor{\bsnm{Glendenning}, \binits{N.K.}}:
\batitle{{First order phase transitions with more than one conserved charge:
  Consequences for neutron stars}}.
\bjtitle{Phys. Rev. D}
\bvolume{46},
\bfpage{1274}--\blpage{1287}
(\byear{1992})
\doiurl{10.1103/PhysRevD.46.1274}
\end{barticle}
\endbibitem

\bibitem[\protect\citeauthoryear{Glendenning}{2001}]{Glendenning:2001pe}
\begin{barticle}
\bauthor{\bsnm{Glendenning}, \binits{N.K.}}:
\batitle{{Phase transitions and crystalline structures in neutron star cores}}.
\bjtitle{Phys. Rept.}
\bvolume{342},
\bfpage{393}--\blpage{447}
(\byear{2001})
\doiurl{10.1016/S0370-1573(00)00080-6}
\end{barticle}
\endbibitem

\bibitem[\protect\citeauthoryear{Bauswein et~al.}{2019}]{Bauswein:2018bma}
\begin{barticle}
\bauthor{\bsnm{Bauswein}, \binits{A.}},
\bauthor{\bsnm{Bastian}, \binits{N.-U.F.}},
\bauthor{\bsnm{Blaschke}, \binits{D.B.}},
\bauthor{\bsnm{Chatziioannou}, \binits{K.}},
\bauthor{\bsnm{Clark}, \binits{J.A.}},
\bauthor{\bsnm{Fischer}, \binits{T.}},
\bauthor{\bsnm{Oertel}, \binits{M.}}:
\batitle{{Identifying a first-order phase transition in neutron star mergers
  through gravitational waves}}.
\bjtitle{Phys. Rev. Lett.}
\bvolume{122}(\bissue{6}),
\bfpage{061102}
(\byear{2019})
\doiurl{10.1103/PhysRevLett.122.061102}
{\href{https://arxiv.org/abs/1809.01116}{{arXiv:1809.01116}}}
{[astro-ph.HE]}
\end{barticle}
\endbibitem

\bibitem[\protect\citeauthoryear{Most et~al.}{2019}]{Most:2018eaw}
\begin{barticle}
\bauthor{\bsnm{Most}, \binits{E.R.}},
\bauthor{\bsnm{Papenfort}, \binits{L.J.}},
\bauthor{\bsnm{Dexheimer}, \binits{V.}},
\bauthor{\bsnm{Hanauske}, \binits{M.}},
\bauthor{\bsnm{Schramm}, \binits{S.}},
\bauthor{\bsnm{St\"ocker}, \binits{H.}},
\bauthor{\bsnm{Rezzolla}, \binits{L.}}:
\batitle{{Signatures of quark-hadron phase transitions in general-relativistic
  neutron-star mergers}}.
\bjtitle{Phys. Rev. Lett.}
\bvolume{122}(\bissue{6}),
\bfpage{061101}
(\byear{2019})
\doiurl{10.1103/PhysRevLett.122.061101}
{\href{https://arxiv.org/abs/1807.03684}{{arXiv:1807.03684}}}
{[astro-ph.HE]}
\end{barticle}
\endbibitem

\bibitem[\protect\citeauthoryear{Moreno et~al.}{2023}]{Moreno:2023xez}
\begin{barticle}
\bauthor{\bsnm{Moreno}, \binits{P.N.}},
\bauthor{\bsnm{Llanes-Estrada}, \binits{F.J.}},
\bauthor{\bsnm{Lope-Oter}, \binits{E.}}:
\batitle{{Ridges in rotating neutron-star properties due to first order phase
  transitions}}.
\bjtitle{Annals Phys.}
\bvolume{459},
\bfpage{169487}
(\byear{2023})
\doiurl{10.1016/j.aop.2023.169487}
{\href{https://arxiv.org/abs/2307.15366}{{arXiv:2307.15366}}}
{[nucl-th]}
\end{barticle}
\endbibitem

\bibitem[\protect\citeauthoryear{Abac et~al.}{2025}]{Abac:2025saz}
\begin{botherref}
\oauthor{\bsnm{Abac}, \binits{A.}}, et al.:
{The Science of the Einstein Telescope}
(2025)
{\href{https://arxiv.org/abs/2503.12263}{{arXiv:2503.12263}}}
{[gr-qc]}
\end{botherref}
\endbibitem

\bibitem[\protect\citeauthoryear{Glendenning and
  Kettner}{2000}]{Glendenning:1998ag}
\begin{barticle}
\bauthor{\bsnm{Glendenning}, \binits{N.K.}},
\bauthor{\bsnm{Kettner}, \binits{C.}}:
\batitle{{Nonidentical neutron star twins}}.
\bjtitle{Astron. Astrophys.}
\bvolume{353},
\bfpage{9}
(\byear{2000})
{\href{https://arxiv.org/abs/astro-ph/9807155}{{arXiv:astro-ph/9807155}}}
\end{barticle}
\endbibitem

\bibitem[\protect\citeauthoryear{Alford et~al.}{2005}]{Alford:2004pf}
\begin{barticle}
\bauthor{\bsnm{Alford}, \binits{M.}},
\bauthor{\bsnm{Braby}, \binits{M.}},
\bauthor{\bsnm{Paris}, \binits{M.W.}},
\bauthor{\bsnm{Reddy}, \binits{S.}}:
\batitle{{Hybrid stars that masquerade as neutron stars}}.
\bjtitle{Astrophys. J.}
\bvolume{629},
\bfpage{969}--\blpage{978}
(\byear{2005})
\doiurl{10.1086/430902}
{\href{https://arxiv.org/abs/nucl-th/0411016}{{arXiv:nucl-th/0411016}}}
\end{barticle}
\endbibitem

\bibitem[\protect\citeauthoryear{Benic et~al.}{2015}]{Benic:2014jia}
\begin{barticle}
\bauthor{\bsnm{Benic}, \binits{S.}},
\bauthor{\bsnm{Blaschke}, \binits{D.}},
\bauthor{\bsnm{Alvarez-Castillo}, \binits{D.E.}},
\bauthor{\bsnm{Fischer}, \binits{T.}},
\bauthor{\bsnm{Typel}, \binits{S.}}:
\batitle{{A new quark-hadron hybrid equation of state for astrophysics - I.
  High-mass twin compact stars}}.
\bjtitle{Astron. Astrophys.}
\bvolume{577},
\bfpage{40}
(\byear{2015})
\doiurl{10.1051/0004-6361/201425318}
{\href{https://arxiv.org/abs/1411.2856}{{arXiv:1411.2856}}}
{[astro-ph.HE]}
\end{barticle}
\endbibitem

\bibitem[\protect\citeauthoryear{Christian et~al.}{2018}]{Christian:2017jni}
\begin{barticle}
\bauthor{\bsnm{Christian}, \binits{J.-E.}},
\bauthor{\bsnm{Zacchi}, \binits{A.}},
\bauthor{\bsnm{Schaffner-Bielich}, \binits{J.}}:
\batitle{{Classifications of Twin Star Solutions for a Constant Speed of Sound
  Parameterized Equation of State}}.
\bjtitle{Eur. Phys. J. A}
\bvolume{54}(\bissue{2}),
\bfpage{28}
(\byear{2018})
\doiurl{10.1140/epja/i2018-12472-y}
{\href{https://arxiv.org/abs/1707.07524}{{arXiv:1707.07524}}}
{[astro-ph.HE]}
\end{barticle}
\endbibitem

\bibitem[\protect\citeauthoryear{Christian and
  Schaffner-Bielich}{2022}]{Christian:2021uhd}
\begin{barticle}
\bauthor{\bsnm{Christian}, \binits{J.-E.}},
\bauthor{\bsnm{Schaffner-Bielich}, \binits{J.}}:
\batitle{{Confirming the Existence of Twin Stars in a NICER Way}}.
\bjtitle{Astrophys. J.}
\bvolume{935}(\bissue{2}),
\bfpage{122}
(\byear{2022})
\doiurl{10.3847/1538-4357/ac75cf}
{\href{https://arxiv.org/abs/2109.04191}{{arXiv:2109.04191}}}
{[astro-ph.HE]}
\end{barticle}
\endbibitem

\bibitem[\protect\citeauthoryear{Seidov}{1971}]{Seidov:1971sv}
\begin{barticle}
\bauthor{\bsnm{Seidov}, \binits{Z.F.}}:
\batitle{{The Stability of a Star with a Phase Change in General Relativity
  Theory}}.
\bjtitle{Sov. Astron.}
\bvolume{15},
\bfpage{347}
(\byear{1971})
\end{barticle}
\endbibitem

\bibitem[\protect\citeauthoryear{Pereira et~al.}{2018}]{Pereira:2017rmp}
\begin{barticle}
\bauthor{\bsnm{Pereira}, \binits{J.P.}},
\bauthor{\bsnm{Flores}, \binits{C.V.}},
\bauthor{\bsnm{Lugones}, \binits{G.}}:
\batitle{{Phase transition effects on the dynamical stability of hybrid neutron
  stars}}.
\bjtitle{Astrophys. J.}
\bvolume{860}(\bissue{1}),
\bfpage{12}
(\byear{2018})
\doiurl{10.3847/1538-4357/aabfbf}
{\href{https://arxiv.org/abs/1706.09371}{{arXiv:1706.09371}}}
{[gr-qc]}
\end{barticle}
\endbibitem

\bibitem[\protect\citeauthoryear{Rau and Sedrakian}{2023}]{Rau:2022ofy}
\begin{barticle}
\bauthor{\bsnm{Rau}, \binits{P.B.}},
\bauthor{\bsnm{Sedrakian}, \binits{A.}}:
\batitle{{Two first-order phase transitions in hybrid compact stars:
  Higher-order multiplet stars, reaction modes, and intermediate conversion
  speeds}}.
\bjtitle{Phys. Rev. D}
\bvolume{107}(\bissue{10}),
\bfpage{103042}
(\byear{2023})
\doiurl{10.1103/PhysRevD.107.103042}
{\href{https://arxiv.org/abs/2212.09828}{{arXiv:2212.09828}}}
{[astro-ph.HE]}
\end{barticle}
\endbibitem

\bibitem[\protect\citeauthoryear{Goncalves and
  Lazzari}{2022}]{Goncalves:2022ymr}
\begin{barticle}
\bauthor{\bsnm{Goncalves}, \binits{V.P.}},
\bauthor{\bsnm{Lazzari}, \binits{L.}}:
\batitle{{Impact of slow conversions on hybrid stars with sequential QCD phase
  transitions}}.
\bjtitle{Eur. Phys. J. C}
\bvolume{82}(\bissue{4}),
\bfpage{288}
(\byear{2022})
\doiurl{10.1140/epjc/s10052-022-10273-5}
{\href{https://arxiv.org/abs/2201.03304}{{arXiv:2201.03304}}}
{[nucl-th]}
\end{barticle}
\endbibitem

\bibitem[\protect\citeauthoryear{Ranea-Sandoval
  et~al.}{2022}]{Ranea-Sandoval:2022izm}
\begin{barticle}
\bauthor{\bsnm{Ranea-Sandoval}, \binits{I.F.}},
\bauthor{\bsnm{Mariani}, \binits{M.}},
\bauthor{\bsnm{Lugones}, \binits{G.}},
\bauthor{\bsnm{Guilera}, \binits{O.M.}}:
\batitle{{Constraining mass, radius, and tidal deformability of compact stars
  with axial wI modes: new universal relations including slow stable hybrid
  stars}}.
\bjtitle{Mon. Not. Roy. Astron. Soc.}
\bvolume{519}(\bissue{2}),
\bfpage{3194}--\blpage{3200}
(\byear{2022})
\doiurl{10.1093/mnras/stac3780}
{\href{https://arxiv.org/abs/2212.10514}{{arXiv:2212.10514}}}
{[astro-ph.HE]}
\end{barticle}
\endbibitem

\bibitem[\protect\citeauthoryear{Lope-Oter and
  Llanes-Estrada}{2022}]{Lope-Oter:2021mjp}
\begin{barticle}
\bauthor{\bsnm{Lope-Oter}, \binits{E.}},
\bauthor{\bsnm{Llanes-Estrada}, \binits{F.J.}}:
\batitle{{Maximum latent heat of neutron star matter independently of General
  Relativity}}.
\bjtitle{Phys. Rev. C}
\bvolume{105},
\bfpage{052801}
(\byear{2022})
\doiurl{10.1103/PhysRevC.105.L052801}
{\href{https://arxiv.org/abs/2103.10799}{{arXiv:2103.10799}}}
{[nucl-th]}
\end{barticle}
\endbibitem

\bibitem[\protect\citeauthoryear{Drischler et~al.}{2020}]{Drischler:2020yad}
\begin{barticle}
\bauthor{\bsnm{Drischler}, \binits{C.}},
\bauthor{\bsnm{Melendez}, \binits{J.A.}},
\bauthor{\bsnm{Furnstahl}, \binits{R.J.}},
\bauthor{\bsnm{Phillips}, \binits{D.R.}}:
\batitle{{Quantifying uncertainties and correlations in the nuclear-matter
  equation of state}}.
\bjtitle{Phys. Rev. C}
\bvolume{102}(\bissue{5}),
\bfpage{054315}
(\byear{2020})
\doiurl{10.1103/PhysRevC.102.054315}
{\href{https://arxiv.org/abs/2004.07805}{{arXiv:2004.07805}}}
{[nucl-th]}
\end{barticle}
\endbibitem

\bibitem[\protect\citeauthoryear{Fujimoto et~al.}{2022}]{Fujimoto:2022ohj}
\begin{botherref}
\oauthor{\bsnm{Fujimoto}, \binits{Y.}},
\oauthor{\bsnm{Fukushima}, \binits{K.}},
\oauthor{\bsnm{McLerran}, \binits{L.D.}},
\oauthor{\bsnm{Praszalowicz}, \binits{M.}}:
{Trace anomaly as signature of conformality in neutron stars}
(2022)
{\href{https://arxiv.org/abs/2207.06753}{{arXiv:2207.06753}}}
{[nucl-th]}
\end{botherref}
\endbibitem

\bibitem[\protect\citeauthoryear{Marczenko et~al.}{2024}]{Marczenko:2023txe}
\begin{barticle}
\bauthor{\bsnm{Marczenko}, \binits{M.}},
\bauthor{\bsnm{Redlich}, \binits{K.}},
\bauthor{\bsnm{Sasaki}, \binits{C.}}:
\batitle{{Curvature of the energy per particle in neutron stars}}.
\bjtitle{Phys. Rev. D}
\bvolume{109}(\bissue{4}),
\bfpage{041302}
(\byear{2024})
\doiurl{10.1103/PhysRevD.109.L041302}
{\href{https://arxiv.org/abs/2311.13401}{{arXiv:2311.13401}}}
{[nucl-th]}
\end{barticle}
\endbibitem

\bibitem[\protect\citeauthoryear{Somasundaram
  et~al.}{2023}]{Somasundaram:2021clp}
\begin{barticle}
\bauthor{\bsnm{Somasundaram}, \binits{R.}},
\bauthor{\bsnm{Tews}, \binits{I.}},
\bauthor{\bsnm{Margueron}, \binits{J.}}:
\batitle{{Investigating signatures of phase transitions in neutron-star
  cores}}.
\bjtitle{Phys. Rev. C}
\bvolume{107}(\bissue{2}),
\bfpage{025801}
(\byear{2023})
\doiurl{10.1103/PhysRevC.107.025801}
{\href{https://arxiv.org/abs/2112.08157}{{arXiv:2112.08157}}}
{[nucl-th]}
\end{barticle}
\endbibitem

\bibitem[\protect\citeauthoryear{Marczenko et~al.}{2023}]{Marczenko:2022jhl}
\begin{barticle}
\bauthor{\bsnm{Marczenko}, \binits{M.}},
\bauthor{\bsnm{McLerran}, \binits{L.}},
\bauthor{\bsnm{Redlich}, \binits{K.}},
\bauthor{\bsnm{Sasaki}, \binits{C.}}:
\batitle{{Reaching percolation and conformal limits in neutron stars}}.
\bjtitle{Phys. Rev. C}
\bvolume{107}(\bissue{2}),
\bfpage{025802}
(\byear{2023})
\doiurl{10.1103/PhysRevC.107.025802}
{\href{https://arxiv.org/abs/2207.13059}{{arXiv:2207.13059}}}
{[nucl-th]}
\end{barticle}
\endbibitem

\bibitem[\protect\citeauthoryear{Jim\'enez et~al.}{2024}]{Jimenez:2024hib}
\begin{barticle}
\bauthor{\bsnm{Jim\'enez}, \binits{J.C.}},
\bauthor{\bsnm{Lazzari}, \binits{L.}},
\bauthor{\bsnm{Gon\c{c}alves}, \binits{V.P.}}:
\batitle{{How the QCD trace anomaly behaves at the core of twin stars?}}
\bjtitle{Phys. Rev. D}
\bvolume{110}(\bissue{11}),
\bfpage{114014}
(\byear{2024})
\doiurl{10.1103/PhysRevD.110.114014}
{\href{https://arxiv.org/abs/2408.11614}{{arXiv:2408.11614}}}
{[hep-ph]}
\end{barticle}
\endbibitem

\bibitem[\protect\citeauthoryear{Cai and Li}{2025}]{Cai:2024oom}
\begin{barticle}
\bauthor{\bsnm{Cai}, \binits{B.-J.}},
\bauthor{\bsnm{Li}, \binits{B.-A.}}:
\batitle{{Unraveling trace anomaly of supradense matter via neutron star
  compactness scaling}}.
\bjtitle{Phys. Rev. D}
\bvolume{112}(\bissue{2}),
\bfpage{023023}
(\byear{2025})
\doiurl{10.1103/3p2p-p3d4}
{\href{https://arxiv.org/abs/2406.05025}{{arXiv:2406.05025}}}
{[astro-ph.HE]}
\end{barticle}
\endbibitem

\bibitem[\protect\citeauthoryear{Komoltsev}{2025}]{Komoltsev:2025vwn}
\begin{botherref}
\oauthor{\bsnm{Komoltsev}, \binits{O.}}:
{Perturbative QCD reveals the softening of matter in the cores of massive
  neutron stars}.
PhD thesis,
Stavanger U.
(2025)
\end{botherref}
\endbibitem

\bibitem[\protect\citeauthoryear{Fortin et~al.}{2016}]{Fortin:2016hny}
\begin{barticle}
\bauthor{\bsnm{Fortin}, \binits{M.}},
\bauthor{\bsnm{Providencia}, \binits{C.}},
\bauthor{\bsnm{Raduta}, \binits{A.R.}},
\bauthor{\bsnm{Gulminelli}, \binits{F.}},
\bauthor{\bsnm{Zdunik}, \binits{J.L.}},
\bauthor{\bsnm{Haensel}, \binits{P.}},
\bauthor{\bsnm{Bejger}, \binits{M.}}:
\batitle{{Neutron star radii and crusts: uncertainties and unified equations of
  state}}.
\bjtitle{Phys. Rev. C}
\bvolume{94}(\bissue{3}),
\bfpage{035804}
(\byear{2016})
\doiurl{10.1103/PhysRevC.94.035804}
{\href{https://arxiv.org/abs/1604.01944}{{arXiv:1604.01944}}}
{[astro-ph.SR]}
\end{barticle}
\endbibitem

\bibitem[\protect\citeauthoryear{Wiringa et~al.}{1988}]{Wiringa:1988tp}
\begin{barticle}
\bauthor{\bsnm{Wiringa}, \binits{R.B.}},
\bauthor{\bsnm{Fiks}, \binits{V.}},
\bauthor{\bsnm{Fabrocini}, \binits{A.}}:
\batitle{{Equation of state for dense nucleon matter}}.
\bjtitle{Phys. Rev. C}
\bvolume{38},
\bfpage{1010}--\blpage{1037}
(\byear{1988})
\doiurl{10.1103/PhysRevC.38.1010}
\end{barticle}
\endbibitem

\bibitem[\protect\citeauthoryear{Kohn and Sham}{1965}]{Kohn:1965zzb}
\begin{barticle}
\bauthor{\bsnm{Kohn}, \binits{W.}},
\bauthor{\bsnm{Sham}, \binits{L.J.}}:
\batitle{{Self-Consistent Equations Including Exchange and Correlation
  Effects}}.
\bjtitle{Phys. Rev.}
\bvolume{140},
\bfpage{1133}--\blpage{1138}
(\byear{1965})
\doiurl{10.1103/PhysRev.140.A1133}
\end{barticle}
\endbibitem

\bibitem[\protect\citeauthoryear{Kouvaris}{2008}]{Kouvaris:2007ay}
\begin{barticle}
\bauthor{\bsnm{Kouvaris}, \binits{C.}}:
\batitle{{WIMP Annihilation and Cooling of Neutron Stars}}.
\bjtitle{Phys. Rev. D}
\bvolume{77},
\bfpage{023006}
(\byear{2008})
\doiurl{10.1103/PhysRevD.77.023006}
{\href{https://arxiv.org/abs/0708.2362}{{arXiv:0708.2362}}}
{[astro-ph]}
\end{barticle}
\endbibitem

\bibitem[\protect\citeauthoryear{de~Lavallaz and
  Fairbairn}{2010}]{deLavallaz:2010wp}
\begin{barticle}
\bauthor{\bsnm{Lavallaz}, \binits{A.}},
\bauthor{\bsnm{Fairbairn}, \binits{M.}}:
\batitle{{Neutron Stars as Dark Matter Probes}}.
\bjtitle{Phys. Rev. D}
\bvolume{81},
\bfpage{123521}
(\byear{2010})
\doiurl{10.1103/PhysRevD.81.123521}
{\href{https://arxiv.org/abs/1004.0629}{{arXiv:1004.0629}}}
{[astro-ph.GA]}
\end{barticle}
\endbibitem

\bibitem[\protect\citeauthoryear{Kouvaris and
  Perez-Garcia}{2014}]{Kouvaris:2014rja}
\begin{barticle}
\bauthor{\bsnm{Kouvaris}, \binits{C.}},
\bauthor{\bsnm{Perez-Garcia}, \binits{M.A.}}:
\batitle{{Can Dark Matter explain the Braking Index of Neutron Stars?}}
\bjtitle{Phys. Rev. D}
\bvolume{89}(\bissue{10}),
\bfpage{103539}
(\byear{2014})
\doiurl{10.1103/PhysRevD.89.103539}
{\href{https://arxiv.org/abs/1401.3644}{{arXiv:1401.3644}}}
{[astro-ph.SR]}
\end{barticle}
\endbibitem

\bibitem[\protect\citeauthoryear{Miao et~al.}{2022}]{Miao:2022rqj}
\begin{barticle}
\bauthor{\bsnm{Miao}, \binits{Z.}},
\bauthor{\bsnm{Zhu}, \binits{Y.}},
\bauthor{\bsnm{Li}, \binits{A.}},
\bauthor{\bsnm{Huang}, \binits{F.}}:
\batitle{{Dark Matter Admixed Neutron Star Properties in the Light of X-Ray
  Pulse Profile Observations}}.
\bjtitle{Astrophys. J.}
\bvolume{936}(\bissue{1}),
\bfpage{69}
(\byear{2022})
\doiurl{10.3847/1538-4357/ac8544}
{\href{https://arxiv.org/abs/2204.05560}{{arXiv:2204.05560}}}
{[astro-ph.HE]}
\end{barticle}
\endbibitem

\bibitem[\protect\citeauthoryear{Yadav et~al.}{2024}]{Yadav:2024xob}
\begin{barticle}
\bauthor{\bsnm{Yadav}, \binits{S.}},
\bauthor{\bsnm{Mishra}, \binits{M.}},
\bauthor{\bsnm{Sarkar}, \binits{T.G.}}:
\batitle{{X-ray emission spectrum for axion\textendash{}photon conversion in
  magnetospheres of strongly magnetized neutron stars}}.
\bjtitle{Eur. Phys. J. C}
\bvolume{84}(\bissue{7}),
\bfpage{687}
(\byear{2024})
\doiurl{10.1140/epjc/s10052-024-13051-7}
{\href{https://arxiv.org/abs/2403.15305}{{arXiv:2403.15305}}}
{[astro-ph.HE]}
\end{barticle}
\endbibitem

\bibitem[\protect\citeauthoryear{Kopp et~al.}{2018}]{Kopp:2018jom}
\begin{barticle}
\bauthor{\bsnm{Kopp}, \binits{J.}},
\bauthor{\bsnm{Laha}, \binits{R.}},
\bauthor{\bsnm{Opferkuch}, \binits{T.}},
\bauthor{\bsnm{Shepherd}, \binits{W.}}:
\batitle{{Cuckoo\textquoteright{}s eggs in neutron stars: can LIGO hear chirps
  from the dark sector?}}
\bjtitle{JHEP}
\bvolume{11},
\bfpage{096}
(\byear{2018})
\doiurl{10.1007/JHEP11(2018)096}
{\href{https://arxiv.org/abs/1807.02527}{{arXiv:1807.02527}}}
{[hep-ph]}
\end{barticle}
\endbibitem

\bibitem[\protect\citeauthoryear{Giangrandi et~al.}{2025}]{Giangrandi:2025rko}
\begin{botherref}
\oauthor{\bsnm{Giangrandi}, \binits{E.}},
\oauthor{\bsnm{Rueter}, \binits{H.}},
\oauthor{\bsnm{Kunert}, \binits{N.}},
\oauthor{\bsnm{Emma}, \binits{M.}},
\oauthor{\bsnm{Abac}, \binits{A.}},
\oauthor{\bsnm{Adhikari}, \binits{A.}},
\oauthor{\bsnm{Dietrich}, \binits{T.}},
\oauthor{\bsnm{Sagun}, \binits{V.}},
\oauthor{\bsnm{Tichy}, \binits{W.}},
\oauthor{\bsnm{Providencia}, \binits{C.}}:
{Numerical Relativity Simulations of Dark Matter Admixed Binary Neutron Stars}.
unpublished
(2025)
\end{botherref}
\endbibitem

\bibitem[\protect\citeauthoryear{Zhang et~al.}{2022}]{Zhang:2020pfh}
\begin{barticle}
\bauthor{\bsnm{Zhang}, \binits{K.}},
\bauthor{\bsnm{Huang}, \binits{G.-Z.}},
\bauthor{\bsnm{Tsao}, \binits{J.-S.}},
\bauthor{\bsnm{Lin}, \binits{F.-L.}}:
\batitle{{GW170817 and GW190425 as hybrid stars of dark and nuclear matter}}.
\bjtitle{Eur. Phys. J. C}
\bvolume{82}(\bissue{4}),
\bfpage{366}
(\byear{2022})
\doiurl{10.1140/epjc/s10052-022-10335-8}
{\href{https://arxiv.org/abs/2002.10961}{{arXiv:2002.10961}}}
{[astro-ph.HE]}
\end{barticle}
\endbibitem

\bibitem[\protect\citeauthoryear{Ellis et~al.}{2018}]{Ellis:2018bkr}
\begin{barticle}
\bauthor{\bsnm{Ellis}, \binits{J.}},
\bauthor{\bsnm{H\"utsi}, \binits{G.}},
\bauthor{\bsnm{Kannike}, \binits{K.}},
\bauthor{\bsnm{Marzola}, \binits{L.}},
\bauthor{\bsnm{Raidal}, \binits{M.}},
\bauthor{\bsnm{Vaskonen}, \binits{V.}}:
\batitle{{Dark Matter Effects On Neutron Star Properties}}.
\bjtitle{Phys. Rev. D}
\bvolume{97}(\bissue{12}),
\bfpage{123007}
(\byear{2018})
\doiurl{10.1103/PhysRevD.97.123007}
{\href{https://arxiv.org/abs/1804.01418}{{arXiv:1804.01418}}}
{[astro-ph.CO]}
\end{barticle}
\endbibitem

\bibitem[\protect\citeauthoryear{Routaray et~al.}{2023}]{Routaray:2022acz}
\begin{barticle}
\bauthor{\bsnm{Routaray}, \binits{P.}},
\bauthor{\bsnm{Quddus}, \binits{A.}},
\bauthor{\bsnm{Chakravarti}, \binits{K.}},
\bauthor{\bsnm{Kumar}, \binits{B.}}:
\batitle{{Probing the impact of WIMP dark matter on universal relations,
  GW170817 posterior, and radial oscillations}}.
\bjtitle{Mon. Not. Roy. Astron. Soc.}
\bvolume{525}(\bissue{4}),
\bfpage{5492}--\blpage{5499}
(\byear{2023})
\doiurl{10.1093/mnras/stad2628}
{\href{https://arxiv.org/abs/2202.04364}{{arXiv:2202.04364}}}
{[nucl-th]}
\end{barticle}
\endbibitem

\bibitem[\protect\citeauthoryear{Nelson et~al.}{2019}]{Nelson:2018xtr}
\begin{barticle}
\bauthor{\bsnm{Nelson}, \binits{A.}},
\bauthor{\bsnm{Reddy}, \binits{S.}},
\bauthor{\bsnm{Zhou}, \binits{D.}}:
\batitle{{Dark halos around neutron stars and gravitational waves}}.
\bjtitle{JCAP}
\bvolume{07},
\bfpage{012}
(\byear{2019})
\doiurl{10.1088/1475-7516/2019/07/012}
{\href{https://arxiv.org/abs/1803.03266}{{arXiv:1803.03266}}}
{[hep-ph]}
\end{barticle}
\endbibitem

\bibitem[\protect\citeauthoryear{Sandin and Ciarcelluti}{2009}]{Sandin:2008db}
\begin{barticle}
\bauthor{\bsnm{Sandin}, \binits{F.}},
\bauthor{\bsnm{Ciarcelluti}, \binits{P.}}:
\batitle{{Effects of mirror dark matter on neutron stars}}.
\bjtitle{Astropart. Phys.}
\bvolume{32},
\bfpage{278}--\blpage{284}
(\byear{2009})
\doiurl{10.1016/j.astropartphys.2009.09.005}
{\href{https://arxiv.org/abs/0809.2942}{{arXiv:0809.2942}}}
{[astro-ph]}
\end{barticle}
\endbibitem

\bibitem[\protect\citeauthoryear{Barbat et~al.}{2024}]{Barbat:2024yvi}
\begin{barticle}
\bauthor{\bsnm{Barbat}, \binits{M.F.}},
\bauthor{\bsnm{Schaffner-Bielich}, \binits{J.}},
\bauthor{\bsnm{Tolos}, \binits{L.}}:
\batitle{{Comprehensive study of compact stars with dark matter}}.
\bjtitle{Phys. Rev. D}
\bvolume{110}(\bissue{2}),
\bfpage{023013}
(\byear{2024})
\doiurl{10.1103/PhysRevD.110.023013}
{\href{https://arxiv.org/abs/2404.12875}{{arXiv:2404.12875}}}
{[astro-ph.HE]}
\end{barticle}
\endbibitem

\bibitem[\protect\citeauthoryear{Hebeler and Schwenk}{2010}]{Hebeler:2009iv}
\begin{barticle}
\bauthor{\bsnm{Hebeler}, \binits{K.}},
\bauthor{\bsnm{Schwenk}, \binits{A.}}:
\batitle{{Chiral three-nucleon forces and neutron matter}}.
\bjtitle{Phys. Rev. C}
\bvolume{82},
\bfpage{014314}
(\byear{2010})
\doiurl{10.1103/PhysRevC.82.014314}
{\href{https://arxiv.org/abs/0911.0483}{{arXiv:0911.0483}}}
{[nucl-th]}
\end{barticle}
\endbibitem

\bibitem[\protect\citeauthoryear{Deliyergiyev
  et~al.}{2019}]{Deliyergiyev:2019vti}
\begin{barticle}
\bauthor{\bsnm{Deliyergiyev}, \binits{M.}},
\bauthor{\bsnm{Del~Popolo}, \binits{A.}},
\bauthor{\bsnm{Tolos}, \binits{L.}},
\bauthor{\bsnm{Le~Delliou}, \binits{M.}},
\bauthor{\bsnm{Lee}, \binits{X.}},
\bauthor{\bsnm{Burgio}, \binits{F.}}:
\batitle{{Dark compact objects: an extensive overview}}.
\bjtitle{Phys. Rev. D}
\bvolume{99}(\bissue{6}),
\bfpage{063015}
(\byear{2019})
\doiurl{10.1103/PhysRevD.99.063015}
{\href{https://arxiv.org/abs/1903.01183}{{arXiv:1903.01183}}}
{[gr-qc]}
\end{barticle}
\endbibitem

\bibitem[\protect\citeauthoryear{Ruester et~al.}{2006}]{Ruester:2005fm}
\begin{barticle}
\bauthor{\bsnm{Ruester}, \binits{S.B.}},
\bauthor{\bsnm{Hempel}, \binits{M.}},
\bauthor{\bsnm{Schaffner-Bielich}, \binits{J.}}:
\batitle{{The outer crust of non-accreting cold neutron stars}}.
\bjtitle{Phys. Rev. C}
\bvolume{73},
\bfpage{035804}
(\byear{2006})
\doiurl{10.1103/PhysRevC.73.035804}
{\href{https://arxiv.org/abs/astro-ph/0509325}{{arXiv:astro-ph/0509325}}}
\end{barticle}
\endbibitem

\bibitem[\protect\citeauthoryear{Fraga et~al.}{2014}]{Fraga:2013qra}
\begin{barticle}
\bauthor{\bsnm{Fraga}, \binits{E.S.}},
\bauthor{\bsnm{Kurkela}, \binits{A.}},
\bauthor{\bsnm{Vuorinen}, \binits{A.}}:
\batitle{{Interacting quark matter equation of state for compact stars}}.
\bjtitle{Astrophys. J. Lett.}
\bvolume{781}(\bissue{2}),
\bfpage{25}
(\byear{2014})
\doiurl{10.1088/2041-8205/781/2/L25}
{\href{https://arxiv.org/abs/1311.5154}{{arXiv:1311.5154}}}
{[nucl-th]}
\end{barticle}
\endbibitem

\bibitem[\protect\citeauthoryear{Narain et~al.}{2006}]{Narain:2006kx}
\begin{barticle}
\bauthor{\bsnm{Narain}, \binits{G.}},
\bauthor{\bsnm{Schaffner-Bielich}, \binits{J.}},
\bauthor{\bsnm{Mishustin}, \binits{I.N.}}:
\batitle{{Compact stars made of fermionic dark matter}}.
\bjtitle{Phys. Rev. D}
\bvolume{74},
\bfpage{063003}
(\byear{2006})
\doiurl{10.1103/PhysRevD.74.063003}
{\href{https://arxiv.org/abs/astro-ph/0605724}{{arXiv:astro-ph/0605724}}}
\end{barticle}
\endbibitem

\bibitem[\protect\citeauthoryear{Dengler et~al.}{2022}]{Dengler:2021qcq}
\begin{barticle}
\bauthor{\bsnm{Dengler}, \binits{Y.}},
\bauthor{\bsnm{Schaffner-Bielich}, \binits{J.}},
\bauthor{\bsnm{Tolos}, \binits{L.}}:
\batitle{{Second Love number of dark compact planets and neutron stars with
  dark matter}}.
\bjtitle{Phys. Rev. D}
\bvolume{105}(\bissue{4}),
\bfpage{043013}
(\byear{2022})
\doiurl{10.1103/PhysRevD.105.043013}
{\href{https://arxiv.org/abs/2111.06197}{{arXiv:2111.06197}}}
{[astro-ph.HE]}
\end{barticle}
\endbibitem

\bibitem[\protect\citeauthoryear{Hippert et~al.}{2023}]{Hippert:2022snq}
\begin{barticle}
\bauthor{\bsnm{Hippert}, \binits{M.}},
\bauthor{\bsnm{Dillingham}, \binits{E.}},
\bauthor{\bsnm{Tan}, \binits{H.}},
\bauthor{\bsnm{Curtin}, \binits{D.}},
\bauthor{\bsnm{Noronha-Hostler}, \binits{J.}},
\bauthor{\bsnm{Yunes}, \binits{N.}}:
\batitle{{Dark matter or regular matter in neutron stars? How to tell the
  difference from the coalescence of compact objects}}.
\bjtitle{Phys. Rev. D}
\bvolume{107}(\bissue{11}),
\bfpage{115028}
(\byear{2023})
\doiurl{10.1103/PhysRevD.107.115028}
{\href{https://arxiv.org/abs/2211.08590}{{arXiv:2211.08590}}}
{[astro-ph.HE]}
\end{barticle}
\endbibitem

\bibitem[\protect\citeauthoryear{Dengler et~al.}{2025}]{Dengler:2025ntz}
\begin{botherref}
\oauthor{\bsnm{Dengler}, \binits{Y.}},
\oauthor{\bsnm{Kulkarni}, \binits{S.}},
\oauthor{\bsnm{Maas}, \binits{A.}},
\oauthor{\bsnm{Radl}, \binits{K.}}:
{Strongly Interacting Dark Matter admixed Neutron Stars}
(2025)
{\href{https://arxiv.org/abs/2503.19691}{{arXiv:2503.19691}}}
{[hep-ph]}
\end{botherref}
\endbibitem

\bibitem[\protect\citeauthoryear{Wellegehausen
  et~al.}{2014}]{Wellegehausen:2013cya}
\begin{barticle}
\bauthor{\bsnm{Wellegehausen}, \binits{B.H.}},
\bauthor{\bsnm{Maas}, \binits{A.}},
\bauthor{\bsnm{Wipf}, \binits{A.}},
\bauthor{\bsnm{Smekal}, \binits{L.}}:
\batitle{{Hadron masses and baryonic scales in $G_2$-QCD at finite density}}.
\bjtitle{Phys. Rev. D}
\bvolume{89}(\bissue{5}),
\bfpage{056007}
(\byear{2014})
\doiurl{10.1103/PhysRevD.89.056007}
{\href{https://arxiv.org/abs/1312.5579}{{arXiv:1312.5579}}}
{[hep-lat]}
\end{barticle}
\endbibitem

\bibitem[\protect\citeauthoryear{Rutherford et~al.}{2023}]{Rutherford:2022xeb}
\begin{barticle}
\bauthor{\bsnm{Rutherford}, \binits{N.}},
\bauthor{\bsnm{Raaijmakers}, \binits{G.}},
\bauthor{\bsnm{Prescod-Weinstein}, \binits{C.}},
\bauthor{\bsnm{Watts}, \binits{A.}}:
\batitle{{Constraining bosonic asymmetric dark matter with neutron star
  mass-radius measurements}}.
\bjtitle{Phys. Rev. D}
\bvolume{107}(\bissue{10}),
\bfpage{103051}
(\byear{2023})
\doiurl{10.1103/PhysRevD.107.103051}
{\href{https://arxiv.org/abs/2208.03282}{{arXiv:2208.03282}}}
{[astro-ph.HE]}
\end{barticle}
\endbibitem

\bibitem[\protect\citeauthoryear{Raaijmakers
  et~al.}{2019}]{Raaijmakers:2019qny}
\begin{barticle}
\bauthor{\bsnm{Raaijmakers}, \binits{G.}}, \betal:
\batitle{{A NICER view of PSR J0030$+$0451: Implications for the dense matter
  equation of state}}.
\bjtitle{Astrophys. J. Lett.}
\bvolume{887}(\bissue{1}),
\bfpage{22}
(\byear{2019})
\doiurl{10.3847/2041-8213/ab451a}
{\href{https://arxiv.org/abs/1912.05703}{{arXiv:1912.05703}}}
{[astro-ph.HE]}
\end{barticle}
\endbibitem

\bibitem[\protect\citeauthoryear{Rutherford et~al.}{2024}]{Rutherford:2024bli}
\begin{botherref}
\oauthor{\bsnm{Rutherford}, \binits{N.}},
\oauthor{\bsnm{Prescod-Weinstein}, \binits{C.}},
\oauthor{\bsnm{Watts}, \binits{A.}}:
{Probing fermionic asymmetric dark matter cores using global neutron star
  properties}.
unpunlished
(2024)
\end{botherref}
\endbibitem

\bibitem[\protect\citeauthoryear{Lopes and Issifu}{2025}]{Lopes:2024ixl}
\begin{barticle}
\bauthor{\bsnm{Lopes}, \binits{L.L.}},
\bauthor{\bsnm{Issifu}, \binits{A.}}:
\batitle{{XTE J1814-338 as a dark matter admixed neutron star}}.
\bjtitle{Phys. Dark Univ.}
\bvolume{48},
\bfpage{101922}
(\byear{2025})
\doiurl{10.1016/j.dark.2025.101922}
{\href{https://arxiv.org/abs/2411.17105}{{arXiv:2411.17105}}}
{[astro-ph.HE]}
\end{barticle}
\endbibitem

\bibitem[\protect\citeauthoryear{Routaray et~al.}{2024}]{Routaray:2023txs}
\begin{barticle}
\bauthor{\bsnm{Routaray}, \binits{P.}},
\bauthor{\bsnm{Das}, \binits{H.C.}},
\bauthor{\bsnm{Pattnaik}, \binits{J.A.}},
\bauthor{\bsnm{Kumar}, \binits{B.}}:
\batitle{{Constraining neutron star properties and dark matter admixture with
  the NITR-I equation of state: Insights from observations and universal
  relations}}.
\bjtitle{Int. J. Mod. Phys. E}
\bvolume{33}(\bissue{11}),
\bfpage{2450052}
(\byear{2024})
\doiurl{10.1142/S0218301324500526}
{\href{https://arxiv.org/abs/2307.12748}{{arXiv:2307.12748}}}
{[math.NA]}
\end{barticle}
\endbibitem

\bibitem[\protect\citeauthoryear{Sagun et~al.}{2023}]{Sagun:2023rzp}
\begin{barticle}
\bauthor{\bsnm{Sagun}, \binits{V.}},
\bauthor{\bsnm{Giangrandi}, \binits{E.}},
\bauthor{\bsnm{Dietrich}, \binits{T.}},
\bauthor{\bsnm{Ivanytskyi}, \binits{O.}},
\bauthor{\bsnm{Negreiros}, \binits{R.}},
\bauthor{\bsnm{Provid\^encia}, \binits{C.}}:
\batitle{{What Is the Nature of the HESS J1731-347 Compact Object?}}
\bjtitle{Astrophys. J.}
\bvolume{958}(\bissue{1}),
\bfpage{49}
(\byear{2023})
\doiurl{10.3847/1538-4357/acfc9e}
{\href{https://arxiv.org/abs/2306.12326}{{arXiv:2306.12326}}}
{[astro-ph.HE]}
\end{barticle}
\endbibitem

\bibitem[\protect\citeauthoryear{Cano and Alarcon}{}]{Alarcon2025}
\begin{botherref}
\oauthor{\bsnm{Cano}, \binits{Y.}},
\oauthor{\bsnm{Alarcon}, \binits{J.M.}}:
Analyzing Fermionic Dark Matter scenarios with the HESS J1731-347 Compact
  Object.
In preparation
\end{botherref}
\endbibitem

\end{thebibliography}

\end{document}